\documentclass{elsarticle}
\usepackage{graphicx}
\usepackage[T1]{fontenc}
\usepackage{color}
\usepackage[bf]{subfigure}
\usepackage{float}
\usepackage{hyperref}
\usepackage{amssymb}
\usepackage{epstopdf}
\usepackage{hypernat}
\usepackage{amsmath}
\usepackage{bm}

\definecolor{greencolor}{rgb}{0,0.5,0.2}
\definecolor{redcolor}{rgb}{1.0,0.,0.}
\definecolor{bluecolor}{rgb}{0,0.,1.}
\definecolor{greycolor}{rgb}{.5,.5,.5}

\RequirePackage[normalem]{ulem} 
\RequirePackage{color}\definecolor{RED}{rgb}{1,0,0}\definecolor{BLUE}{rgb}{0,0,1} 

\begin{document}

\title{The Kuramoto model in complex networks}

\author[icmc]{Francisco A. Rodrigues}
   \ead{francisco@icmc.usp.br}
    \author[ifsc,pik]{Thomas K. DM. Peron\corref{cor1}}
   \ead{thomaskaue@gmail.com}
  \author[pik,ph]{Peng Ji\corref{cor1}}
  \ead{pengji@pik-potsdam.de}
    \author[pik,ph,aber,nizhny]{J\"urgen Kurths}
    \ead{kurths@pik-potsdam.de}

\cortext[cor1]{Corresponding authors}

\address[icmc]{Departamento de Matem\'{a}tica Aplicada e Estat\'{i}stica, Instituto de Ci\^{e}ncias Matem\'{a}ticas e de Computa\c{c}\~{a}o, Universidade de S\~{a}o Paulo, Caixa Postal 668, 13560-970 S\~{a}o Carlos,  S\~ao Paulo, Brazil}
\address[ifsc]{Instituto de F\'{\i}sica de S\~{a}o Carlos, Universidade de S\~{a}o Paulo, Caixa Postal 369, 13560-970, S\~{a}o Carlos,  S\~ao Paulo, Brazil}
\address[pik]{Potsdam Institute for Climate Impact Research (PIK), 14473 Potsdam, Germany}
\address[ph]{Department of Physics, Humboldt University, 12489 Berlin, Germany}
\address[aber]{Institute for Complex Systems and Mathematical Biology, University of Aberdeen, Aberdeen AB24 3UE, United Kingdom}
\address[nizhny]{Department of Control Theory, Nizhny Novgorod State University, Gagarin Avenue 23, Nizhny Novgorod
606950, Russia}

\date{\today}

\begin{abstract}
Synchronization of an ensemble of oscillators is an emergent phenomenon present in several complex systems, ranging from social and physical to biological and technological systems. The most successful approach to describe how coherent behavior emerges in these complex systems is given by the paradigmatic Kuramoto model. This model has been traditionally studied in complete graphs. However, besides being intrinsically dynamical, complex systems present very heterogeneous structure, which can be represented as complex networks. This report is dedicated to review main contributions in the field of synchronization in networks of Kuramoto oscillators. In particular, we provide an overview of the impact of network patterns on the local and global dynamics of coupled phase oscillators. We cover many relevant topics, which encompass  a description of the most used analytical approaches and the analysis of several numerical results. Furthermore, we discuss recent developments on variations of the Kuramoto model in networks, including the presence of noise and inertia. The rich potential for applications is discussed for special fields in engineering, neuroscience, physics and Earth science. Finally, we conclude by discussing problems that remain open after the last decade of intensive research on the Kuramoto model and point out some promising directions for future research.
\end{abstract}

\maketitle

\tableofcontents
\section*{List of Abbreviations}

\begin{tabular}{ l l  }	
 BA & Barab\'asi-Albert\\
 CM & configuration model\\ 
 ER & Erd\H{o}s-R\'enyi\\	 
 FPE & Fokker-Planck equation\\
 FSS & finite-size scaling\\
 GA  & Gaussian approximation\\
 MFA & mean-field approximation\\
 MSF & master stability function\\
 OA & Ott-Antonsen\\
 PA & preferential attachment\\
 SF & scale-free\\
 SW & small-world
\end{tabular}

\section*{List of Symbols}

\begin{tabular}{ l l  }
 $\mathbf{A}$ & Adjacency matrix \\
 $\mathcal{A}$ & Assortativity\\
      $\mathcal{BS}$ & Basin stability\\
  $\alpha$ & Damping constant \\
  $\beta$ & Critical exponent of the phase transition\\
  $\bar{\nu}$ & Finite-size scaling exponent\\
  $cc_i$ & Local clustering coefficient\\
  $D$ & Noise strength\\
  $\sigma^{\rm{A}}$ & Adjacency matrix eigenvalues\\
  $\sigma^{\rm{L}}$ & Laplacian matrix eigenvalues\\
 $E$ & Number of edges\\
  $k$ & Degree\\
  $\left\langle k \right\rangle$ & Network average degree\\ 
  $k_{\min}$ & Minimum degree\\
  $k_{\max}$ & Maximum degree \\
  $P(k)$ & Degree distribution\\ 
    $\mathbf{L}$ & Laplacian matrix\\
    $\lambda$ & Coupling strength\\
    $\lambda_c$ & Critical coupling strength\\
    $\lambda_c^{\rm{I}}$ & Critical coupling strength for increasing coupling branch\\
    $\lambda_c^{\rm{D}}$ & Critical coupling strength for decreasing coupling branch\\
      $\left\langle \cdot \right\rangle$ & Ensemble average  \\
     $\left\langle \cdot \right\rangle_t$ & Temporal average\\ 
     $N$ & Network size\\
     $\nu$ & Frequency \\
 $\theta$ & Phase \\
 $\phi$ & Phase in the rotating frame\\
 $t$ & Time \\
 $\omega$ & Natural frequency\\
 $g(\omega)$ & Natural frequency distribution\\
 $\psi$ & Mean phase\\
 $\Omega$ & Locking frequency\\
 $R$ & Kuramoto order parameter\\ 
 $r$ & Global order parameter accounting the mean-field of uncorrelated networks\\
 $r_i$ & Local order parameter\\
 $r_i^{\rm{T}}$ & Local order parameter taking into account time fluctuations.\\
 $r_{\rm{lock}}$ & Contribution of locked oscillators to the order parameter\\
 $r_{\rm{drift}}$ & Contribution of drifting oscillators to the order parameter\\
 $r^{\rm{D}}$ & Order parameter associated to the decreasing branch\\
 $r^{\rm{I}}$ & Order parameter associated to the increasing branch\\
 $\tau$ & Time delay\\
 $\tau_r$ & Relaxation time\\
 $\chi$ & Susceptibility\\ 
 $\mathcal{T}$ & Transitivity
\end{tabular}

\section{Introduction}
\label{sec:introduction}

Synchronization phenomena are ubiquitous in nature, science, society, and technology.  Examples of oscillators are fireflies, lasers, neurons and heart cells~\cite{pikovsky2003synchronization}. Among  the many models proposed for a description of synchronization~\cite{pikovsky2003synchronization}, the \textit{Kuramoto model} is the most popular nowadays. It describes self-sustained phase oscillators rotating at heterogeneous intrinsic frequencies coupled through the sine of their phase differences. This model exhibits a phase transition at a critical coupling, beyond which a collective behavior is achieved. Since its original formulation 40 years ago~\cite{kuramoto1975,kuramoto1984chemical}, several variations, extensions and applications of the Kuramoto model have been documented in the literature. In 2005, Acebr\'on et al.~\cite{acebron2005kuramoto} published the first survey addressing the Kuramoto model, discussing the main works available at that time. 


In parallel with the advances in the study of the traditional Kuramoto model, over almost 
the last two decades one has witnessed the rapid development of the new field
of network science~\cite{barabasi1999emergence}, which not only brought new insights into the characterization of real networks, but also introduced a new dimension
in the study of dynamical systems~\cite{barabasi2011NetworkTakeover,bianconi2015interdisciplinary}. Researchers were puzzled by the 
question of how the connectivity pattern between elements in a network
can influence the performance of dynamical processes, such  as  epidemic 
spreading, percolation, diffusion, opinion formation and synchronization. This 
apparently simple question has motivated a lot of studies comprised in several reviews (e.g.~\cite{newman2003StructureFunctionFunctionComplexNetworks,
dorogovtsev2002evolution,albert2002statistical,boccaletti2006complex, costa2007characterization,szabo2007evolutionary,arenas2008synchronization,costa2011analyzing,saberi2015recent,
kivela2014multilayer,boccaletti2014structure,porter2014DynamicalSystems}) and books (e.g.~\cite{strogatz2003sync,bornholdt2006handbook,
pastor2007evolution,dorogovtsev2013evolution,
cohen2010complex,newman2010networks,estrada2010network}) on the topic. 
Curiously, the rise of network science is intimately related to the study of synchronization among coupled oscillators. As described in~\cite{strogatz2003sync}, Watts and Strogatz conceived the idea of including shortcuts between oscillators connected as a regular graph to analyze how crickets synchronize their chirps. It turns out that the simple inclusion of a few 
shortcuts greatly reduces the average topological distance between the 
oscillators, improving the synchronous behavior between 
them~\cite{watts1999small,strogatz2001exploring}. In this way, through this process, the so-called small-world phenomenon was formalized and quantified in the context of networks. This analysis is  a milestone in the study of complex systems triggering an overwhelming number of papers. In 2004, the Kuramoto model was generalized to 
scale-free networks~\cite{moreno2004synchronization} to address the role played by highly connected nodes (hubs) in network dynamics. After that, most of the focusing has mainly aimed at determining how network structure influences the onset of 
synchronization. 

In 2006, Boccaletti et al.~\cite{boccaletti2006complex} 
provided the first review encompassing structural and dynamical properties 
of complex networks, where the first theoretical 
approaches to the Kuramoto model in networks were 
reviewed. However, the study of the interplay between network structure and 
dynamics was still in its infancy. In the following years this study rapidly 
evolved in a way that in 2008 Arenas et al.~\cite{arenas2008synchronization} 
published a survey in Physics Reports devoted to the analysis of synchronization in 
networks. The authors focused on two main topics, namely the study of 
synchronization in the framework of the master stability function (MSF) and 
networks of Kuramoto oscillators. Regarding the latter subject, 
significant new analytical and numerical findings were revised, mainly studies on the relation between synchronization and network structure.

However, several important new results on the Kuramoto model have been published 
since then. In particular, the development of new network models has enabled the study of how different network properties affect 
synchronization. More specifically, the early works on the Kuramoto model in 
networks have naturally focused on the influence of topological properties 
present in the traditional models, such as the presence of shortcuts
and hubs. In the last years new classes of random network models that go 
beyond the reproduction of the degree distribution of real-world networks 
have been proposed. Basic topological properties such as the occurrence of triangles, 
emergence of communities, degree-degree correlations and distribution of 
subgraphs were incorporated in variations of the traditional configuration
model allowing the investigation of the effect of these 
properties on dynamical processes. Not only these non-trivial properties encountered in real networks, but also new types of network representations 
have been incorporated in the investigations. Namely, in the last couple of years the so-called \textit{multilayer networks} have been attracting the attention of network researchers and, as we shall see, neglecting the multilayer character can greatly alter the synchronous behavior between the oscillators~\cite{kivela2014multilayer,boccaletti2014structure}.

It is also important to emphasize that until 2011 only continuous synchronization transitions in the Kuramoto model in networks were reported. The discovery of first-order phase transitions to synchronization (also named as ``Explosive Synchronization''~\cite{gomez2011explosive}) as a consequence of the correlation between structure and local dynamics has triggered several investigations. Moreover, the study of temporal networks, whose structures change in time, has provided new versions of the Kuramoto model. The exploration of several properties in the model, such as the inclusion of stochastic fluctuations, time-delay and repulsive couplings, is also a new tendency in the analysis of the Kuramoto model in networks. 

The study of the Kuramoto model in complex networks has also been boosted thanks to findings of new synchronization phenomena, such as the emergence of chimera states in which networks of identical oscillators can split into synchronized and desynchronized subpopulations~\cite{panaggio2015ChimeraStates}. The ansatz proposed by E. Ott and T. Antonsen~\cite{ott2008low,ott2009long}, which allows a dimensional reduction to a small number of coupled differential equations, is another recent remarkable result that has attracted the interest of researchers. This ansatz has been receiving great attention since 2008 along with its generalizations to networks of Kuramoto oscillators.

Finally, the Kuramoto model in complex networks has been used in several applications, such as  modeling neuronal activity and power grids. In particular, the latter system can be suitably modeled by a second-order Kuramoto model, a fact that motivated many other works aiming at generalizing the model to complex networks. In terms of even large perturbations, the recently proposed concept of basin stability has been proved to deepen insights not only on the stability of real power grids, but also on other dynamical systems~\cite{menck_how_2013}. 

After the survey published in 2008, a large number of papers 
considering the networked Kuramoto model have been published. Noteworthy, recently D\"orfler and Bullo provided a comprehensive survey focused on the control of synchronization of phase oscillators applied to technological networks~\cite{dorfler2014synchronization}. However, this survey does not cover the main recent works related to the Kuramoto model.
For this reason, it is timely to provide a survey about the Kuramoto model in complex networks to contextualize the fundamental works and 
enable the advance of the field.

\subsection{Outline}

This report is organized as follows. We begin our review by analysing important aspects of the Kuramoto model in well-known network models in Sec.~\ref{sec:traditional}. There, we discuss the first numerical investigations of the model in complex topologies and also introduce analytical treatments that will be used throughout 
the text. Finite-size effects and the transient dynamics are 
also examined.  In Section~\ref{sec:different_topologies} we describe the dynamics
of Kuramoto oscillators coupled in networks that mimic typical properties
of real-world structures. In particular, we review the works devoted to analyzing the influence of non-vanishing clustering-coefficient, degree-degree correlations and presence of nodes grouped into communities on 
network synchronization. \textcolor{black}{Effects of time-delay and adaptive couplings are studied in Sec.~\ref{sec:general_coupling}.} Section~\ref{sec:explosive_sync} reviews the very recent works on 
the correlation between natural frequencies and local topology and other
conditions that are known to yield discontinuous phase transitions in networks made up of Kuramoto oscillators. Section~\ref{sec:stochastic} is concerned with the 
recent developments on the stochastic Kuramoto model in networks. The second-order Kuramoto is presented in Sec.~\ref{sec:second_KM}, where we discuss the recently introduced concept of basin stability.\textcolor{black}{ Section~\ref{sec:optimization} discusses extensive numerical simulations on the optimization of synchronization in networks}. \textcolor{black}{In Sec.~\ref{sec:applications} we summarize relevant applications of the first- and second-order Kuramoto model in real complex systems, such as power-grids, neuronal systems, networks of semiconductor junctions and seismology.} Finally, 
in Section~\ref{sec:conclusions} we present our perspectives and conclusions. 


\section{First-order Kuramoto on traditional network models}
\label{sec:traditional}
 
Although the first report on a synchronization phenomenon dates back to Huygens in the 17th century~\cite{pikovsky2003synchronization}, the topic of spontaneous emergence 
of collective behavior in large populations of oscillators
was only brought to higher attention after the work by 
Wiener~\cite{wiener1958nonlinear,wiener1961cybernetics}. Wiener was interested in the generation of alpha
rhythms in the brain and his guess was that this 
particular phenomenon was somehow related with the same
mechanism that yields coherent behavior in other biological 
systems, such as in the synchronous flash of fireflies. Wiener's 
idea was interesting and anticipating, but at the same time too complex to get 
analytical insights from it. A more promising approach was later developed by Winfree~\cite{winfree1967biological,winfree1980geometry}, who was the first to properly state the problem 
of collective synchronization mathematically. He proposed a model 
of a large population of interacting phase oscillators with 
distributed natural frequencies. By simulating his model, he 
found that spontaneous synchronization emerges as a threshold 
process, a phenomenon akin to a phase transition. His main finding was: if the spread of 
the frequencies is higher compared to the coupling between the 
oscillators, then each oscillator would run at its own natural 
frequency, causing the population to behave incoherently. On the 
other hand, in case that the coupling is sufficiently strong to overcome 
the heterogeneity in the frequencies, then the system spontaneously
locks into synchrony~\cite{winfree1967biological}. 

Deeply motivated by these results, Kuramoto simplified Winfree's approach to obtain an analytically tractable model, which at the same time would preserve the fundamental assumptions of having oscillators with distributed frequencies interacting through a collective rhythm produced by the rest of the population. 
The Kuramoto model consists of a population of $N$ phase oscillators whose evolution is dictated by the governing equations~\cite{kuramoto1975,kuramoto1984chemical} 
\begin{equation}
\dot{\theta}_i = \omega_i + \frac{\lambda}{N}\sum_{j=1}^N \sin(\theta_j - \theta_i), \quad i=1\ldots, N, 
\label{eq:KM}
\end{equation}
where $\theta_i$ denotes the phase of the $i$th oscillator, $\lambda$ is the coupling strength and $\omega_i$ 
the natural frequencies, which are distributed according
to a given probability density $g(\omega)$. In his original approach, Kuramoto considered $g(\omega)$ to be
unimodal and symmetric centered at $\omega = \bar{\omega}$, which can be assumed to be $\bar{\omega} = 0$ after a shift. Henceforth, throughout this review, we consider the mean frequency as $\bar{\omega}=0$ to have $g(\omega) = g(-\omega)$, always when the distribution $g(\omega)$ is even and symmetric, without loss of generality. Kuramoto further introduced the order parameter~\cite{kuramoto1975,kuramoto1984chemical}
\begin{equation}
R e^{i\psi(t)} = \frac{1}{N} \sum_{j=1}^N e^{i\theta_j(t)},
\label{eq:KM_ORDER_PARAMETER}
\end{equation}
in order to quantify the overall synchrony of the population. This quantity has the interesting interpretation of being the centroid of a set of $N$ points $e^{i\theta_j}$ distributed in the unit circle in the complex plane. If the phases are 
uniformly spread in the range $[0,2\pi]$ then $R \approx 0$ meaning that there is no synchrony among the oscillators. On the other hand, when all the oscillators rotate grouped into a synchronous cluster with the same average phase $\psi(t)$ we have
$R \approx 1$. We can rewrite the set of Eqs.~\ref{eq:KM} using the mean-field quantities $R$ and $\psi$ by multiplying both sides of Eq.~\ref{eq:KM_ORDER_PARAMETER} by $e^{-i\theta_i}$ and equating the imaginary parts
to obtain
\begin{equation}
\dot{\theta}_i = \omega_i + \lambda R\sin(\psi - \theta_i).
\label{eq:KM_uncoupled}
\end{equation}
In this formulation, the phases $\theta_i$ seem to evolve independently from each other, but the interaction is actually set through $R$ and $\psi$. Furthermore, note that the effective coupling is now proportional to the order parameter $R$, creating a feedback relation between coupling and synchronization. More specifically, small increments in the order parameter $R$ end up by increasing the effective coupling in Eq.~\ref{eq:KM_uncoupled} attracting, in this way, more oscillators to the synchronous group. From this process, a self-consistent 
relation between the phases $\theta_i$ and the mean-field is found, i.e. $R$ and $\psi$ will define the evolution of 
$\theta_i$, but at the same time, the phases $\theta_i$ self-consistently yield the mean-field through Eq.~\ref{eq:KM_ORDER_PARAMETER}.

In the limit of infinite number of oscillators, the system can be described by the probability density $\rho(\theta,\omega,t)$ so that $\rho(\theta,\omega,t)d\theta$ gives
the fraction of oscillators with phase betweeen 
$\theta$ and $\theta + d\theta$ at time $t$ for a given 
natural frequency $\omega$. Since $\rho$ is nonnegative and $2\pi-$periodic in $\theta$, we have that it 
satisfies the normalization condition 
\begin{equation}
\int_{-\pi}^{\pi} \rho(\theta,\omega,t)d\theta = 1.
\label{eq:KM_normalization_condition}
\end{equation} 
Furthermore, the density $\rho$ should obey the continuity
equation 
\begin{equation}
\frac{\partial \rho}{\partial t} + \frac{\partial}{\partial \theta}(\rho v) = 0, 
\label{eq:KM_continuity_equation}
\end{equation}
where $v(\theta,\omega,t) = \omega + \lambda R\sin(\psi - \theta)$ is the angular velocity of a given oscillator with phase $\theta$ and natural frequency $\omega$ at time $t$. In the continuum limit, 
the order parameter $R$ and the average phase $\psi$ defined in Eq.~\ref{eq:KM_ORDER_PARAMETER} are determined in terms of the probability density $\rho(\theta,\omega,t)$ as
\begin{equation}
R e^{i\psi(t)} = \int_{-\pi}^{\pi} \int_{-\infty}^{\infty} e^{i\theta} \rho(\theta,\omega,t)g(\omega) d\omega d\theta.
\label{eq:KM_ORDER_PARAMETER_CONTINUUM}
\end{equation}  
Equations~\ref{eq:KM_continuity_equation} and~\ref{eq:KM_ORDER_PARAMETER_CONTINUUM} admit the trivial solution $R=0$, which corresponds
to the stationary distribution $\rho = 1/2\pi$, characterizing 
the incoherent state. In the partial synchronized state ($0 < R < 
1$), the continuity equation~(\ref{eq:KM_continuity_equation}) yields 
in the stationary regime ($\partial \rho / \partial t = 0$)
\begin{equation}
\rho(\theta,\omega)=\begin{cases}
\delta\left[\theta-\psi-\arcsin\left(\frac{\omega}{\lambda R}\right)\right], & \textrm{ if }|\omega|\leq \lambda R  \\
\frac{\sqrt{\omega^{2}-(\lambda R)^{2}}}{2\pi|\omega-\lambda R\sin(\theta-\psi)|} & \textrm{ otherwise.}
\end{cases}
\label{eq:KM_stationary_distribution}
\end{equation}
The solutions for the stationary distribution assert that in 
the partial synchronized state the oscillators are divided 
into two groups. Specifically, those with frequencies $|\omega| \leq \lambda R$ correspond to the oscillators entrained by mean-field, i.e. the oscillators that evolve locked in a common average phase $\psi(t) = \Omega t$, where $\Omega$ is the average frequency of the population. On the other hand, the oscillators 
with $|\omega | > \lambda R$ (referred as drifting oscillators) 
rotate incoherently. Inserting the stationary distributions in Eq.~\ref{eq:KM_ORDER_PARAMETER_CONTINUUM}, one obtains the 
self-consistent equation for $R$ 
\begin{equation}
R = \lambda R \int_{-\pi/2}^{\pi/2} \cos^2 \theta g(\lambda R \sin\theta) d\theta, 
\label{eq:KM_SELF_EQUATION}
\end{equation} 
where the integral corresponds to the contribution of drifting oscillators vanished due to $g(\omega) = g(-\omega)$ and the symmetry $\rho(\theta + \pi, -\omega) = \rho(\theta,\omega)$. By letting $R \rightarrow 0^+$ in Eq.~\ref{eq:KM_SELF_EQUATION}, we get 
\begin{equation}
\lambda_c^{\rm{KM}} = \frac{2}{\pi g(0)},
\label{eq:KM_CRITICAL_COUPLING}
\end{equation}
which is the critical coupling strength for the onset of synchronization firstly obtained by Kuramoto~\cite{kuramoto1975,kuramoto1984chemical}. Moreover, 
expanding the right-hand side of Eq.~\ref{eq:KM_SELF_EQUATION} in 
powers of $\lambda R$, given that $g''(0) < 0$, yields 
\begin{equation}
R \sim \sqrt{\frac{-16 (\lambda - \lambda_c)}{\pi \lambda_c^4g''(0)}}, 
\label{eq:KM_order_parameter_near_onset}
\end{equation}
for $\lambda \rightarrow \lambda_c$. Thus, near the transition point, the order parameter yields the form $R \sim (\lambda - 
\lambda_c)^\beta$ with $\beta = 1/2$, clearly showing an analogy to a second-order phase transition observed in magnetic 
systems.

Besides being a timely contribution for the understanding 
of how synchronization in large populations of mutually
coupled oscillators sets in, Kuramoto's analysis also 
established a link between mean-field techniques in 
statistical
physics and nonlinear dynamics. However, although 
undoubtedly brilliant and insightful, his approach was not 
rigorous and left many questions
that puzzled the researchers throughout the following 
years~
\cite{strogatz2000FromKuramotoToCrawford,acebron2005kuramoto} 
making it still matter of fundamental research in recent 
times~
\cite{mirollo2005SpectrumLockedState,mirollo2007SpectrumPartiallyLockedState,ott2008low,lee2010VorticesTwoDimensionalKuramotoModel,chiba2015AproofKuramotoConjecture,dietert2014StabilityBifurcationKuramotoModel,fernandez2014LandauDampingKuramotoModel}.   

In parallel with the studies on the 
traditional Kuramoto model, one 
has witnessed the emergence of an
overwhelming number of works
focused on particular effects caused by the 
introduction of heterogeneous connection patterns, letting the 
interactions to be no longer only restricted to global coupling. 
In this section we will analyse the results uncovered for standard
network topologies, such as in Erd\H{o}s-R\'enyi (ER), scale-free (SF) and small-world (SW) networks. In order to do so, we will firstly discuss the required approximations to analytically treat 
the problem in uncorrelated networks, i.e. networks in which the degree of connected nodes are not correlated (see Appendix~\ref{sec:appendix}). Subsequently, we discuss the scaling with the system 
size and finally we study the relaxation dynamics of the model.

The generalization of the Kuramoto model in complex networks is 
obtained by including the connectivity in the coupling term as~\cite{arenas2008synchronization,barrat2008DynamicalProcessesOnComplexNetworks}
\begin{equation}
\dot{\theta}_i = \omega_i + \sum_{j=1}^N \lambda_{ij} 
A_{ij} \sin(\theta_j - \theta_i), 
\label{eq:KMNETWORKS}
\end{equation}
where $\lambda_{ij}$ is the coupling strength between nodes $i$ 
and $j$ and $A_{ij}$ the elements of the adjacency matrix $\mathbf{A}$ ($A_{ij}=1$ if the re is a connection between nodes $i$ and $j$, or $A_{ij}=0$, otherwise). The 
definition of the model in Eq.~\ref{eq:KMNETWORKS} already brings 
the first problem when treating the model in complex topologies: the choice of $\lambda_{ij}$. In the fully connected graph the coupling $\lambda_{ij} = \lambda/N$ is 
adopted so that the model is well behaved in the thermodynamic 
limit $N\rightarrow \infty$, since in this case the connectivity 
of each oscillators grows linearly with the system size. Thus,
the normalization factor for the coupling strength in networks
should be defined in a way to incorporate the same scaling observed in the dependence
of the connectivity of the nodes on the system size. However, 
different network models generate different scalings, which 
make the choice for an intensive coupling $\lambda_{ij}$ to be
not unique. This led many researchers to simply adopt a constant coupling $\lambda_{ij} = \lambda$ $\forall i,j$ without
any dependence on $N$, i.e.~\cite{arenas2008synchronization}
\begin{equation}
\dot{\theta}_i = \omega_i + \lambda \sum_{j=1}^N A_{ij}\sin(\theta_j - \theta_i).
\label{eq:KMNETWORKS_CONSTANT_COUPLING}
\end{equation} 
 Setting the interaction between the oscillators  to be through a constant coupling strength without any further normalization factor seems to be more appropriate when comparing the synchronization of different networks, since the total number of connections can scale differently with the network size, depending on the topology considered.  The choice for an intensive coupling is indeed an important issue regarding the formulation of the Kuramoto model in complex networks, specially when concerning the determination of the onset of synchronization. Arenas et al.~\cite{arenas2008synchronization} provide an interesting discussion about the different prescriptions for intensive couplings and also the corresponding consequences of each choice in the network dynamics. Here we discuss the results with the normalization terms used in the original papers, commenting, when possible, the limiting cases of different definitions for the coupling strength $\lambda_{ij}$.

\subsection{Early works}
\label{subsec:early_works}

\begin{figure}[!t]
\begin{center}
\includegraphics[width=0.65\linewidth]{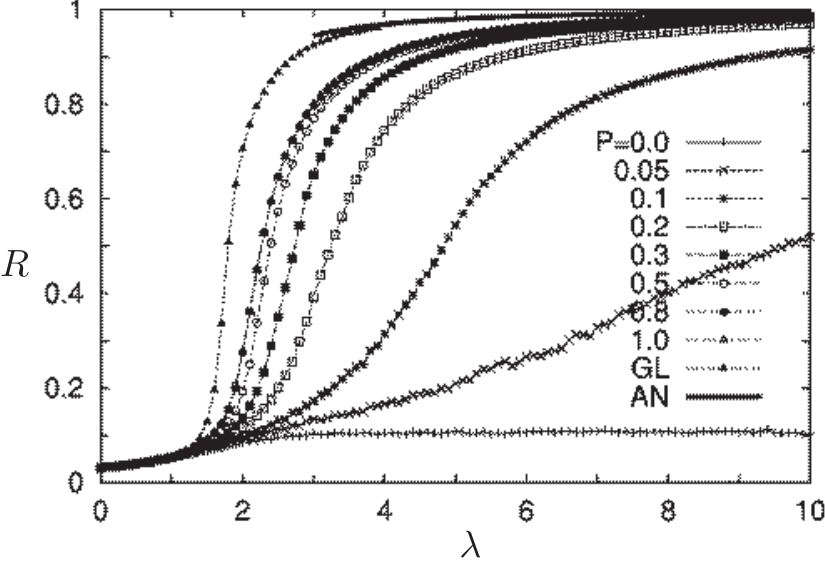}
\end{center}
\caption{ Order parameter $R$ (Eq.~\ref{eq:KM_ORDER_PARAMETER}) as a function of coupling $\lambda$ of SW networks with different values of the rewiring probability $p$. GL and AN stand for globally coupled network and its respective analytical solution, respectively. Adapted with permission from~\cite{hong2002synchronization}. Copyrighted by the American Physical Society.}
\label{fig:hong2002fig1}
\end{figure}

The first works on the Kuramoto model in complex networks aimed at quantifying the influence of the 
SW phenomenon on the overall network synchronization~\cite{watts1999small,hong2002synchronization,moreno2004synchronization,moreno2004fitness}. 
This issue was systematically investigated in~\cite{hong2002synchronization} by
considering oscillators with natural frequencies distributed according to a Gaussian distribution coupled in SW networks originated from one-dimensional regular lattices. By numerically evolving the equations (\ref{eq:KMNETWORKS}) with $\lambda_{ij} = \lambda/\left\langle k \right\rangle$ in order to obtain the dependence of the order parameter $R$ (Eq.~\ref{eq:KM_ORDER_PARAMETER}) on the coupling strength for different values of the rewiring probability $p$ (see Appendix~\ref{sec:appendix}), 
it was found that a small percentage of shortcuts is able to 
dramatically improve network synchronization in comparison to the
completely regular case. Interestingly, this enhancement in the
coherence between the oscillators was verified to saturate for 
intermediate values of the rewiring probability (see Fig.~\ref{fig:hong2002fig1}). In other words, 
for $p \gtrsim 0.5$, the synchronization of SW networks exhibit
no significant difference than the fully random case ($p=1$). This shows that, in an application context where the optimization of the network synchronization is sought, no improvement in the collective
behavior between oscillators is obtained beyond a critical value of
the rewiring probability, leading to a save of resources in cases in which rewirings have costs associated.  

These results thus suggest that the critical coupling $\lambda_c$ 
for the onset of synchronization should be a decreasing function of 
$p$. However, to thoroughly evaluate this dependence, finite-size
effects should be taken into account. The order parameter was then proposed to take the usual scaling form as~\cite{hong2002synchronization,hong2007FiniteSizeInComplexNetworks}
\begin{equation}
R = N^{-\frac{\beta}{\bar{\nu}}} F\left[(\lambda - \lambda_c)N^{\frac{1}{\bar{\nu}}} \right], 
\label{eq:hong2002_scaling}
\end{equation}
where $F\left[ \cdot \right]$ is a scaling function. At $\lambda = \lambda_c$ the
function $F$ becomes independent of $N$ and, by plotting $R N^{\beta/\bar{\nu}}$ for different values of $N$, the ratio $\beta/\bar{\nu}$ is then determined by the value that yields the best matching between the curves at $\lambda_c$. 
Once $\beta/\bar{\nu}$ is calculated, one can determine $\bar{\nu}$ through
\begin{equation}
\ln\left[\left.\frac{d R}{d\lambda}\right|_{\lambda_{c}}\right]=\frac{1-\beta}{\bar{\nu}}\ln N+\mbox{const}, 
 \label{eq:hong2002_scaling_2}
\end{equation} 
which together with $\beta/\bar{\nu}$ gives the exponents $\beta$ and $
\bar{\nu}$. Despite the sparse number of connections, the 
dynamics in SW networks apparently exhibited the same scaling with the system 
size as the model in the fully connected graph~\cite{daido1987scalingBehaviorAtTheOnset}, namely with $\beta 
\sim 1/2$ and $\bar{\nu} \sim 2$, and consequently the same mean-field 
character of the scaling near 
the critical coupling $R \sim (\lambda - \lambda_c)^\beta$~\cite{hong2002synchronization}. However, new results on the finite-size scaling (FSS) of Kuramoto model in the fully connected graph~\cite{hong2007EntrainmentTransitionInPopOfRandomFreqOsc,choi2013ExtendedFSS} and in SW networks~\cite{lee2014FiniteTimeFiniteSize} verified that the correct critical exponent should be $\bar{\nu} = 5/2$ instead of $\bar{\nu} = 2$ for both topologies (see also~\cite{brede2008SynchronizationDirectedFeedForward} for results on the FSS of directed SW networks). We shall discuss finite-size effects in more details in Sec.~\ref{subsec:finite_size_effects}. 

Moreno and Pacheco addressed the same questions for
SF networks generated by the Barab\'asi-Albert (BA) model (see Appendix~\ref{sec:appendix})~\cite{moreno2004synchronization} with uniform
frequency distributions.  
Surprisingly, a similar dependence of the order parameter $R$ 
on the coupling strength $\lambda$ was verified. In fact, the obtained critical exponent, $\beta \sim 0.46$, suggested that
SF networks also exhibit the same square-root behavior 
of the mean-field synchronization transition observed in the 
standard Kuramoto model~\cite{moreno2004synchronization}. Interestingly, the same choice of uniform
frequency distribution is known to yield a discontinuity in the
order parameter as a function of coupling $\lambda$ in the 
fully connected graph~\cite{pazo2005thermodynamic} (see Sec.~\ref{sec:explosive_sync}), in
striking contrast with the result in SF networks~\cite{moreno2004synchronization}.



The first numerical results left many questions to be solved regarding the onset of synchronization specially due to the finite value of the critical coupling $\lambda_c$ found
in SF networks~\cite{moreno2004synchronization,arenas2008synchronization}. The reason for this unexpected behavior
resides in the fact that critical properties of other dynamical
processes, such as epidemic spreading and percolation, were
predicted to vanish as a consequence of the high degree of heterogeneity found in these networks~\cite{barrat2008DynamicalProcessesOnComplexNetworks}.

In contrast with the formulation in the fully connected graph, 
the Kuramoto model has no exact solution in heterogeneous networks. 
Unfortunately, in the latter case, the system of equations (\ref{eq:KMNETWORKS_CONSTANT_COUPLING}) 
cannot be exactly decoupled by a global mean-field as in Eq.~\ref{eq:KM_uncoupled} 
and approximations need to be considered. 

Before presenting the most adopted mean-field approach, let us first discuss the so-called \textit{time-average} approximation. Restrepo et al.~\cite{restrepo2005onset} defined a local order parameter in a way to take explicitly into account the
contribution of time fluctuations. More specifically, the 
local mean-field of the neighborhood of node $i$ is given by~\cite{restrepo2005onset,restrepo2006SynchronizationLargeDirected}
\begin{equation}
r_i^{\rm{T}} e^{i\psi_i} = \sum_{j=1}^N A_{ij} \left\langle e^{i\theta_j} \right\rangle_t, 
\label{eq:restrepo_rn}
\end{equation}
where $\left\langle \cdots \right\rangle_t$ is a time average. Using Eq.~\ref{eq:restrepo_rn}, it is possible to write Eq.~\ref{eq:KMNETWORKS_CONSTANT_COUPLING} as 
\begin{equation}
\dot{\theta}_i = \omega_i + \lambda r_i^{\rm{T}} \sin(\psi_i - \theta_i) - \lambda h_i(t),
\label{eq:restrepo_rn_equations_of_motion_with_hn}
\end{equation}
where $h_i(t) = \textrm{Im}\{e^{-i\theta_i}\sum_{j=1}^N A_{ij}(\left\langle e^{i\theta_j} \right\rangle_t - e^{i\theta_j}) \}$ is the term that evaluates the contribution of time fluctuations. Beyond the onset of synchronization it is expected that the oscillators are locked in a common phase
making the order parameter to be of order $r_i^{\rm{T}} \sim k_i$. Since $h_i(t)$ is a sum of independent terms we then expect $h_i(t) \sim \sqrt{k_i}$~\cite{restrepo2005onset}. Thus, for networks in the limit of large average degrees, the term $h_i(t)$ can be neglected in comparison with the magnitude of $r_i^{\rm{T}}$
leading to
\begin{equation}
\dot{\theta}_i = \omega_i + \lambda r_i^{\rm{T}} \sin(\psi_i - \theta_i).
\label{eq:restrepo_rn_equations_of_motion_with_hn}
\end{equation}
In the time-independent regime $\dot{\theta}_i=0$, the locked oscillators have their phases given by
$\sin(\theta_i - \psi_i) = \omega_i/\lambda r_i^{\rm{T}}$. In this way, the order parameter (\ref{eq:restrepo_rn}) can be written 
as
\begin{equation}
r_{i}^{\rm{T}}=\sum_{j=1}^N A_{ij}\langle e^{i(\theta_{j}-\psi_{i})}\rangle_{t}=\sum_{|\omega_{j}|\leq\lambda r_{j}^{\rm{T}}}A_{ij}e^{i(\theta_{j}-\psi_{i})}+\sum_{|\omega_{j}|>\lambda r_{j}^{\rm{T}}}A_{ij}\langle e^{i(\theta_{j}-\psi_{i})}\rangle_{t}, 
\label{eq:restrepo_rn_sync_n_async}
\end{equation}
where the two terms above stand for the contribution 
of synchronous and drifting oscillators, respectively. The latter
can be computed by noting that the time average in Eq.~\ref{eq:restrepo_rn_sync_n_async} is given by~\cite{restrepo2005onset}
\begin{equation}
\langle e^{i\theta_{j}}\rangle_{t}=\int_{-\pi}^{\pi}\rho(\theta|\omega_j,r_j^{\rm{T}})d\theta,
\label{eq:retrepo_average_theta}
\end{equation}
where $\rho(\theta|\omega_j,r_j^{\rm{T}})d\theta$ is the probability
of finding the phase $\theta_j$ between $\theta$ and $\theta + d\theta$ for a given $\omega_j$ and $r_j^{\rm{T}}$. As 
$\rho \sim 1/|\dot{\theta}|$ we have that  
\begin{equation}
\rho(\theta|\omega_j,r_j^{\rm{T}})=\frac{\sqrt{\omega_{j}^{2}-\lambda^{2}(r_{j}^{\rm{T}})^{2}}}{2\pi|\omega_{j}+\lambda r_{j}^{{\rm T}}\sin(\psi_{j}-\theta)|}.
\label{eq:restrepo_Pj}
\end{equation}
Using Eq.~\ref{eq:restrepo_Pj} and~\ref{eq:retrepo_average_theta} we obtain the contribution of drifting oscillators, i.e.
\begin{eqnarray*}
\sum_{|\omega_{j}|>\lambda r_{j}^{\rm{T}}} A_{ij}\langle e^{i\theta_{j}}\rangle_{t} & = & \sum_{|\omega_{j}|>\lambda r_{j}^{\rm{T}}} A_{ij}\lambda r_{j}^{\rm{T}}\sqrt{\omega_{j}^{2}-\lambda (r_{j}^{\rm{T}})^{2}}{\rm sgn}(\omega_{j})\frac{1}{2\pi}\\
 &  & \times\int_{-\pi}^{\pi}\frac{e^{i\theta}\sin(\theta-\psi_{j})d\theta}{\omega_{j}^{2}-\lambda^{2}(r_{j}^{\rm{T}})^{2}\sin^{2}(\theta-\psi_{j})}
\label{eq:restrepo_contribution_drifting}
\end{eqnarray*}
Assuming that the variables $r_i^{\rm{T}}$, $\psi_i$ and $\omega_i$ are statistically 
independent and considering that the frequency distribution $g(\omega)$ is symmetric, we have that summation in Eq.~\ref{eq:restrepo_contribution_drifting} is of the order $\sqrt{k_i}$ and can be neglected in comparison with the contribution of locked oscillators~\cite{restrepo2005onset}. Under these conditions Eq.~\ref{eq:restrepo_rn_sync_n_async} is reduced to
\begin{equation}
r_{i}^{\rm{T}}=\sum_{|\omega_{j}|\leq\lambda r_{j}^{\rm{T}}}A_{ij}\cos(\psi_{j}-\psi_{i})\sqrt{1-\left(\frac{\omega_{j}}{\lambda r_{j}^{\rm{T}}}\right)^{2}}
\label{eq:restrepo_rn_sync_n_async_reduced}
\end{equation}
The previous self-consistent equation for the order parameter $r_i^{\rm{T}}$ is valid 
if the onset of synchronization is reached, i.e., for $\lambda > \lambda_c$. Furthermore, the minimal 
value for $\lambda_c$ is obtained for $\cos(\psi_i - \psi_j)=1$, which yields~\cite{restrepo2005onset}
\begin{equation}
r_{i}^{\rm{T}}=\sum_{|\omega_{j}|\leq\lambda r_{j}^{\rm{T}}}A_{ij}\sqrt{1-\left(\frac{\omega_{j}}{\lambda r_{j}^{\rm{T}}}\right)^{2}}
\label{eq:restrepo_rn_sync_n_async_reduced}
\end{equation}
Thus, once the adjacency matrix $\mathbf{A}$ and the natural frequencies $\omega_i$ of the whole population are known, 
Eq.~\ref{eq:restrepo_rn_sync_n_async_reduced} can be numerically solved in order to obtain 
the dependence of $r_i^{\rm{T}}$ on $\lambda$. Since the time fluctuations can be neglected for connected networks with sufficient large average degrees, from now on we abandon the upper script and the time average in Eq.~\ref{eq:restrepo_rn} and define the local order parameter simply by $r_i e^{i\psi_i(t)} = \sum_{j=1}^N A_{ij}e^{i\theta_j(t)}$. In this formulation, the overall network synchronization 
can be quantified through averaging the local order parameters $r_i$ as~\cite{restrepo2005onset}
\begin{equation}
z = re^{i\psi(t)} = \frac{\sum_{j=1}^N r_j}{\sum_{j=1}^N k_j}.
\label{eq:restrepo_r_rn}
\end{equation} 
Suppose now that the oscillators are not described by a particular sequence of natural frequencies but rather
by frequencies distributed according to some function $g(\omega)$. Then, Eq.~\ref{eq:restrepo_rn_sync_n_async_reduced} can be formulated as 
\begin{equation}
r_{i}=\sum_{j}A_{ij}\int_{-\lambda r_{j}}^{\lambda r_{j}}g(\omega)\sqrt{1-\left(\frac{\omega}{\lambda r_{j}}\right)^{2}}d\omega=\lambda\sum_{j}A_{ij}r_j\int_{-1}^{1}g(x\lambda r_{j})\sqrt{1-x^{2}}dx,
\label{eq:resptrepo_frequency_approximation}
\end{equation}
where $x = \omega/\lambda r_j$. Equation~\ref{eq:resptrepo_frequency_approximation} is also known as the \textit{frequency approximation}~\cite{restrepo2005onset}. Near the onset of synchronization, $r_j \rightarrow 0^{+}$, one can use the first-order approximation $g(x\lambda r_j)\approx g(0)$ to obtain
\begin{equation}
r_j^{(0)} = \frac{\lambda}{\lambda^{\rm{KM}}_c} \sum_{j=1}^N A_{ij} r_j^{(0)}.
\label{eq:restrepo_first_order_approximation}
\end{equation}
The smallest value of $\lambda$ that satisfies the previous equations is precisely 
the critical coupling $\lambda_c$, which is identified to be dependent on the
largest eigenvalue $\sigma_{\max}^{\rm{A}}$ of $\mathbf{A}$~\cite{restrepo2005onset}:
\begin{equation}
\lambda_c = \frac{\lambda_c^{\rm{KM}}}{\sigma_{\max}^{\rm{A}}}.
\label{eq:restrepo_critical_coupling_perturbation_theory}
\end{equation}
This result can be complemented by the estimation  
of $\sigma_{\max}^{\rm{A}}$ for different network models. In particular, for uncorrelated networks with a given degree distribution we have that~\cite{chung2003spectra} 
\begin{equation}
\sigma_{\max}^{\rm{A}}\sim\begin{cases}
\frac{\left\langle k^{2}\right\rangle }{\left\langle k\right\rangle } & \textrm{{\rm  if }}\frac{\left\langle k^{2}\right\rangle }{\left\langle k\right\rangle }>\sqrt{k_{\max}}\ln N\\
\sqrt{k_{\max}} & \textrm{{\rm if }}\sqrt{k_{\max}}>\frac{\left\langle k^{2}\right\rangle }{\left\langle k\right\rangle }\ln^{2}N, 
\end{cases}
\label{eq:adjacency_matrix_largest_eigenvalue}
\end{equation} 
where $k_{\max}$ is the network largest degree. Note that the critical coupling $\lambda_c$ in Eq.~\ref{eq:restrepo_critical_coupling_perturbation_theory}
takes naturally into account the finite size $N$ of the network. For this reason, although it provides accurate results for the critical coupling in cases where other 
approaches fail~\cite{restrepo2005onset}, Eq.~\ref{eq:restrepo_critical_coupling_perturbation_theory} 
predicts a vanishing onset of synchronization in the 
thermodynamic limit of SF networks. For instance, for 
networks with $P(k) \sim k^{-\gamma}$ with $\gamma \leq 
3$, the largest degree scales with system size as 
$k_{\max} \sim N^{1/(\gamma - 1)}$, making $\sigma_{\max}^{\rm{A}} 
\sim N^{1/2(\gamma -1)}$, which diverges for $N\rightarrow 
\infty$ and thus leading to the well-known result of 
vanishing $\lambda_c \rightarrow 0 $, where no phase 
transition is expected~
\cite{ichinomiya2004frequency,restrepo2005onset,arenas2008synchronization}. 
This is also true
for SF networks with $\gamma > 3$. Specifically, in this 
case, for sufficient large $N$, the largest eigenvalue 
will almost surely scale as $\sigma_{\max}^{\rm{A}} \sim 
\sqrt{k_{\max}}$, which also diverges in the limit $N 
\rightarrow \infty$ predicting, in this way, the absence 
of a critical coupling for the onset of synchronization.  
In contrast to Eq.~\ref{eq:restrepo_critical_coupling_perturbation_theory}, the critical coupling predicted using the mean-field approximation (MFA) remains finite for SF networks with $\gamma > 3$, as we shall soon discuss.

Equation~\ref{eq:resptrepo_frequency_approximation} can be further explored in order to obtain the
dependence of the order parameter near the critical point. A second-order expansion of the term $g(x\lambda r k_j)$ gives
\begin{equation}
r_{i}=\lambda\sum_{j=1}^{N}A_{ij}r_{j}\int_{-1}^{1}\left(g(0)+\frac{1}{2}g''(0)(x\lambda r_{j})^{2}\right)\sqrt{1-x^{2}} dx.
\label{eq:restrepo_expansion_perturbation_theory}
\end{equation}
By considering perturbations in the local order parameter in the previous equation, it follows that for $\lambda \rightarrow \lambda_c$
the total order parameter $r$ defined in Eq.~\ref{eq:restrepo_r_rn} is given by~\cite{restrepo2005onset}
\begin{equation}
r^{2}=\left(\frac{\eta_{1}}{c(\lambda_{c}^{{\rm KM}})^2}\right)\left(\frac{\lambda}{\lambda_{c}}-1\right)\left(\frac{\lambda}{\lambda_{c}}\right)^{-3},
\label{eq:restrepo_r_perturbation_theory}
\end{equation}
where $c = -\pi g''(0)\lambda_c^{\rm{KM}}/16$ and
\begin{equation}
\eta_{1}=\frac{\left\langle \upsilon\right\rangle ^{2}(\sigma_{\max}^{\rm{A}})^{2}}{N\left\langle k\right\rangle ^{2}\left\langle \upsilon^{4}\right\rangle },
\label{eq:restrepo_eta_perturbation_theory}
\end{equation}
with $\upsilon$ being the normalized eigenvector associated to $\sigma_{\max}^{\rm{A}}$ of $\mathbf{A}$. 

One of the most employed approach in the
analytical treatment of dynamical processes in
networks is to consider MFAs~\cite{barrat2008DynamicalProcessesOnComplexNetworks,dorogovtsev2008criticalPhenomenaInComplexNetworks}. 
Such an approximation scheme relies on the assumption that 
the dynamical state of a given node $i$ depends on a global 
common field that is equally felt by all individuals 
in the network. This is translated in the dynamics 
of the Kuramoto model in networks by considering that
the oscillators interact through a global field $re^{i\psi}
$ that is related with the local mean-field $r_j e^{i
\psi_j(t)}$ as
\begin{equation}
re^{i\psi} = \frac{1}{k_j} r_je^{i\psi_j} = \frac{1}{k_j} 
\sum_{j=1}^N A_{ij}e^{i\theta_j},
\label{eq:mean_field_approximation}
\end{equation}
i.e., Eq.~\ref{eq:mean_field_approximation} 
basically states that the local mean-field felt by a node 
should be proportional to a global mean-field weighted by
the local connectivity, i.e. $r_j = r k_j$. This assumption
is reasonable to be adopted only if the network is well 
connected (sufficient large average degree) without the 
presence of communities, i.e. a portion of the network that is
more connected within itself than with the other nodes~\cite{arenas2008synchronization,restrepo2005onset}. Using this approximation for the local order parameter the
equations governing the phases evolution are decoupled as
\begin{equation}
\dot{\theta}_i = \omega_i +\lambda r k_i \sin(\psi - \theta_i). 
\label{eq:mean_field_approximation_eq_decoupled}
\end{equation}
Note that the effective coupling of node $i$ has a term 
proportional to its local topology, in contrast with the
case of the fully connected graph. For this reason the 
effects of the network topology on dynamics can be hidden 
if the coupling strength in Eq.~\ref{eq:KMNETWORKS} is 
chosen as $\lambda_{ij} = \lambda/k_i$, where the 
normalization by the degree neutralizes the heterogeneity
in the field initially imposed by the network topology~\cite{arenas2008synchronization}. It is interesting to remark that the MFA in Eq.~\ref{eq:mean_field_approximation} is 
precisely identical to the so-called \textit{annealed 
network approximation}~\cite{dorogovtsev2008criticalPhenomenaInComplexNetworks}. There, one 
substitutes $A_{ij}$ by its expected value over an ensemble average of uncorrelated
networks with a given degree sequence $\{k_i\}$, $i=1,...,N$. In other words, the original problem of a network defined by the adjacency matrix $A_{ij}$ is now mapped 
into fully connected weighted network described by the 
adjacency matrix $\tilde{\mathbf{A}}$ with
\begin{equation}
\tilde{A}_{ij} = \frac{k_i k_j}{N\left\langle k \right\rangle}.
\label{eq:mean_field_annealed_approximation}
\end{equation} 
Noteworthy, $\tilde{A}_{ij}$ is also the probability 
in the configuration model that nodes $i$ and $j$ are 
connected~\cite{newman2010networks}. Replacing $A_{ij}$ by $\tilde{A}_{ij}$
in Eq.~\ref{eq:KMNETWORKS_CONSTANT_COUPLING} we get 
\begin{equation}
\dot{\theta}_i = \omega_i + \lambda k_i \sum_{j=1}^N \frac{k_j}{N\left\langle k \right\rangle} \sin(\theta_j - \theta_i).  
\label{eq:mean_field_annealed_approximation_eq_of_motion}
\end{equation}
In order to decouple Eqs.~\ref{eq:mean_field_annealed_approximation_eq_of_motion}, this formulation motivates the following definition of the global order parameter: 
\begin{equation}
r e^{\psi(t)}= \frac{1}{N \left\langle k \right\rangle}\sum_{j=1}^N k_j e^{i\theta_j(t)}, 
\label{eq:mean_field_annealed_order_parameter}
\end{equation}
which is equivalent to the original definition for the
MFA in Eq.~\ref{eq:mean_field_approximation} and leads precisely 
to the same set of equations as in~(\ref{eq:mean_field_approximation_eq_decoupled}).

After this detour to show the equivalence of different
treatments, let us return to the analysis of Eq.~\ref{eq:resptrepo_frequency_approximation} in the frequency approximation scheme. Applying the MFA to $r_j$ and summing over $i$ in both sides in Eq.~\ref{eq:resptrepo_frequency_approximation} we obtain~\cite{ichinomiya2004frequency,restrepo2005onset,lee2005synchronization}
\begin{equation}
\sum_{j=1}^{N}k_{j}=\lambda\sum_{j=1}^{N}k_{j}^{2}\int_{-1}^{1}g(x\lambda rk_{j})\sqrt{1-x^{2}}dx.
\label{eq:restrepo_mean_field_approximation}
\end{equation}
Tending $r \rightarrow 0^{+}$ one finally gets $\lambda_c$
within the MFA
\begin{equation}
\lambda_c = \lambda_c^{\rm{KM}} \frac{\left\langle k \right\rangle}{\left\langle k^2 \right\rangle}. 
\label{eq:mean_field_critical_coupling}
\end{equation}
This equation is one of the most known results related to the dynamics of Kuramoto oscillators in networks. It asserts that the value of the critical coupling for the onset of synchronization in the fully connected graph is rescaled by the ratio $\left\langle k \right\rangle /\langle k^2 \rangle$ of the first two moments of the degree distribution $P(k)$. Therefore, according to Eq.~\ref{eq:mean_field_critical_coupling}, the more heterogeneous the network, the weaker the coupling strength required 
to synchronize its oscillators. This highlights the role
played by the hubs in network dynamics acting improving 
the overall collective behavior. Furthermore, in contrast 
to $\lambda_c$ predicted by the frequency approximation (Eq.~\ref{eq:restrepo_critical_coupling_perturbation_theory}), the mean-field scheme gives a finite $\lambda_c$ for SF networks with 
$\gamma > 3$ in the thermodynamic limit $N \rightarrow \infty$, in agreement with simulations of sufficient large networks~\cite{moreno2004synchronization,arenas2008synchronization}. However, 
problems arise for more heterogeneous networks with $2 < \gamma < 3$. In principle, one would expect partial synchronization to emerge for any $\lambda > 0$, since $\left\langle k^2 \right\rangle \rightarrow \infty$ for this range of $\gamma$. However, as extensive simulations show~\cite{arenas2008synchronization}, that seems to be not the case even if the finite number of nodes is taken into account in the estimation of $\lambda_c$ using Eq.~\ref{eq:mean_field_critical_coupling}. More specifically, for $\gamma = 3$, the second moment of the degree distribution scales with the system size as $\left\langle k^2 \right\rangle \sim \ln N$, leading to $\lambda_c \sim 1/\ln N$ ~\cite{dorogovtsev2008criticalPhenomenaInComplexNetworks,arenas2008synchronization,dorogovtsev2010lectures}. Although a very high number of oscillators is indeed a limiting factor, simulations with reasonably large networks already present discrepancies with this estimative of $\lambda_c$. As previously mentioned, evidences show that in fact $\lambda_c$ for SF networks with $2\leq \gamma \leq 3$ seems to converge to a constant value as the
system size $N$ is increased, in striking contrast with 
the prediction of the MFA. Therefore, the question left is what is the source of the disagreement between the result predicted in Eq.~\ref{eq:mean_field_critical_coupling} and the results observed in simulations. Much has been conjectured about this puzzle~\cite{arenas2008synchronization,barrat2008DynamicalProcessesOnComplexNetworks,dorogovtsev2008criticalPhenomenaInComplexNetworks,dorogovtsev2010lectures} but the problem remains open.


Nevertheless, despite the inconsistency in determining $\lambda_c$ in the limit of large $N$, further developments can be made using the MFA. Similarly as before, one can expand $g(x\lambda r k_j) \approx g(0) + \frac{1}{2} g''(0)(x\lambda r k_j)^2$ to get
\begin{equation}
r^{2}=\left(\frac{\eta_{2}}{c(\lambda_{c}^{{\rm KM}})^2}\right)\left(\frac{\lambda}{\lambda_{c}}-1\right)\left(\frac{\lambda}{\lambda_{c}}\right)^{-3},
\label{eq:restrepo_mean_field_r_near_onset}
\end{equation}
where 
\begin{equation}
\eta_{2}=\frac{\left\langle k^{2}\right\rangle ^{3}}{\left\langle k^{4}\right\rangle \left\langle k\right\rangle ^{2}}, 
\label{eq:restrepo_eta2_mean_field}
\end{equation}
which holds once $\left\langle k^4 \right\rangle$ remains finite~\cite{restrepo2005onset}. 

The mean-field result for $\lambda_c$ (Eq.~\ref{eq:mean_field_critical_coupling}) was firstly obtained
by Ichinomiya~\cite{ichinomiya2004frequency} using the continuum limit of Eqs.~\ref{eq:KMNETWORKS_CONSTANT_COUPLING}. Similarly as
considered in the original approach of Kuramoto, 
in the limit $N\rightarrow \infty$, the population of oscillators can be
described by the density $\rho(\theta,t|\omega,k)$ of oscillators that have phase $\theta$ at time $t$ for a given
frequency $\omega$ and degree $k$. It is further assumed 
that, for a given $\omega$ and $k$, $\rho(\theta,t|\omega,k)$ is normalized as
\begin{equation}
\int_0^{2\pi} \rho(\theta,t|\omega,k) d\theta = 1.
\label{eq:Ichinomiya_normalization}
\end{equation}
Considering an 
uncorrelated network with a given degree distribution 
$P(k)$, the probability that a randomly selected edge has 
at its end a node with phase $\theta$ at time $t$ for a 
given degree $k$ and frequency $\omega$ is given by
\begin{equation}
\frac{k P(k)}{\left\langle k \right\rangle }g(\omega)
\rho(\theta,t|\omega,k). 
\label{eq:Ichinomiya_probability}
\end{equation}
The equations for the phases evolution in the continuum 
limit are then 
obtained by replacing the sum by the average using Eq.~\ref{eq:Ichinomiya_probability} in the right-hand side of 
Eq.~\ref{eq:KMNETWORKS_CONSTANT_COUPLING}, i.e.~
\cite{ichinomiya2004frequency} 
\begin{equation}
\dot{\theta} = \omega + \frac{\lambda k}{\left
\langle k \right\rangle} \int d\omega' \int dk' \int d
\theta' g(\omega')P(k')k'\rho(\theta',t|\omega',k')
\sin(\theta' - \theta), 
\label{eq:Ichinomiya_continuumlimit}
\end{equation}
where now the network dynamics is described by the average
phases $\theta(t)$. In order to decouple Eqs.~\ref{eq:Ichinomiya_continuumlimit} one can define the global order parameter as
\begin{equation}
re^{i\psi(t)} = \frac{1}{\left\langle k \right\rangle} \int d\omega \int dk \int d\theta P(k) k g(\omega) \rho(\theta,t|\omega,k) e^{i\theta(t)}. 
\label{eq:Ichinomiya_order_parameter_continuum_limit}
\end{equation}
Note that Eq.~\ref{eq:Ichinomiya_order_parameter_continuum_limit} is exactly the continuum limit version of the order parameter 
introduced in the annealed network approximation (Eq.~\ref{eq:mean_field_annealed_order_parameter}). Writing Eq.~\ref{eq:Ichinomiya_order_parameter_continuum_limit} in terms of the order parameter in the continuum limit, distribution $\rho(\theta,t|\omega,k)$ should then
obey the continuity equation 
\begin{equation}
\frac{\partial\rho(\theta,t|\omega,k)}{\partial t}=-\frac{\partial}{\partial\theta}\left\{ \rho(\theta,t|\omega,k)\left[\omega+\lambda kr\sin(\psi-\theta)\right]\right\}, 
\label{eq:Ichinomiya_continuity_equation}
\end{equation}
whose solutions in the stationary regime assuming $\psi = 0$, without loss of generality, are given by
\begin{equation}
\rho(\theta|\omega,k)=\begin{cases}
\delta\left[\theta-\arcsin\left(\frac{\omega}{\lambda kr}\right)\right] & \mbox{if } \left|\omega\right|\leq \lambda kr\\
\frac{C(\omega,k)}{\left|\omega-\lambda kr\sin\theta\right|} & \mbox{otherwise,}
\end{cases}, 
\label{eq:Ichinomiya_solutions_cont_equation}
\end{equation}
where $C(\omega,k) = \sqrt{\omega^2 - (\lambda k r)^2}/2\pi$ and $\delta(\cdot)$ is the Dirac delta function. 
The first and second terms correspond to the density of synchronous and drifting oscillators, respectively. In particular, the latter corresponds to the distribution of drifting oscillators in Eq.~\ref{eq:restrepo_Pj} formulated in terms of the phases $\theta(t)$. Substituting Eq.~\ref{eq:Ichinomiya_solutions_cont_equation} in Eq.~\ref{eq:Ichinomiya_order_parameter_continuum_limit} one gets~\cite{ichinomiya2004frequency}
\begin{equation}
\int dkP(k)k=\lambda\int dkP(k)k^{2}\int_{-1}^{1}dxg(x\lambda rk)\sqrt{1-x^{2}},
\label{eq:Ichinomiya_last_eq}
\end{equation}
which tending $r\rightarrow 0^{+}$ leads to the same critical coupling obtained in Eq.~\ref{eq:mean_field_critical_coupling}. 

\begin{figure}[!t]
\begin{center}
\includegraphics[width=0.8\linewidth]{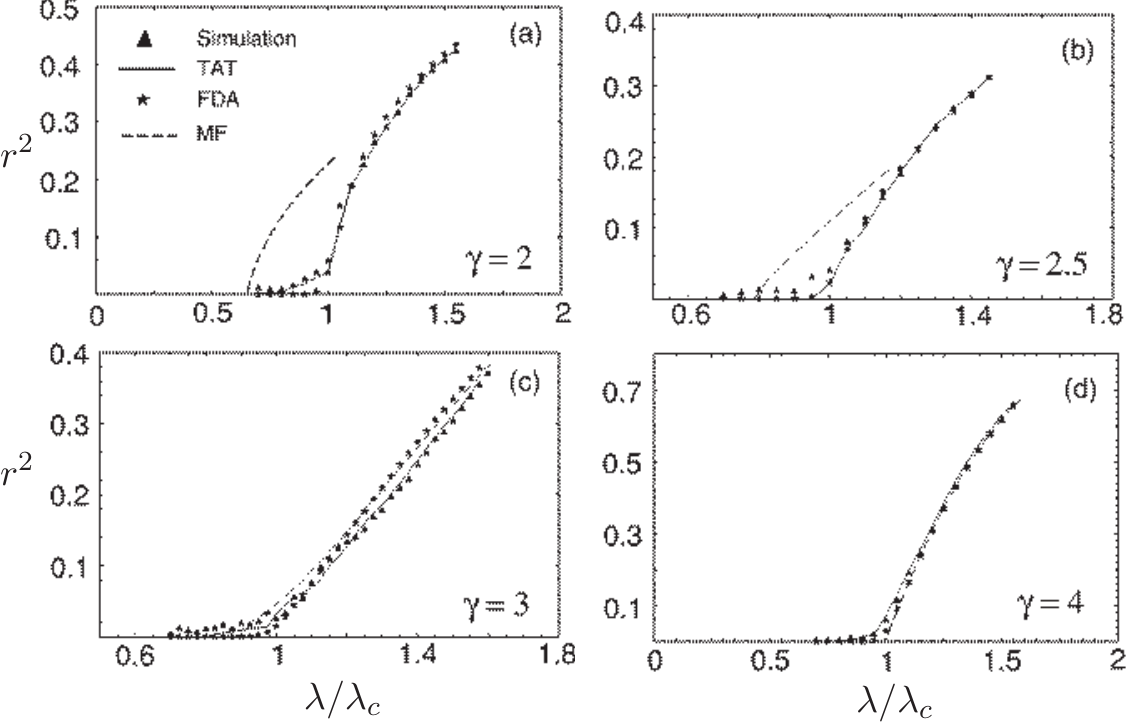}
\end{center}
\caption{ Comparison of order parameter $r^2$ obtained from simulations and 
theory for SF networks with power-law exponent (a) $\gamma = 2$, (b) $\gamma = 
2.5$, (c) $\gamma = 3$ and (d) $\gamma = 4$. In all cases the degrees are bounded as $50 
\leq k 
\leq 2000$~\cite{restrepo2005onset}. TAT, FDA and MF stand for time-average theory (Eq.~\ref{eq:restrepo_rn_sync_n_async_reduced}), frequency distribution 
approximation (Eq.~\ref{eq:resptrepo_frequency_approximation}) and mean-field (Eq.~\ref{eq:restrepo_mean_field_r_near_onset}), respectively. 
Adapted with permission from~\cite{restrepo2005onset}. Copyrighted by the American Physical Society.}
\label{fig:restrepo2005fig1}
\end{figure}

The comparison of the aforementioned approximations with simulations is shown in Fig.~\ref{fig:restrepo2005fig1} by plotting $r^2$ as a function of $\lambda/\lambda_c$ for SF networks with different $\gamma$. As we can see, the time-averaged approximation provides the best agreement with the simulation results for all $\gamma$ considered in Fig.~\ref{fig:restrepo2005fig1}. Furthermore, the mean-field technique completely fails to determine the onset
of synchronization in networks with $\gamma = 2$ (Fig.~\ref{fig:restrepo2005fig1}(a)) and $2.5$ (Fig.~\ref{fig:restrepo2005fig1}(b)), clearly showing
the limitations of such an approximation in highly heterogeneous networks. On the other hand, for more homogeneous networks, as for $\gamma = 4$ (Fig.~\ref{fig:restrepo2005fig1}(d)), the mean-field solution approaches the results provided by the frequency distribution approximation, which are in better agreement with the simulations. 

So far we analyzed the onset of synchronization in networks in which 
the oscillators are symmetrically coupled. A natural and interesting 
question would then be how the network dynamics is affected by the
introduction of asymmetric interactions. Given an undirected network, 
one way to introduce asymmetry in the couplings between the oscillators
is to consider normalization factors in $\lambda_{ij}$ in Eq.~\ref{eq:KMNETWORKS} that depend on local topological properties of node 
$i$. This scenario was investigated in~\cite{oh2007SynchronizationTransitionHeterogeneously}, as follows  
\begin{equation}
\dot{\theta}_i = \omega_i + \frac{\lambda}{k_i^{1 - \eta}}\sum_{j=1}^N A_{ij}\sin(\theta_j - \theta_i).
\label{eq:oh2007_eq_of_motion}
\end{equation}
Rewriting the equations in terms of $r$ in the
MFA we get
\begin{equation}
\dot{\theta}_i = \omega_i + \lambda k_i^{\eta}r\sin(\psi - \theta_i).
\label{eq:oh2007_eq_of_motion_decoupled}
\end{equation}
In SF networks the coupling $\lambda_{ij} = \lambda/k_i^{1-\eta}$ has 
the interesting property of tuning the influence of hubs on the 
network dynamics. More specifically, depending on the choice of $\eta$, the contribution of high degree nodes to the heterogeneous mean-field can significantly change the onset of synchronization. By developing an analogous self-consistent analysis within the MFA as in Eq.~\ref{eq:restrepo_mean_field_approximation}, the effect of $\eta$ on the network dynamics can be summarized by calculating the $\lambda_c$, which assumes the following forms~\cite{oh2007SynchronizationTransitionHeterogeneously}
\begin{equation}
\lambda_{c}=\begin{cases}
\lambda_{c}^{{\rm KM}}\frac{\left\langle k\right\rangle }{\left\langle k^{1+\eta}\right\rangle } & \textrm{ if }\gamma>3\eta+2\\
\lambda_{c}^{{\rm KM}}\frac{\zeta(\gamma-1)}{\zeta(\gamma-\eta-1)} & \textrm{ if }\eta+2<\gamma<3\eta+2\\
\sim N^{(\eta-\gamma+2)/(\gamma-1)} & \textrm{ if }\eta+1<\gamma<\eta+2,
\end{cases}
\label{eq:oh2007_critical_couplings}
\end{equation}       
where $\zeta$ is the Hurwitz zeta function. For $\eta=1$ we recover the symmetric 
coupling case for which the MFA predicts $\lambda_c \rightarrow 0 $ in the thermodynamic limit for networks with $2< \gamma < 3$. For a general $\eta$, the regime in which $\lambda_c$ is expected to vanish is obtained for SF networks in the range $\eta + 1 < \gamma < \eta + 2$.  On the other hand, for $\gamma> \eta + 2$, $\lambda_c$ remains finite as $N \rightarrow \infty$. Note that $\eta = 0$ in Eq.~\ref{eq:oh2007_critical_couplings} completely removes the dependence of $\lambda_c$ on the network topology, since it leads to the coupling normalization $\lambda_{ij} = \lambda/k_i$ as previously discussed. 

One can further show that the nature of the phase synchronization transition
crucially depends on the exponent $\eta$. More specifically, near the onset of synchronization, $r\sim (\lambda - \lambda_c)^{\beta}$, with $\beta$ given by
\begin{equation}
\beta=\begin{cases}
\frac{1}{2} & \textrm{ if }\gamma>3\eta+2\\
\eta/(\gamma-2-\eta) & \textrm{ if }\eta+2<\gamma<3\eta+2\\
\eta/(\eta-\gamma+2) & \textrm{ if }\eta+1<\gamma<\eta+2\\
(\gamma-1)/(\eta-\gamma+2) & \textrm{ if }\gamma<\eta+1
\end{cases}.
\label{eq:oh2007_exponent_beta}
\end{equation}
The result above generalizes the scaling of $r$ 
for different exponents $\eta$. More specifically, in the absence of a normalization factor ($\eta =1$), the mean-field transition
with $\beta = 1/2$ is only obtained for SF networks with $\gamma = 5$. However, by decreasing $\eta$, one obtains mean-field transitions with $\beta=1/2$ 
for broader degree distributions, a fact that shows how $\eta$ acts on the network dynamics by attenuating the effect of hubs.    


\subsection{Finite-size effects}
\label{subsec:finite_size_effects}

We discussed the main early contributions to the
theoretical analysis of the Kuramoto model networks, including the 
approximations used to obtain the dependence of the order parameter on the 
coupling strength and the related critical exponents. However, the numerical
determination of $\lambda_c$ essentially relies on
simulations of networks that are inherently consisted of a finite number of 
oscillators. Therefore, the effects of such a limitation should be 
considered in the analysis. Furthermore, as we shall see, another important
feature should be considered for the correct estimation of the critical
exponents of the phase transition, namely the sample-to-sample fluctuations 
generated by different realizations of random natural frequencies as well as
randomness introduced by fluctuations in the network topology. 

We begin by first analyzing the scaling of the $r$ 
in uncorrelated SF networks following closely~\cite{hong2007FiniteSizeScalingOfSynchronized}. Considering that
the natural frequencies are distributed by a unimodal and even 
distribution $g(\omega)$, it is convenient to write the self-consistent
mean-field equation for $r$ in the symbolic form $r = \widetilde{\mathcal{R}}(\lambda r)$, with
\begin{equation}
\widetilde{\mathcal{R}}(\lambda r) \equiv \frac{1}{N}\sum_{|\omega_j|\leq k_j \lambda r}\frac{k_{j}}{\left\langle k\right\rangle }\sqrt{1-\left(\frac{\omega_{j}}{k_{j}\lambda r}\right)^{2}}
\label{eq:hong2007FiniteSize_Tilde_Psi_Def}
\end{equation}
In the limit $N \rightarrow \infty$, Eq.~\ref{eq:hong2007FiniteSize_Tilde_Psi_Def} should converge to
\begin{equation}
\mathcal{R}(\lambda r) \equiv \lim_{N\rightarrow\infty}\widetilde{\mathcal{R}}(\lambda r)=\left\langle \widetilde{\mathcal{R}}(\lambda r)\right\rangle =\frac{1}{\left\langle k\right\rangle }\int dkP(k)ku(k \lambda r), 
\label{eq:hong2007FiniteSizeSF_Psi_Def}
\end{equation} 
where $u(y) = \int_{-y}^y d\omega g(\omega) \sqrt{1 - \omega^2/y^2}$. It is also convenient to define~\cite{hong2007FiniteSizeScalingOfSynchronized}
\begin{equation}
\hat{u}(y) = \frac{\pi}{2} g(0) y- u(y).
\label{eq:hong2007FiniteSizeSF_u_hat}
\end{equation}
Using Eq.~\ref{eq:hong2007FiniteSizeSF_u_hat} we can then rewrite
$\mathcal{R}(\lambda r)$ as
\begin{equation}
\mathcal{R}(\lambda r)=\frac{\pi\left\langle k^{2}\right\rangle }{2\left\langle k\right\rangle }g(0) \lambda r -\hat{\mathcal{R}}(\lambda r),
\label{eq:hong2007FiniteSizeSF_Psi_approx}
\end{equation} 
where $\hat{\mathcal{R}}(\lambda r) = \frac{1}{\left\langle k \right\rangle}\int dk P(k) k \hat{u}(k \lambda r)$. For SF networks with $\gamma > 5$, $\langle k^4 \rangle $ remains finite yielding $\hat{\mathcal{R}}$ for small $r$: 
\begin{equation}
\hat{\mathcal{R}}(\lambda r)\backsimeq c_0\frac{\left\langle k^{4}\right\rangle }{\left\langle k^{2}\right\rangle }(\lambda r)^{3},  
\label{eq:hong2007FiniteSizeSF_Psi_Hat_Gamma_l_5}
\end{equation} 
where $c_0 = -\pi g''(0)/16$ is a positive constant. If $3 < \gamma < 5$, Eq.~\ref{eq:hong2007FiniteSizeSF_Psi_Hat_Gamma_l_5} no longer holds, since
in this case the moment $\left\langle k^{4}\right\rangle $ diverges. Using
the definition of $\hat{\mathcal{R}}(\lambda r)$ and supposing that $P(k) = C k^{-\gamma}$, we then obtain 
\begin{equation}
\hat{\mathcal{R}}(\lambda r)=\frac{1}{\left\langle k\right\rangle }\int_{0}^{\infty}dkP(k)k\hat{u}(k\lambda r)=c_{1}(\lambda r)^{\gamma-2}, 
\label{eq:hong2007FiniteSizeSF_Psi_Hat_2}
\end{equation} 
where $c_1 = C \left\langle k \right\rangle^{-1}\int_0^{\infty}dy y^{-\gamma + 1}\hat{u}(y)$. 

Having calculated these quantities we are able to estimate the contributions 
of sample-to-sample fluctuations in the FSS. Such fluctuations are quantified
by $\delta \tilde{\mathcal{R}}(\lambda r) \equiv \tilde{\mathcal{R}}(\lambda r) - \mathcal{R}(\lambda r)$, which 
basically calculates the deviations of the function $\tilde{\mathcal{R}}(\lambda r)$ corresponding to a single realization from the ensemble average $\mathcal{R}(\lambda r)$ defined in Eq.~\ref{eq:hong2007FiniteSizeSF_Psi_Def}. Futhermore, $\tilde{\mathcal{R}}(\lambda r)$ can be seen as the mean value of the random variable   
\begin{equation}
\vartheta(\omega,k)=\frac{k}{\left\langle k\right\rangle }\sqrt{1-\left(\frac{\omega}{k\lambda r}\right)^{2}}\Theta\left(1-\frac{|\omega|}{k\lambda r}\right),
\label{eq:hong2007FiniteSizeSF_Chi_Def}
\end{equation} 
where $\Theta(\cdot)$ is the Heaviside function. In this way, using the central limit theorem, we expect $\delta \tilde{\mathcal{R}}(\lambda r)$ to be Gaussian distributed with zero mean and with the variance
\begin{equation}
\langle(\delta\tilde{\mathcal{R}})^{2}\rangle=\frac{1}{N}(\langle\vartheta^{2}\rangle-\langle\vartheta\rangle^{2})\equiv\frac{f(r)}{N}.
\label{eq:hong2007FiniteSizeSF_Variance}
\end{equation}
Similarly as before, $\left\langle \vartheta^2 \right\rangle$ will remain finite for networks with $\gamma > 4$, hence $f(r)$ can be straightforwardly calculated for small $r$ as~\cite{hong2007FiniteSizeScalingOfSynchronized}
\begin{equation}
f(r) \backsimeq \langle \vartheta^2\rangle \backsimeq \frac{4 \left\langle k^3 \right\rangle}{3 \left\langle k \right\rangle^2}g(0) \lambda r, 
\label{eq:hong2007FiniteSizeSF_D_r_gamma_l_4}
\end{equation} 
where the contribution of $\left\langle \vartheta \right\rangle$ was neglected 
since $\left\langle \vartheta \right\rangle \sim r$ near $\lambda_c$. For networks with $3 < \gamma < 4$, one can
evaluate the contribution analogously as performed in Eq.~\ref{eq:hong2007FiniteSizeSF_Psi_Hat_2} to obtain 
\begin{equation}
f(r) \backsimeq d_1 (\lambda r)^{\gamma - 3},
\label{eq:hong2007FiniteSizeSF_D_r_3_4}
\end{equation}
where $d_1 = C \left\langle k \right\rangle^{-2}\int_0^{\infty} dy y^{2-\gamma}\int_{-y}^y d\omega g(\omega)(1 - \omega^2/y^2)$ is also a positive constant.

\begin{figure}[!t]
\begin{center}
\includegraphics[width=0.6\linewidth]{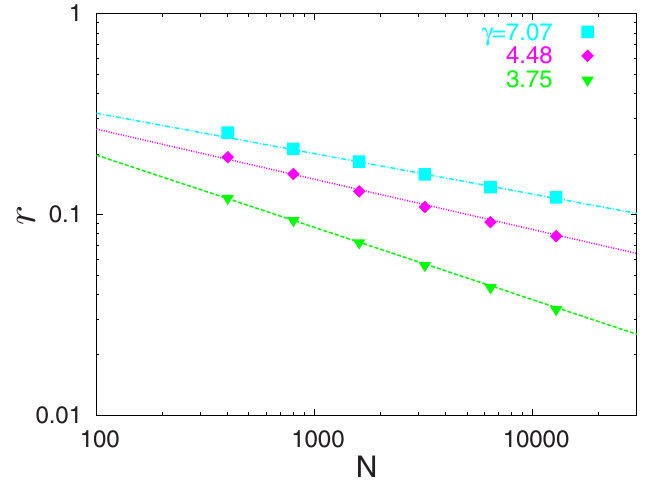}
\end{center}
\caption{ Size dependence of the order parameter $r$ for SF networks with $\gamma = 7.07, 4.48$ and 3.75. The lines $r \sim N^{-\beta/\bar{\nu}}$ are obtained using the exponents given in Eq.~ \ref{eq:hong2007fssonnetworks_main_result_exponents}. Adapted with permission from~\cite{hong2007FiniteSizeScalingOfSynchronized}. Copyrighted by the American Physical Society.}
\label{fig:hong2007fig1_FSS_SF}
\end{figure}

All the expansions near the onset of synchronization of the function $r = \tilde{\mathcal{R}}(\lambda r)$ can be summarized into the general form~\cite{hong2007FiniteSizeScalingOfSynchronized}
\begin{equation}
r=\frac{\lambda}{\lambda_{c}}r-c(\lambda r)^{a}+d(\lambda r)^{b/2}N^{-1/2}\xi, 
\label{eq:hong2007FiniteSizeSF_scaling_r}
\end{equation}
where $c$ and $d$ are positive constants, $\lambda_c = 2\left\langle k \right\rangle /\pi g(0) \left\langle k^2 \right\rangle$, $\xi$ a Gaussian random variable with zero mean and unit variance, and $a$ and $b$ constants that depend on $\gamma$. Finally, the scaling relation in Eq.~\ref{eq:hong2002_scaling_2} can be used to identify
($\beta,\bar{\nu}$) for uncorrelated SF networks~\cite{hong2007EntrainmentTransitionInPopOfRandomFreqOsc,hong2007FiniteSizeScalingOfSynchronized}
\begin{equation}
(\beta,\bar{\nu}) = \left\{
\begin{array}{cccl}
\frac{1}{2}, &\frac{5}{2}, & \qquad\mbox{$\gamma>5$;} \\
\frac{1}{\gamma-3}, &\frac{2\gamma-5}{\gamma-3}, & \qquad\mbox{$4<\gamma<5$;}\\
\frac{1}{\gamma-3}, &\frac{\gamma-1}{\gamma-3}, & \qquad\mbox{$3< \gamma < 4$.}
\end{array}
\right. \label{eq:hong2007fssonnetworks_main_result_exponents}
\end{equation}
Figure~\ref{fig:hong2007fig1_FSS_SF} shows the scaling of $r$ on $N$ at $\lambda_c$ estimated by the mean-field method together with the expected behavior $r \sim N^{-\beta/\bar{\nu}}$ using the exponents given by Eq.~\ref{eq:hong2007fssonnetworks_main_result_exponents}.

It is important to remark that Lee~\cite{lee2005synchronization} provided
the first estimation for the critical exponents associated to the size dependence in SF networks using a different approach, namely by analyzing the size of the largest synchronous component. However, the results in~\cite{lee2005synchronization} only agree with those in Eq.~\ref{eq:hong2007fssonnetworks_main_result_exponents}
in the prediction of $\beta$ for networks with
with $\gamma > 5$. The source of the discrepancy between
these two results is precisely the consideration of sample-to-sample fluctuations. Interestingly, not only in complex topologies 
the critical exponents have their values corrected, but also well-known results on the model in the fully connected graph had their estimations revisited by the inclusion of such fluctuations. Specifically, 
the exponent $\bar{\nu}$ no longer assumes the usual mean-field estimation $\bar{\nu} = 2$~\cite{kuramoto1984cooperativeDynamicsOscillatorCommunity,daido1987scalingBehaviorAtTheOnset} in the globally coupled system submitted to quenched randomness in the frequencies, but rather $\bar{\nu} = 5/2$ if the averages over different realizations are taken into account~\cite{hong2007FiniteSizeScalingOfSynchronized,choi2013ExtendedFSS,lee2014FiniteTimeFiniteSize} . The same effect is also observed in SW networks. In Sec.~\ref{subsec:early_works} we saw that such structures were firstly reported
to have the same scaling as the fully connected graph, i.e., $(\beta,\bar{\nu}) 
= (1/2,2)$~\cite{hong2002synchronization}. Nevertheless, the statement that the synchronization phase 
transitions in the fully connected graph and in SW networks have the same 
critical exponents still holds, as recent studies showed that the dependence on 
the system size in the latter is, in fact, better described by exponents $
(\beta,\bar{\nu}) = (1/2,5/2)$~\cite{lee2014FiniteTimeFiniteSize}, in agreement with 
the revisited
calculation for the fully connected graph~
\cite{hong2007EntrainmentTransitionInPopOfRandomFreqOsc,choi2013ExtendedFSS}. 
Noteworthy, as it will be discussed in Sec.~\ref{sec:stochastic}, in 
the presence of noise, the order parameter is shown to exhibit the usual 
mean-field scaling characterized by $(\beta,\bar{\nu})=(1/2,2)$~\cite{sonnenschein2012onset}. 

Sample-to-sample fluctuations arise in the dynamics of globally coupled oscillators due to the random disorder introduced by the different frequency assignments between realizations. If the oscillators are now coupled through 
a network another source of disorder is included, i.e., the randomness associated
to the different link configuration (or also ``link-disorder fluctuations'')~\cite{hong2013LinkDisorderFluctuationsEffects}. Thus, given the fluctuations induced by frequency and link assignments, one should study the isolated effect of each kind of quenched disorder to the scaling with the system
size. In order to only analyze the influence of link-disorder, the sequence of natural frequencies $\{\omega_i\}$ ($i=1,...,N$) can be deterministically assigned according to~\cite{hong2013LinkDisorderFluctuationsEffects}  
\begin{equation}
\frac{i-0.5}{N} = \int_{-\infty}^{\omega_i} g(\omega) d\omega,
\label{eq:hong2013LinkDisorder_freq_assignment}
\end{equation}
which removes the disorder due to frequencies, since then $\{\omega_i\}$ is uniquely set over different realizations. Considering the particular case of ER networks, one can show by using the procedure in~\cite{hong2007FiniteSizeScalingOfSynchronized} that for $\lambda \rightarrow \lambda_c$, $r$ is given by the following self-consistent equation    
\begin{equation}
r = \frac{\lambda}{\lambda_c} r - c(\lambda r)^3 + d(\lambda r)^{1/2} N^{-1/2}\xi,
\label{eq:hong2013LinkDisorder_scaling_r}
\end{equation}
where in this case $c = -[\pi g''(0)\left\langle k^4 \right\rangle/16 \left\langle k \right\rangle]$, $d = \sqrt{[4g(0)\left\langle k^3 \right\rangle/3\left\langle k^2 \right\rangle]}$ and $\xi$ is a Gaussian random variable with zero mean and unity variance, similarly as in Eq.~\ref{eq:hong2007FiniteSizeSF_scaling_r}. Hence, Eq.~\ref{eq:hong2013LinkDisorder_scaling_r} is used to determine
the scaling of $r$~\cite{hong2013LinkDisorderFluctuationsEffects}:
\begin{equation}
r = N^{-1/5}F[(\lambda - \lambda_c)N^{2/5}], 
\label{eq:hong2013LinkDisorder_scaling_r_2}
\end{equation}
leading to $(\beta,\bar{\nu}) =  (1/2,5/2)$. Interestingly, the 
scaling induced by link-disorder is precisely the same as induced by quenched 
frequency disorder in the fully connected graph. Furthermore, if one relaxes
the condition imposed by Eq.~\ref{eq:hong2013LinkDisorder_freq_assignment} 
and randomly distribute the frequencies according to $g(\omega)$, the result
$(\beta,\bar{\nu}) =  (1/2,5/2)$ is again obtained~\cite{hong2013LinkDisorderFluctuationsEffects}. This suggests that the 
scaling is dominated by the fluctuations in the network connectivity, regardless 
of the frequency disorder. Note that this independence 
of the scaling on the frequency disorder is not observed in the fully connected 
graph, where $\bar{\nu}$ is reduced to $\bar{\nu} = 5/4$ in case the
frequencies are regularly distributed as in Eq.~\ref{eq:hong2013LinkDisorder_freq_assignment}~\cite{hong2015FiniteSizeScalingDynamicsFluctuationHyper}

The next is step is then verifying how the synchronization
phase transition changes if the link-disorder is removed, while keeping 
the fluctuations in the frequency. More specifically, in this case, the connections between oscillators are kept constant over different realizations of the network dynamics (networks whose adjacency matrices are fixed over time or over different realizations are also referred as \textit{quenched networks}~\cite{um2014NatureOfSynchronizationTransitions,castellano2010ThresholdsForEpidemicSpreading}). At first, one could expect that the phase transition remains unchanged except from a shift in the coupling strength. However, the scaling of $r$ can significantly change whether the networks are quenched or annealed~\cite{um2014NatureOfSynchronizationTransitions}. In order to illustrate this phenomenon, let us consider ER networks with uniform natural frequency distribution given by
\begin{equation}
g(\omega)=\begin{cases}
\frac{1}{2\epsilon}, & \textrm{ if }|\omega|\leq \epsilon\\
0 & \textrm{otherwise.}
\end{cases}
\label{eq:um2014NatureOfSync_freq_dist}
\end{equation}
Substituting Eq.~\ref{eq:um2014NatureOfSync_freq_dist} into 
Eq.~\ref{eq:restrepo_mean_field_approximation} we obtain  
\begin{equation}
r=\frac{\pi\lambda r}{4\epsilon\left\langle k \right\rangle}\left[\left\langle 
k^{2}\right\rangle -\int_{k>\epsilon/\lambda r} 
dkP(k)k^{2}f\left(\frac{\epsilon}{k\lambda r}\right)\right], 
\label{eq:um2014NatureOfSync_order_parameter_uniform_freq}
\end{equation}
where 
\begin{equation}
f(x) = 1- \frac{2}{\pi}\left[ \arcsin x + x \sqrt{1 - x^2}\right].
\label{eq:um2014NatureOfSync_funcf}
\end{equation}
Using mean-field analysis one can straightforwardly show that the 
onset of synchronization is characterized by $\lambda_c = 
\frac{4\epsilon \left\langle k \right\rangle }{\pi \left\langle k^2 
\right\rangle}$. Furthermore, for $\lambda \gtrsim \lambda_c$ we have
\begin{equation}
\frac{\lambda}{\lambda_c} - 1 \approx \frac{1}{\left\langle k^2 
\right\rangle} \int_{k>\epsilon/\lambda r} dk P(k) k^2 f \left( 
\frac{\epsilon}{k \lambda_c r} \right) .
\label{eq:um2014NatureOfSync_approx1}
\end{equation}
For small values of $r$, only high degree nodes are taken into account in the integral above. However, for ER networks, $P(k) = \left\langle k \right\rangle^k e^{-\left\langle k \right\rangle}/k!$ decays exponentially fast in a way that only degrees $k \gtrsim \epsilon/(\lambda_c r)$ effectively contribute to the integral in Eq.~\ref{eq:um2014NatureOfSync_approx1}. Hence, expanding $f(x)$ for $x \lesssim 1$, 
we obtain
\begin{equation}
\frac{\lambda}{\lambda_c} - 1 \sim \frac{1}{\left\langle k^2 \right\rangle} \sqrt{\frac{\epsilon}{\lambda_c r} } P\left( \frac{\epsilon}{\lambda_c r}\right), 
\label{eq:um2014NatureOfSync_approx2}
\end{equation}  
which yields the following logarithmic scaling
\begin{equation}
r \approx \frac{\epsilon}{\lambda_c}\left| \ln \left(\frac{\lambda - \lambda_c}{\lambda_c} \right)\right|^{-1}.
\label{eq:um2014NatureOfSync_logarithmic_scaling}
\end{equation}

Although the scaling in Eq.~\ref{eq:um2014NatureOfSync_logarithmic_scaling} leads to an abrupt logarithmic increase of $r$, the mean-field approach predicts that the transition to the synchronous state in ER networks with uniform frequency distributions is still continuous. 
\begin{figure}[!t]
\begin{center}
\includegraphics[width=1.0\linewidth]{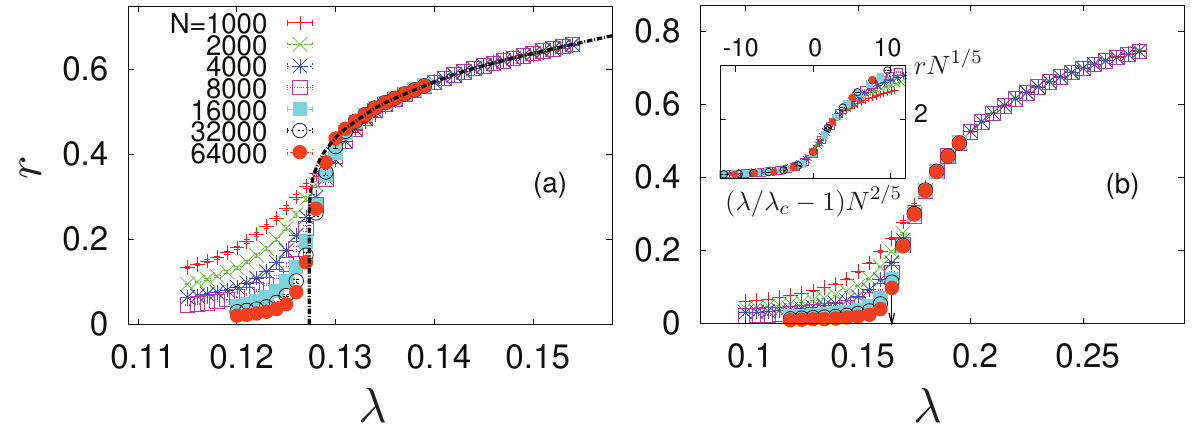}
\end{center}
\caption{ Order parameter $r$ as a function of coupling $\lambda$ for (a) annealed and (b) quenched ER networks with $\langle k \rangle = 4$ for different system sizes $N$. In panel (a) different network realizations are considered, yielding a good agreement with the the logarithm scaling in Eq.~\ref{eq:um2014NatureOfSync_logarithmic_scaling}. A different scaling with is obtained for the quenched network, as depicted in the inset in panel (b). Adapted with permission from~\cite{um2014NatureOfSynchronizationTransitions}. Copyrighted by the American Physical Society.}
\label{fig:um2014NatureOfSync_fig1}
\end{figure} 
Figure~\ref{fig:um2014NatureOfSync_fig1}(a) shows the dependence of $r$ on 
$\lambda$ for various $N$ in annealed ER networks. We see 
a good agreement with the logarithm scaling 
predicted by the MFA. On the other hand, as shown in Figure~
\ref{fig:um2014NatureOfSync_fig1}(b), the phase transition assumes the usual 
mean-field behavior characterized by the exponents $(\beta,\bar{\nu}) = (1/2,5/2)$, 
as in other topologies previously discussed in this section. Therefore, besides having the value of $\lambda_c$ shifted (see Fig.~\ref{fig:um2014NatureOfSync_fig1}), the 
nature of the synchronization phase transition and, consequently, its dependence  
on $N$ are also drastically changed if the network structure is quenched, regardless of $g(\omega)$. This phenomenon can be explained by comparing the effective frequencies $\langle \dot{\theta}_i \rangle_t$, in annealed and quenched topologies. Precisely, in annealed networks, the effective frequencies are shown to converge in the incoherent state to the natural frequencies $\omega_i$, whereas the distribution of $\langle \dot{\theta}_i \rangle_t$ tends to a $\delta$ for $\lambda_c$. Curiously, 
this effect seems to not be present in quenched networks~\cite{um2014NatureOfSynchronizationTransitions}, where in the synchronous state the distribution of $\langle \dot{\theta}_i \rangle_t$ significantly differs from the original 
frequency distribution $g(\omega)$ in both incoherent and synchronous state, leading in this way to deviations in the prediction by the mean-field calculation~\cite{um2014NatureOfSynchronizationTransitions}.

\subsection{Relaxation dynamics}
\label{subsec:relaxation}

Most of the studies on network synchronization focus on effects
of network topology on the dynamics in the stationary regime. 
However, 
the question of how fast the network converges to the equilibrium 
state is equally important. Instead of asking about 
the necessary coupling strength required to synchronize the 
oscillators, the question could be, given certain conditions, how 
long does a given network take to reach the synchronized state or 
fall into incoherence. In real 
applications, important questions would then be which kind of 
network topology promotes the fastest time scale to reach 
synchronization or, given that coherent motion is already attained, 
how robust such a state is against perturbations quantified in 
terms of the time required to return to the synchronized state. 

\subsubsection{Small-world networks}

%

\begin{figure}[!t]
\begin{center}
\includegraphics[width=1.0\linewidth]{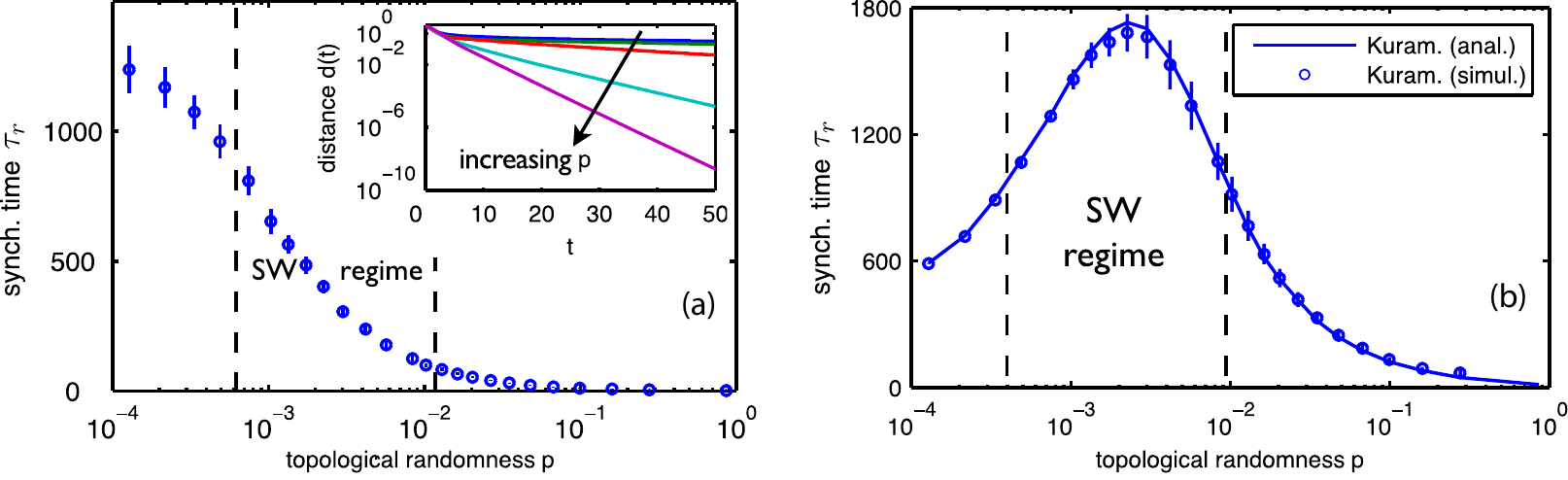}
\end{center}
\caption{ Relaxation time measured by Eq.~\ref{eq:grabow2010DoSWSyncFastest_distance2} for SW networks ensembles with (a) 
fixed average degree $\left\langle k \right\rangle = 4$ and (b) with fixed average shortest path 
length $\ell = 4$. Inset of panel (a) depicts the 
distance $d$ (Eq.~\ref{eq:grabow2010DoSWSyncFastest_distance}) over time. Solid line is calculated using Eq.~\ref{eq:grabow2010Fastest_tau_r_sigma_2} Adapted from~\cite{grabow2010small}}.
\label{fig:grabow2010SoSmallFastest_fig1}
\end{figure}

In~\cite{grabow2010small} the relaxation time $\tau_r$ of SW networks made 
up of identical 
oscillators subjected to coupling $\lambda_{ij} = \lambda/k_i$ in 
Eq.~\ref{eq:KMNETWORKS} was thoroughly  investigated by numerically 
calculating $\tau_r$, i.e. the time needed for 
the network to reach the stationary state. Specifically, 
considering the distance
\begin{equation}
d(t) = \max_{i,j} \textrm{ dist } (\theta_i(t),\theta_j(t)),
\label{eq:grabow2010DoSWSyncFastest_distance}
\end{equation}
if the network is connected and formed by a single giant component, $\tau_r$ is derived through~\cite{grabow2010small} 
\begin{equation}
d(t) \sim \exp(-t/\tau_r).
\label{eq:grabow2010DoSWSyncFastest_distance2}
\end{equation}
Fixing $N$ and $\left\langle k \right\rangle$ one clearly
sees the effect of increasing the heterogeneity on $\tau_r$ in Fig.~
\ref{fig:grabow2010SoSmallFastest_fig1}(a).
Similar results were also obtained in~\cite{hong2002synchronization,son2008relaxation}. However, a surprising effect emerges if the average shortest path length $ \ell $ (see Appendix~\ref{sec:appendix}) is 
fixed. Figure~\ref{fig:grabow2010SoSmallFastest_fig1}(b) shows the dependence 
of $\tau_r$ on the rewiring probability $p$ of networks constructed with varying $\left
\langle k(p) \right\rangle$ so that $\ell$ remains unchanged throughout the whole range 
of $p$. Unexpectedly, besides having a non-monotonic dependence on $p$, the peak 
in the time needed to the networks synchronize all its oscillators is precisely 
in the SW regime. Therefore, in terms of $\tau_r$, the inclusion of
shortcuts ends up by delaying the onset of synchronization, a non-intuitive 
effect given the fact that SW networks are able to synchronize at lower coupling
strengths in comparison to regular structures~
\cite{watts1998collective,hong2002synchronization,arenas2008synchronization}. 
These results were verified to hold not only for networks of Kuramoto oscillators, 
but also for chaotic and pulse-coupled systems~
\cite{grabow2010small,grabow2011speed}. Further insights can be gained by linearizing Eq.~\ref{eq:KMNETWORKS} for $\omega_i = \omega$ $\forall i$ around the synchronized state $\theta_1(t) = \cdots = \theta_N = \theta(t)$ to obtain 
the evolution for the phase perturbations $\phi_i(t) = \theta_i(t) - \theta(t)$:
\begin{equation}
\dot{\phi}_i = \sum_{j=1}^N \mathcal{L}_{ij} \phi_j, 
\label{eq:grabow2010Fastest_linearizing_eq_mov}
\end{equation}
where $\mathcal{L}_{ij}$ are the elements of the weighted Laplacian 
$\boldsymbol{\mathcal{L}}$ defined as
\begin{equation}
\mathcal{L}_{ij} = \frac{\lambda}{k_i}(1 - \delta_{ij}) - \lambda \delta_{ij},
\label{eq:grabow2010Fastest_laplacian_elements}
\end{equation}
where $\delta_{ij}$ is the Kronecker delta. Near the invariant trajectory, the second smallest eigenvalue $\sigma_2$ 
of $\mathcal{L}$ dominates the asymptotic decay leading 
to~\cite{grabow2010small,grabow2011speed}
\begin{equation}
\tau_r =  - \frac{1}{\textrm{Re } \sigma_2}.
\label{eq:grabow2010Fastest_tau_r_sigma_2}
\end{equation}
This equation can be used together with the results for the spectra of SW networks~\cite{grabow2012SmallWorldNetworkSpectraInMeanFieldTheory,grabow2015CollectiveRelaxationDynamicsSmallWorldNetworks}  in order to analytically determine $\tau_r$, as shown 
in Fig.~\ref{fig:grabow2010SoSmallFastest_fig1}(b). 

\subsubsection{Scale-free networks}
\label{subsubsec:relaxation_dynamics_SF}


Analytical progress on the temporal behavior of $r$
is impracticable with the approaches presented in Sec.~\ref{subsec:early_works}, 
since in this case the dynamics in the stationary regime is targeted. However, 
thanks to the theory introduced by Ott and Antonsen (OA)~\cite{ott2008low,ott2009long} one can assess the 
full relaxation dynamics by reducing the system's dimension to a smaller set of 
differential equations that describes the temporal evolution of $r$. To describe the potential of the theory in~\cite{ott2008low,ott2009long}, consider an uncorrelated network whose
node dynamics is described by Eqs.~\ref{eq:Ichinomiya_continuumlimit} in the continuum limit. Expanding the time-dependent density of the oscillators $\rho(\theta,t|\omega,k)$ in a Fourier series in $\theta$ we have  
\begin{equation}
\rho(\theta,t|\omega,k)=\frac{1}{ 2\pi }\left\{ 1+\left[\sum_{n=1}^{\infty}\hat{\rho}_n(\omega,k,t) e^{in\theta}+\textrm{(c.c)}\right]\right\}, 
\label{eq:yoon2015CriticalBehavior_expanding_rho}
\end{equation}
where c.c. stands for the complex conjugate. The core of the OA theory is the ansatz~\cite{ott2008low,ott2009long} 
\begin{equation}
\hat{\rho}_n(\omega,k,t) = \left[\hat{\rho}(\omega,k,t) \right]^n.
\label{eq:yoon2015CriticalBehavior_OA_ansatz}
\end{equation}
Substituting Eq.~\ref{eq:yoon2015CriticalBehavior_expanding_rho} into 
Eq.~\ref{eq:Ichinomiya_continuity_equation} we get
\begin{equation}
\frac{\partial\hat{\rho}}{\partial t}+i\omega\hat{\rho}+\frac{\lambda k}{2}(z\hat{\rho}^{2}-z^*)=0
\label{eq:Yoon2014CriticalBehavior_alpha_eq}
\end{equation}
Furthermore, substituting Eq.~\ref{eq:yoon2015CriticalBehavior_expanding_rho} 
into Eq.~\ref{eq:Ichinomiya_order_parameter_continuum_limit} we obtain the evolution of $z = r e^{i\psi}$
\begin{equation}
z(t) = \frac{1}{\left\langle k \right\rangle}  \int dk P(k) k \int d\omega g(\omega)\hat{\rho}^*(\omega,k,t).
\label{eq:Yoon2014CriticalBehavior_order_parameter}
\end{equation}
In the stationary regime $\partial \hat{\rho}/\partial t=0$, Eq.~\ref{eq:Yoon2014CriticalBehavior_alpha_eq} admits the solutions
\begin{equation}
\hat{\rho}_0(\omega,k) = \begin{cases} -\frac{i \omega}{\lambda k r} +\sqrt{1-\left(\frac{\omega}{\lambda k r}\right)^2}, &  |\omega|\leq \lambda k r \\
-\frac{i\omega}{\lambda k r}\Bigl[1 - \sqrt{ 1-\left(\frac{\lambda k r}{\omega}\right)^2} \,\,\Bigr] & \text{otherwise}, 
\end{cases}
\label{eq:yoon2014CriticalBehavior_alpha0}
\end{equation}
which if substituted into Eq.~\ref{eq:Yoon2014CriticalBehavior_order_parameter} 
and considering a symmetric $g(\omega)$ recovers Eq.~\ref{eq:Ichinomiya_last_eq}. 

In order to estimate $\tau_r$, we consider small perturbations around the stationary state in Eq.~\ref{eq:yoon2014CriticalBehavior_alpha0}~\cite{yoon2015critical}: 
\begin{eqnarray}
z(t) &=& z_0 + \delta z(t),  \nonumber \\
\hat{\rho}(t) &=& \hat{\rho}_0(\omega,k) + \delta\hat{\rho}(\omega,k,t), 
\label{eq:yoon2014CriticalBehavior_perturbations}
\end{eqnarray}
where $z_0 = z(t = 0)$ and $\delta z (t = 0) \ll z_0$. Substituting Eq.~\ref{eq:yoon2014CriticalBehavior_perturbations} into Eq.~\ref{eq:Yoon2014CriticalBehavior_alpha_eq} and~\ref{eq:Yoon2014CriticalBehavior_order_parameter}, and ignoring second
order terms we obtain the evolution equation for $\delta \hat{\rho}$:  
\begin{equation}
\delta\dot{\hat{\rho}}+i\omega\delta \hat{\rho}+\frac{\lambda k}{2}(2z_{0} \hat{\rho}_{0}\delta \hat{\rho}+\hat{\rho}_{0}^{2}\delta z-\delta z^{*})=0,
\label{eq:yoon2014CriticalBehavior_alpha_perturbation}
\end{equation}
which, analogously to Eq.~\ref{eq:Yoon2014CriticalBehavior_alpha_eq}, needs 
to be solved with 
\begin{equation}
\delta z(t) = \frac{1}{\left\langle k \right\rangle}\int dk P(k) k \int d \omega  g(\omega) \delta \hat{\rho}^*(\omega,k,t).
\label{eq:yoon2014CriticalBehavior_z_perturbation}
\end{equation}
Taking the Laplace transform and integrating both sides of 
Eq.~\ref{eq:yoon2014CriticalBehavior_alpha_perturbation} one
obtains~\cite{yoon2015critical}
\begin{equation}
\delta\hat{\rho} (s,\omega,k) = \frac{\delta \hat{\rho}(t =0) + \frac{\lambda k}{2}(\delta z^*(s)-\hat{\rho}_0^2 \delta z(s))}{s +  i \omega + \lambda k z_0 \hat{\rho}_0}.
\label{eq:yoon2014CriticalBehavior_alpha_laplacian_transform}
\end{equation}
Substituting Eq.~\ref{eq:yoon2014CriticalBehavior_alpha_laplacian_transform} into 
Eq.~\ref{eq:yoon2014CriticalBehavior_z_perturbation} gives~\cite{yoon2015critical}
\begin{equation}
\delta z(s) = \frac{B(s)}{1-\frac{\lambda}{\langle k \rangle}\int dk  P(k) k^2 J(s,k)},
\label{eq:yoon2014CriticalBehavior_delta_z_final}
\end{equation}
where 
\begin{eqnarray}
\!\!\!\!\!\! B(s) &{=}& \frac{1}{\langle k \rangle} \!\! \int \!\! dk P(k) k^2 \int\!\! d\omega \frac{g(\omega)\delta \hat{\rho}(\omega,k,t{=}0)}{s {+}  i \omega {+} \lambda k r \hat{\rho}_0},
\label{eq:yoon2014CriticalBehavior_B} \\
\!\!\!\!\!\!J(s,k) &{=}& \frac{1}{2}\int_{-\infty}^{\infty} d\omega \frac{g(\omega)[1-\hat{\rho}_0^2(\omega,k)]}{s +  i \omega + \lambda k r\hat{\rho}_0(\omega,k)}.
\label{eq:yoon2014CriticalBehavior_A}
\end{eqnarray}
Therefore, with Eqs.~\ref{eq:yoon2014CriticalBehavior_delta_z_final}-\ref{eq:yoon2014CriticalBehavior_A} one is able to assess the 
relaxation dynamics of any uncorrelated network. However, in order to do so, two regimes should be considered separately, namely $\lambda < \lambda_c$
and $\lambda > \lambda_c$. (i) In the incoherent regime ($\lambda < \lambda_c$) we have that
$r = 0$, making $J$ in Eq.~\ref{eq:yoon2014CriticalBehavior_A} to be independent of $k$. In this way, the poles of Eq.~\ref{eq:yoon2014CriticalBehavior_delta_z_final} are simply given by~\cite{yoon2015critical}
\begin{equation}
\frac{\langle k^2 \rangle J(s)}{\langle k \rangle} = \frac{1}{\lambda}.
\label{eq:yoon2014CriticalBehavior_poles_delta_z_K_leq_Kc}
\end{equation}
Considering the particular case of a Lorentzian
$g(\omega) = [\pi(\omega^2 + 1)]^{-1}$, Eq.~\ref{eq:yoon2014CriticalBehavior_poles_delta_z_K_leq_Kc} yields
\begin{equation}
\frac{\left\langle k^2 \right\rangle }{2\left\langle k \right\rangle (1 + s) } = \frac{1}{\lambda}, 
\label{eq:yoon2014CriticalBehavior_withlorentzian_1}
\end{equation}
which written in terms of the critical coupling $\lambda_c = 2\langle k \rangle/\pi g(0)\langle k^2 \rangle$ is
\begin{equation}
s_0 =  - \frac{\lambda_c - \lambda}{\lambda_c}
\label{eq:yoon2014CriticalBehavior_s0}
\end{equation}
Taking the inverse Laplace transform of Eq.~\ref{eq:yoon2014CriticalBehavior_delta_z_final} we obtain 
\begin{equation}
\delta z(t) \sim \exp(-t/\tau_r),  
\label{eq:yoon2014CriticalBehavior_delta_z_t}
\end{equation}
where $\tau_r = - s_0^{-1}$. Therefore, the relaxation time is then given 
by
\begin{equation}
\tau_r = \frac{\lambda}{\lambda - \lambda_c}.
\label{eq:yoon2014CriticalBehavior_relaxation_time_K_leq_Kc}
\end{equation}
The result above is valid for any uncorrelated network that 
has a finite second moment $\left\langle k^2 \right\rangle$, which is the case
of SF networks with $\gamma > 3$. Note that the relaxation 
time $\tau_r$ tends to infinity as $\lambda \rightarrow \lambda_c$.

ii) For $\lambda > \lambda_c$, Eq.~\ref{eq:yoon2014CriticalBehavior_A} is no longer
independent of $k$, since $r>0$, and particular effects of the degree distribution should be evidenced in the estimation of $\tau_r$.  In this case, and again for $g(\omega) = [\pi(\omega^2 + 1)]^{-1}$, one can show that~\cite{yoon2015critical} 
\begin{equation}
J(s,k)=\frac{\sqrt{1+(\lambda k r)^2}-1}{(\lambda k r)^2 [\sqrt{1+(\lambda k r)^2}+s]}, 
\label{eq:yoon2014CriticalBehavior_A_K_geq_Kc}
\end{equation}
where the poles of Eq.~\ref{eq:yoon2014CriticalBehavior_delta_z_final} are
now calculated by
\begin{equation}
\frac{1}{\langle k \rangle \lambda^2 r^2}\int dk  P(k) \frac{\sqrt{1+(\lambda k r)^2}-1}{\sqrt{1+(\lambda k r)^2}+s} = \frac{1}{\lambda}.
\label{eq:yoon2014CriticalBehavior_poles_K_leq_Kc}
\end{equation}
Considering SF networks with $\gamma > 5$, so that the fourth moment $\langle k^4 \rangle$ is finite, and expanding Eq.~\ref{eq:yoon2014CriticalBehavior_poles_K_leq_Kc} near the onset of synchronization yields
\begin{equation}
\tau_{r} = - \frac{1}{s_0} \simeq \frac{2 \langle k^2\rangle}{\langle k^4 \rangle \lambda^2 r^2} = \frac{\lambda_c}{2(\lambda - \lambda_c)}, 
\label{eq:yoon2014CriticalBehavior_relaxation_time_gamma_geq_5}
\end{equation}
which has the same form as encountered in the fully connected graph~\cite{yoon2015critical,daido2015SusceptibilityLargePopulations}. Interestingly, it can be further shown that networks with $3 < \gamma < 5$ have
the similar dependence $\tau \sim (\lambda - \lambda_c)^{-1}$ for $\tau_r$~\cite{yoon2015critical}. 

\begin{center}
\begin{table}[t]
\caption{Dependence of the order parameter $r$, relaxation rate $\tau_r^{-1}$ and susceptibility $\chi$ on the coupling $\lambda$ near the onset of synchronization at zero field $a = 0$, and as a function of the field amplitude $a$ at $\lambda = \lambda_c$ for networks with $P(k) \sim k^{-\gamma}$. From~\cite{yoon2015critical}.}
\begin{tabular}{cccc}
\hline 
Coupling & $r$ & $\tau_{r}^{-1}$ & $\chi$\tabularnewline
\hline 
\hline 
 &  & $\gamma>5$ & \tabularnewline
$\lambda<\lambda_{c}$ & 0 & $1-\frac{\lambda}{\lambda_{c}}$ & $\frac{1}{2}\left(1-\frac{\lambda}{\lambda_{c}}\right)^{-1}$\tabularnewline
$\lambda>\lambda_{c}$ & $\left(\frac{\lambda}{\lambda_{c}}-1\right)^{1/2}$ & $2\left(\frac{\lambda}{\lambda_{c}}-1\right)$ & $\frac{1}{4}\left(\frac{\lambda}{\lambda_{c}}-1\right)^{-1}$\tabularnewline
$\lambda=\lambda_{c}$ & $\propto a^{1/3}$ & $\propto a^{2/3}$ & $\propto a^{-2/3}$\tabularnewline
 &  & $3<\gamma\leq$5 & \tabularnewline
$\lambda<\lambda_{c}$ & 0 & $1-\frac{\lambda}{\lambda_{c}}$ & $\frac{1}{2}\left(\frac{\lambda}{\lambda_{c}}-1\right)^{-1}$\tabularnewline
$\lambda>\lambda_{c}$ & $\left(\frac{\lambda}{\lambda_{c}}-1\right)^{1/(\gamma-3)}$ & $\propto\left(\frac{\lambda}{\lambda_{c}}-1\right)$ & $\frac{1}{2(\gamma-3)}\left(\frac{\lambda}{\lambda_{c}}-1\right)^{-1}$\tabularnewline
$\lambda=\lambda_{c}$ & $\propto a^{1/(\gamma-2)}$ & $\propto a^{(\gamma-3)/(\gamma-2)}$ & $\propto a^{-(\gamma-3)/(\gamma-2)}$\tabularnewline
\hline 
\end{tabular}
\label{table:yoon2014critical}
\end{table}
\end{center}

Another interesting case is when the model is under the 
influence of an external field acting on the oscillators' 
phases. This scenario is particular appealing 
for the study of relaxation dynamics of phase oscillators, 
since many insights can be gained from known results 
relating the relaxation rate and susceptibility in the 
statistical mechanics of magnetic systems~
\cite{stanley1987introduction}.  
The phase evolution in the presence of a uniform local field can be formulated as
\begin{equation}
\dot{\theta}_i = \omega_i + \lambda \sum_{j=1}^N A_{ij}\sin(\theta_j - \theta_i) - a \sin\theta_i,
\label{eq:yoon2015CriticalBehavior_external_field}
\end{equation}
The model in~\ref{eq:yoon2015CriticalBehavior_external_field} was 
firstly studied by Shinomoto and Kuramoto~
\cite{shinomoto1986phase} and is relevant in the modeling of 
excitable systems and Josephson junctions~
\cite{acebron2005kuramoto}. For this reason, constant $a$ 
can be referred as the amplitude of a periodic force~
\cite{shinomoto1986phase,sakaguchi1988CooperativePhenomenaCoupledOscillator,strogatz1989CollectiveDynamicsRandomPinning,coolen2003partially} 
or the excitation threshold of the model~
\cite{tessone2007theory,sonnenschein2013excitable,sonnenschein2014cooperative} 
(see Sec.~\ref{sec:stochastic} for investigations 
of this model in the presence of stochastic fluctuations). 
In the same way that in magnetic systems the response to 
changes in the magnetic field can be quantified by the 
magnetic susceptibility, it it possible to define the 
correspondent susceptibility $\chi$ of the order parameter 
$r$ with respect to small variations of $a$ as~\cite{yoon2015critical,daido2015SusceptibilityLargePopulations} 
\begin{equation}
\chi = \frac{dr(a)}{da}.
\label{eq:Susceptibility}
\end{equation}
Similar definitions for the susceptibility as in Eq.~\ref{eq:Susceptibility} can be found in classical XY 
models~\cite{stanley1987introduction}. By using the same mean-field treatment used 
to derive the relaxation time, in~\cite{yoon2015critical} 
it was shown that $\chi$ and $\tau_r$ are also related 
in SF networks, exhibiting the same universality class
as in the Ising model. The effects of network heterogeneity on the scaling near the onset of synchronization of the variables $r$, $\tau_r$ and $\chi$
are summarized in Table~\ref{table:yoon2014critical}.

\subsection{\textcolor{black}{Further approaches}}
\begin{figure}[!t]
\begin{center}
\includegraphics[width=0.75\linewidth]{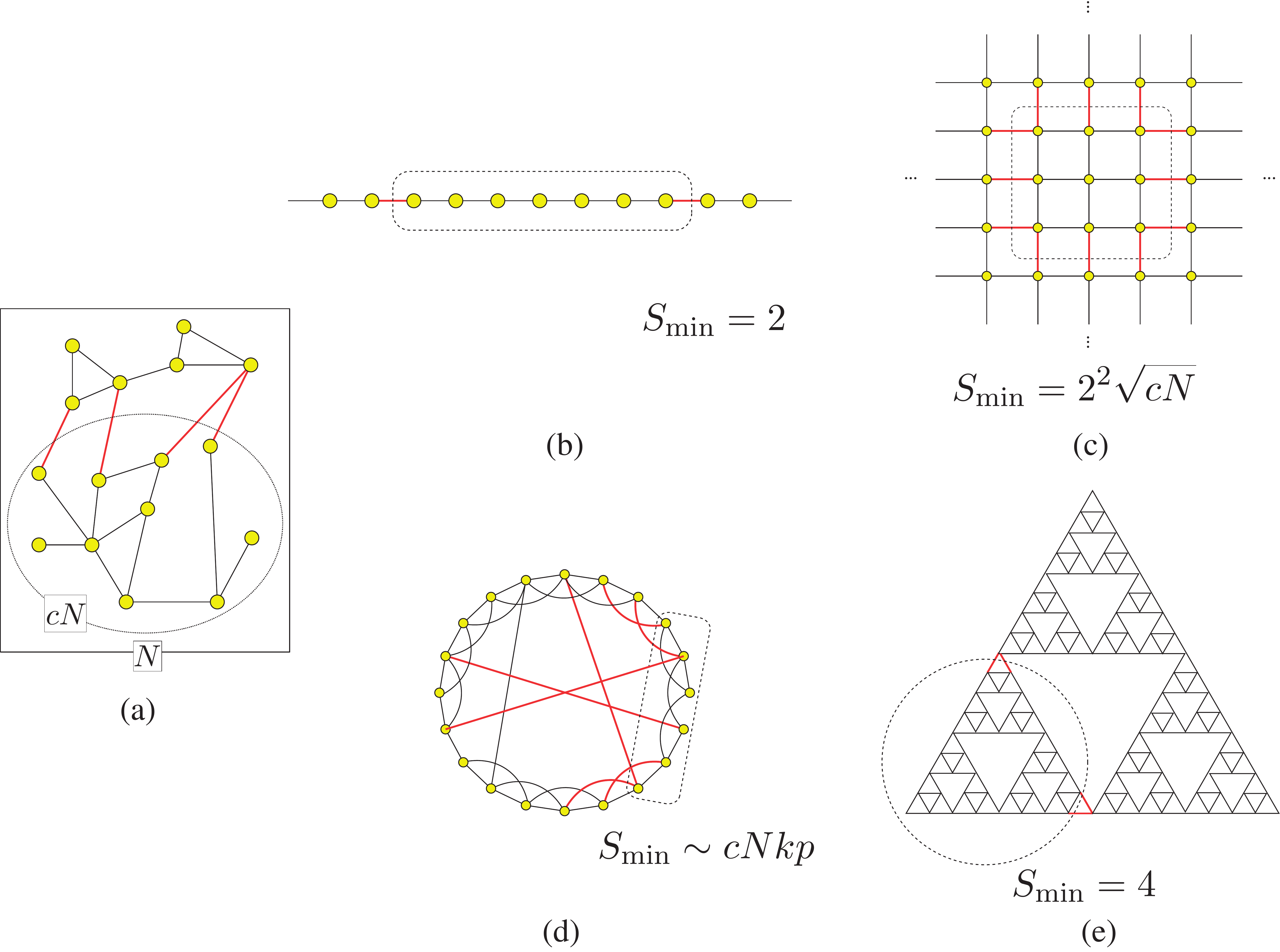}
\end{center}
\caption{(a) Illustration of the definition of surface are. Given $cN$ ($0<cN<1$) randomly chosen nodes, the surface of the set $\{cN\}$ is defined as the set of links connecting nodes inside $\{cN\}$ with those outside of $\{cN\}$. The links belonging to the surface of $\{cN\}$ are depicted in red. Minimum surface area $S_{\min}$ of (b) one-dimensional lattice ($S_{\min} = 2$), (c) two-dimensional lattice ($S_{\min}=2^2 \sqrt{cN}$), (d) SW network ($S_{\min}\sim cNkp$) and (e) Sierpinski gasket ($S_{\min} = 4$). }
\label{fig:mori2010}
\end{figure} 

\textcolor{black}{
As we shall see throughout the text, great part of the
investigations of the Kuramoto model in networks has been
concentrated in uncovering local or global synchronization 
properties as a function, e.g., of the coupling 
strength or other parameters of interest. There are, however, 
some very interesting works that address these and other issues through different perspectives. It is thus worth discussing briefly some of these approaches.}

\textcolor{black}{
Instead of asking how strong should be the coupling 
strength between the oscillators in order to achieve synchronization, one could formulate the problem in a different way by 
asking what are the necessary conditions that must be satisfied
for a given network to exhibit partial or global 
synchronization. This kind of approach is actually usually
explored in the context of the MSF formalism whereby precise
conditions, which depend on the dynamics under consideration and on the 
network structure, are obtained for the stability of the completely 
synchronized state. On the other hand, conditions such as these are 
rarely addressed when studying Kuramoto oscillators in networks. Seeking 
to fill this gap, 
Mori and Odagaki~\cite{mori2010NecessaryConditionFreqSync,mori2009SynchronizationSW} 
carried out an interesting analysis through which necessary conditions 
for frequency synchronization were 
derived. The analysis in~\cite{mori2010NecessaryConditionFreqSync} is based on the concept of \textit{surface 
area} of sets of nodes in a network. More specifically, given a set of 
$cN$ connected nodes, where $0<c<1$, the surface area 
$S(\{cN\})$ is defined as the number of links that connect nodes
belonging to the set $\{cN\}$ with the rest of the nodes outside 
$\{cN\}$.  This concept is illustrated in Fig.~\ref{fig:mori2010}(a). Provided that the 
frequency distribution $g(\omega)$ has a finite variance, Mori derived 
that a necessary condition for a network to sustain complete 
synchronization is given by~\cite{mori2010NecessaryConditionFreqSync}  }
\textcolor{black}{\begin{equation}
\lim_{N\rightarrow \infty } \frac{S(\{cN\})}{\sqrt{N}}>0\textrm{ for any set $\{cN\}$}.
\label{eq:mori_condition}
 \end{equation}}
\textcolor{black}{Therefore, according to the above condition, if the minimum possible surface area $S_{\min}$ grows more rapidly with the system size than $\sqrt{N}$ then the fully synchronized state is achievable. Figure~\ref{fig:mori2010} shows examples
of applications of condition (\ref{eq:mori_condition}). In the one-dimensional lattice (Fig.~\ref{fig:mori2010}(b)), $S_{\min} = 2$ for any set $\{cN\}$ meaning that complete synchronization is not supported in this topology. On the other hand, the two-dimensional lattice (Fig.~\ref{fig:mori2010}(c)) reaches the fully synchronized state, since $S_{\min} = 2^2 \sqrt{N}$. For the SW network (Fig.~\ref{fig:mori2010}(d)), $S_{\min}$ can be approximate as the balance between ingoing and outgoing rewired links related to a set of $cN$ connected nodes, i.e. $S(\{cN\})=cNkp(1-c) + (1-c)Nkpc = 2c(1-c)Nkp$. For small $c$, $S_{\min} \sim cNkp$, which satisfies condition (\ref{eq:mori_condition}), implying that complete synchronization is attainable in SW networks. Furthermore, a similar condition as in Eq.~\ref{eq:mori_condition} can be derived for partial synchronization. Interestingly, in this case one can show that
the Sierpinski gasket network (Fig.~\ref{fig:mori2010}(e)) does not support neither complete nor partial synchronization~\cite{mori2010NecessaryConditionFreqSync}.}

\textcolor{black}{The necessary conditions obtained in~\cite{mori2010NecessaryConditionFreqSync} consist in an alternative and interesting approach to assess how particular topological structures can affect network synchronization. However, it remains to be shown whether such conditions are necessary and sufficient to assure that complete or partial synchronization are attainable. Furthermore, it would be interesting to relate these findings with the recently observed phenomenon of \textit{erosion of synchronization}, which consists in the loss of perfect synchronization due to coupling frustration~\cite{skardal2015ErosionOfSyncInNet,skardal2015ErosionofSync2}.}     

%

\textcolor{black}{
Finally, we remark that necessary conditions for the 
stability of the phase-locked state were recently derived in terms 
of the Coates graph of the Jacobian matrix~\cite{do2012GraphicalNotation,epperlein2013meso}.} 

%


\section{First-order Kuramoto model on different types of networks}
\label{sec:different_topologies}

Great part of the works developed on synchronization of Kuramoto 
oscillators in the past few years has aimed at understanding of how the heterogeneity in the connectivity pattern impacts on the overall network dynamics with the hope that it would bring insights into the dynamics of real systems as well. 
Early studies focused mainly on the influence of random connections, inclusion of shortcuts, and presence of highly connected nodes (hubs), 
which are properties of traditional random network models. 
However, these topological properties do not reflect main 
structures observed in real-world networks. It is important to emphasize that most of analytical 
approaches are based on MFAs that are only valid
for uncorrelated networks in the limit of large populations of oscillators
and sufficiently high average degree. Obviously this imposes a constraint 
to the thorough comprehension of synchronization of real networks, 
since they are finite and often exhibit sparsity, degree-degree correlations, 
presence of loops, community structure, and other properties that make the 
mean-field calculations no longer valid~\cite{melnik_unreasonable_2011,gleeson2012accuracy,newman2010networks}. While there is still an ongoing effort to generalize MFAs for more sophisticated topologies~\cite{restrepo2014mean}, many numerical studies have extensively investigated synchronization of Kuramoto oscillators in networks with properties observed in real structures. In this section we will discuss these results. Noteworthy, the analysis
of the Kuramoto model in modular networks is often closely related to the 
development of methods of community detection. Here we also discuss some 
of the main approaches on this regard.

\subsection{Networks with non-vanishing transitivity}
\label{sec:different_topologies_clustering}

One of the simplest topological properties of real-world networks that traditional random models fail to reproduce in the limit of large networks is \textit{transitivity} (or \textit{clustering})~\cite{newman2010networks}, which indicates the probability that two neighbours of a common node are also connected with each other, forming a triangle (or a cycle of order three)~\cite{newman2001random,newman2010networks}. The occurrence of triangles in the network topology can be quantified either locally or globally. The latter    
is expressed in terms of the transitivity $\mathcal{T}$ (\textit{or global clustering coefficient}) defined as~\cite{newman2010networks,costa2007characterization} (see also Appendix) 
\begin{equation}
\mathcal{T} = \frac{3 \times \mbox{(number of triangles in the network)}}{\mbox{(number of connected triples)}} = \frac{3N_{\triangle}}{N_3}, 
\label{eq:transitivity_T}
\end{equation}
where 
\begin{equation}
N_\triangle = \frac{1}{3} \sum_{i,j,k} A_{ij}A_{jk}A_{ki} \mbox{ and }N_3 = \sum_{i,j,k} A_{ki}A_{kj}.
\label{eq:N_triangle_and_N3}
\end{equation}
Similarly, the \textit{local clustering coefficient} $cc_i$ of node $i$ is expressed as~\cite{newman2010networks,costa2007characterization}
\begin{equation}
cc_i = \frac{1}{k_i(k_i-1)} \sum_{j,k=1}^{N} A_{ij}A_{jk}A_{ki}.
\label{eq:clustering_coefficient_ci}
\end{equation}
Alternatively, the global clustering of a network can be calculated 
by the respective average over all nodes $\left\langle cc_i \right\rangle = N^{-1}\sum_{i=1}^N cc_i$.  For $N\rightarrow \infty$, it can be shown that networks constructed via the configuration model~\cite{newman2001random,molloy1995critical} have locally a tree-like
structure, i.e., networks in which $\mathcal{T}\rightarrow 0$~\cite{newman2010networks}. On the other hand, real-world networks have strongly clustered structures~\cite{watts1998collective,newman2010networks,serrano2006clusteringI}. 

\begin{figure}[!t]
\begin{center}
\includegraphics[width=0.5\linewidth]{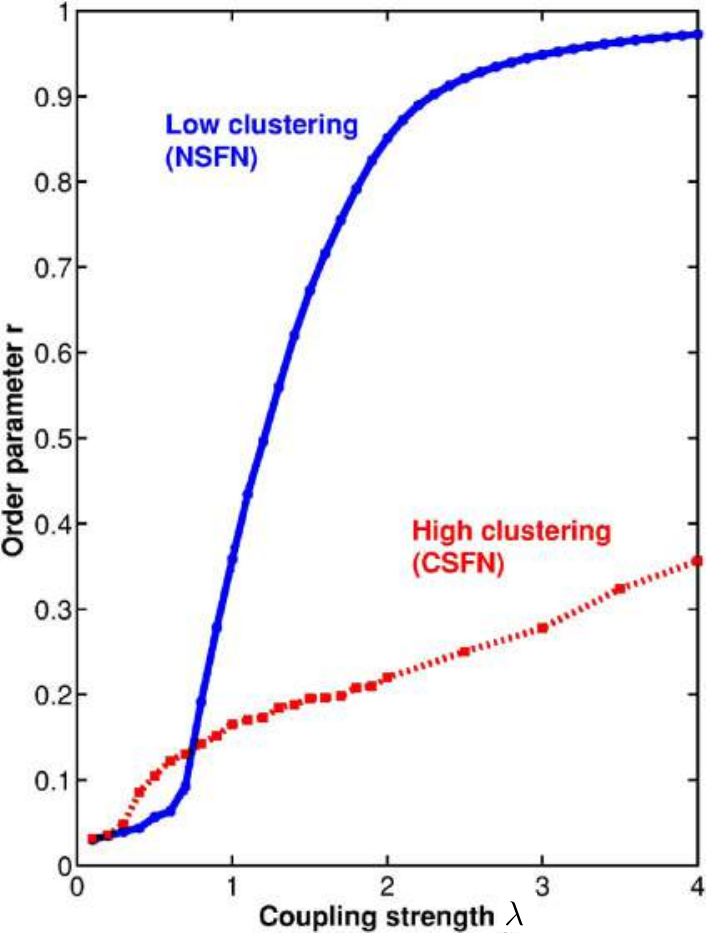}
\end{center}
\caption{ Order parameter $r$ as a function of the coupling strength $\lambda$ for networks with low (circles) and high (squares) clustering. Reprinted with permission from~\cite{mcgraw2007analysis}. Copyright 2007 by the American Physical Society.}
\label{fig:mcgraw2007fig1}
\end{figure}

McGraw and Menzinger~\cite{mcgraw2005clustering,mcgraw2007analysis,mcgraw2008laplacian} investigated synchronization of Kuramoto oscillators in networks with non-vanishing 
clustering 
coefficients. 
In order to precisely compare the coherence of
clustered networks with that of non-clustered networks, the authors adopted the stochastic 
rewiring algorithm proposed by Kim~\cite{kim2004performance}. It consists of
randomly selecting
two edges and rewiring the connection of the associated nodes, accepting 
the new configuration in case the number of triangles in the network is 
increased. This procedure is then successively repeated until the desired 
transitivity is reached. The key feature of this process is that the 
degree sequence and consequently the degree distribution of the original 
network remain unchanged~\cite{kim2004performance}. By applying this 
methodology to compare clustered with non-clustered 
networks, it was found that the coherence between 
oscillators is generally 
suppressed if the number of triangles is increased, 
regardless of whether the 
network topology is  ER random or SF~
\cite{mcgraw2005clustering,mcgraw2007analysis,mcgraw2008laplacian}. 
Yet, clustered SF topologies exhibit an interesting 
behavior that is absent in ER networks. Namely, in 
clustered SF networks, the synchronization at low values 
of the coupling strength
is enhanced when compared with unclustered networks with 
the same degree distribution~
\cite{mcgraw2005clustering,mcgraw2007analysis,mcgraw2008laplacian} 
(Fig.~\ref{fig:mcgraw2007fig1}). The same effect of 
clustering on the onset of synchronization in SF networks 
was 
reported in~\cite{gomez2007synchronizationlocal}. 
Evidence is shown that this particular behavior of the order 
parameter as a function of the coupling strength in 
clustered networks may be related to the following effect: 
in the process of stochastically rewiring the connections, 
not only the clustering of the networks is being changed, 
but also other properties, including the average shortest 
path length and the network modularity. In particular, 
changes in the local connections in order to increase the 
number of triangles contribute to increase the 
topological distance between the nodes and, in some cases, 
yielding different communities. Thus, for weak couplings, 
partial synchronization is favored by these newly formed local 
connections, while higher couplings are required to 
overcome the long distances created so that $r \sim 1$ is 
reached~\cite{mcgraw2005clustering,mcgraw2007analysis,mcgraw2008laplacian,gomez2007synchronizationlocal}.  
It is also important to remark that the emergence of non vanishing synchronization for small coupling strengths might be due to the existence of relatively constant local mean-fields produced by nodes evolving around periodic stable orbits, a phenomenon called \textit{collective almost synchronization}~\cite{baptista2012CollectiveAlmostSync}.  

These findings do not only show how the network dynamics is influenced by properties not encountered in traditional random models, but also demonstrate how difficult it is to disentangle the impact of different topological properties. More specifically, as previously mentioned, although the degree distribution of the networks is preserved, 
other network properties are modified along with the increasing of cycles of 
order three. In particular, the algorithm in~\cite{kim2004performance} is 
known to strongly change the network assortativity and, in some cases, to 
induce the emergence of communities~\cite{mcgraw2005clustering,mcgraw2007analysis,mcgraw2008laplacian,xulvi2004reshuffling,arruda2013InfluenceNetworkProperties}. Therefore, the influence of 
triangles on network synchronization is hard to be distinguished from these 
side effects generated by stochastic rewiring algorithms. Another limitation of using stochastic rewiring models to generate clustered networks resides 
in their analytic intractability, limiting the approaches to numerical calculations.

\begin{figure}[!t]
\begin{center}
\includegraphics[width=1.0\linewidth]{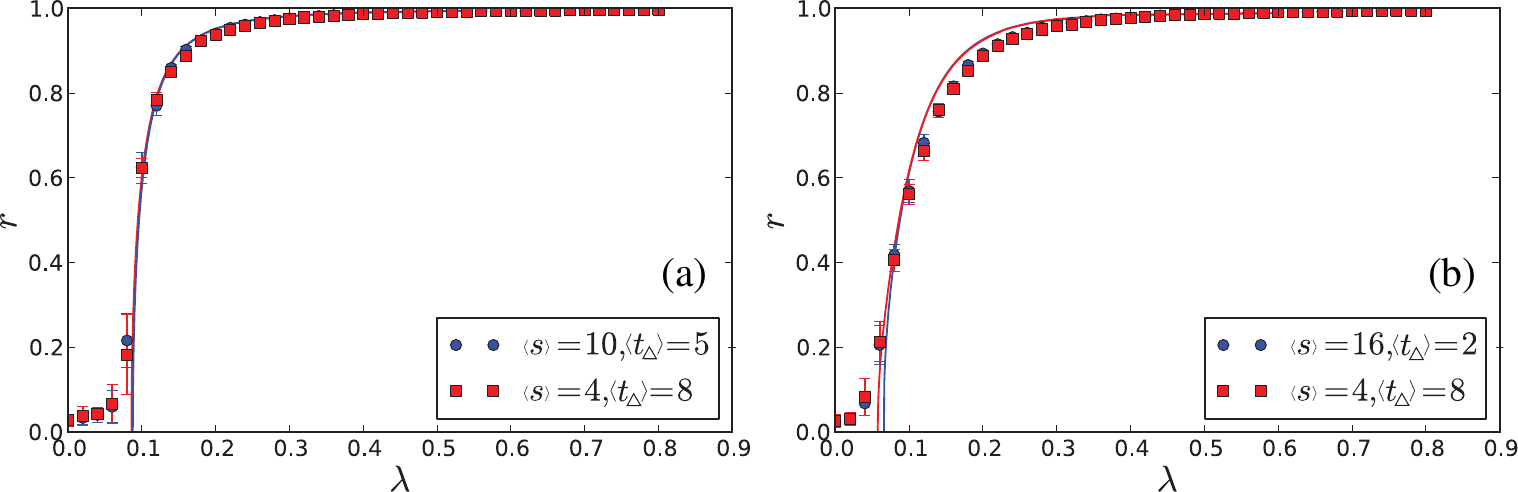}
\end{center}
\caption{ Synchronization diagram for networks constructed with the 
configuration model for clustered networks considering (a) $P(s,t_
\triangle)$ as a double Poisson degree distribution (Eq.~\ref{eq:peron2013synchronization_doublePoisson}) and (b) double 
SF degree distribution $P(s,t_\triangle) \propto 
s^{\gamma_s}t_{\triangle}^{-\gamma_{t}}$, where $\gamma_s = 
\gamma_{t} = 3$. Lines are obtained by numerically solving 
Eq.~\ref{eq:peron2013synchronization_implicit_equation} for different 
values of $\lambda$ and dots are obtained numerically evolving Eqs.~\ref{eq:KMNETWORKS_CONSTANT_COUPLING}. Each point is an average over 10 different 
realizations. Other parameters: $N = 10^3$ and $\left\langle k \right\rangle =20$. Adapted from~\cite{peron2013synchronization}. Copyrighted by the American Physical Society.}
\label{fig:peron2013synchronization_fig1}
\end{figure}

In order to overcome this difficulty and untangle the effects of triangles from other topological properties, Peron et al.~\cite{peron2013synchronization} analytically and numerically studied synchronization of Kuramoto oscillators in a class 
of random graph models that yields clustered networks, while keeping assortativity close to zero. Specifically, they considered the model proposed independently by Newman~\cite{newman2009random} and Miller~\cite{miller2009percolation}. It can be seen as a generalization of the standard configuration model for clustered random networks. Specifically, instead of setting the degree distribution $P(k)$ by drawing a single degree sequence $\{k_i\}$, the model proposed in~\cite{newman2009random,miller2009percolation} sets two different degree sequences. The edges that do not participate in triangles, called single edges, are specified by the sequence $\{s_i\} = \{s_1,s_2,...,s_N\}$, where $s_i$ is the number of single edges attached to node $i$. Similarly, the sequence of triangles $\{t_{\triangle i}\} = \{t_{\triangle 1}, t_{\triangle 2},...,t_{\triangle N}\}$ will dictate the number of triangles associated to each node in the network~\cite{newman2009random,miller2009percolation}. The two sequences define the joint degree sequence $\{s_i,t_{\triangle i}\}$ from which it is convenient to define the  joint degree distribution $P(s,t_\triangle)$. The standard degree of node $i$ is obtained by $k_i = s_i + 2 t_{\triangle i}$ and the relation between the degree distribution $P(k)$ and $P(s,t_{\triangle})$ is given by
\begin{equation}
P(k) = \sum_{s,t_{\triangle}=0}^\infty P(s,t_\triangle) \delta_{k,s+2t_\triangle},
\label{eq:P_k_Pst_newman_clustered_random_model}
\end{equation}
Furthermore, it is useful to define the probability density  $\rho(\theta,t|\omega,s,t_{\triangle})$ of nodes with phase $\theta$ at time $t$ for a given frequency $\omega$ with $s$ single edges and $t_\triangle$ triangles. With these quantities, the equations of motion in the continuum limit are written as~\cite{peron2013synchronization} 
\begin{eqnarray}\nonumber
\dot{\theta} & = & \omega+\frac{\lambda}{\left\langle k \right\rangle}\int ds'\int dt'_{\triangle}\int d\omega'\int d\theta'g(\omega')P(s',t'_{\triangle})(s'+2t'_{\triangle})\\
 &  & \times\rho(\theta',t|\omega',s',t'_{\triangle})\sin(\theta'-\theta), 
\label{eq:peron2013synchronization_equatios_of_motion}
\end{eqnarray}
where $g(\omega)$ is the frequency distribution. It can be shown that for Gaussian frequency distributions $g(\omega) = (\sqrt{2\pi})^{-1} e^{-\omega^2/2}$, the following implicit equation for the order parameter $r$ is obtained~\cite{peron2013synchronization}: 
\begin{eqnarray}\nonumber
\lambda & = & \sqrt{\frac{8}{\pi}}\left\langle k\right\rangle \left\{ \int ds\int dt_{\triangle}(s+2t_{\triangle})^{2}P(s,t_{\triangle})e^{-\lambda^{2}(s+2t_{\triangle})^{2}r^{2}/4}\right.\\
 &  & \times\left.\left[I_{0}\left(\frac{\lambda^{2}(s+2t_{\triangle})^{2}r^{2}}{4}\right)+I_{1}\left(\frac{\lambda^{2}(s+2t_{\triangle})^{2}r^{2}}{4}\right)\right]\right\} ^{-1},
\label{eq:peron2013synchronization_implicit_equation}
\end{eqnarray}
where $I_0$ and $I_1$ are the modified Bessel functions of first kind. 
Solving the equation above for different values of $\lambda$ one can 
then uncover the dependence $r=r(\lambda)$. Moreover, by fixing the 
total average degree $\left\langle k \right\rangle = \left\langle s 
\right\rangle  +\left\langle 2t_{\triangle} \right\rangle $ and 
varying the average number of triangles $\left\langle t_\triangle 
\right\rangle $, it is possible to analytically quantify the influence 
of triangles on the network synchronization. Figure~
\ref{fig:peron2013synchronization_fig1}(a) shows the synchronization 
diagram with a double Poisson degree distribution, i.e., 
\begin{equation}
P(s,t_{\triangle})=e^{-\left\langle s\right\rangle }\frac{\left\langle 
s\right\rangle ^{s}}{s!}e^{-\left\langle t_{\triangle}\right\rangle }
\frac{\left\langle t_{\triangle}\right\rangle ^{t_{\triangle}}}
{t_{\triangle}!}.
\label{eq:peron2013synchronization_doublePoisson}
\end{equation}
Interestingly, the critical coupling does 
not suffer significant changes as $\left\langle t_\triangle \right
\rangle$ is varied (Fig.~\ref{fig:peron2013synchronization_fig1}). This also holds for networks with a double SF distribution $P(s,t_\triangle) \propto s^{-\gamma_s} t_
\triangle^{-\gamma_t}$ (Fig.
\ref{fig:peron2013synchronization_fig1}(b)). These results suggest that 
the presence of triangles in the topology poorly affects the network 
dynamics, since the dependence on $r$ can be described by MFAs developed for local tree-like networks~\cite{peron2013synchronization}. It is noteworthy that similar findings were reported on the performance of other dynamical processes in clustered networks, such as bond percolation, $k$-core size percolations, and epidemic spreading~\cite{melnik_unreasonable_2011,gleeson2012accuracy}. As demonstrated in~\cite{melnik_unreasonable_2011,gleeson2012accuracy}, mean-field theories for local tree-like networks yield remarkably accurate results even for networks with high values of clustering coefficient if the average shortest path is sufficiently small. 

Although the results in~\cite{peron2013synchronization} contribute with more evidences that  cycles of order three do not play an important role in network dynamics, it is important to specify the limitations of the configuration model with non-vanishing clustering to evaluate the contributions of triangles on the dynamics. The model in~\cite{newman2009random,miller2009percolation} is limited to generate networks in the so-called low-clustering regime in which the network transitivity has an upper bound  $\mathcal{T}_{\max}= 1/( \left\langle k \right\rangle - 1)$~\cite{serrano2006clusteringI,serrano2006clusteringII}. The reason for this bound resides in the fact that the model does not allow the creation of overlapping triangles. In other words, a given edge is restricted to participate in a single triangle;  limiting the clustering coefficient of a node with degree $k$ to $cc(k) \leq 1/(k-1)$. Moreover, the configuration model for clustered networks is not completely independent from effects of degree-degree correlations. It is possible to show that the assortativity $\mathcal{A}$ (see Appendix~\ref{sec:appendix} for definition) of the model in~\cite{newman2009random,miller2009percolation} as a function of $\mathcal{T}$ is given by~\cite{huang2013robustness}
\begin{equation}
\mathcal{A} = \frac{\mathcal{T} - \mathcal{T}^2 - \left\langle k \right\rangle \mathcal{T}^2 }{1 - \mathcal{T} + \left\langle k \right\rangle \mathcal{T} - 2\left\langle k \right\rangle \mathcal{T}^2}.
\label{eq:}
\end{equation}
However, in contrast to networks generated with 
stochastic 
rewiring algorithms, $\mathcal{A}$ can be attenuated 
either by increasing $\left\langle k \right\rangle$ or by 
setting
the average number of triangles $\left\langle 
t_{\triangle} \right
\rangle$ in order to obtain transitivity values close to $
\mathcal{T}_{\max}$, since $\mathcal{A}(\mathcal{T}_{\max}) = 0$. Unfortunately, 
the strategy of increasing $\left\langle k \right\rangle$ comes with the price of 
decreasing $\mathcal{T}_{\max}$. Nevertheless, 
even though $\mathcal{T}$ achieved for high $\left\langle k \right\rangle$ is not as 
high as the ones achieved by stochastic rewiring algorithms~
\cite{kim2004performance,holme2002growing,klemm2002highly,serrano2005tuning,bansal2009exploring}, 
it is possible to obtain a significantly higher clustering than those 
obtained in random network models~\cite{newman2009random,miller2009percolation}.

There are, however, other network models~\cite{serrano2006clusteringI,serrano2006clusteringII,newman2009random,shi2007networks,gleeson2009bond,karrer2010random,zlatic2012networks} that go beyond the low-clustering regime. For instance, an interesting generalization of the 
clustered random network
\cite{newman2009random,miller2009percolation} was introduced in~\cite{karrer2010random}, where networks can be 
constructed not only by single-edges and triangles, but also with 
arbitrary distributions of different kinds of subgraphs. In principle, 
one could mimic the subgraph structure of real-world networks using the 
model in~\cite{karrer2010random}. However, the implementation and 
analytical tractability of the model greatly increase as the 
connectivity pattern of the subgraphs becomes more complex. Another 
interesting model that can be suitably used to evaluate the dynamics of 
networks with similar topology as real structures was 
proposed in~\cite{zlatic2012networks}. Instead of 
focusing on how many triangles are attached to a given node, the model 
in \cite{serrano2006clusteringI,serrano2006clusteringII,zlatic2012networks} is based on the concept of edge multiplicity, 
which is the number of triangles that a given edge participates. More 
specifically, each node is described by a $(N-1)$-dimensional vector $
\boldsymbol{k}_i = (k_i^{(0)},k_i^{(1)},...,k_i^{(M)}   )$, where 
$k^{(l)}$ is the number of edges with multiplicity $l$ attached to node 
$i$ and $M = N - 2$ is the maximum possible edge multiplicity. The total 
degree is then obtained by summing the contribution of all multiplicities, i.e., $k_i = \sum_{m=0}^M k_i^{(m)}$. The great 
advantage of the model based on edge multiplicities over other random 
models is that, while the evaluation of subgraph distributions in real 
networks is  a potentially expensive task depending on the number of 
nodes, the distribution of edge multiplicities is easily calculated from 
real data, which can then be used as an input in the random model~\cite{zlatic2012networks}. The 
analysis of the Kuramoto model and the development of MFAs in these and other random network models~\cite{allard2012BondPercolationOnAClassOfCorrelated,allard2009HeterogeneousBondPercolationMultiple,berchenko2009EmergenceAndSizeOfTheGiant,ghoshal2009RandomHypergraphsAndApplications,dufresne2013PercolationOnRandomNetworksKCore,newman2003PropertiesOfHighlyClusteredNetworks,serrano2006PercolationAndEpidemicThresholdsClustered,vazquez2006SpreadingDynamicsOnHeterogeneousPops,vazquez2003ResilienceDamageGraphsDegreeCorrelations,sendina2015AssortativityAndLeadership} for clustered networks are 
promising directions for future research.

\subsection{Assortative networks}
\label{subsec:assortative}

As discussed in Sec. \ref{sec:different_topologies_clustering}, the presence of high 
$\mathcal{T}$ is inherently related with the emergence of non-vanishing 
$\mathcal{A}$. In fact, it is possible to express 
$\mathcal{A}$ in terms of $\mathcal{T}$ 
for general networks~\cite{estrada2011combinatorial,ramos2013random}.
Thus it comes with no surprise the fact that stochastic algorithms 
designed to yield degree-degree correlations while keeping the degree 
distribution fixed also lead to strongly clustered structures~
\cite{xulvi2004reshuffling}. Therefore, it is expected that populations 
of oscillators coupled through networks constructed via stochastic 
rewiring models that prioritize clustering and assortativity end up a similar dynamical behavior~	
\cite{mcgraw2005clustering,mcgraw2007analysis,mcgraw2008laplacian}.

The influence of degree-degree correlations on network dynamics has been extensively investigated in the context of the 
MSF~
\cite{arenas2008synchronization,motter2005NetworkSyncDiffParadoxHeterogeneity,
chavez2006degreeMixing,
diBernardo2007effectsSyncNetworkOscillators,sorrentino2007synchronizability}. Curiously, the thorough analysis of the 
influence of assortative mixing in the dynamics of Kuramoto oscillators has been only addressed 
very recently and mostly in the context of correlation between natural frequencies and degrees~\cite{brede2010SynchronizationTransitionsCorrelatedOscillators,zhu2013criterion,li2013effect,liu2013effects,li2013reexamination,sendina2015effects} (See Sec.~\ref{sec:explosive_sync}).  

Despite the great interest in the dynamics of networks with
assortative mixing, there is a lack of theoretical approaches
to tackle the problem. However,  an important 
step has been taken towards filling this gap. Very recently, Restrepo and 
Ott~\cite{restrepo2014mean} generalized the mean-field formulation of 
uncorrelated networks in order to account for  directed connections and 
  degree-degree correlations. Considering a directed network 
characterized by the degree distribution $P(\boldsymbol{k})$, where $
\boldsymbol{k} = (k^{\rm{in}},k^{\rm{out}})$, the 
assortativity function $q(\boldsymbol{k}' \rightarrow \boldsymbol{k})$ 
is defined as the probability of having an outgoing edge from a node with 
degree $\boldsymbol{k}'$ reaching a node with degree $
\boldsymbol{k}$. For uncorrelated directed networks, 
$q(\boldsymbol{k}' \rightarrow \boldsymbol{k})$ is reduced to 
$q(\boldsymbol{k}' \rightarrow \boldsymbol{k}) = k^{'\rm{out}}
k^{\rm{in}}/(N\left\langle k \right\rangle)$. In the limit of large
networks ($N\rightarrow \infty$) the population of oscillators 
can be described by the density $\rho(\theta,t,\omega|\boldsymbol{k})$ of oscillators with phase $\theta$ at time $t$ for a given degree $\boldsymbol{k}$ and frequency $\omega$. In this limit, the order parameter
$z = re^{i\psi}$ (Eq.~\ref{eq:restrepo_r_rn})
can be written in terms of the distribution $P(\boldsymbol{k}')$ and $\rho(\theta,t,\omega|\boldsymbol{k})$ as
\begin{equation}
z(\boldsymbol{k},t) = \sum_{\boldsymbol{k}'} P(\boldsymbol{k}')q(\boldsymbol{k}' \rightarrow \boldsymbol{k})\int d\omega' \int \frac{d\theta'}{2\pi} \rho(\theta',t,\omega'|\boldsymbol{k}')e^{i\theta'}
\label{eq:restrepo2014epl_localmeanfield_continuum}, 
\end{equation}
whereby the following continuity equation is derived
\begin{equation}
\frac{\partial\rho(\theta,t,\omega|\boldsymbol{k})}{\partial t}+\frac{\partial}{\partial\theta}\left\{ \left[\omega+\lambda\textrm{Im}(e^{-i\theta}z(\boldsymbol{k},t))\right]\rho(\theta,t,\omega|\boldsymbol{k})\right\} =0.
\label{eq:restrepo2014_continuityequation}
\end{equation}
Seeking to analyze the time-dependent behavior of the model, they~\cite{restrepo2014mean} employed the OA ansatz~\cite{ott2008low,ott2009long}, which allows the following expansion
\begin{equation}
\rho(\theta,t,\omega|\boldsymbol{k})=g(\omega|\boldsymbol{k})\left\{ 1+\left[\sum_{n=1}^{\infty}\left[\hat{\rho}(\omega,\boldsymbol{k},t)\right]^{n}e^{in\theta}+\textrm{(c.c)}\right]\right\}, 
\label{eq:restrepo2014_OAansatz}
\end{equation}
where $g(\omega|\boldsymbol{k})$ is the natural frequency distribution given $\boldsymbol{k}$ and $\textrm{(c.c)}$ denotes the corresponding complex conjugate. Substituting Eq.~\ref{eq:restrepo2014_OAansatz} into Eq.~\ref{eq:restrepo2014_continuityequation}, one finds that the coefficients $\hat{\rho}(\omega,\boldsymbol{k},t)$ satisfy 
\begin{equation}
\frac{\partial \hat{\rho}}{\partial t} -i\omega \hat{\rho} + \frac{\lambda}{2}(z^* \hat{\rho}^2 - z) = 0.
\label{eq:restrepo2014_equationforb}
\end{equation}
Furthermore, substituting Eq.~\ref{eq:restrepo2014_OAansatz} into Eq.~\ref{eq:restrepo2014epl_localmeanfield_continuum}, we can write parameter $z(\boldsymbol{k},t)$ as a function of coefficients $\hat{\rho}(\omega,\boldsymbol{k},t)$, i.e.
\begin{equation}
z(\boldsymbol{k},t)=\sum_{\boldsymbol{k}'}P(\boldsymbol{k}')q(\boldsymbol{k}'\rightarrow\boldsymbol{k})\int g(\omega'|\boldsymbol{k}')\hat{\rho}(\omega',\boldsymbol{k'},t)d\omega'.
\label{eq:restrepo2014_R_with_b}
\end{equation}
Finally, considering Lorentzian frequency distributions as 
\begin{equation}
g(\omega|\boldsymbol{k}) = \frac{1}{\pi}\frac{\Delta(\boldsymbol{k})}{\left[\omega - \omega_0(\boldsymbol{k}) \right]^2 + \Delta^2(\boldsymbol{k})},
\label{eq:restrepo2014_freqdist}
\end{equation}
the whole network dynamics is exactly described
by~\cite{restrepo2014mean}
\begin{eqnarray}\nonumber
\left\{ \frac{\partial}{\partial t}+\left[-i\omega_{0}(\boldsymbol{k})+\Delta(\boldsymbol{k})\right]\right\} \tilde{\rho}(\boldsymbol{k},t)\\
+\frac{\lambda}{2}\sum_{\boldsymbol{k}'}P(\boldsymbol{k}')q(\boldsymbol{k}' & \rightarrow\boldsymbol{k})\left[\tilde{\rho}(\boldsymbol{k}',t)^{*}\tilde{\rho}^{2}(\boldsymbol{k},t)-\tilde{\rho}(\boldsymbol{k}',t)\right]=0,
\label{eq:restrepo2014_hat_b}
\end{eqnarray}
where $\tilde{\rho}(\boldsymbol{k},t)\equiv \hat{\rho}(\omega_{0}
(\boldsymbol{k})+i\Delta(\boldsymbol{k}),\boldsymbol{k},t)
$. Note that the frequency distribution is correlated with 
the local topology of the nodes through the parameters $
\omega_0(\boldsymbol{k})$ and $\Delta(\boldsymbol{k})$ (a 
similar case was considered in~
\cite{sonnenschein2013networks}, see also 
Sec.~\ref{sec:stochastic}). With Eq.~
\ref{eq:restrepo2014_hat_b} the dimension of the system is 
exactly reduced and the complete dynamics is now described 
by the evolution of 
the coefficients $\hat{\rho}(\boldsymbol{k},t)$. Hence, one is left 
with a set of equations with dimension equal to the number of 
different degrees $\boldsymbol{k}$, which can be further reduced 
by approximating the 
summation over $\boldsymbol{k}$~\cite{restrepo2014mean}. By using 
the MFA for assortative networks combined 
with the OA theory, the authors in~\cite{restrepo2014mean} were 
able to uncover new sequences of bifurcations in strongly 
assortative networks. Specifically, besides the transition from 
the incoherence to a steady state, bifurcations between the latter 
and oscillatory regimes were also observed, constituting an effect 
induced by the assortative mixing in the network 
structure~\cite{restrepo2014mean} that was unseen in previous 
numerical works on assortative 
networks~\cite{mcgraw2005clustering,mcgraw2007analysis,mcgraw2008laplacian}. 
The technique developed in~\cite{restrepo2014mean} was recently 
generalized to account general correlations between neighbours' 
frequencies, where it was found that chaos can be induced in 
network dynamics for sufficiently assortative frequency 
assignments~\cite{skardal2015frequency}. Furthermore, 
in~\cite{brede2010SynchronizationTransitionsCorrelatedOscillators} 
it was numerically verified that the relaxation time of ER 
networks is not affected by such frequency-frequency correlations. 
It would be interesting to combine the approaches presented in 
this section with the one in  
Sec.~\ref{subsubsec:relaxation_dynamics_SF} in order to analytically verify the findings in~\cite{brede2010SynchronizationTransitionsCorrelatedOscillators}  
as well as extend the results to SF networks.
    
\subsection{Networks with community structure}
\label{subsec:community}

\textcolor{black}{
The detection and analysis of communities is undoubtedly one
of most active topics in network science~\cite{newman2012communities,fortunato2010community}. The observation 
of such topological patterns in a wide range of real-world networks (e.g.~\cite{girvan2002community,ravasz2002hierarchical,fletcher2013network,zhao2011community,palla2005uncovering}) motivated an overwhelming number of works aimed at developing 
or improving methods to partition networks~\cite{newman2012communities,fortunato2010community}. All these efforts to uncover and characterize the modular structure of networks naturally raised questions about the role played by the presence of communities in dynamical processes~\cite{barrat2008DynamicalProcessesOnComplexNetworks}. Regarding synchronization of phase oscillators, one of the first studies to investigate the collective dynamics of modular networks made up of Kuramoto oscillators is by Oh et al.~\cite{oh2005modular}. By analyzing real and synthetic topologies, it was verified that the organization of the nodes into communities tends to hinder network synchronization. This effect is already expected, since the sparser the connection between nodes in different communities, the harder for these nodes to lock in a common phase~\cite{moreno2004fitness,arenas2006SynchronizationRevealsTopological,guan_transition_2008}. Furthermore, as shown in~\cite{oh2005modular}, the dependence of the order parameter on the coupling strength and its scaling with the system size turn out to be significantly affected by intermodular pattern of connections. More specifically, while the critical coupling for the onset of synchronization in random modular networks can be directly evaluated by a finite-size analysis, the determination of this quantity is not straightforward in networks generated through hierarchical models~\cite{oh2005modular,ravasz2003hierarchical}.}
\textcolor{black}{
The dynamics of Kuramoto oscillators can also reveal in great detail the hierarchical structure of modular networks. In particular, the case of identical oscillators is of special interest. Since there the dynamics has only one attractor, the system will reach the complete synchronous state regardless of the coupling strength, making the time scale necessary to lock all oscillators to be entirely dependent on the community structure of the network~\cite{arenas2008synchronization}. Based on these ideas, Arenas et al. devoted a series of papers~\cite{arenas2006SynchronizationRevealsTopological,arenas2006synchronizationProcessesInComplexNetworks,arenas2007synchronizationAndModularityInComplexNetworks} to characterize modular networks using dynamics. Basically, in networks with well defined modular structure, nodes are expected to lock their phases first with neighbours belonging to the same community and then, subsequently as the dynamics evolves, more and more nodes belonging to different clusters become entrained in the mean-field. Thus, by starting with different
initial conditions, one can unveil the impact of modularity on network synchronization by tracing the emergence of different synchronous components at different time scales. In order to address this task, the average correlation between the phases of pairs of oscillators can be quantified by the local parameter~\cite{arenas2006SynchronizationRevealsTopological}
\begin{equation}
\rho_{ij}(t) = \left\langle \cos\left[\theta_i(t) - \theta_j(t)\right]\right\rangle,
\label{eq:arenas_rho_ij}
\end{equation} 
where $\left\langle \cdot \right\rangle$ here denotes averages over 
different realizations of the initial conditions. Furthermore,
one can define the \textit{dynamic connectivity matrix}
\begin{equation}
{\cal D}_{t}(\varepsilon)_{ij}=\begin{cases}
1 & \mbox{ if }\rho_{ij}(t)>\varepsilon\\
0 & \mbox{ if }\rho_{ij}(t)<\varepsilon,
\end{cases}
\label{eq:arenas_Dt_ij}
\end{equation}
where $\varepsilon$ is a given threshold. The matrix $\boldsymbol{\mathcal{D}}_{t}$ gives the precise information about the emergence of connected synchronous components as the system evolves. More specifically, at a given  time $t$, for sufficiently high $\varepsilon$, $\boldsymbol{{\cal D}}_{t}$ will depict the connections only between the locked oscillators located at the core of the communities, whereas as 
$\varepsilon$ is decreased the hierarchical structure is revealed. Moreover, as shown in~\cite{arenas2006SynchronizationRevealsTopological,arenas2006synchronizationProcessesInComplexNetworks,arenas2007synchronizationAndModularityInComplexNetworks}, besides visual inspection, one can precisely obtain the complete information about the merge of
different synchronous components over time by simply analyzing
the spectrum of $\boldsymbol{{\cal D}}_{t}$. This approach proved to be accurate in the determination of the modular organization of networks with well defined communities and with a homogeneous degree distribution. However, for modular networks with heterogeneous degrees, the hierarchical structure uncovered 
might be not as clear as in the homogeneous degree case due to the different 
time scales
at which hubs lock their phases with the mean-
field}~\cite{arenas2008synchronization,arenas2006SynchronizationRevealsTopological,arenas2006synchronizationProcessesInComplexNetworks,arenas2007synchronizationAndModularityInComplexNetworks,pereira2010HubSynchronization}.  

\textcolor{black}{
The aforementioned results have not only shed light 
on the microscopic mechanism related to local synchronization 
in modular networks, but also highlighted the potential of 
the Kuramoto model to probe network topology through its dynamics. 
In other words, these findings showed that the phases evolution could now be 
used as a signature whereby communities could be detected. This 
kind of approach to uncover modules in networks fits the class
of \textit{dynamical clustering algorithms}~\cite{fortunato2010community} and is conceptually
different from traditional methods of community detection. 
Specifically, the latter are solely based on the information 
provided by the network topology (e.g. betweenness 
centrality~\cite{girvan2002community,newman2004findingAndEvaluatingCommunity}, network spectrum~\cite{newman2006FindingCommunityStructureUsingEigen} and optimization of the modularity function~\cite{fortunato2010community}), whereas the former considered features 
obtained from a given dynamical process taking place in the network, which are then used to 
define the community membership for each node~\cite{fortunato2010community}. A method that exemplifies the application of these concepts is the one by Wang et al.~\cite{wang2009extractingHierarchical}, which consists in 
a modification of the approach by Arenas et al.~\cite{arenas2006SynchronizationRevealsTopological,arenas2006synchronizationProcessesInComplexNetworks,arenas2007synchronizationAndModularityInComplexNetworks} so that 
communities can be automatically assigned. Namely, instead
of using $\boldsymbol{\mathcal{D}}_t(\varepsilon)$
to classify the nodes, Wang et al.~\cite{wang2009extractingHierarchical} define the \textit{sync-affinity
matrix} $\boldsymbol{S}$, whose elements are given by
\begin{equation}
S_{ij}(\varepsilon) = T_{ij}(\varepsilon)/t_{\max},  
\label{eq:wang2009extracing_sync_affinity_matrix}
\end{equation}
where $T_{ij}$ is the time required to synchronize the phases
of nodes $i$ and $j$, and with $t_{\max} = \max\{T_{ij}(\varepsilon)\}$ being the longest time needed for two oscillators to synchronize. After evolving the dynamics and computing the sync-affinity matrix, a hierarchical clustering algorithm is then employed using matrix $\boldsymbol{S}$ and through which the communities are assigned to the nodes~\cite{wang2009extractingHierarchical}. }
  
\textcolor{black}{  
Although the method in~\cite{wang2009extractingHierarchical} considers the Kuramoto model in its methodology, the network partition task is still performed by hierarchical clustering algorithms, 
similar as other methods~\cite{fortunato2010community}. However, it is also possible to uncover communities in networks by agglomerating nodes through adaptive processes using dynamics akin to the Kuramoto model~\cite{boccaletti2007detectingComplexNetworkModularity,oh2008ModularSynchronizationGaugeKuramoto}. For instance, inspired by a modification of the Kuramoto model designed to model opinion dynamics - the Opinion Changing Rate model (OCR) -, Boccaletti et al.~\cite{boccaletti2007detectingComplexNetworkModularity} proposed a method in which nodes within the same community tune their instantaneous frequency to a common value. This effect is obtained by considering that 
each node $i$ ($i=1,...,N$) is characterized by an unbounded real variable $x_i$~\cite{boccaletti2007detectingComplexNetworkModularity} that evolves according to
\begin{equation}
\dot{x}_i=\omega_{i}+\frac{\lambda}{\sum_{j=1}^{N}A_{ij}b_{ij}^{\alpha(t)}}\sum_{j=1}^{N}A_{ij}b_{ij}^{\alpha(t)}\sin(x_{j}-x_{i})c e^{-c|x_{j}-x_{i}|},
\label{eq:boccaletti2007detection_equations_of_motion}
\end{equation}
where $b_{ij}$ is the betweenness centrality of the edge connecting nodes $i$ and $j$, $\alpha(t)$ is a time-dependent exponent and 
$c$ a tuning exponential factor. The method to find the best network partition works as follows: given a network, the coupling strength $\lambda$ is set to an arbitrary constant in way that in the unweighted case ($\alpha = 0$) the network is in the fully synchronous state. Considering random initial conditions for $x_i$, random frequency distribution and $\alpha(t=0)=0$, equations (\ref{eq:boccaletti2007detection_equations_of_motion}) are numerically integrated and, as the dynamics evolves, $\alpha(t)$ is progressively decreased by the amount $\delta \alpha$ according to the rule $\alpha(t_{j+1}) = \alpha(t_j) - \delta \alpha$ for $t_{j+1}>t>t_j$, where $t_{j+1} - t_j = T$, with $T$ being a tunable parameter. Links connecting nodes belonging to different
communities are expected to present higher values of betweenness centrality in comparison with links attached to nodes of one single community. Thus, as $\alpha(t)$ is decreased, the coupling strengths between nodes inside the same community are strengthened forcing them to synchronize their phases; whereas 
couplings between different communities are weakened. Nodes in this process that are identified with the same instantaneous frequency $\dot{x}$ are assigned to the same community. 
The best network partition is then selected to the one corresponding to the value of $\alpha$ that maximizes the modularity function~\cite{boccaletti2007detectingComplexNetworkModularity}. The OCR model for community detection presented 
a better performance of correct classification of nodes in synthetic modular networks than the traditional method in~\cite{newman2004findingAndEvaluatingCommunity} and the modularity optimization algorithm introduced in~\cite{newman2004fastAlgorithmForDetectingCommunityStructure}. Moreover, by including a slight modification in the dynamics, the authors were able to significantly improve the performance of the algorithm. Precisely,  at a given time $t$, the neighbors of node $i$ having phases in the range $[\theta_i - \epsilon,\theta_i + \epsilon]$ are considered to be 
its ``compatible'' neighbors. In the next step, $\omega_i(t+\Delta t)$ is set as the average over the natural frequencies $\omega_i$ of these compatible nodes~\cite{boccaletti2007detectingComplexNetworkModularity} selected in the previous time step. The modified dynamics significantly increased the accuracy of classification compared with others~\cite{boccaletti2007detectingComplexNetworkModularity}.
Although less accurate than methods based on simulated annealing~\cite{danon2005comparing}, the classification method using the OCR model has a significant
speed advantage scaling with $\mathcal{O}(N^2)$, offering a good
trade-off between accuracy and computation time compared with others~\cite{boccaletti2007detectingComplexNetworkModularity}.} 

 
\textcolor{black}{ 
In some cases, the modular structure of real-world networks may not 
be clearly defined in a way that attributing to a node the 
membership of only one community might lead to a loss of 
information when describing the system under investigation. This 
certainly introduces a great limitation to most of community identification methods, since nodes that belong to more than 
one community are impossible to be detected using such approaches~\cite{fortunato2010community}.
Thus, networks with overlapping communities require 
special methods that are able to quantify the special role played 
by nodes lying in the interfaces between two or more communities.
Among the different approaches developed to fill the need for this more accurate description of modules, it is also possible to find
methods that attribute the community membership of each node
based on the network collective dynamics~\cite{fortunato2010community}. This is the case
of the method introduced in~\cite{li2008synchronizationInterfaces}. The approach relies on the
assumption that nodes placed at the interfaces of different 
functional clusters shall present a peculiar dynamical behavior 
that could be detected by a suitable algorithm through which the 
membership of each community is then quantified. This is illustrated
in~\cite{li2008synchronizationInterfaces} by considering the 
special case of a random network made up of two well defined 
communities, denoted by labels $A$ and $B$, where the level of 
overlap of the nodes belonging to the interface is controllable. 
The natural frequencies are assigned according to different 
distributions for each community, namely $g_A(\omega)$ and 
$g_B(\omega)$ with means $\bar{\omega}_A$ and $\bar{\omega}_B$, 
respectively. As shown in~\cite{li2008synchronizationInterfaces}, 
in the long time dynamics, nodes belonging to the communities have 
their instantaneous frequencies $\dot{\theta}_i$ converged to the 
corresponding mean frequencies ($\bar{\omega}_A$ and $\bar{\omega}
_B$). On the other hand, nodes located at the overlapping region 
between communities exhibit an oscillating instantaneous frequency
around the network mean frequency $\bar{\omega} = (\bar{\omega}_A +
\bar{\omega}_B)/2$. Having observed this phenomenon, the authors 
introduced the node index $C_i = \mbox{sgn}[\dot{\theta}_i - 
\bar{\omega}]\min_t\{|\dot{\theta}_i - \bar{\omega}|\}$ in order to 
quantify the participation of nodes in each community. In 
particular, $C_i$ approaches $0$ as node $i$ becomes equally 
connected with communities $A$ and $B$, whereas nodes outside 
the overlapping region are assigned extreme values of the measurement~
\cite{li2008synchronizationInterfaces}.  }

\textcolor{black}{
While interesting from a theoretical point of view and insightful 
regarding the dynamics of overlapping modular networks, the method in~\cite{li2008synchronizationInterfaces} exhibits particular 
limitations not faced by other methods~\cite{fortunato2010community}. For instance, 
although the identification of overlapping nodes is possible, its application is only feasible given a prior 
knowledge about the modular structure so that the natural frequencies can be suitably assigned. Moreover, the identification of the oscillators lying in the interfaces crucially depends on the coupling strength for which moderate 
values must be selected in order to not lock all oscillators in one single frequency, otherwise nodes connecting communities cannot be distinguished. 
The method in~\cite{li2008synchronizationInterfaces} was later extended~\cite{almendral2010dynamicsOfoverlappingStructuresModularNetworks} to networks with an arbitrary number of overlapping communities, yet the generalized algorithm also suffers from the same  abovementioned limitations. Nevertheless, as demonstrated in~\cite{sendina2011UnveilingProteinFunctions}, the methods presented in~\cite{li2008synchronizationInterfaces,almendral2010dynamicsOfoverlappingStructuresModularNetworks} can be jointly employed with other traditional algorithms of community detection in order to gain further insight into the functional role of modules in real networks. }

\textcolor{black}{
The limitation of assigning natural frequencies according to a pre-defined 
modular structure was overcome in a modified approach~\cite{wu2012OverlappingCommunityDetectionNetworkDynamics,wu_clustering_2014}. Specifically,  the following model was considered
\begin{equation}
\dot{\theta}_i = \omega_i + \frac{\lambda_{+}}{N}\sum_{j=1}^N A_{ij} \sin(\theta_j - \theta_i) + \frac{\lambda_{-}}{N}\sum_{j=1}^N (1 - A_{ij})\sin(\theta_j - \theta_i),
\label{eq:wu2012Overlapping_equations_of_motion}
\end{equation}
where $\lambda_+ > 0$, $\lambda_-\leq 0$ and the natural frequencies $\omega_i$ are randomly drawn according to a uniform distribution. With 
the introduction of positive and negative couplings, the phases of connected
nodes will attract each other, whereas unconnected nodes will be repelled. 
Thus, as shown in~\cite{wu2012OverlappingCommunityDetectionNetworkDynamics,wu_clustering_2014}, the long-time behavior of phases $\theta_i(t)$ 
will quantify the membership of each node, and those one lying with 
intermediate values are identified as belonging to interfaces between 
two or more communities. Furthermore, in contrast to~\cite{li2008synchronizationInterfaces,almendral2010dynamicsOfoverlappingStructuresModularNetworks} the accuracy 
of the generalized method no longer depends on the scale of the coupling strength, since the emergence of phase lag between nodes belonging 
to different communities is guaranteed by the presence of negative couplings~\cite{wu2012OverlappingCommunityDetectionNetworkDynamics,wu_clustering_2014}. }

%
\textcolor{black}{
It is worth mentioning that many papers analytically investigated 
the low-dimensional behavior of Kuramoto oscillators coupled in fully 
connected graphs but organized into subpopulations characterized by different 
frequency and couplings 
distributions~\cite{montbrio2004synchronizationOfTwoInteractingPopulations,pikovsky2008partially,skardal2012HierarchicalSynchrony,barreto2008synchronizationNetworksofNetworks,laing2009chimeraStatesInHeterogeneousNetworks,kawamura2010phaseSynchronizationBetweenCollective,martens2010chimerasInANetworkOfThreeOscillatorPop}. 
Extensions of these studies accounting modular heterogeneous networks are 
promising topics for future works.}

\textcolor{black}{Finally, we point out that the
multilayer character~\cite{kivela2014multilayer,boccaletti2014structure} of real-world networks has recently started
to be incorporated in the study of phase oscillators~\cite{um2011SynchronizationInterdependent,louzada2013BreathingSynchronizationInterconnected,nicosia2014SpontaneousSyncDrivenByEnergy,zhang2015explosive}}

\section{General couplings}
\label{sec:general_coupling}

\subsection{Time-delayed couplings}
\label{sec:time_delay}


Networks of Kuramoto oscillators with time-delayed couplings attract lots of attention with applications to neural networks, arrays of lasers and microwave oscillators~\cite{schuster_mutual_1989, yeung_time_1999,earl_synchronization_2003}.   
A time delay naturally appears in the interaction between oscillators due to  the finite speed  propagation of a signal.  
One typical example is the biological evidence of the delayed system in the plasmodium of the slime mold, Physarum polycephalum, where  the time delay is evaluated  related to the propagation velocity between two plasmodimal oscillators coupled  by the protoplasmic streaming  \citep{takamatsu_time_2000}.  
A question then arises of how important time delays are on the dynamics of a system.  Izhikevich~\citep{izhikevich_phase_1998} addressed this question and found that 
the time delay starts to play a significant role when the ratio of its absolute value to the oscillating  period is sufficiently large~\cite{izhikevich_phase_1998}. 
In particular, in weakly connected oscillators, the time delay can be neglected when its value is comparable with one or few periods, otherwise, it gives rise to rich and complicated phenomena even for two-coupled oscillators~\citep{schuster_mutual_1989,izhikevich_phase_1998,takamatsu_time_2000,dhuys_synchronization_2008}. 

In a simple system of two Kuramoto oscillators, Schuster and Wagner~\cite{schuster_mutual_1989} firstly showed that time delays induce a multitude of synchronized solutions. 
One typical biological example is that, in a living two-coupled system with the plasmodium of Physarum polycephalum, time delays induce various oscillation states, including unentrained, antiphase, and in-phase oscillations~\citep{takamatsu_time_2000}. 
Additionally, the stability diagram for the two mutually coupled Kuramoto oscillators was further  investigated under the influence of feedback~\citep{dhuys_synchronization_2008} and noise~\citep{dhuys_stochastic_2014}.

In networked delay-coupled oscillators, research has been focusing on, e.g., uniform  time  delay~\citep{yeung_time_1999,montbrio_time_2006,earl_synchronization_2003,dhuys_synchronization_2008,
nordenfelt_bursting_2013,nordenfelt_frequency_2014,nordenfelt_cyclic_2014,
choi_synchronization_2000,ares_collective_2012,louzada2013BreathingSynchronizationInterconnected}, uniform time delay either with phase shift~\citep{jorg_synchronization_2014} or with noise~\citep{montbrio_time_2006} or both of them~\citep{yeung_time_1999,dhuys_stochastic_2014}, and distance-dependent time delays  \citep{zanette_propagating_2000,ko_effects_2007, lee_large_2009,jeong_time-delayed_2002,ko_wave_2004,eguiluz_structural_2011}. 
The time delay yields rich phenomena including bistability between 
synchronized and incoherent states,  unsteady solutions with 
time-dependent order 
parameters~\cite{yeung_time_1999,dhuys_synchronization_2008}, and 
multistabilities where synchronized states coexist with stable 
incoherent states~\cite{choi_synchronization_2000}, and it may  
result in suppression of the collective 
frequency~\cite{choi_synchronization_2000}. 
Additionally the time-delay related system was extended to high 
dimensional systems~\citep{jeong_time-delayed_2002,ares_collective_2012}. 
Ares et al.~\citep{ares_collective_2012} introduced a continuum 
description of long-wavelength modes of spatially extended 
systems of $d$-dimensional coupled oscillators with time delays and implemented this approach to the segmentation clock of vertebrate embryos. They found 
that the interplay of moving boundaries and time delays also leads 
to non-local effects. 

In this section we shall examine the 
effects of time-delays in the dynamics of Kuramoto oscillators by first employing stability analysis in small-sized graphs and, subsequently, we consider large populations of oscillators in random networks.

%
%
%
%
%
%
%
%
%
%
%
%
%
%
%

\subsubsection{Two limit cycle oscillators}

Let us first consider the simplest system with two mutually delay-coupled Kuramoto oscillators, which was originally studied by Schuster and Wagner~\citep{schuster_mutual_1989}. The governing dynamics follows 
\begin{equation}
\begin{split}
\dot{\theta}_1(t) = \omega_1 + \lambda \sin(\theta_2(t-\tau) - \theta_1(t)),  \\
\dot{\theta}_2(t) = \omega_2 + \lambda \sin(\theta_1(t-\tau) - \theta_2(t)),
\end{split}
\label{Eq:schuster_wagner_two_mutually_coupled_oscillators}
\end{equation}
where the two oscillators interact with each other after a retardation time $\tau$.  

To look for the most general synchronized solutions of this system, providing that the two oscillators are synchronized with  a common frequency $\Omega$ and a constant phase shift $\varphi$, and their phases correspondingly follow 
\begin{equation}
\theta_{1,2}(t) = \Omega t \pm \varphi/2.
\label{Eq_Schuster_wagner_theta_1_2}
\end{equation} 
The stability of this synchronized solution can be evaluated either from the Lyapunov exponents in terms of small perturbations or from the new concept of basin stability~\citep{menck_how_2013} given large perturbations (see Sec. \ref{sec:second_KM}).

Inserting the solution into the governing dynamics~(\ref{Eq:schuster_wagner_two_mutually_coupled_oscillators}) yields 
\begin{equation}
\Delta \omega = 2 \lambda \sin(\varphi) \cos(\Omega \tau), 
\label{Eq:time_delay_two_mutually_Delta_omega}
\end{equation}
and
\begin{equation}
\Omega = \overline{\omega} - \lambda \sin(\Omega \tau) \cos(\varphi), 
\label{Eq:time_delay_two_mutually_Omega}
\end{equation}
where $\Delta \omega$ denotes the natural frequency difference as $\Delta \omega = \omega_1 - \omega_2$ and $\overline{\omega}$ denotes the average natural frequency defined as  $\overline{\omega}=\frac{\omega_1+\omega_2}{2}$.  

The phase difference $\varphi$ as a function of $\Omega$ is calculated directly from  Eq. (\ref{Eq:time_delay_two_mutually_Delta_omega}) as 
\begin{equation}
\begin{split}
& \varphi=\arcsin(\Delta \omega / 2 \lambda \cos(\Omega \tau)) \quad \text{if} \quad \cos(\Omega \tau)>0,\\
& \varphi=\pi- \arcsin(\Delta \omega / 2 \lambda \cos(\Omega \tau))  \quad \text{otherwise}.
\end{split}
\label{Eq:schuster_wager_phase_shift}
\end{equation}
When oscillators are identical, e.g., $\omega_1=\omega_2=1$ without loss of generality,  the phase difference $\varphi$ is either $0$ or $\pi$ (from Eq.~(\ref{Eq:schuster_wager_phase_shift})).  
In the absence of the coupling delay, i.e., $\tau=0$, if $\lambda>0$, the oscillators will approach to a symmetric state with $\theta_1(t) = \theta_2(t)$; if $\lambda<0$, they are coupled repulsively and evolve to an antiphase state with $\theta_1(t) = \theta_2(t)+\pi$~\citep{dhuys_synchronization_2008}.

Inserting the solutions (\ref{Eq:schuster_wager_phase_shift}) into Eq.~(\ref{Eq:time_delay_two_mutually_Omega}) and eliminating $\varphi$ yields   
stable phase locked solutions of the common frequency $\Omega$ depending on the sign of the coupling \citep{schuster_mutual_1989,takamatsu_time_2000,dhuys_synchronization_2008} 
\begin{equation}
\begin{split}
0=\overline{\omega}-\Omega  - \lambda \tan(\Omega \tau) \sqrt{\cos^2(\Omega \tau) - \Delta \omega^2/4\lambda^2 },  \quad \text{if} \quad \lambda>0 \\
0=\overline{\omega}-\Omega  + \lambda \tan(\Omega \tau) \sqrt{\cos^2(\Omega \tau) - \Delta \omega^2/4\lambda^2 }.  \quad \text{if} \quad \lambda<0
\end{split}
\end{equation}

In the presence of the time delay, the system illustrates two solutions: frequency-locked symmetric if $\lambda >0$ and antisymmetric states if $\lambda <0$.  
For identical oscillators, e.g., $\omega_1=\omega_2=1$, in the case of symmetric solutions, the frequency is determined, i.e., $\Omega = 1 - \lambda \sin(\Omega \tau)$, and the in-phase state is stable if and only if $\lambda \cos(\Omega \tau) >0$. In the case of antisymmetric states, the frequency is determined by $\Omega = 1 + \lambda \sin(\Omega \tau)$ and the antiphase solution is stable if and only if $\lambda \cos(\Omega \tau) <0$~\citep{dhuys_synchronization_2008}. 

\begin{figure}
 \begin{center}
\includegraphics[width=0.6\linewidth]{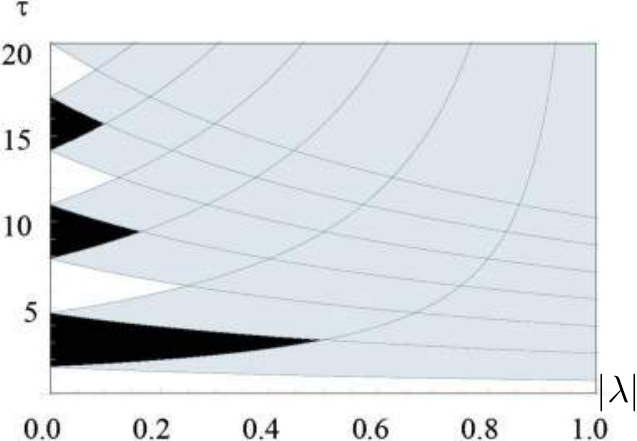} 
 \end{center}
\caption{Stable regions in parameter space of $|\lambda|$ and $\tau$ In the black regions, in-phase solutions exist and are  linearly stable; in the white regions, anti-phase solutions exist and are linearly stable; both states coexist in the gray regions; in the dark gray regions, stable in-phase solutions coexist with oscillatory solutions.  Adapted with permission from~\cite{dhuys_synchronization_2008}. Copyright 2008, AIP Publishing LLC.}
\label{Fig:dhuys_synchronization_2008_fig_3}
\end{figure}

In Fig.~\ref{Fig:dhuys_synchronization_2008_fig_3}, the stability diagram of the two mutually coupled Kuramoto models is shown for in-phase and antiphase solutions  as a function of the coupling strength $|\lambda|$ and the time delay $\tau$. For sufficiently small coupling strength, the system will jump from one state to another with the increases in  $\tau$. While, for strong coupling strength, multiple states coexist and the time delay therein induces multistability~\citep{dhuys_synchronization_2008}.
 
Under the influence of independent noise sources, the dynamics of  the two mutually delay-coupled oscillators  follows~\citep{dhuys_stochastic_2014}
\begin{equation}
\begin{split}
\dot{\theta}_1(t) = \omega_1 + \lambda \sin(\theta_2(t-\tau) - \theta_1(t)) + \xi_1(t),  \\
\dot{\theta}_2(t) = \omega_2 + \lambda \sin(\theta_1(t-\tau) - \theta_2(t)) + \xi_2(t), 
\end{split}
\label{Eq_Dhuys_et_al_stochastic_switching_time_delay_two_oscillators}
\end{equation}
 subject to Gaussian white noise $\xi_i(t)$ with $\left\langle \xi_i(t) \right\rangle=0$ and $\left\langle \xi_i(t)\xi_j(t') \right\rangle = 2D\delta_{ij}\delta(t-t')$ (see Sec.~\ref{sec:stochastic} for effects of noise). 
 
\begin{figure}
\begin{center}
\includegraphics[width=0.8\linewidth]{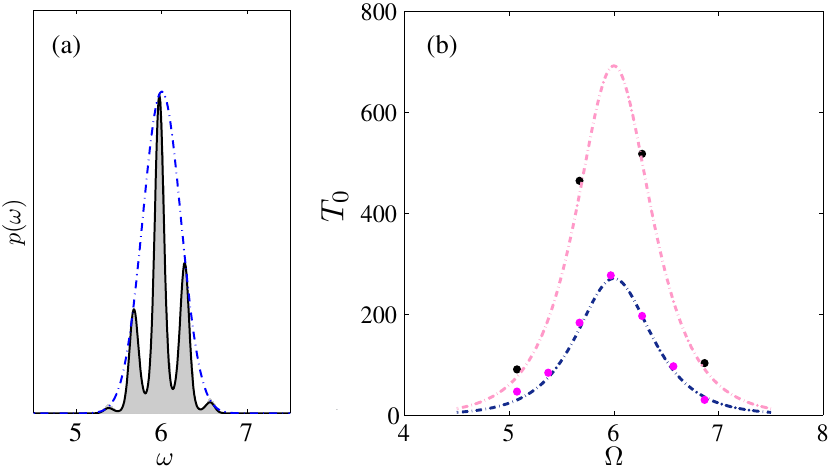}
\end{center}
\caption{(a) The frequency distribution for two identical coupled oscillators. The analytical approximation obtained from Eq.~\ref{Eq:Dhuys_omega_distribution_identical_two_oscillators} is plotted in black and the corresponding Gaussian envelop is shown in blue dashdotted lines.  
(b) Mean residence time of the orbits versus their frequency for one (upper black dots and upper dashed pink curve) and two identical oscillators. Pink dots represent numerical results and the blue curve is obtained from Eq.~\ref{Eq:Dhuys_stochastic_2014_residence_two_oscillators}. Coupling values: (a) $\lambda=2$ and (b) $\lambda=3$. Remaining parameters: $\tau=10$, $\omega_0=6$, and $D=0.5$. Adapted with permission from \citep{dhuys_stochastic_2014}. Copyrighted by the American Physical Society.}
\label{Fig:dhuys_stochastic_2014_fig_3_b_and_fig_4_b}
\end{figure}

Delayed interactions can induce multistability (Fig.~\ref{Fig:dhuys_synchronization_2008_fig_3}) and multiple periodic orbits coexist with different frequencies. Under the influence of noise, the two mutually coupled oscillators can switch between coexistent orbits (jump between frequencies). 
Furthermore, there are two main characteristics to be considered: the distribution of frequencies and the residence times of the orbits. Considering that the system is governed by the two driving terms $x_1(t)$ and $x_2(t)$, defined as $x_{1,2}(t) = [\theta_{1,2}(t) - \theta_{2,1}(t-\tau)]$, and assuming that the oscillators are locked to the same frequency but differ from the noise term, yields 
\begin{equation}
\dot{\theta}_{1,2}(t-\tau) \approx \frac{x_1(t) +x_2(t)}{2\tau} + \xi_{1,2}(t-\tau).
\end{equation}
Therefore, the system can be rewritten using a function of a two-dimensional potential: 
\begin{equation}
\begin{split}
& \dot{x}_{1,2}(t) = - \frac{\partial V(x_1,x_2)}{\partial x_{1,2}} + \tilde{\xi}_{1,2}(t),\\
& V(x_1,x_2)=\frac{1}{4\tau}(2x_0-x_1-x_2)^2 + \frac{\varphi}{2}(x_1-x_2) - \lambda (\cos(x_1)+\cos(x_2)),
\end{split}
\end{equation}
where $x_0=\frac{\omega_1+\omega_2}{2} \tau$ and $\tilde{\xi}_{1,2}(t) = \xi_{1,2}(t) - \xi_{2,1}(t-\tau)$. The potential is $4\pi$-periodic~\citep{dhuys_stochastic_2014}.   
For identical oscillators with $\omega_1=\omega_2=\omega_0$ and therein $\varphi=0$, 
the frequency distribution can be obtained~\citep{dhuys_stochastic_2014}
\begin{equation}
p(\overline{\omega}) \propto e^{-\frac{\tau}{2D}(\overline{\omega}-\omega_0)^2} I_0(\lambda \cos(\overline{\omega} \tau)/D),
\label{Eq:Dhuys_omega_distribution_identical_two_oscillators}
\end{equation}
where $I_0(y)$ is a modified Bessel function defined as $I_0(y)=\sum \frac{y^{2n}}{2^{2n}(n!)^2}$ and $\overline{\omega}=\frac{x_1(t)+x_2(t)}{2\tau}$ is the average frequency of the system. The Bessel function $I_0(y)$ is symmetric and its height is determined by the average frequency $\omega_0$.  The Gaussian envelope of the frequency distribution can also be obtained correspondingly.  Fig.~\ref{Fig:dhuys_stochastic_2014_fig_3_b_and_fig_4_b} shows that the theoretical approximation is in good agreements with numerical results of the complete original system (\ref{Eq_Dhuys_et_al_stochastic_switching_time_delay_two_oscillators}).  

Given the potential model, in the limit of low noise, strong coupling and large delay, the average residence time can be approximated as a function of the common frequency $\Omega$ as follows~\citep{dhuys_stochastic_2014} 
\begin{equation}
T_0(\Omega) \approx \frac{\pi}{2\lambda} \frac{e^{\frac{\lambda}{D} + \frac{•\pi^2}{8\tau D}}}{\cosh[\frac{\pi(\Omega-\omega_0)}{2D}]}. 
\label{Eq:Dhuys_stochastic_2014_residence_two_oscillators}
\end{equation}
The average residence time increases with the coupling strength $\lambda$ and decreases with the increases in the noise strength $D$. For fixed frequency $\Omega$, effects of the delay on the residence time are limited and the dependency vanishes for long delays. The approximation is validated by numerical results in Fig.~\ref{Fig:dhuys_stochastic_2014_fig_3_b_and_fig_4_b} (b).

Another interesting related work considers the influence of  feedback.  
With an additional feedback, the system has a stronger symmetry than the one without feedback (\ref{Eq:schuster_wagner_two_mutually_coupled_oscillators}). The corresponding dynamics with feedback  is governed by~\cite{dhuys_synchronization_2008} 
\begin{equation}
\begin{split}
\dot{\theta}_1(t) = 1 + \frac{\lambda}{2} \sin(\theta_2(t-\tau) - \theta_1(t)) + \sin(\theta_1(t-\tau) - \theta_1(t)),  \\
\dot{\theta}_2(t) = 1 + \frac{\lambda}{2} \sin(\theta_1(t-\tau) - \theta_2(t)) + \sin(\theta_2(t-\tau) - \theta_2(t)).
\end{split}
\label{Eq:two_mutually_coupled_oscillators_feedback}
\end{equation}

The system also admits in general in-phase and anti-phase solutions. 
 A linear stability analysis for the in-phase solutions leads to the same stability criterion as without feedback.  But for the anti-phase solutions, the linear stability analysis becomes complicated. In this case, the only locking frequency is $\Omega = 1$ and the anti-phase states lose and regain stability in a Hopf bifurcation~\cite{dhuys_synchronization_2008}.

\begin{figure}
\begin{center}
\includegraphics[width=0.6\linewidth]{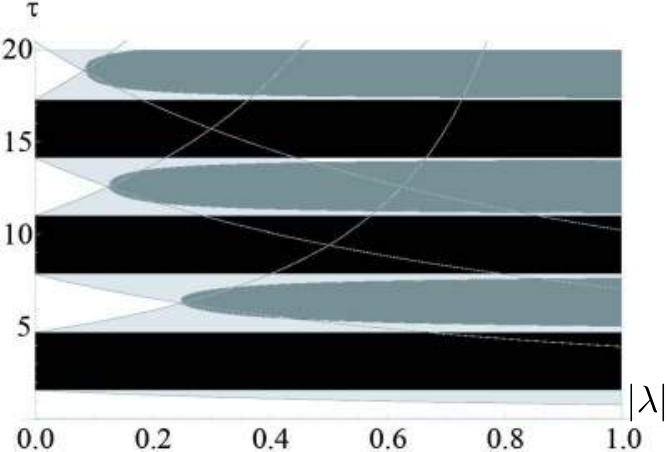}
\end{center}
\caption{Stable regions in parameter space of $|\lambda|$ and $\tau$   with feedback. In the black regions, in-phase solutions exist and are  linearly stable; in the white regions, anti-phase solutions exist and are linearly stable; both states coexist in the gray regions; in the dark gray regions, stable in-phase solutions coexist with oscillatory solutions.  Adapted with permission~\cite{dhuys_synchronization_2008}. Copyright 2008, AIP Publishing LLC.}
\label{Fig:dhuys_synchronization_2008_fig_6}
\end{figure}

The stability diagram is plotted in Fig.~\ref{Fig:dhuys_synchronization_2008_fig_6}. In-phase and antiphase solutions  alternate with each other for increasing $\tau$, like in the case without feedback.  However, the parameter combinations of the stable in-phase solutions enlarge compared to the case without feedback. Stable antiphase solutions exist for small time delays $\tau$ and small coupling strengths $|\lambda|$. For high $|\lambda|$, oscillatory solutions emerge and coexist with in-phase solutions.

\subsubsection{Networks of oscillators with uniform time delay}

Next, let us consider networks of oscillators with the same constant time delay for all interactions~\cite{yeung_time_1999,montbrio_time_2006,choi_synchronization_2000,ares_collective_2012,jorg_synchronization_2014,
nordenfelt_frequency_2014}.  
Recall that time delays induce various solutions, e.g., bistability between synchronized and incoherent states,  unsteady solutions with time-dependent order parameters~\cite{yeung_time_1999,dhuys_synchronization_2008}, and multistabilities where synchronized states coexist with stable incoherent states~\cite{choi_synchronization_2000}.  

In terms of the solutions subjected to small perturbations, Yeung and Strogatz~\cite{yeung_time_1999} provided a detailed analysis on the mean-field Kuramoto model, and derived exact formulas of the stability boundaries of the incoherent and synchronized states for identical oscillators. 
For bimodal natural frequency distributions, Montbri\'o et al.~\cite{montbrio_time_2006}  found that bimodality induces a quasiperiodic pattern thanks to the existence of a new time scale. Moreover, the stability boundaries of the incoherent state have a completely different structure in the parameter space in contrast to the case with unimodal distribution~\cite{yeung_time_1999}.  

To investigate effects of time delays on collective synchronization of coupled oscillators, the self-consistent equations for the order parameter in the stationary state in terms of a Fokker-Planck equation can be derived for a system of globally coupled oscillators with distributed natural frequencies \cite{choi_synchronization_2000}. 
Due to the time delay, the nonzero synchronization frequency breaks the symmetry of the integration interval of the natural frequency distribution and therein the drifting as well as the synchronous oscillators contribute to the phase coherence. 
 The transition between incoherent states and coherent states is in general discontinuous instead of continuous and the synchronization frequency is suppressed by the time delay. In particular, the synchronization frequency of oscillators in a coherent state is found to decrease with the increases in  the time delay.  
  Corresponding curves for the synchronization frequency and phase synchronization with respect to the time delay could also be observed in Rulkov and Hindmarsh-Rose neuron networks with time-delayed chemical synapses~\cite{nordenfelt_bursting_2013}. 
  The interplay between the time delay and time-dependent parameters gives rise to low-dimensional echo effects, where time delays strongly affect the magnitude and shape of mean-field oscillations for slow variations of natural frequencies  \cite{barabash_homogeneous_2014}. 
Additionally, the inclusion of noise to the delayed system induces mode hoppings between coexistent attractors, and the reduction of this system to a nondelayed Langevin equation allows the calculation of analytical solutions of the common frequency distribution and their corresponding residence times~\citep{dhuys_stochastic_2014}.

In networks of identical phase oscillators with delayed coupling, one of the simplest model is given by the following equation~\cite{earl_synchronization_2003,jorg_synchronization_2014}
\begin{equation}
\dot{\theta}_i(t) = \omega_0 + \frac{\lambda}{k_i} \sum^{N}_{j=1}A_{ij} \sin(\theta_j(t-\tau) - \theta_i(t) ),
\label{Eq:TDC_original_systesm}
\end{equation}
where each oscillator $i$ receives signals from $k_i$ others, $\tau$ is the delay and the adjacency matrix $A_{ij}$ encodes the connection topology of a  graph with $A_{ij}=1$ if oscillator $j$ sends signals to oscillator $i$ and $A_{ij}=0$, otherwise. Consider oscillators with uniform in-degree $k$, i.e., $k_i=k, \forall i=1,2, \ldots, N$. 
The synchronized state in which all oscillators move in phase at a fixed common frequency $\Omega$ is given by   
\begin{equation}
\theta_i(t) = \Omega t,
\label{Eq:TDC_synchronized_state_Omega}
\end{equation}
where $\Omega$ is determined by the following algebraic equation \cite{niebur_collective_1991,earl_synchronization_2003}
\begin{equation}
\Omega = \omega_0 + \lambda \sin(-\Omega \tau ),
\label{TDC_synchronized_state_Omega_governed}
\end{equation}
 which is obtained by substituting Eq.~(\ref{Eq:TDC_synchronized_state_Omega}) into the original dynamics~(\ref{Eq:TDC_original_systesm}). 
To determine the local stability of the synchronous state~(\ref{Eq:TDC_synchronized_state_Omega}), consider a linear stability analysis by adding a small perturbation near the synchronous state
\begin{equation}
\theta_i(t) = \Omega t + \varepsilon \phi_i(t) ,
\label{Eq:TDC_theta_perturbation}
\end{equation}
where $0<\varepsilon  \ll  1$.  
Providing $\phi_i = v_i e^{\mu t}$ and substituting Eq.~(\ref{Eq:TDC_theta_perturbation}) into the original  dynamics~(\ref{Eq:TDC_original_systesm}), Earl and Strogatz~\cite{earl_synchronization_2003} obtained the first-order approximation of the solution 
\begin{equation}
\mathbf{A}v = \sigma^{\rm{A}} v,
\end{equation}
with 
\begin{equation}
\sigma^{\rm{A}} = \frac{k e^{\mu \tau} \left[ \mu + \lambda \cos(\Omega \tau) \right]}{\lambda \cos(\Omega \tau)},
\end{equation}
where $\sigma^{\rm{A}}$ is an eigenvalue of $\mathbf{A}$.  
 The synchronized state is linearly stable with $\mathrm{Re}(\sigma^{\rm{A}})<0$ if and only if~\cite{yeung_time_1999} 
\begin{equation}
\lambda \cos(\Omega \tau)>0.
\end{equation}
 This stability criterion depends on the coupling strength $\lambda$ and on the time delay $\tau$, and it is  independent of  the network topology. In particular, the criterion of the synchronous state applies to any network in which each oscillator has a uniform in-degree, independent of all other network properties. 
However, the ability of a network to achieve the full synchronous state also depends on the occurrence and stability of non-synchronous solutions~\citep{dhuys_synchronization_2008}. These states are influenced by the network topology and the delay, and the underlying coupling topology plays an essential role in stability criteria. 

To deepen the understanding of effects of time delays on general networks, it is insightful to firstly analyse different types of motifs, e.g., open chains of Kuramoto oscillators, bidirectionally coupled rings,  and  unidirectionally coupled rings \citep{dhuys_synchronization_2008}.  
Depending on networks possessing different symmetries, the delay behaves differently on the bifurcation diagram: (i) in an open chain topology of Kuramoto oscillators, the mathematical analysis is similar to that of two mutually  coupled oscillators, i.e., oscillators move either in-phase with each other or   anti-phase with their neighbors~\citep{dhuys_synchronization_2008}.

(ii) In a unidirectional ring, where oscillator $i$ is coupled to the oscillator $i+1$, the system allows $N$ different frequency-locked solutions. In particular, all oscillators move either in-phase solutions or exhibit spatiotemporal symmetries. Such solutions are stable if and only if
 $\lambda \cos(\Delta \phi - \Omega \tau)>0$, where $\Delta\phi = \frac{2 j \pi}{N}$ with
 $0 \leq j < N$.  Moreover, the system also has other unlocked solutions, which could be found numerically. 
 The time delay plays a vital role in inducing multiple solutions without enhancing their linear  stability~\citep{dhuys_synchronization_2008}.  Under influence of noise, oscillators can switch between different solutions and spend equally much time in different oscillation patterns~\citep{dhuys_stochastic_2014}. 

(iii) In a bidirectional ring, where each Kuramoto oscillator is coupled to its two neighbors,  the stability of $N$ frequency-locked solutions was conducted following the same analysis in unidirectional rings~\citep{dhuys_synchronization_2008}. In particular, two special solutions were addressed where each oscillator moves either in-phase or antiphase with its second nearest neighbor with an even number of oscillators, and  the in-phase solutions are stable if and only if $\lambda \cos(\Omega \tau)<0$, whereas solutions with $\Delta \phi = \pm \pi/2$ are always unstable for nonzero delay. As a sufficient stability criterion for other solutions, a state $\theta_n(t) = \Omega t + n \Delta \phi$ is stable if $\lambda \cos(\Omega \tau + \Delta \phi) >0$ and $\lambda \cos(\Omega \tau -   \Delta  \phi) >0$.  
 In contrast to that of the unidirectional ring, the delay of bidirectional ring exhibits different influences on the stability of the in-phase and out-of-phase states~\cite{dhuys_synchronization_2008}.

(iv) In this line of research, the relative amount of short cyclic motifs on different classes of networks  with the same average degree as the governing topological factor were also considered in networks of identical time-delayed Kuramoto oscillators~\citep{nordenfelt_cyclic_2014}. The patterns occurring from the cyclic motifs are distinguishable particularly in terms of the momentary frequency dispersion $S^2$, which is defined as follows 
\begin{equation}
S^2 \equiv \frac{1}{NT} \int_0^T \sum\limits_i [\dot{\theta}_i(t) - \Omega(t)]^2 dt,
\end{equation}
where $\Omega(t) = \sum_i \dot{\theta_i}(t) / N$ and $T$ denotes a long time intervals. One interesting conclusion is that bidirectional random networks with short cyclic motifs and unidirectional random networks are two opposite ends of a spectrum of different classes of networks~\citep{nordenfelt_cyclic_2014}. 

Note that there are two different terms considered: the momentary frequency dispersion $S^2$ and the sustained frequency dispersion $\sigma^2$ \citep{nordenfelt_frequency_2014}. The sustained frequency dispersion is defined as follows 
\begin{equation}
\sigma^2 \equiv \langle (\Omega_i - \Omega )^2 \rangle,
\end{equation}
where $\Omega_i = \frac{1}{T} \int^T_0 \dot{\theta}_i(t) dt $ and $\Omega $ denotes the average frequency of all oscillators.  
The two terms exhibit  different information. In particular, the momentary frequency dispersion emerges from the cyclic motifs~\citep{nordenfelt_cyclic_2014}.    
The pronounced frequency dispersion emerges in networks with binomial degree distribution, along with a relatively smooth transition from synchronization to incoherence with the increases in the time delay $\tau$, in contrast to that of networks with identical degree  distributions~\citep{nordenfelt_frequency_2014}. 
Frequency dispersion can also be induced by the time delay in networks with a binomial degree distribution instead of networks with identical degree, i.e. $k_i = k$~\cite{nordenfelt_frequency_2014}.

Additional to the time delay, it has been shown that phase shifts can show similar features as time  delays, e.g., inducing an additional effective phase shift~\citep{yeung_time_1999,panaggio2015ChimeraStates}.  
The interplay between phase shifts and time delays can lead to different oscillator dynamics~\citep{jorg_synchronization_2014}. 
This raises the questions of whether time delays and phase shifts play a similar role in networks of coupled Kuramoto oscillators.  
Consider the following dynamics as a function of a uniform time delay $\tau$  and phase shift $\varphi$~\citep{jorg_synchronization_2014} 
\begin{equation}
\dot{\theta}_i(t) = \omega_0 + \frac{\lambda}{k_i} \sum \limits^{N}_{j=1} A_{ij} \sin(\theta_j(t-\tau) - \theta_i(t) - \varphi). 
\label{Eq:Jorg_Synchronization_2014_orignal_system}
\end{equation}
The normalization of the coupling strength $k_i$ allows in-phase synchronized states with
$\theta_i(t) = \Omega t$. 
 The collective frequency $\Omega$ therein follows the equation 
\begin{equation}
\Omega = \omega_0 - \lambda \sin(\Omega \tau + \varphi).
\end{equation}
To address the effects of the interplay of time delays and phase shifts on synchronization, the collective frequency is kept constant as 
$\Omega = \Omega_0$ by varying $\tau$ and adjusting $\varphi(\tau)$ correspondingly 
    \begin{equation}
  \varphi(\tau) = \Psi - \Omega_0 \tau,
  \label{Eq:Jorg_synchronization_2014_alpha_tau}  
  \end{equation} 
where $\Psi$ is a constant that determines the value of a constant collective frequency $\Omega_0$, and fulfills the following condition
\begin{equation}
\Omega_0 = \omega_0 + \lambda \sin(-\Psi). 
\end{equation}
Although the collective frequency $\Omega$ is  kept constant,  the relaxation rate $\tau_r^{-1}$ (see Sec.~\ref{subsec:relaxation} for definition) does change~\citep{jorg_synchronization_2014}.

\begin{figure}
\begin{center}
\includegraphics[width=0.75\linewidth]{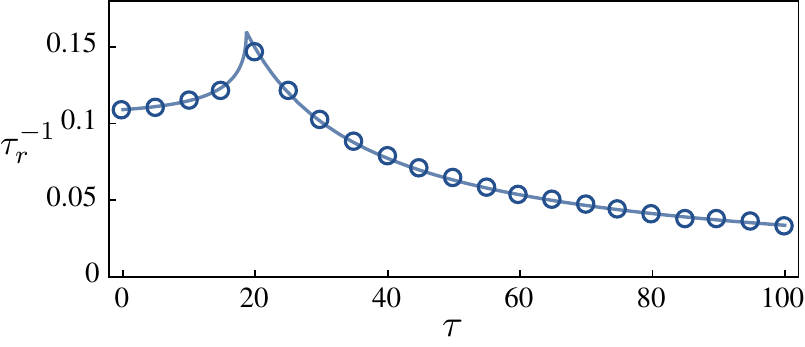}
\end{center}
\caption{ Slowest relaxation rate $\tau_r^{-1}$ as a function of the coupling delay $\tau$ for a globally coupled system. Numerical simulations of Eqs. \ref{Eq:Jorg_Synchronization_2014_orignal_system} and \ref{Eq:Jorg_synchronization_2014_alpha_tau} are in circles. Analytical approximation is obtained from Eq.~\ref{Eq:Jorg_synchronization_2014_lambda_0}. Here, $N=40$, $\omega=1$, $\Psi=5.5$, and $\lambda=0.15$. Adapted with permission from~\cite{jorg_synchronization_2014}.}
\label{Fig:Jorg_synchronization_2014_fig_1}
\end{figure}

To calculate the synchronization rate $\tau_r^{-1}$ numerically, initially oscillators are synchronized, and  system is then subjected to small random perturbations. When the Kuramoto order parameter approaches $r \sim 1$,  relaxation becomes exponential for large times, and $\tau_r^{-1}$ is approximated to fit this exponential~\citep{jorg_synchronization_2014} (see Sec.~\ref{subsec:relaxation}). 
It was observed numerically that the $\tau_r^{-1}$ as a function of the time delay $\tau$ displays a characteristic peak with a maximum in globally coupled networks (shown in  Fig.~\ref{Fig:Jorg_synchronization_2014_fig_1}(a))  and in nearest-neighbor coupled networks~\cite{jorg_synchronization_2014}.

To deepen the insights into the relaxation dynamics as a function of time delays and phase shifts,    a linearized method was performed near the synchronized state, $\theta_i(t) = \Omega t +  \varepsilon \phi_i(t)$ with $\varepsilon \ll 1$. 
The synchronized state is stable if and only if $a>0$, where $a \equiv \lambda \cos(\Psi)$.  
 Substituting this linear approximation into the original system (\ref{Eq:Jorg_Synchronization_2014_orignal_system}) yields the first order in $\phi_i$ as 
\begin{equation}
\dot{\phi}_i(t) = a \sum_j b_{ij} [\phi_j(t-\tau) - \phi_i(t)],
\end{equation}
where $b_{ij} \equiv A_{ij}/k_i$. Here, only stable states are considered. Via introducing the collective relaxation modes $\eta_i(t) = \sum_j (d^{-1}_{ij} \phi_j(t))$, where $d_{ij}$ follows $\sum_{jk} d^{-1}_{ij} b_{jk} d_{kl} = \mu_i \delta_{il}$ and $\mu_i$ are eigenvalues of the normalized matrix $b_{ij}$, and given the ansatz $\eta(t) = e^{-\sigma t}$ yield the characteristic equation 
\begin{equation}
a - \sigma = a \mu e^{\sigma \tau},
\end{equation}
where the index $i$  is dropped for notational simplicity. The solutions follow the Lambert $W$ function, i.e., $W(z) e^{W(z)} = z$ for $z \in \mathbb{C}$. This function has discrete branches $W_n(z)$ and each branch corresponds to one relaxation rate $\tau^{-1}_{r_n} = \text{Re}(\sigma_n)$. The slowest relaxation rate $\tau_r^{-1}$ is shown as \citep{jorg_synchronization_2014}
\begin{equation}
\sigma_0 = a - \frac{1}{\tau} W_0(z_\tau),
\label{Eq:Jorg_synchronization_2014_lambda_0}
\end{equation}
where $z_\tau \equiv \mu a \tau e^{a \tau}$. 

%
%
 
 Consider two cases $\mu >0$ and $\mu <0$, which exhibit the Fourier modes in nearest-neighbor coupled oscillators with long wavelengths and short wavelengths, respectively. When $\mu >0$, collective models are the Fourier modes with long wavelengths. In this case, one can show that $\tau_r^{-1}$ satisfies~\cite{jorg_synchronization_2014}
\begin{equation}
\frac{d \tau_r^{-1}}{d \tau}  = - \frac{\tau_r^{-1}}{ \tau + (\lambda \cos\Psi - \tau_r^{-1} )^{-1}} <0, 
\end{equation}
Note that $\tau_r^{-1}$ decreases monotonically,  and approaches   
$0$ for large $\tau$.  Moreover, given two eigenvalues if $\mu_1 
> \mu_2$, their relative rates obey $\tau_{r1}^{-1}>\tau_{r2}^{-1}$, which is validated numerically. 

For $\mu <0$, as suggested by Fig.~\ref{Fig:Jorg_synchronization_2014_fig_1}, the relaxation rate displays a peak and its  maximum is reached at $\tau^{*}$, where  
\begin{equation}
\tau^{*} \equiv \frac{1}{\lambda \cos\Psi } W_0 \left(\frac{e^{-1}}{|\mu|}\right). 
\end{equation}

As exhibited numerically, the signs of the derivation of the relaxation rate before and after the threshold are opposite. 
For $\tau < \tau^{*}$, one can analytically show that $\frac{d \tau^{-1}_r}{ d \tau} >0$ from Eq.~(\ref{Eq:Jorg_synchronization_2014_lambda_0}).
For $\tau > \tau^{*}$,  $\frac{d \tau^{-1}_r}{ d \tau} <0$.  The system synchronizes slower as the increases in the time delay.  
 For globally coupled networks, eigenvalues $ \mu = (N-1)^{-1}$ and the maximal synchronization rate $\tau^{-1}_r$ is located at $\tau^{*} = W_0(e^{-1}(N-1))/\lambda \cos \Psi$. 
In the limit of large time delays,  $\tau^{-1}_r \approx - \ln (|\mu|) / \tau  $, which indicates that the synchronization rates with eigenvalues $\mu$ and $-\mu$ approach the same asymptotic behavior. 

\subsubsection{Networks of oscillators with distance-dependent time delays}

At last of this subsection, we consider an ensemble of coupled Kuramoto oscillators embedded in a metric space and subjected to (distance-dependent) time-delayed  interactions~\citep{zanette_propagating_2000,jeong_time-delayed_2002,ko_wave_2004,eguiluz_structural_2011}. 
 The time delay $\tau_{ij}$ is the time required to transfer a signal from node $j$ to $i$ depending on, for example, the distance between $j$ and $i$~\cite{zanette_propagating_2000}. When oscillators are sparsely and randomly connected, short delays could induce frequency synchronization~\cite{zanette_propagating_2000}. Even in a general coupling form,  a small fraction of connections with time delays can destabilize synchronous states~\cite{ko_effects_2007}.  

Time delays, proportional to the Euclidean distances between interacting oscillators, yield various spatial patterns including traveling rolls, squarelike and rhombuslike patterns, spirals, and targets~\cite{jeong_time-delayed_2002}. Distance-dependent time delays can also induce near regular waves even through oscillators that are randomly coupled~\cite{ko_wave_2004}. Surprisingly, it is also possible to induce frequency synchronization and  allow propagating structures in one-dimensional globally coupled systems~\citep{zanette_propagating_2000}. Jeong et al.~\citep{jeong_time-delayed_2002} investigated further effects of the distance-dependent time delay in a two-dimensional array of coupled Kuramoto oscillators, where each oscillator is coupled to oscillators within a finite radius. They found that time delays yield various well defined spatio-temporal patterns, including travelling rolls, squarelike and rhombuslike patterns, spirals, and targets. 
Moreover, distance-dependent time delays are found to induce traveling waves even with long-range interactions in regular topology~\citep{zanette_propagating_2000,jeong_time-delayed_2002}. This line of research was also extended to irregular topologies~\citep{ko_wave_2004}.  In the presence of time delays and phase lag,  pulse-coupled networks  exhibit analogous behaviour, like chimera states~\cite{wildie_metastability_2012}. 

The following system with randomly coupled identical Kuramoto oscillators is generally considered  with time delays proportional to Euclidean distances:
\begin{equation}
\dot{\theta}_i(t) = \omega_0 + \frac{\lambda}{k_i} \sum \limits_j A_{ij} \sin[\theta_j(t-\tau_{ij}) - \theta_i(t)], 
\label{Eq:Distributed_time_delay_dynamics}
\end{equation}
where the time delay $\tau_{ij} = \frac{d_{ij}}{v}$. The coupling from oscillators $j$ to $i$ is assumed to be propagated along the distance $d_{ij}$ with finite constant speed $v$. Assume that the oscillators  are equally spaced on a ring with circumference $L$ for simplicity and the position $x_i$ of the $i$-th oscillator is therein $x_i = \frac{L}{N}i$. The distance $d_{ij}$ is taken as the shortest Euclidean distance between oscillators $i$ and $j$, i.e., $d_{ij} = \text{min}\{|x_j-x_i|, L-|x_j-x_i|\}$.  
Note that here identical oscillators are mainly considered. The system of nonidentical oscillators has also attracted much  attention~\cite{yeung_time_1999,montbrio_time_2006,choi_synchronization_2000,ares_collective_2012}.

Model~(\ref{Eq:Distributed_time_delay_dynamics}) with different 
natural frequencies $\omega_i$ can be 
derived from neural networks to investigate the effect of 
the large-scale structural connectivity on neural dynamics at 
the neural population level~\cite{cabral2011RoleLocalNetworkOscillationsResting}. The network describes a periodic 
trajectory (a limit cycle) in phase space  and its dynamics can 
be approximated by the phase $\theta_i$ on this limit cycle. 
$\omega_i$ is the intrinsic frequency of node $i$ and  is drawn 
from a fixed Gaussian distribution with mean frequency 
$\bar{\omega} = 60 \mathrm{Hz}$ and standard derivation. The 
coupling term between one node and another is written as a 
$2\pi$-periodic sin function of their phase difference and is 
taken into account the interaction delay $\tau_{ij}$. The delay 
$\tau_{ij}$ from node $j$ to $i$ is calculated using $\tau_{ij} = 
\left\langle \tau \right\rangle L_{ij}/\left\langle L 
\right\rangle$, where $\left\langle \tau \right\rangle$ denotes 
the mean delay, $L$ is the fiber length matrix and $\left\langle 
L \right\rangle$ is the mean fiber length. The biologically-based 
Kuramoto model in the presence of time-delayed interactions gives 
rise to the emergence of slow neural activity fluctuations with 
their patterns significantly correlated with the empirically 
measured functional connectivity (see also 
Sec.~\ref{subsec:neuroscience} for more applications of neuronal 
models).

Noteworthy, assuming a fixed amplitude and taking only the phase dynamics into account, the dynamics of the full Stuart-Landau equation with feedback can be reduced to a delay-coupled Kuramoto model~\cite{dhuys_amplitude_2010} and its linear stability analysis can also be derived via a similar mathematical process~\cite{earl_synchronization_2003,dhuys_synchronization_2008}. 

\begin{figure}
\begin{center}
\includegraphics[width=0.6\linewidth]{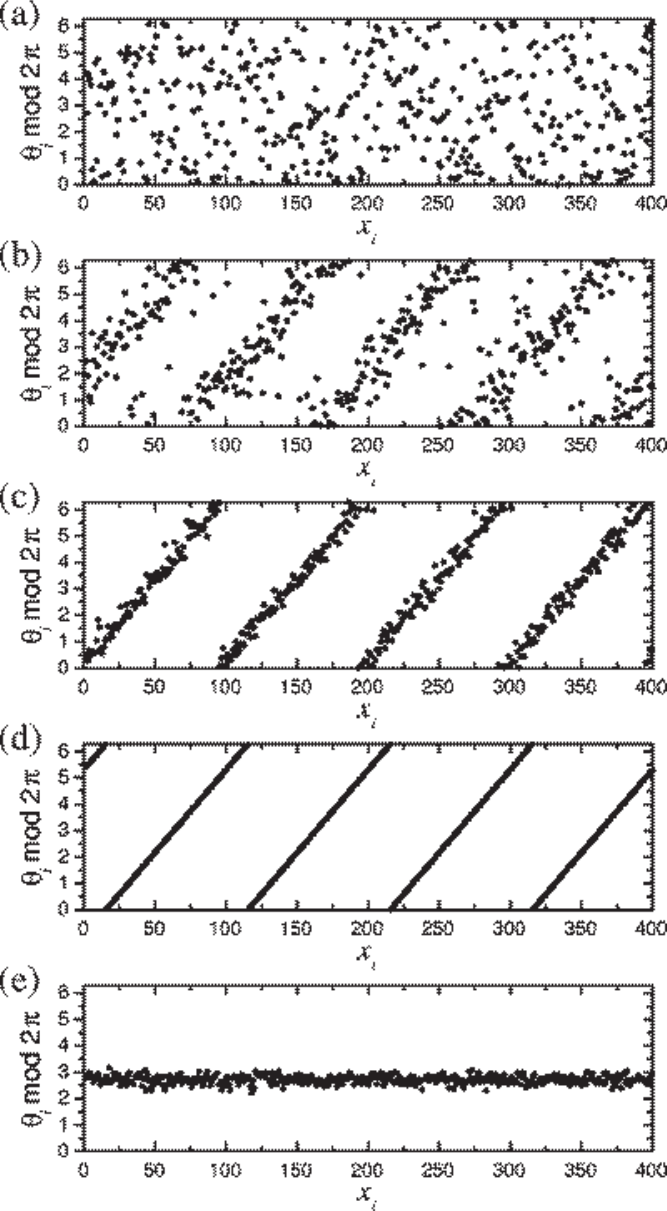}
\end{center}
\caption{Phase of the oscillators with  $\omega_0 = \pi/10$,$N=400$, $K=0.4$, $1/v=0.22$, and $L=400$. (a) The average degree $\langle k \rangle = 4$, the average frequency $\Omega \approx 0.31$, and the frequency dispersion $\sigma \approx 0.05$. (b) $\langle k \rangle = 8$: a wave state with winding number $n=4$, $\Omega \approx 0.295$,  and $\sigma \approx 0.015$. (c)  $\langle k \rangle = 20$: a wave state with $m=4$, $\Omega \approx 0.2933$ and $\sigma = 4\times 10^{-4}$.  (d) All-to-all coupling: a wave state with $m=4$, $\Omega = 0.291$, and $\sigma = 10^{-5}$. (e) $\langle k \rangle = 20$: a near in-phase synchronous state, $\Omega = 0.0424$, and $\sigma < 10^{-5}$. Reprinted with permission from \citep{ko_wave_2004}. Copyright 2011 by the American Physical Society.}
\label{Figure_Ko_wave_2004_figure_1}
\end{figure}

Wave formation is observed in randomly coupled oscillators 
\citep{ko_wave_2004}, where, for each oscillator $i$, $\langle k  
\rangle /2$ oscillators are randomly chosen to couple to $i$ 
bidirectionally. Moreover, the system shows different behaviors, 
including wave states and in-phase state, in terms of different 
topologies as shown in Fig.~\ref{Figure_Ko_wave_2004_figure_1} 
with the difference of panels from (a) to (e) in the coupling 
topology and the initial phase configurations. 
To quantify the oscillating behaviors, the average frequency 
$\Omega = \frac{1}{N} \sum_i \dot{\theta}_i$ and the frequency 
dispersion $\sigma = \sqrt{\frac{1}{N} \sum_i(\dot{\theta}_i - 
\Omega)}$ are defined. 
With the increases of the average degree $\langle k \rangle$ (as 
shown from Fig.~\ref{Figure_Ko_wave_2004_figure_1}(a) to (c)), 
the system exhibits wave states similar to that of all-to-all 
coupling case (Fig.~\ref{Figure_Ko_wave_2004_figure_1}(d)).  
Additionally to the wave states in 
Fig.~\ref{Figure_Ko_wave_2004_figure_1}(c), the system can also 
exhibit near in-phase synchronous states in 
Fig.~\ref{Figure_Ko_wave_2004_figure_1}(e). 

In the case when phase differences remain typically small in the system, 
Eguiluz et al.~\cite{eguiluz_structural_2011} considered a linearization of a set of coupled Kuramoto oscillators. In this case, the corresponding governing dynamics becomes 
\begin{equation}
\dot{\theta}_i(t) = \omega_i + \lambda \sum^{N}_{j=1}A_{ij} \left[\theta_j(t-\tau_{ij}) - \theta_i(t) \right]. 
\label{Eq:DTC_Eguluz_orignal_dynamics}
\end{equation}
All oscillators rotate at constant speed $\Omega$ with sufficiently small phase differences and their phases are represented by 
\begin{equation}
\theta_i(t) = \Omega t + \phi_i,
\end{equation}
where $\phi_i$ is the initial phase of the $i$-th oscillator.
The dependence of the locking frequency $\Omega$ on the network parameters and on the time delays is given by~\cite{eguiluz_structural_2011}
\begin{equation}
\Omega = \frac{\left\langle \boldsymbol{\omega} \right\rangle}{1+ \left\langle \mathbf{T} \right\rangle},
\label{Eq:DTC_Omega}
\end{equation}
and the phases 
\begin{equation}
\frac{\omega_i - \left\langle \boldsymbol{\omega} \right\rangle - (\omega_i \left\langle \mathbf{T} \right\rangle - \left\langle \boldsymbol{\omega} \right\rangle T_i)}{1+ \left\langle \mathbf{T} \right\rangle } =  (\mathbf{L}\boldsymbol{\phi})_i,
\label{Eq:DTC_phi}
\end{equation}
where $\pmb{\omega}$ is the frequency vector with components $\omega_i$, $\mathbf{T}$ is a vector of components $T_i = \sum_j A_{ij} \tau_{ij}$ indicating the total delay on node $i$, 
$\mathbf{L}$ is the Laplacian matrix with $L_{ij} = k_{i}^{\mathrm{in}} \delta_{ij} - A_{ij}$, $k_{i}^{\mathrm{in}}$ is the in degree of the node $i$, $\delta_{ij}$ is the Kronecker delta,
$\pmb{\phi}$ is the phase vector with components $\phi_i$, 
 $\left\langle \pmb{x} \right\rangle = \sum_i c_i x_i$ and $\boldsymbol{x}$ is normalized with $\sum_i c_i =1$, with $c_i$ being the components of the left eigenvector of $\mathbf{L}$ with eigenvalue 0. 
 Without loss of generality, the coupling strength is recalled, i.e. $\lambda =1$. If all nodes of the network are reached by at least one node, Eq.~\ref{Eq:DTC_Eguluz_orignal_dynamics} has a unique phase locked solution given by Eqs.~(\ref{Eq:DTC_Omega}) and (\ref{Eq:DTC_phi}) \cite{eguiluz_structural_2011}.



%
%

For uncorrelated and undirected networks of identical oscillators with constant time delay, i.e. $\omega_i = \omega_0$ and $\tau_{ij} = \tau$,  the analytic formulas of the mean phases $\Phi_k$ of nodes with degree $k$ and the locking frequency $\Omega$  are obtained as~\cite{eguiluz_structural_2011} 
\begin{equation}
\Phi_{{k}} = \frac{\Omega \left\langle k \right\rangle \tau }{k}  +a,
\end{equation}
and 
\begin{equation}
\Omega = \frac{\omega_0}{1 + \left\langle k \right\rangle  \tau},
\end{equation}
where $a$ is an arbitrary constant.   
This indicates that in uncorrelated networks the phase difference of nodes depends on their degree difference. In this limit, the locking frequency depends on the average degree but is independent of the degree distribution. 

%
%
%
%
%
%

Along the lines of these topics, 
considering the limit of $N \rightarrow \infty$, Ko and Ermentrout~\citep{ko_effects_2007} analyzed effects of axonal time delay on synchronization and wave formation using an approximate equation, which for discrete oscillators is equivalent to an equation for a continuum of oscillators as follows
\begin{equation}
\frac{\partial \theta(x,t)}{\partial t} = \omega_0 + \frac{\lambda}{\overline{P}} \int^{L/2}_{-L/2} P(x,x+y) \sin(\theta(x+y,t) - \theta(x,t) - 2\pi\tau|y|) dy,
\label{Eq:Ko_Ermentrout_continuum_equation_1}
\end{equation}
where $\tau $ is the relative unit time delay, $P(x,x+y)$ is the connecting probability between oscillators at $x$ and $x+y$,  oscillators are located uniformly along a ring with circumference $L$, $x_i = L i /N$, and $\overline{P}$ is the average connecting probability defined by $\overline{P} = \int^{L/2}_{-L/2} \int^{L/2}_{-L/2} P(x,y) dy dx$. To simplify the notation, without loss of generality, $L$ is set unity, i.e., $L=1$. For randomly coupled oscillators with perfect synchronous solutions or traveling wave solutions, $P(x,x+y)$ is a constant. Therefore, the continuum equation can be simplified and is exactly the form as that of globally coupled case as follows  
\begin{equation}
\frac{\partial \theta(x,t)}{\partial t} = \omega_0  + \lambda \int^{1/2}_{-1/2} \sin(\theta(x+y,t) - \theta(x,t) - 2\pi \tau |y|) dy. 
\label{Eq:Ko_Ermentrout_continuum_equation_2}
\end{equation}

\begin{figure}
\begin{center}
\includegraphics[width=0.7\linewidth]{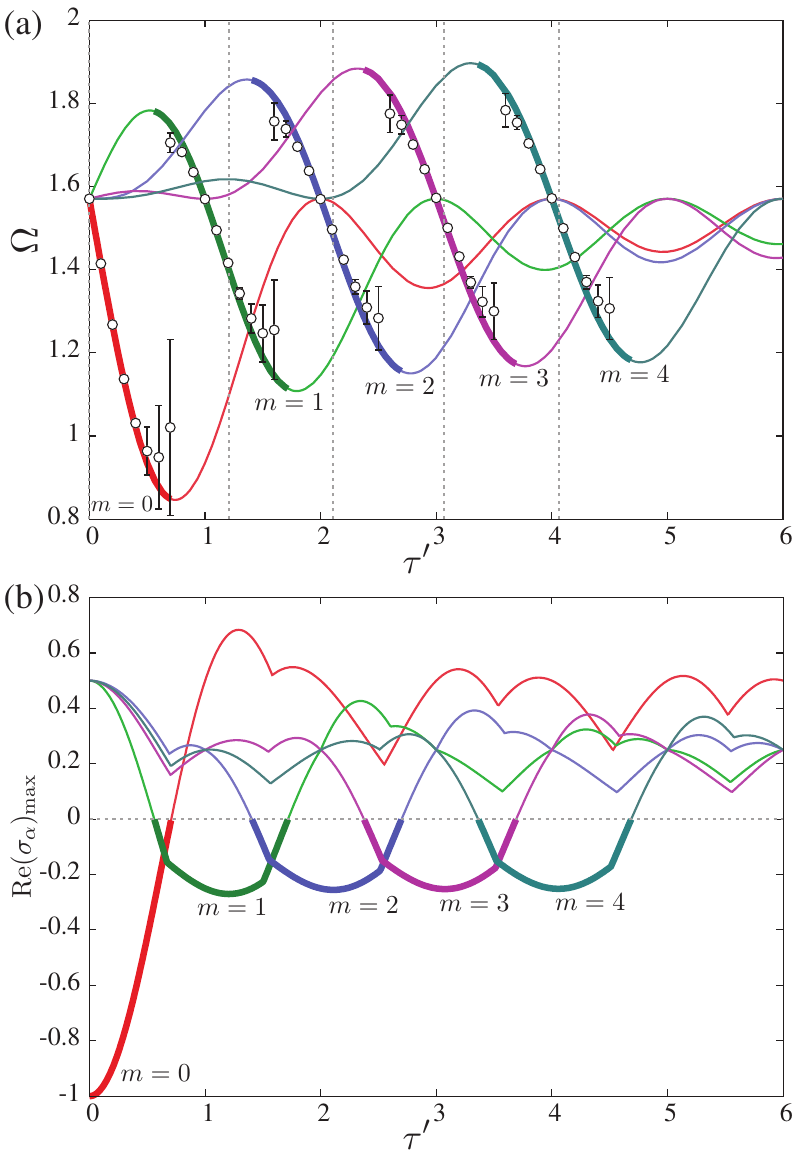}
\end{center}
\caption{States and their linear stability as a function of the relative unit time delay $\tau$ with $\omega_0 = \pi/2$ and $\lambda =1$. Thick curves indicate stable states identified by the linear stability analysis of the system (\ref{Eq:Ko_Ermentrout_continuum_equation_2}). (a) Synchronization frequency $\Omega$. Symbols and error bars denote the time average of $\Omega $ and the frequency dispersion calculated numerically with $N=1600$ and $ \langle k \rangle = 40$.  Dashed vertical lines are located at the minimum of $\text{Re}(\sigma_a)_{\text{max}}$. (b) The maximum real part $\text{Re}(\sigma_a)_{\text{max}}$ of eigenvalue $\sigma_a$ given by Eq.~\ref{Eq:Ko_Ermentrout_real_eigenvalues}. Adapted with permission from~\citep{ko_effects_2007}. Copyrighted by the American Physical Society.}
\label{Fig:Ko_Ermentrout_Effects_2007_Figure_4_a_and_b}
\end{figure}

Providing that the possible solutions obey the following form $\theta(x,t) = \Omega_q t + q x$, where $\Omega_q$ is the synchronization frequency, $q = 2m\pi $ is the wave vector and  $m = 0, \pm 1, \pm 2, \ldots$. Substituting these solutions into Eq.~\ref{Eq:Ko_Ermentrout_continuum_equation_2} yields 
\begin{equation}
\Omega_q = \omega_0 + \lambda \left \{ \int^{1/2}_{0} \sin(qy-2\pi\tau y) d y + \int^{0}_{-1/2} \sin(qy + 2\pi\tau y) d y \right \}. 
\label{Eq:Omega_k_tau_m}
\end{equation}. 
In Figure \ref{Fig:Ko_Ermentrout_Effects_2007_Figure_4_a_and_b}(a), the synchronization frequency $\Omega$ as a function of the relative unit time delay $\tau$ and winding number $m$ is obtained from Eq.~\ref{Eq:Omega_k_tau_m}. Symbols and error bars are obtained numerically considering randomly coupled oscillators. Numerical and theoretical results fit well to each other. Thick parts of the curves are identified by the linear stability analysis.  Dashed vertical lines indicate the minimum of the maximal real values of eigenvalues and correspond to the most locally stable states.  The frequency dispersions are small around the dashed vertical lines and enlarge when states move away from these lines. Order parameter also reflects this trend~\citep{ko_effects_2007}.

Consider the stability of this possible solutions and let  $\theta(x,t) = \Omega_q t+ q x + \varepsilon \phi(x,t)$, where $\varepsilon  \ll 1$ and $\phi(x,t) = e^{i a x } e^{\sigma_a t}$ with $a$ following the form $2\pi q$ and $q = 0,\pm 1, \pm 2,\ldots$. Via inserting these quantities into Eq.~\ref{Eq:Ko_Ermentrout_continuum_equation_2}, the real part of $\sigma_a$  follows
\begin{equation}
\text{Re}(\sigma_a) = \lambda \int^{1/2}_{-1/2} \cos(qy-2\pi\tau |y|) [\cos(a y) -1] dy. 
\label{Eq:Ko_Ermentrout_real_eigenvalues}
\end{equation}
As shown in Fig.~\ref{Fig:Ko_Ermentrout_Effects_2007_Figure_4_a_and_b}(b), maximal values of $\text{Re}(\lambda_a)_{\text{max}}$ as a function of $\tau$ are shown and their negative values correspond to the stability region as thick curves. Bifurcations occur at the end points of the thick curves. The results are even good for very sparsely coupled cases with $k_i \ll N$. 

Additionally to the distance-dependent time delays, time delay $\tau$ could vary according to a  distribution~\cite{zanette_propagating_2000,ko_effects_2007, lee_large_2009, eguiluz_structural_2011,ko_wave_2004,lee_dynamics_2011}. 
In the continuum limit, where $N \rightarrow \infty$, Lee et al.~\cite{lee_large_2009} provided a framework for the study of delay heterogeneity on the stability of incoherent states and the dynamics transitions from stable incoherent states to stable coherent states, using the OA ansatz~\citep{ott2008low}. In comparison to the case with uniform time delay, spread in the delay distribution function greatly affects the system dynamics, e.g., the critical coupling.

It is worth noting that the influence of time delays was also investigated under the presence of time-varying parameters~\citep{barabash_homogeneous_2014}, a model that is relevant for the study of effects of anaesthesia on  macroscopic dynamics of the  brain~\citep{sheeba_neuronal_2008}.  
Specifically, instead of solely having time-dependent coupling 
function, natural frequencies and coupling strength are also allowed to vary over time~\cite{barabash_homogeneous_2014}. The problem was tackled using the OA ansatz~\citep{ott2008low,barabash_homogeneous_2014} .

\subsection{Time-varying coupling}
\label{sec:time_varying_coupling}

Due to the ubiquity of synchronization in complex systems, it is interesting to understand effects of time-varying interactions on collective behaviors of networked oscillators. 
Adaptive mechanisms can be found in biological systems~\citep{ermentrout_adaptive_1991} and neural systems~\citep{bechhoefer_feedback_2005}, e.g., the synchronization of pacemaker cells  in the heart,  evolving patterns of neuronal synchronization in the brain~\citep{taylor_spontaneous_2010}, and a typical Asian firefly specie, Pteroptyx malaccae, with adaptive frequency to achieve perfect synchronization with stimulating frequency~\citep{ermentrout_adaptive_1991} (see Sec.~\ref{sec:second_KM}). 

 Feedback theory  is  commonly used for time-varying coupling, and it is one of the most important ideas developed in the last century, with fundamental implications especially for biological neural  systems~\cite{bechhoefer_feedback_2005}.  For example in neuroscience, spike-timing dependent plasticity (STDP) enhances synchronization of neural activity via enlarging the synchronization zones, decreasing the relaxation time, and increasing its robustness against  perturbations~\cite{nowotny_enhancement_2003,zhigulin_robustness_2003}. In networked Kuramoto oscillators,  STDP can give rise to multistability~\cite{maistrenko2007MultistabilityKMSynapticPlasticity}.

Adaptive scheme can suppress the negative effect of the heterogeneity in oscillator networks~\cite{nishikawa_heterogeneity_2003,ren_adaptive_2014,lu2008SynchronizationDiscreteTime,zhou_dynamical_2006}.   
Even in a system of identical Kuramoto oscillators, the presence of coupling plasticity induces the occurrence of multi-stable states, i.e., the existence of a desynchronized state and a state consisting of two anti-phase clusters~\citep{chandrasekar_adaptive_2014}. 
In adaptive and multilayer networks in the absence of  frequency-degree corrections, explosive synchronization occurs \citep{zhang2015explosive} (see Sec.~\ref{sec:explosive_sync}).  
 Additional to adaptive weights (coupling), adaptive natural frequency induces special features, including long waiting times before synchronization and three stability regimes \cite{taylor_spontaneous_2010,skardal2014ComplexMacroscopic}. 

Here we aim at presenting recent results of synchronization on networks of Kuramoto oscillators, where coupling strengths slowly adapt. We will also discuss effects of different kinds of adaptive coupling schemes on global and local synchronization.

The system consists of an ensemble of $N$ phase oscillators and its governing dynamics is given by 
\begin{equation}
\dot{\theta}_i = \omega_i +\lambda \sum_{j \in \mathcal{N}_i} w_{ij}(t) \sin(\theta_j - \theta_i), 
\label{Eq:gutierrez_emerging_2011_theta}
\end{equation}
where each oscillator $i$ has the randomly assigned natural frequency 
$\omega_i$ and interacts with $K$ randomly selected oscillators, which 
form the set $\mathcal{N}_i$. Coupling $w_{ij}$ is the non-negative and 
time-dependent weight of a directed link from node $j$ to $i$ in 
~\citep{gutierrez_emerging_2011,assenza_emergence_2011}. 
The adaptive evolution of the weights $w_{ij}$  is 
defined  as follows  \citep{gutierrez_emerging_2011}
\begin{equation}
\dot{w}_{ij} (t) = p_{ij} - \left( \sum_{l \in \mathcal{N}_i} p_{il} \right) w_{ij}(t), 
\label{Eq:gutierrez_emerging_2011_w_ij}
\end{equation}
where $p_{ij}$ quantifies the local synchronization between nodes $i$ and $j$ over a characteristic memory time $T$, and it is governed by 
\begin{equation}
p_{ij} = \frac{1}{T} \left | \int^{t}_{-\infty} e^{-(t-t')/T} e^{i[\theta_i(t') - \theta_j(t')]} dt' \right |. 
\label{Eq:quitierrez_emerging_2011_p_ij}
\end{equation}
The sum of the incoming weights of each nodes is kept constant at all times, i.e., $\sum_{j \in \mathcal{N}_i} w_{ij}=1$. 
In directed networks  with the same incoming and outgoing links, the coupling weights $w_{ij}$ can also bes adjusted as  \citep{assenza_emergence_2011}
\begin{equation}
\dot{w}_{ij}(t) = w_{ij}(t) \left[ s_i p_{ij} - \sum_j w_{ij}(t) p_{ij}  \right], 
\label{Eq:Assenza_emergence_2011_w_ij}
\end{equation}
where $s_i$ denotes the sum of the total incoming strength, i.e., $s_i = \sum^{N}_{j=1} w_{ij}$,  and $p_{ij}$ denotes the average phase correlation between nodes $i$ and $j$ over the time interval $[t-T,t]$ as 
\begin{equation}
p_{ij} = \frac{1}{T} \left | \int^{t}_{t-T} e^{i[\theta_j(t') -\theta_i(t')]} dt'\right |. 
\label{Eq:Assenza_emergence_2011_p_ij}
\end{equation}
Note that here memory dependent adaptation mechanisms are considered. The local phase correlation $p_{ij}$ could  also be instantaneous ~\citep{avalos-gaytan_assortative_2012}.

The adaptive scheme  has the form of the replicator equation and exhibits the feature of homophily and homeostasis, accounted for by, respectively, the first and the second term in the right-hand side of Eq.~\ref{Eq:gutierrez_emerging_2011_w_ij} and~\ref{Eq:Assenza_emergence_2011_w_ij}.
When the first term is larger  than the second one, the adaptive mechanism will enhance their coupling strength, which exhibits effects of homophily. On the other hand, the mechanism will depress their coupling strength, as effects of homoeostasis~\citep{avalos-gaytan_assortative_2012}.

The influence of adaptive coupling on global synchronization $r$ and local synchronization  $r_{\text{link}}$ between connected links  is investigated with the respective definitions  \citep{gutierrez_emerging_2011}
\begin{equation}
r=\lim_{\Delta t \rightarrow \infty} \frac{1}{\Delta t N} \int^{t_s + \Delta t}_{t_s} \left| \sum^N_{j=1} e^{i\theta_j(t')}  \right| d t', 
\end{equation}
and 
\begin{equation}
r_{\text{link}} = \frac{1}{N} \sum^N_{i=1} \sum^N_{j \in \mathcal{N}_i}w_{ij} \lim_{\Delta t \rightarrow  \infty} \frac{1}{\Delta t} \left | \int^{t_s + \Delta t}_{t_s} e^{i[\theta_j(t') - \theta_i(t')]} d t' \right | , 
\label{Eq:guitterrez_emerging_2011_r_link}
\end{equation}
where  $t_s$ identifies the beginning of averaging the global and local 
synchronization over a suitably long time interval denoted by  $\Delta 
t$. Intuitively, $r$ quantifies the global time averaged order parameter 
and $r_{\text{link}}$  the time averaged local synchronization between 
connected nodes. The two measures coincide with each other for very small 
or very high coupling strengths but become different with the emergence 
of modularity, where $r_{\text{link}}>r$ and only local synchrony occurs.

\begin{figure}
\begin{center}
\includegraphics[width=0.7\linewidth]{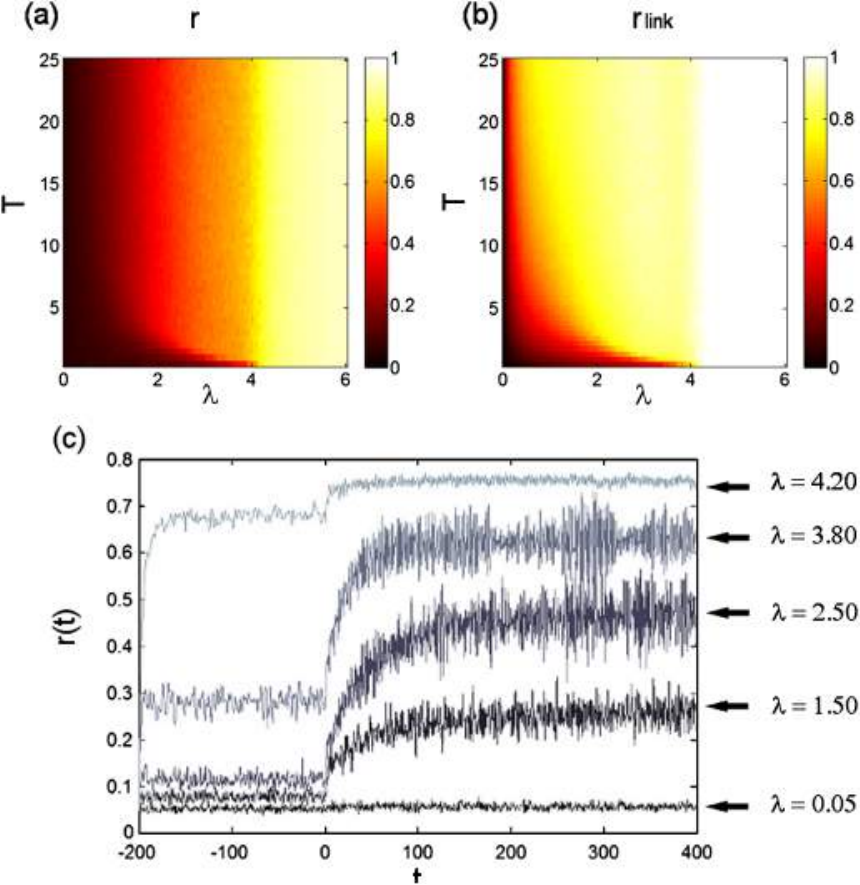}
\end{center}
\caption{Projection of (a) global synchronization $r$ and (b) local synchronization $r_{\text{link}}$ over the parameter space of $(T, \lambda)$ in a directed weight-adapted network. (c) Time evolution of global order parameter $r(t)$ considering $T = 15$ for different couplings. Remaining parameters: $t_s=2000$, $\Delta t = 1000$ and $T = 15$. From~\cite{gutierrez_emerging_2011}.}
\label{Fig:Gutierrez_emerging_2011_Fig_1}
\end{figure}

(i) Let us consider first the adaptive coupling in Eq.~\ref{Eq:gutierrez_emerging_2011_w_ij}. In simulations, the system is integrated with initial values $w_{ij}=1/K$ and without the adaptive coupling during the first $200$ time units.  Subsequently, at the time instant defined as $t=0$, the adaptation mechanism (\ref{Eq:gutierrez_emerging_2011_w_ij}) is switched on. After $t_s$ integrating time units, $r$ and $r_{\rm{link}}$ are monitored over another $\Delta t$ time periods.  
The time evolution of the instantaneous order parameter $r(t)$ is shown in Fig.~\ref{Fig:Gutierrez_emerging_2011_Fig_1} (c)   via varying the value of $\lambda$ and fixing $T=15$, showing the striking result that  the adaptive mechanism generally enhances global synchronization $r(t)$ for $t>0$ in comparison to that for $t<0$. This result was also observed with different adaptive mechanisms \cite{ren_adaptive_2007}.

Figure~\ref{Fig:Gutierrez_emerging_2011_Fig_1} plots the global synchronization $r$ in (a) and the local synchornization $r_{\text{link}}$ (b) with respect to  $\lambda$ and $T$. For  large values of $T$, $r$ and $r_{\text{link}}$  solely depend on the coupling $\lambda$ up to the threshold $\lambda_c$, which is determined as in the classical Kuramoto model~\citep{strogatz2000FromKuramotoToCrawford}, i.e., $\lambda_c^{\rm{KM}}=\frac{2}{\pi g(0)}$ (Eq.~\ref{eq:KM_CRITICAL_COUPLING}). Within this period, due to the emergence of modular structure, $r_{\text{link}}$ grows much faster than $r$, persisting as such for a wide region in the plane (for $\lambda <1$ and $T>1$).  With increasing $\lambda$ beyond $\lambda_c$, higher global and local synchronizations are achieved. In this case, $p_{ij} \approx 1$ and $w_{ij} \approx 1/K$ (one can also  approximate its value by setting $\dot{w}_{ij}=0$ in Eq.~\ref{Eq:gutierrez_emerging_2011_w_ij}).  

\begin{figure}
\begin{center}
\includegraphics[width=0.7\linewidth]{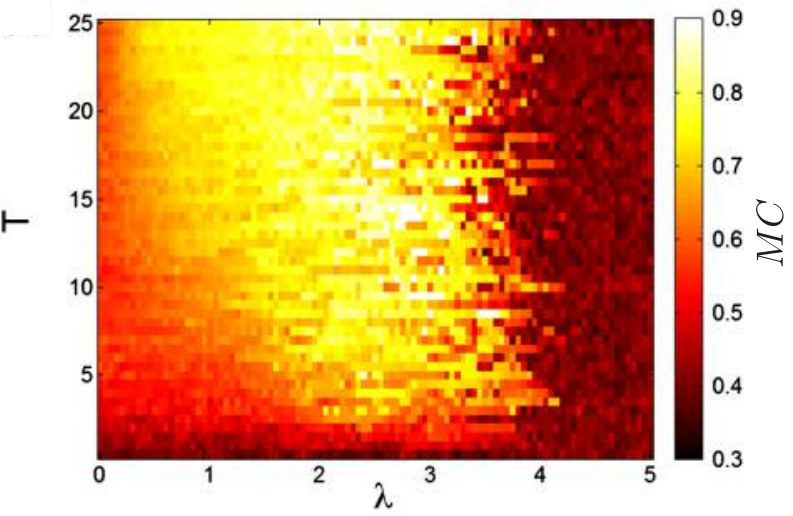}
\end{center}
\caption{Modularity cohesion $MC$ (Eq.~\ref{Eq:Gutierrez_emergencing_2011_MC}) as a function coupling $\lambda$ and characteristic memory time $T$ for networks consisting of $N = 300$ oscillators. From~\cite{gutierrez_emerging_2011}. }
\label{Fig:Gutierrez_emergencing_2011_Fig_2_a}
\end{figure}

To deepen the understanding of this structure, the modularity of the partition of  networks is measured by its \textit{modular cohesion} $MC$ as follows~\cite{gutierrez_emerging_2011}
\begin{equation}
MC = \frac{1}{N} \sum^{M}_{\mu=1} \sum_{i,j \in C_\mu} w_{ij},
\label{Eq:Gutierrez_emergencing_2011_MC}
\end{equation}
where the fast community detection   algorithm~\citep{newman_fast_2004,duch_community_2005} allows us to split  the network into  $M$ non-overlapping communities and $C_\mu$ stands for the $\mu$th community. $MC$ quantifies the degree of the partition, takes values within $[0,1]$, and $MC = 1$, when the network is split into disconnect modules. Interestingly, the network is separated into components when $\lambda$ is relatively low and $T$ is relatively high (Fig.~\ref{Fig:Gutierrez_emergencing_2011_Fig_2_a}). 
At small values of $\lambda$, the adaption mechanism redistributes the weights, which leads to a highly heterogeneous topology. With the increases of $\lambda$, more and more distinct modules occur and $MC$ increases respectively. 
When $\lambda > \lambda_c$, the coupling weights are persistent at a constant value $1/K =0.05$ (the initial value) and all oscillators belong to a single synchronous component. In this case, the adaptation mechanism is negligible.

%
%

(ii) Secondly, we consider directed networks with the same governing dynamics (\ref{Eq:gutierrez_emerging_2011_theta}) but different adaptive-coupling schemes  (\ref{Eq:Assenza_emergence_2011_w_ij}). 
Following the same simulation process, the time 
evolution of $r$ can exhibit quasi-periodic oscillations for some special coupling strengths~\cite{assenza_emergence_2011}. Interestingly, the network dynamics with competitive interactions (Eqs.~\ref{Eq:gutierrez_emerging_2011_theta},~\ref{Eq:Assenza_emergence_2011_w_ij} and~\ref{Eq:Assenza_emergence_2011_p_ij}) presented similar behaviors as in the above analysis of Fig.~\ref{Fig:Gutierrez_emerging_2011_Fig_1}. Specifically,
the projections of $r$ and $r_{\rm{link}}$ on the plane $(\lambda,T)$ also exhibit the presence of incoherent, partially ordered and totally synchronized regimes. 
In the incoherent state, a modular structure emerges, and a large internal synchronization of each modular is achieved but without global and local synchronization. At low coupling strength, a distribution of weights can be fitted by a power-law distribution.  
With increasing $\lambda$, the system comes to the partially ordered phase, where $r$ oscillates. Meanwhile, multiple components coexist and within each component perfect local synchronization is achieved. 
For high $\lambda$, the network dynamics ends up in a fully synchronized state and the stationary network with $\dot{w}_{ij}=0$ in~(\ref{Eq:Assenza_emergence_2011_w_ij}) is very similar to the initial regular network without adaptation, where each link shares the same weights, i.e., $w_{ij}=1/K$.

\begin{figure}
\begin{center}
\includegraphics[width=\linewidth]{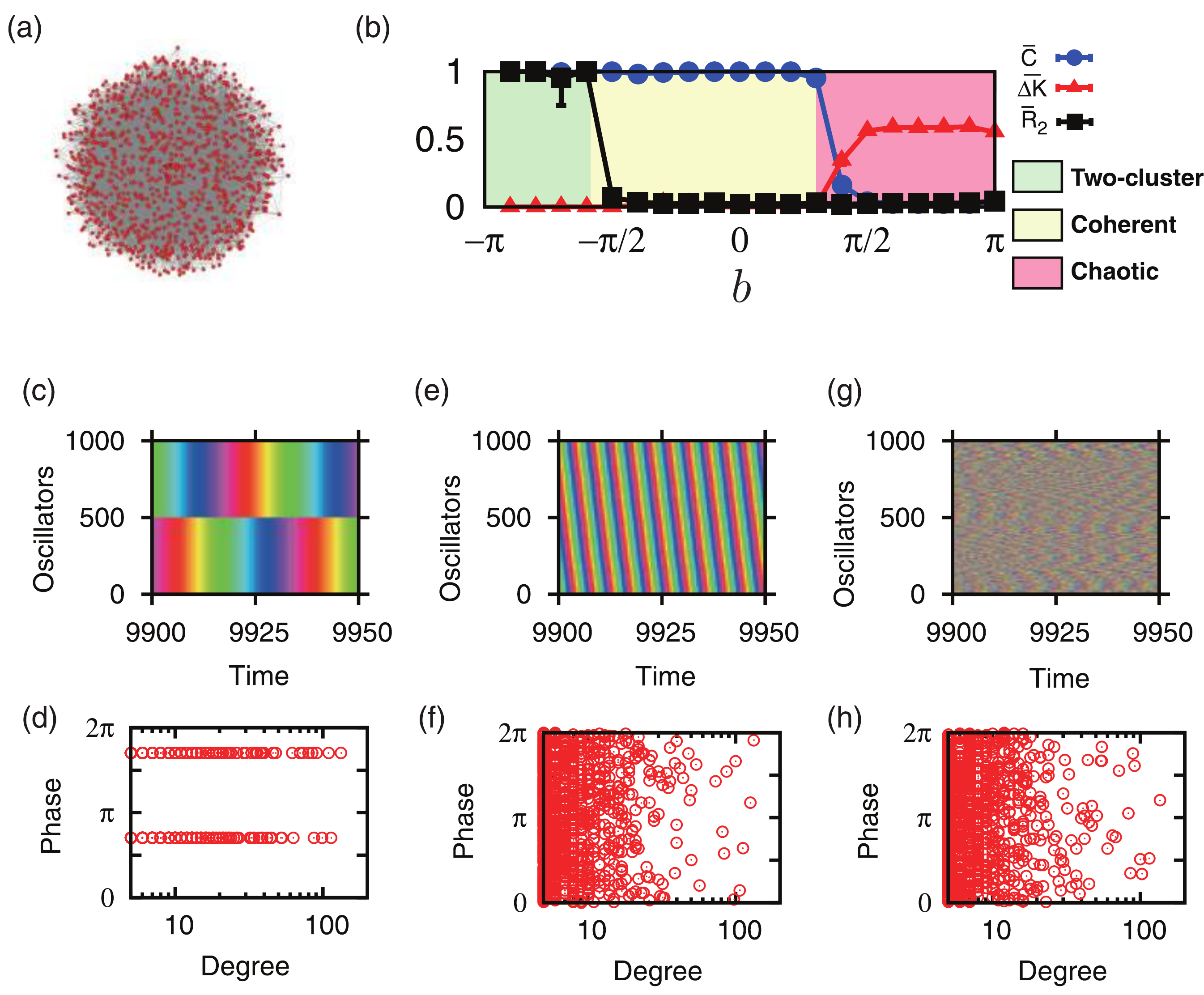}
\end{center}
\caption{ Coevolving dynamics on a scale-free network and its topology is shown in (a). Three regimes are exhibited in (b) as a function of the characteristic $b$ in terms of  the second order parameter $\overline{R}_2$, the long-time correlation $\overline{C}$ and the rate of
change of weights $\overline{\Delta K}$ with $ a = 0.3\pi$ and $\varepsilon= 0.005$. Time evolution of phase and phase vs degree are shown in (c-d) for the two-cluster state with $b=-0.7 \pi$,  (e-f) for the coherent state with $b=-0.1 \pi$, and (g-h) the chaotic state with $b=0.7 \pi$. Adapted with permission from~\citep{aoki_self-organized_2011}. Copyrighted by the American Physical Society.}
\label{Fig:Aoki_Aoyagi_self_organized_2011_Fig_11}
\end{figure}

The above results are induced by the competitive mechanisms (\ref{Eq:gutierrez_emerging_2011_w_ij}) and (\ref{Eq:quitierrez_emerging_2011_p_ij}) or   (\ref{Eq:Assenza_emergence_2011_w_ij}) and (\ref{Eq:Assenza_emergence_2011_p_ij}). 
Other adaptive mechanisms proportional to the instantaneous phase difference between the oscillators have also been considered~\cite{ren_adaptive_2007,ren_adaptive_2014,sendina2008PhaseLockingSFTopologies}. For instance, Aoki and Aoyagi~\citep{aoki_co-evolution_2009} investigated coevolving dynamics in a weighted network of identical oscillators  as follows 
\begin{equation}
\dot{\theta}_i = \omega_i - \frac{1}{N} \sum_j  w_{ij} \sin(\theta_i - \theta_j +\varphi), 
\end{equation}
where the natural frequency is identical with $\omega_i=\omega_0=1$, without loss of generality. The constant phase shift denoted by $\varphi$  is induced by a short delay of the coupling and the coupling weight from the $j$th to the $i$th oscillator is denoted by $w_{ij}$.  
The dynamical model for the coupling weights $w_{ij}$  depends on the relative timing of phase difference as follows 
\begin{equation}
\dot{w}_{ij} = - \varepsilon \sin(\theta_i - \theta_j + b),
\end{equation}
 where $\varepsilon$ determines the time scale of the evolution of  
 the coupling weights with $\varepsilon \ll 1$ 
 and $b$ a different phase shift. 
The coupling $w_{ij}$ is bounded within the range of $[-1,1]$, and outside the boundary, $w_{ij}$ is immediately set to the approximate bounded value ($1$ or $-1$). An alternative method can be implemented by including a nonlinear term similar as in~\cite{ren_adaptive_2007} and the results are qualitatively the same (see Fig.~\ref{Fig:Aoki_Aoyagi_self_organized_2011_Fig_11}).
With this coevolving dynamical coupling, the system has three distinct types of self-organized phase patterns (Fig. \ref{Fig:Aoki_Aoyagi_self_organized_2011_Fig_11} (b)): a two-cluster state, a coherent state, and a chaotic state,  depending on $\varphi$ and $b$. The results are independent of the network parameter values, and are robust against the topological differences between the SF and the all-to-all connections. 

A generalization of the order parameter 
(Eq.~\ref{eq:KM_ORDER_PARAMETER}) can be defined as
\begin{equation}
R_m = \left | \frac{1}{N} \sum^{N}_{j=1} e^{im \theta_j} \right |, 
\end{equation}
where $m=1$ or $2$. 
$|R_1|$ quantifies the degree of asymmetry in clustering (refer to other sections), while  
$|R_2|$ quantifies the degree of cluster synchrony in the system~\cite{skardal_cluster_2011} and converges to $1$ when a two-cluster state  with anti-phase synchronization emerges. 

The normalized rate of change of the weights averaged over all connections is given by~\citep{aoki_self-organized_2011} 
\begin{equation}
\Delta K (t) = \frac{1}{N (N-1)} \sum_{i \neq j} \frac{|w_{ij}(t) - w_{ij} (t-\delta) | }{\delta},
\end{equation}
where $\delta$ is a sampling interval $\delta \approx  \frac{2 \pi}{\omega_0} \ll 1/\varepsilon$. In the region $b \in [-\pi, 0]$, the time evolution of coupling weights exhibits the Hebbian-like characteric and this is a positive feedback. 

In the two-cluster state in Fig. 
\ref{Fig:Aoki_Aoyagi_self_organized_2011_Fig_11} (b) and (c-d), 
$R_1$ is kept close to zero, $R_2$ converges to $1$ with anti-phase synchronization. The rate $\Delta K(t)$ becomes zero 
if the coupling weights are in frozen states, where 
oscillators within the same cluster run at the same frequency. The ratio of the populations in the two clusters depends on the initial conditions. 
On the other hand, in the 
coherent state in Fig. 
\ref{Fig:Aoki_Aoyagi_self_organized_2011_Fig_11} (b) and (e-f), 
the system is in the steady state and the rate of change of 
weights $\Delta K$ remains close to zero. To characterize it, the autocorrelation function $C(\tau) = 
\left\langle |\frac{1}{N} \sum_j e^{i\theta_j(t)} e^{-
i\theta_j(t-\tau)} | \right\rangle$ is defined  showing that 
$C(\tau)$ remains close to $1$, i.e., the oscillators are in the 
steady state. In right parameter region of Fig. 
\ref{Fig:Aoki_Aoyagi_self_organized_2011_Fig_11} (b), where both 
the two-cluster state and the coherent state become unstable in  
Fig. \ref{Fig:Aoki_Aoyagi_self_organized_2011_Fig_11} (g-h), the 
system behaves chaotically with positive Lyapunov exponents and the 
number of positive Lyapunov exponents is a linear function of the 
number of degrees of freedom of the system. Furthermore, $\Delta K$ does not decay and 
$C(\tau)$ converges to zero. 
Additional to the two-cluster state, the system can be designed 
to exhibit an $m$-cluster state by choosing an appropriate 
coupling evolution function, e.g. $ \dot{w}_{ij} = 
\cos[(m-1) \theta]$~\citep{aoki_self-organized_2011}. 

\section{Correlations between dynamics and topology (Explosive synchronization)}
\label{sec:explosive_sync}


Several works have verified the occurrence of discontinuous synchronization transitions in the tradition Kuramoto model since 90's~\cite{bonilla1992nonlinear}. Regarding the model without inertia, in 1992, Bonilla et al.~\cite{bonilla1992nonlinear} provided one of the first studies on first-order transitions in complete graphs. The authors further showed that such transitions could emerge when bimodal frequency distributions are considered. After this work, only in 2005 the subject of abrupt transitions in the first-order Kuramoto model was brought to attention again~\cite{pazo2005thermodynamic}. Analyzing the model with uniform frequency distributions, Paz\'o~\cite{pazo2005thermodynamic} showed that suddenly after a given critical coupling $\lambda_c$, all the population of oscillators is entrained in the mean-field, making the order parameter $r$ to ``jump''
from $r \sim 0$ to $r_c \lesssim 1$ with the dependence $r-r_c \propto (\lambda - \lambda_c)^{2/3}$, for $\lambda > \lambda_c$. The work by Paz\'o triggered further investigations devoted to the analysis of first-order transitions~\cite{basnarkov2007phase,basnarkov2008kuramoto,filatrella2007generalized}. In particular, Basnarkov and Urumov~\cite{basnarkov2007phase} analyzed the phase transition in the Kuramoto model and also generalized the results to unimodal frequency distributions composed by a plateau followed by tails in the form $|\omega - \omega_0|^m$. They observed discontinuous transitions for $m>0$. The same authors later reported that asymmetric frequency distributions could also lead to this dependency of the order parameter as a function of coupling~\cite{basnarkov2008kuramoto}.

\begin{figure}[!t]
\begin{center}
\includegraphics[width=0.7\linewidth]{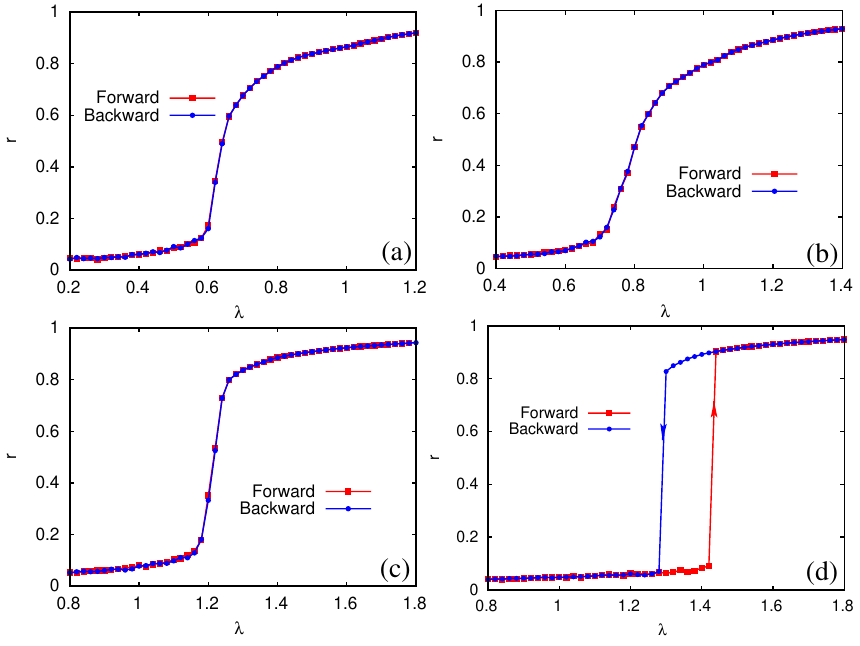}
\end{center}
\caption{Synchronization diagram $r(\lambda)$ for networks generated from the model proposed in~\cite{gomez2006scale} for (a) $\alpha = 1$ (ER), (b) $\alpha = 0.6$, (c) $\alpha = 0.2$ and (d) $\alpha = 0$ (BA). All networks have $N = 10^3$ and $\left\langle k \right\rangle = 6$. Reprinted with permission from~\cite{gomez2011explosive}. Copyright
2011 by the American Physical Society.}
\label{fig:gomezgardenez_explosive_fig1}
\end{figure}

\begin{figure}[!t]
\begin{center}
\includegraphics[width=0.7\linewidth]{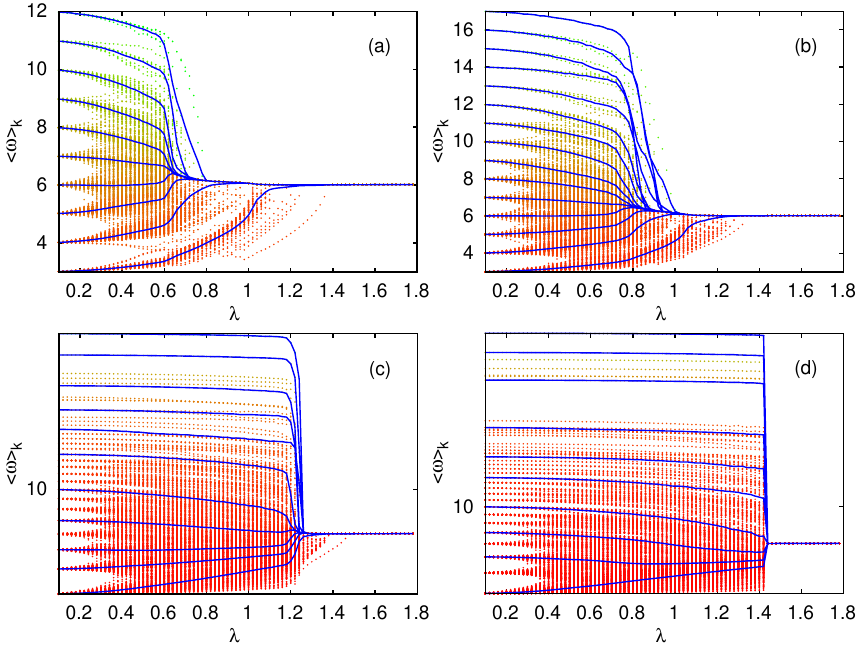}
\end{center}
\caption{ Effective frequencies $\left\langle \omega \right\rangle_k$ as a function of coupling $\lambda$ for networks using model~\cite{goltsev2006k} with (a) $\alpha = 1$ (ER), (b) $\alpha = 0.6$, (c) $\alpha = 0.2$ and (d) $\alpha = 0$ (BA). All networks have $N = 10^3$ and $\left\langle k \right\rangle = 6$. Reprinted with permission~\cite{gomez2011explosive}. Copyright 2011 by the American Physical Society.}
\label{fig:gomezgardenez_explosive_fig2}
\end{figure}

Further works verified that the same conditions to yield discontinuous transitions in fully connected graph do not necessarily have the same effect in heterogeneous topologies~\cite{moreno2004synchronization,arenas2008synchronization}. However, in 2011 G\'omez-Garde\~nez et al.~\cite{gomez2011explosive} firstly demonstrated the occurrence of a discontinuous transition -- or ``explosive synchronization'' (ES) as named by them -- in SF networks. The authors suggested that the correlation between topology and dynamics is the mechanism responsible for such a transition. More specifically, they considered the natural frequencies to be positively correlated with the degrees in the form $\omega_i = k_i$, where $k_i = \sum_{j=1}^N A_{ij}$ is the degree of node $i$. By using a random network model~\cite{gomez2006scale} able to tune continuously from ER to BA networks via controlling a single parameter $\alpha \in [0,1]$ in the model, the authors showed that explosive synchronization only happens for $\alpha = 0$, i.e., when the network is SF with a degree distribution given by $P(k) \sim k^{-\gamma}$, as shown in Fig.~\ref{fig:gomezgardenez_explosive_fig1}~\cite{gomez2011explosive}. Furthermore, to analyze the network dynamics locally, the authors also considered the effective frequency
\begin{equation}
\omega_i^{\rm{eff}} = \frac{1}{T} \int_t^{t + T} \dot \theta_i (\tau) d\tau. 
\label{eq:effect_freq_gomez_PRL_2011}
\end{equation} 
Figure~\ref{fig:gomezgardenez_explosive_fig2} shows the corresponding dependency of $\omega_i^{\textrm{eff}}$ for the forward propagation of coupling $\lambda$ along with the effective 
frequency for each group of nodes with degree $k$, $\left\langle \omega \right\rangle_k = \sum_{[i|k_i = k]} \omega_i^{\rm{eff}}/(NP(k))$~\cite{gomez2011explosive}. As we can see, for Fig.~\ref{fig:gomezgardenez_explosive_fig2}(a)-(c) the groups of nodes join smoothly the synchronous component as the coupling is increased. However, for $\alpha  = 0$ (Fig.~\ref{fig:gomezgardenez_explosive_fig2}(d)) all nodes join the giant synchronous component abruptly at $\lambda_c \approx 1.4$, evolving with $\left\langle \omega \right\rangle_k  = \left\langle k \right\rangle = 6$~\cite{gomez2011explosive}.

These findings led the authors to conjecture that the necessary condition for the emergence of an abrupt variation in the order parameter is the correlation between the dynamics and the network structure. Other dynamical processes in networks share similar characteristics, such as the case of percolation and epidemic spreading, where there is also a microscopic mechanism that leads to the abrupt transition~\cite{achlioptas2009explosive,da2010explosive,da2014solution,waagen2014given,gomez2015AbruptTransitionsReinfections}.

\subsection{Analytical approaches}
\label{subsec:explosive_sync_analytical_approaches}

The first work on explosive synchronization focused mostly on the numerical analysis of the model in complex networks and the theoretical calculation of the critical coupling for a star graph. Naturally the discovery of such transitions raised many questions 
on the subject, a fact that motivated many following works~\cite{gomez2011explosive,peron2012determination,peron2012explosive,ji2013cluster,zhang2013explosive,su2013explosive,leyva2012explosive_chaotic,skardal2014disorder,coutinho2013kuramoto,zhu2013criterion,li2013effect,liu2013effects,chen2013explosive,leyva2013explosive,zhu2013explosive,sonnenschein2013networks,li2013reexamination,skardal2013effects,zou2014basin,zhang2014explosive}.
The first analytical approach to the model in networks without degree-degree correlations was carried out by Peron and Rodrigues~\cite{peron2012determination}. More specifically, the authors considered the equations of motion (Eq.~\ref{eq:Ichinomiya_continuumlimit}) in the continuum limit for the case in which $\omega = k$, i.e., 
\begin{equation}
\dot{\theta} = k + \lambda k \int dk' \int d\theta' \frac{k'P(k')}{\left\langle k \right\rangle} \rho(\theta',t|k') \sin(\theta' - \theta), 
\label{eq:eq_motion_cont_limit_peron_rodrigues}
\end{equation}
where $\rho(\theta,t|k)$ is the density of oscillators with phase $\theta$ at time $t$ for a given degree $k$, similarly as before. Using the corresponding definition for the order parameter $r$ in the continuum limit
\begin{equation}
r e^{i\psi(t)} = \frac{1}{\left\langle k \right\rangle} \int dk \int d\theta kP(k) \rho(\theta,t|k) e^{i\theta} 
\label{eq:order_parameter_peron_rodrigues1}
\end{equation}
and rewriting the equations in the rotating frame it is possible to show that the real part of the order parameter is given by~\cite{peron2012determination}
\begin{equation}
\left\langle k \right\rangle r = \int_{\left\langle k \right\rangle/(1 + \lambda r)}^{\left\langle k \right\rangle/(1-\lambda r)} dk P(k) k \sqrt{1 - \frac{1}{\lambda^2 r^2} \left( \frac{k - \left\langle k \right\rangle}{k}\right)^2}.
\label{eq:order_parameter_peron_rodrigues2}
\end{equation}
Changing the variable in Eq.~\ref{eq:order_parameter_peron_rodrigues2} and letting $r\rightarrow 0^+$ we obtain~\cite{peron2012determination}
\begin{equation}
\lambda_c = \frac{2}{\pi \left\langle k \right\rangle P(\left\langle k \right\rangle)},
\label{eq:critical_coupling_peron_rodrigues}
\end{equation} 
which is the critical coupling for the onset of synchronization. 
It is interesting to note that for the case in which natural frequencies are equal to degrees, the critical coupling no longer depends on the ratio between the first moments of the degree distribution $\left\langle k \right\rangle /\left\langle k^2 \right\rangle$, in contrast to the case where the frequencies are drawn from a unimodal and even distribution~\cite{ichinomiya2004frequency}. Noteworthy, strictly speaking, the result
in Eq.~\ref{eq:critical_coupling_peron_rodrigues} is valid for networks
with symmetric $P(k)$, although it provides good estimates for $\lambda_c$ in SF networks with reasonable size~\cite{peron2012determination}. Following a similar analysis, the expression for $\lambda_c$ 
was recently generalized for networks with $P(k) \sim k^{-\gamma}$ and it is given by~\cite{pinto2015ExplosiveSyncPartialCorr}
\begin{equation}
\lambda_c = \frac{2\frac{(\gamma - 1)^{\gamma-2}}{(\gamma - 2)^{\gamma-1}}}{\sqrt{\pi^2 + \left[\log(\gamma - 2) - \sum_{i=1}^{\gamma - 3} \frac{1}{i} \left( \frac{\gamma - 1}{\gamma - 2}\right)^{i} \right]^2}}, 
\label{eq:lambda_c_pinto_saa}
\end{equation} 
which is valid for any integer $\gamma > 2$.

\begin{figure}[!t]
\begin{center}
\includegraphics[width=0.7\linewidth]{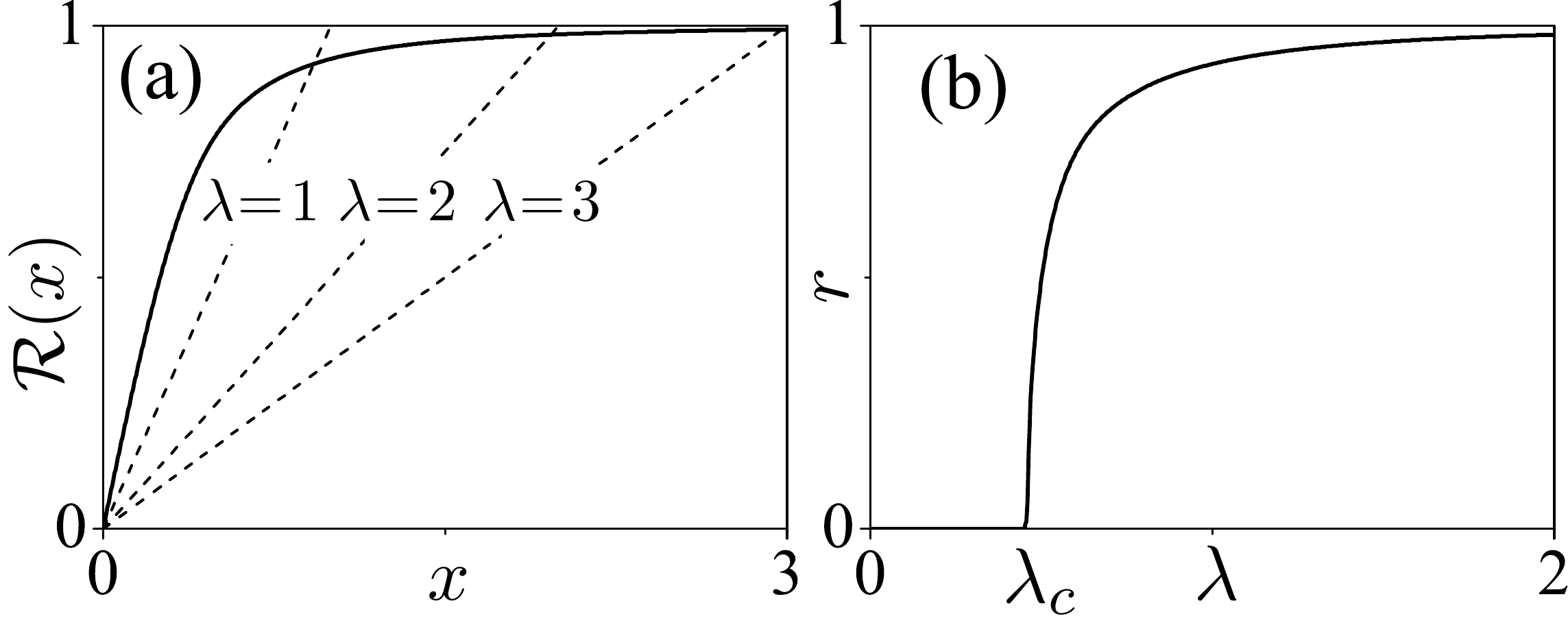}
\end{center}
\caption{ (a) Example of solution of Eq.~\ref{eq:coutinho_order_parameter_final} for an ER network with $\left\langle k \right\rangle = 10$. The dashed lines correspond to the curves $x/\lambda$, whose intersection with $\mathcal{R}(x)$ yield the solutions of Eq.~\ref{eq:coutinho_system2} leading to (b) the order parameter $r$ as a function of $\lambda$. Adapted with permission from~\cite{coutinho2013kuramoto}. Copyrighted by the American Physical Society.}
\label{fig:coutinho_kuramoto_fig1}
\end{figure} 

A different analysis of explosive synchronization was carried out by Coutinho et al.~\cite{coutinho2013kuramoto}, where further properties of the model were uncovered. Noting that in the rotating frame the equations of motion are written as $\dot\theta_i - \Omega = (k_i - \Omega) + \lambda r k_i \sin(\psi - \theta_i)$, with $\Omega$ being the average frequency, it is possible to rewrite the order parameter as 
\begin{equation}
re^{i\psi(t)}=\frac{1}{N\left\langle k\right\rangle }\sum_{\left|k_{j}-\Omega\right|\leq k_{j}\lambda r}k_{j}e^{i\theta_j}+\frac{1}{N\left\langle k\right\rangle }\sum_{\left|k_{j}-\Omega\right|>k_{j}\lambda r}k_{j}e^{i\theta_{j}},
 \label{eq:coutinho_r_discrete}
\end{equation}
where the first and second terms are the contributions due to locked and 
drifting oscillators, respectively. Taking the continuum limit of Eq.~\ref{eq:coutinho_r_discrete} we can write each contribution as
\begin{equation}
r_{{\rm {lock}}}=\frac{1}{\left\langle k\right\rangle }\int_{|k-\Omega|\leq\lambda kr}dkP(k)k\sqrt{1-\left(\frac{k-\Omega}{\lambda kr}\right)^{2}}
\label{eq:lock_coutinho}
\end{equation}
and
\begin{equation}
r_{{\rm {drift}}}=\frac{1}{\left\langle k\right\rangle }\int_{|k-\Omega|>\lambda kr}dkP(k)\frac{k-\Omega}{\lambda r}\left[1-\sqrt{1-\left(\frac{\lambda kr}{k-\Omega}\right)^{2}}\right].
\label{eq:drift_coutinho}
\end{equation}
Defining $x \equiv \lambda r$ and substituting Eq.~\ref{eq:lock_coutinho} and~\ref{eq:drift_coutinho} into Eq.~\ref{eq:coutinho_r_discrete} ($r = r_{\rm{lock}} + r_{\rm{drift}}$) we obtain
\begin{equation}\label{eq:coutinho_system1}
\left\langle k\right\rangle -\Omega=\int_{|k-\Omega|>xk}dkP(k)(k-\Omega)\sqrt{1-\left(\frac{xk}{k-\Omega}\right)^{2}},
\end{equation}
\begin{equation} \label{eq:coutinho_system2}
\mathcal{R}(x) = \frac{x}{\lambda}, 
\end{equation}
where $\mathcal{R}(x)$ is given through Eq.~\ref{eq:coutinho_r_discrete}, as follows
\begin{equation}
\mathcal{R}(x)=\frac{1}{\left\langle k\right\rangle }\int_{|k-\Omega(x)|\leq xk}dkP(k)k\sqrt{1-\left(\frac{k-\Omega(x)}{xk}\right)^{2}}
\label{eq:coutinho_order_parameter_final}.
\end{equation}
The set of Eqs.~\ref{eq:coutinho_system1},~\ref{eq:coutinho_system2} and~\ref{eq:coutinho_order_parameter_final} gives the full recipe to uncover the dependency of $r = r(\lambda)$. First, the average frequency $\Omega = \Omega(x)$ is obtained by solving Eq.~\ref{eq:coutinho_system1} for variable $x$. Then the result is inserted into Eq.~\ref{eq:coutinho_order_parameter_final} to get $\mathcal{R}=\mathcal{R}(x)$. Finally, using Eq.~\ref{eq:coutinho_system2} we yield $r = r(\lambda)$ by calculating the intersections between the curves $\mathcal{R}(x)$ and $\frac{x}{\lambda}$ for different values of $\lambda$.  

\begin{figure}[!t]
\begin{center}
\includegraphics[width=0.7\linewidth]{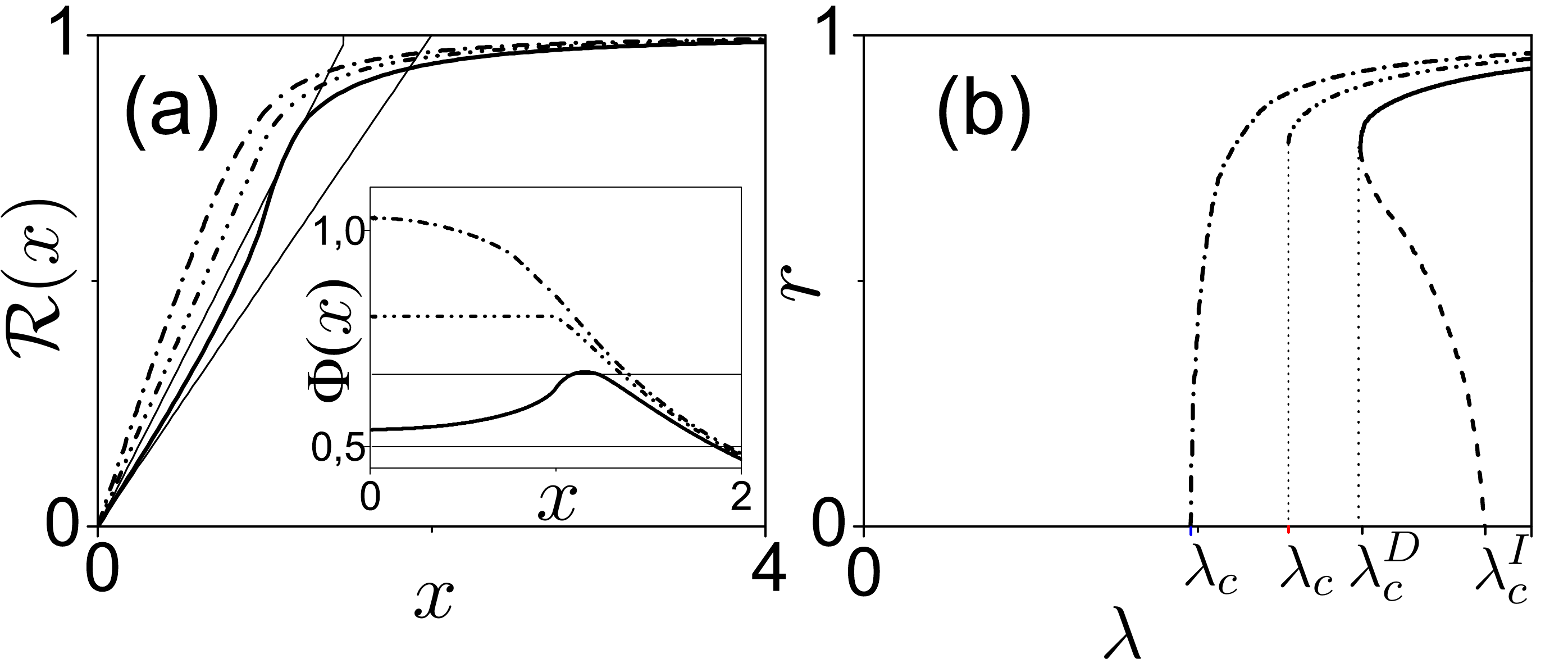}
\end{center}
\caption{ (a) Solutions of Eq.~\ref{eq:coutinho_order_parameter_final} for SF networks considering degree distribution
exponents $\gamma = 3.2$ (dash-dotted), $\gamma = 3$ (dash-dot-dotted) and $\gamma = 2.8$ (solid lines). The inset shows function $\Phi(x)$ for SF networks with degree distribution
exponents $\gamma = 3.2$ (dash-dotted), $\gamma = 3$ (dash-dot-dotted) and $\gamma = 2.8$ (solid lines). The thin solid lines correspond to the curves $\lambda = 1.5$ and $\lambda = 2$, respectively, whose intersections yield the order parameter $r$ as a function of coupling $\lambda$. (b) The order parameter $r$ vs. $\lambda$ for SF networks with $\gamma = 3$ (dash-dot-dotted) and $\gamma = 2.8$ (solid lines). Adapted with permission from~\cite{coutinho2013kuramoto}. Copyrighted by the American Physical Society.}
\label{fig:coutinho_kuramoto_fig2}
\end{figure}

Now we discuss this behavior for basic network classes: i) Figure~\ref{fig:coutinho_kuramoto_fig1}(a) shows $\mathcal{R}(x)$ (Eq.~\ref{eq:coutinho_system2}) for an ER network with the degree distribution $P(k) = \left\langle k \right\rangle^k e^{-\left\langle k \right\rangle}/k!$. The intersection between the solid  (Eq.~\ref{eq:coutinho_order_parameter_final}) and the dashed lines (Eq.~\ref{eq:coutinho_system2}) in Fig. ~\ref{fig:coutinho_kuramoto_fig1}(a) yields the curve in Fig.~\ref{fig:coutinho_kuramoto_fig1}(b). ii) The same process can be applied to networks with SF distributions $P(k) \sim k^{-\gamma}$. Fig.~\ref{fig:coutinho_kuramoto_fig2} shows the solution of Eqs.~\ref{eq:coutinho_system1} and~\ref{eq:coutinho_system2} for different values of $\gamma$. As we can see, for $\gamma = 3.2$ the phase transition
is continuous and is characterized by a critical coupling $\lambda_c$. For $\lambda < \lambda_c$ the only possible solution is the trivial one given by $r = 0$. The scenario changes for $2 < \gamma < 3$. As shown in Fig.~\ref{fig:coutinho_kuramoto_fig2}, for these values of $\gamma$, discontinuous transitions emerge for which a hysteresis, characterized by the two critical couplings, $\lambda_{c}^{\rm{D}}$ and $\lambda_{c}^{\rm{I}}$, takes place. Interestingly, as shown in~\cite{coutinho2013kuramoto}, for $\gamma = 3$ the phase transition of the order parameter $r$ as a function of the coupling $\lambda$ is discontinuous, however with the absence of hysteresis. This particular kind of phase transitions is known as \textit{hybrid phase transition} and is observed
in other dynamical processes, such as in $k$-core~\cite{dorogovtsev2006k,goltsev2006k} and bootstrap percolation~\cite{baxter2011heterogeneous} and in avalanches in interdependent networks~\cite{baxter2012avalanche} as well.

\begin{figure}[!t]
\begin{center}
\includegraphics[width=0.7\linewidth]{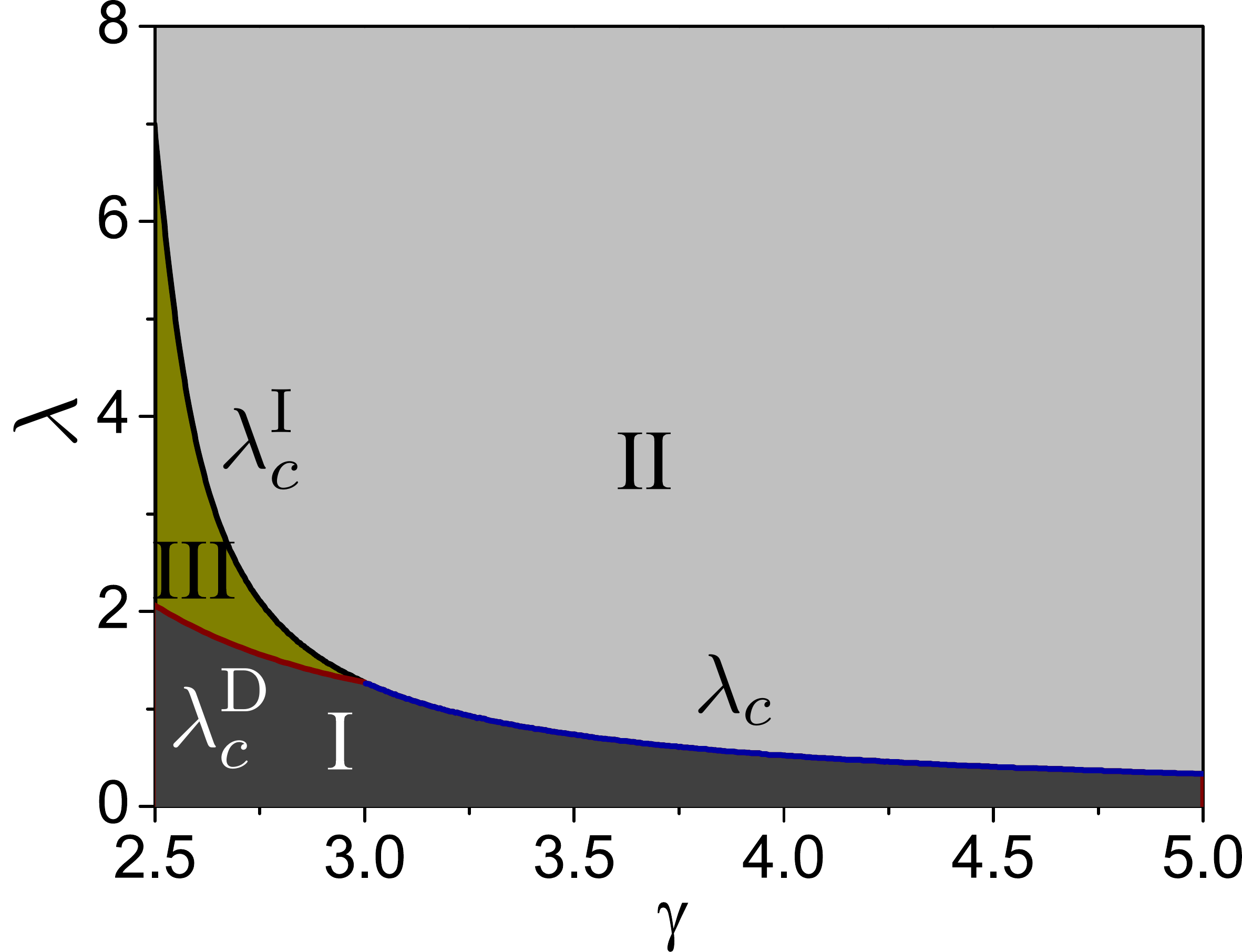}
\end{center}
\caption{ Phase diagram of the Kuramoto model with degree-correlated frequencies in SF networks. Area I denotes the region with incoherence, i.e., $r =0$. Synchronization emerges for parameters belonging to area II. The system will present hysteresis in area III, where there is bistability between the incoherent and synchronous state. Adapted with permission from~\cite{coutinho2013kuramoto}. Copyrighted by the American Physical Society.}
\label{fig:coutinho_kuramoto_fig3}
\end{figure}

To precisely analyze the nature of the phase transitions for the problem 
of correlated frequencies and degrees, Coutinho et al. defined the function~\cite{coutinho2013kuramoto}
\begin{equation}
\Phi(x)=\frac{1}{\lambda},
\label{eq:coutinho_function_Phi_I}
\end{equation}
where $\Phi(x) \equiv \mathcal{R}(x)/x$ and is given by
\begin{eqnarray}\nonumber
\Phi(x) & = & (\gamma-2)\left(\frac{\Omega(x)}{k_{\min}}\right)^{2-\gamma}\int_{-1}^{1}dy(1-x y)^{\gamma-3}\\
 &  & \times\sqrt{1-y^{2}}\Theta(1-x y)\Theta\left(x y-\frac{k_{\min}-\Omega(x)}{k_{\min}}\right),
 \label{eq:coutinho_function_Phi_II}
\end{eqnarray}   
where $y\equiv [k - \Omega(x)]/(x k )$. The horizontal lines in the inset of Fig.~\ref{fig:coutinho_kuramoto_fig2} are given by $1/\lambda$ and their intersection with $\Phi(x)$ gives the solutions of the order parameter $r$. To obtain a discontinuous transition, the function $\Phi(x)$
must have a maximum for $x \neq 0$. Since $\lim_{x\rightarrow \infty}\Phi(x) = 0$, $\Phi(x)$ must be an increasing function at $x =0$, i.e. it has its maximum value at $x \neq 0$. Thus, since $\left.\partial\Phi(x)/\partial x\right|_{x=0}=0$, a sufficient condition for the existence of a first-order transition is $\left.\partial^{2}\Phi(x)/\partial x^{2}\right|_{x=0}>0$, which is equivalent to~\cite{coutinho2013kuramoto}
\begin{equation}
\frac{(\gamma-4)(\gamma-3)}{4(\gamma-2)}-\frac{\Omega''(0)}{\Omega(0)}>0,
\label{eq:coutinho_condition_ES}
\end{equation}
where $\Omega''(0) = \left.\partial^2 \Omega(x)/\partial^2 x \right|_{x=0}$. Although there is no explicit solution of $\Omega$ as a function of $x$, it is possible to determine its dependence on 
exponent $\gamma$ for $x$. This dependence is given by~\cite{coutinho2013kuramoto}
\begin{equation}
\frac{\Omega(0)}{k_{\min}}-2=\frac{\pi^{2}}{4}(\gamma-3),
\label{eq:coutinho_Omega0}
\end{equation}
\begin{equation}
\frac{\Omega''(0)}{k_{\min}}\simeq1.71(\gamma-3)
\label{eq:coutinho_Omegaddot0}
\end{equation}
Inserting Eqs.~\ref{eq:coutinho_Omega0} and~\ref{eq:coutinho_Omegaddot0}
into Eq.~\ref{eq:coutinho_condition_ES} we obtain the phase
diagram depicted in Fig.~\ref{fig:coutinho_kuramoto_fig3}. Notice that three regions are uncovered. Region I is marked by the presence of incoherence, i.e., no synchronous behavior emerges for $\lambda < \lambda_c$. Synchronization ($r>0$) takes place in region II, where $\lambda > \lambda_c^{\rm{I}}$. Finally, the system will generate a hysteresis for parameters belonging to region III in which one stable and one metastable solution are found. Furthermore, by expanding the function $\Phi(x)$ using Taylor series around $x$ one finds that, 
for $\gamma = 3$, the order parameter follows $r - r_c \propto (\lambda - \lambda_c)^\beta$, where $\beta = 2/3$, which is the same critical exponent found by Paz\'o~\cite{pazo2005thermodynamic} for uniform frequency distributions~\cite{coutinho2013kuramoto}.

The findings in~\cite{coutinho2013kuramoto} yield important insights into the dynamics of Kuramoto oscillators with frequencies correlated with degrees in networks without degree-degree correlations. Specifically, it predicts that SF networks with $\gamma > 3$ undergo a second-order transition, changing to a hybrid transition at $\gamma = 3$ with no hysteresis and finally becoming a first-order transition for $\gamma < 3$.
However, as remarked by the authors in~\cite{coutinho2013kuramoto}, the result for $\gamma > 3$ is in apparent contradiction
to those presented by Gom\'ez-Garde\~nes et al.~\cite{gomez2011explosive}. More specifically, in~\cite{gomez2011explosive} the authors numerically found that SF networks with $\gamma = 3.3$ constructed through
the CM undergo a first-order transition, in contrast 
to what is predicted by the calculations in~\cite{coutinho2013kuramoto}. 
This inconsistency between both results
may be due to finite-size effects presented by networks of size $N = 10^3$ 
considered in~\cite{gomez2011explosive}. It is important to note that networks with finite-size 
constructed by the CM might exhibit non-vanishing values for the clustering-coefficient and also can have degree-degree correlations~\cite{newman2010networks}. Thus, this makes the assumption of 
networks in the absence of degree-degree correlations no longer valid leading, in this way, to a different 
result as the one predicted by diagram in 
Fig.~\ref{fig:coutinho_kuramoto_fig3}. As we shall see in the next sections, non-vanishing values of the assortativity coefficient can significantly change the synchronization properties of networks under correlation between frequencies and degrees. 

Besides analyzing the Kuramoto model for the frequency assignment $\omega_i = k_i$, Coutinho et al. also studied 
the model in star graphs~\cite{coutinho2013kuramoto}, i.e., networks made up of a central node connected to $K$ peripheral nodes with degree $k_j=1$, for $j=1,...,K$, where $K$ is the total number of peripheral nodes.
The equations for the model in the star graph are given by
\begin{equation}
\dot\theta_H = \omega_H + \lambda \sum_{j=1}^{K} \sin(\theta_j - \theta_H),
\label{eq:coutinho_hub_star_eq_motion}
\end{equation}
\begin{equation}
\dot\theta_j = \omega_j + \lambda \sin(\theta_H - \theta_j),
\label{eq:coutinho_peripheral_star_motion}
\end{equation}
where $\theta_H$ and $\theta_j$ are the phases of the central (or hub) node and the $j$-th peripheral node. By appropriately defining the order parameter as
\begin{equation}
r e^{i\psi} = \frac{1}{K}\sum_{j=1}^K e^{i\theta_j}, 
\label{eq:coutinho_order_parameter_star}
\end{equation} 
it is possible to decouple the Eqs~\ref{eq:coutinho_hub_star_eq_motion} and~\ref{eq:coutinho_peripheral_star_motion}  as 
\begin{equation}
\dot{\theta}_{H}-\Omega=(\omega_{H}-\Omega)-\lambda Kr\sin(\theta_{H}-\psi) \mbox{ and}
\label{eq:coutinho_eq_mov_dec_star_1}
\end{equation}
\begin{equation}
\dot{\theta}_{j}-\dot{\theta}_{H}=(\omega_{j}-\dot{\theta}_{H})-\lambda\sin(\theta_{j}-\theta_{H}).
\label{eq:coutinho_eq_mov_dec_star_2}
\end{equation}
The hub will be locked when $\dot \theta_H = \Omega$, satisfying
\begin{equation}
\omega_H - \Omega = \lambda K r \sin (\theta_H - \psi), 
\label{eq:coutinho_hub_star_locked}
\end{equation}
where $\Omega$ is the common frequency. Furthermore, the locked solution associated to 
Eq.~\ref{eq:coutinho_peripheral_star_motion} is given by
\begin{equation}
\omega_j - \Omega = \lambda \sin(\theta_H - \theta_j), 
\label{eq:coutinho_peripheral_star_locked}
\end{equation}
for $\left| \omega_j - \Omega \right| < \lambda$. On the other hand, if $\left| \omega_j - \Omega \right| > \lambda$ then 
the $j$-th peripheral node is not entrained by the mean-field. In this way, similarly as in Eq.~\ref{eq:coutinho_r_discrete}, the contributions 
to the order parameter can be explicitly written as~\cite{coutinho2013kuramoto}
\begin{equation}
re^{i(\psi-\theta_{H})}=\frac{1}{K}\sum_{|\omega_{j}-\Omega|\leq\lambda r}e^{i(\theta_{j}-\theta_{H})}+\frac{1}{K}\sum_{|\omega_{j}-\Omega|>\lambda r}e^{i(\theta_{j}-\theta_{H})}.
\label{eq:coutinho_order_parameter_star}
\end{equation} 
Considering that the peripheral nodes have identical frequencies so that frequencies and degrees are correlated in the star gaph, i.e. with frequency distribution
\begin{equation}
g(\omega) = \frac{1}{K}\sum_{j=1}^K \delta(\omega - \omega_j), 
\label{eq:coutinho_freq_dist_star}
\end{equation}
one can derive the following implicit equations for $\Omega$ and $r$~\cite{coutinho2013kuramoto},
\begin{equation}
r^{2}=\left(\frac{\Omega-\omega_{H}}{K\lambda}\right)^{2}+\left[\int_{-\lambda}^{+\lambda}d\omega g(\omega+\Omega)\sqrt{1-\left(\frac{\omega}{\lambda}\right)^{2}}\right]^2,
\label{eq:coutinho_r_star1}
\end{equation}
\begin{equation}
\frac{\Omega-\omega_{H}}{K}=\int_{|\omega|>\lambda}d\omega g(\omega+\Omega)\omega\left[1-\sqrt{1-\left(\frac{\lambda}{\omega}\right)^{2}}\right]. 
\label{eq:coutinho_r_star2}
\end{equation}
The average frequency $\Omega$ should be first determined through Eq.~\ref{eq:coutinho_r_star2}, which can then
be used to calculate the order parameter $r$ via Eq.~\ref{eq:coutinho_r_star1}. As mentioned in~\cite{coutinho2013kuramoto}, if $\omega_H - \left\langle \omega_j \right\rangle < \omega_c$, where $\left\langle \omega_j \right\rangle$ is the average frequency of the peripheral nodes and $\omega_c$ a certain threshold, the phase transition is continuous and $r$ increases as a function of $\lambda$. On the order hand, if $\omega_H - \left\langle \omega_j \right\rangle > \omega_c$ then the transition becomes discontinuous with the presence of hysteresis in the plane $\lambda-r$. It is interesting to note that the mechanism behind the emergence of abrupt transitions in star graphs is a 
sufficient large frequency mismatch between the hub and its neighbors. This gives also insights into the dynamics of SF networks with correlation between frequencies and degrees. In their topology hubs can locally
form star-like structures leading to a frequency mismatch between 
oscillators and, consequently, yielding a first-order transition~\cite{coutinho2013kuramoto,zou2014basin}. 

\begin{figure}[!t]
\begin{center}
\includegraphics[width=0.7\linewidth]{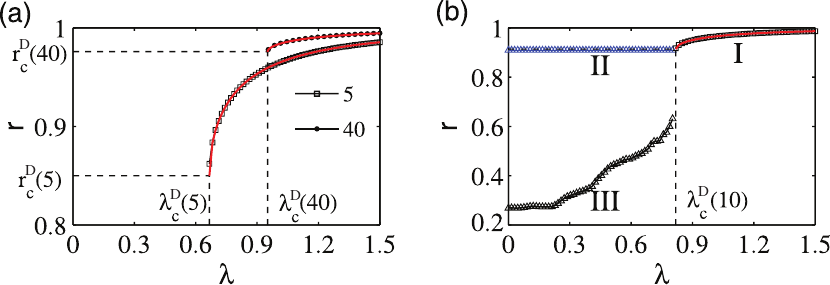}
\end{center}
\caption{(a) The order parameter $r$ as a function of coupling $\lambda$ for star-graphs with $K = 5$ and $40$  and  (b)  $K = 10$. Branch II corresponds to the dynamics without the inclusion of disorder $\zeta_j$, whereas I and III the frequencies of the peripheral nodes are given by $\omega_j = \omega + \zeta_j$, where $\zeta_j \in [-0.05, 0.05]$. Adapted from~\cite{zou2014basin}.}
\label{fig:zou_basin_fig1}
\end{figure}

Zou et al.~\cite{zou2014basin} also addressed the relation 
between the dynamics of star and SF networks
using a different approach, namely by analyzing the basin 
of attraction of the synchronized state. The 
authors defined first the state space 
of Eq.~\ref{eq:coutinho_hub_star_eq_motion} and~
\ref{eq:coutinho_peripheral_star_motion} as a $K + 1$ 
dimensional torus $\mathbb{T}^{K+1}$~\cite{zou2014basin} , 
where $K$ is the number of peripheral nodes in the star 
networks, as previously defined. By setting $\omega_H = K
\omega$ and $\omega_j = \omega$ ($j=1,...,K$)  in Eqs.~
\ref{eq:coutinho_hub_star_eq_motion} and~
\ref{eq:coutinho_peripheral_star_motion}; and defining $
\boldsymbol{\theta}=(\theta_1,..., \theta_K,\theta_H)$, $\boldsymbol{\omega} 
= (\omega,...,\omega,K\omega)$ and the function $
\mathbf{H}: \mathbb{T}^{K+1} \rightarrow \mathbb{T}^{K+1}$ 
given by $\left( \sin(\theta_H - \theta_1), \sin(\theta_H 
- \theta_2), ..., \sin(\theta_H - \theta_K), \sum_{j=1}^K 
\sin(\theta_j - \theta_H)\right)$, the Eqs.~
\ref{eq:coutinho_hub_star_eq_motion} and~
\ref{eq:coutinho_peripheral_star_motion} can be rewritten as
\begin{equation}
\dot{\boldsymbol{\theta} } = \boldsymbol{\omega} + \lambda \mathbf{H}(\boldsymbol{\theta} ). 
\label{eq:zou_basin_eq_mov}
\end{equation}
Therefore, the locking manifold is given by
\begin{equation}
M_a := \{ \boldsymbol{\theta}  \in \mathbb{T}^{K+1}: \theta_1 = \cdots = \theta_K \mbox{ and } \theta_{H} - \theta_1 = a  \}, 
\label{eq:zou_basin_locking_manifold}
\end{equation}
admitting the solutions $\dot{\boldsymbol{\theta} } = \boldsymbol{\omega} - \lambda \mathbf{H}(\mathbf{a)}$, where $\mathbf{a} = c(1, ..., 1) + (0, ..., 0, a)$, with $c$ being a constant real number, leading to $\boldsymbol{\theta} (t) = \left[\boldsymbol{\omega} - \lambda \mathbf{H}(\mathbf{a}) \right]t + \boldsymbol{\theta}_0$, where $\boldsymbol{\omega}_0 \in M_a$, and satisfying
\begin{equation}
-(K-1) \omega + \lambda(K+1)\sin a = 0. 
\label{eq:zou_basin_condition_backward}
\end{equation}
Equation~\ref{eq:zou_basin_condition_backward} is valid for $(K-1)\omega/(K+1)\lambda \leq 1$, which yields 
the critical coupling $\lambda_c^{\rm{D}}$ for the backward continuation~\cite{zou2014basin}:
\begin{equation}
\lambda_c^{\rm{D}} = \frac{(K-1)\omega}{K+1}.
\label{eq:zou_basin_critical_coupling_backward}
\end{equation}
For $\lambda> \lambda_c^{\rm{D}}$ and using the solutions provided by the locked manifold in Eq.~\ref{eq:zou_basin_locking_manifold}, one can get the full dependence of the order parameter
$r$ as a function of $\lambda$, which is given by
\begin{equation}
r^2=\frac{K^{2}+1}{(K+1)^{2}}+\frac{2K}{(K+1)^{2}}\sqrt{1-\left[\frac{(K-1)\omega}{(K+1)\lambda}\right]^{2}}, 
\label{eq:zou_basin_r2}
\end{equation}
which for $\lambda = \lambda_c^{\rm{D}}$ yields~\cite{zou2014basin}
\begin{equation}
r_c^{\rm{D}} = \frac{\sqrt{K^2+1}}{K + 1}. 
\label{eq:zou_basin_rcD}
\end{equation}
Equation~\ref{eq:zou_basin_rcD} gives the critical value 
of the order parameter at the transition point for the 
backward propagation. The comparison between the 
theoretical results in Eq.~\ref{eq:zou_basin_critical_coupling_backward} and~\ref{eq:zou_basin_rcD} and numerical simulations are shown 
in Fig.~\ref{fig:zou_basin_fig1}. Noteworthy, the 
expression for $r_c^{\rm{D}}$ obtained in~\cite{zou2014basin} 
differs from the one 
calculated in the original paper by G\'omez-Garde\~nez et al.~\cite{gomez2011explosive}, where the authors
obtained $r_c^{\rm{D}} = K/(K+1)$, considering $\omega = 1$. 
However,  if we tend the number of peripheral nodes
to zero ($K \rightarrow 0$), one expects to obtain perfect synchronization 
in the system composed only by a central node. This condition is included 
by the order parameter in Eq.~\ref{eq:zou_basin_rcD}, but not by the result 
presented in~\cite{gomez2011explosive}. 

In order to determine the conditions for synchronization for the case
in which the coupling strength is adiabatically increased, one should 
notice that the locking manifold $M_a$ in Eq.~\ref{eq:zou_basin_locking_manifold} is locally attractive for 
$\lambda \in [\lambda_c^{\rm{D}},\lambda_c^{\rm{I}}]$. Thus, as the coupling increases, $M_a$ becomes globally attractive for $\lambda > \lambda_c^{\rm{I}}$, i.e., any incoherent state for $\lambda > \lambda_c^{\rm{I}}$ is attracted to the manifold $M_a$. Using this information and the theory developed in~\cite{pereira2013connectivity,pereira2014TowardsATheoryDiffusive}, one gets that the critical coupling for the forward propagation 
is given by~\cite{zou2014basin} 
\begin{equation}
\lambda_c^{\rm{I}} \thickapprox \left( \frac{K-1}{B\sqrt{K}}\right)\omega,
\label{eq:zou_basin_critical_coupling_forward}
\end{equation} 
where $B$ is a constant that depends on the initial conditions~\cite{zou2014basin}. As we can see,  Eq.~\ref{eq:zou_basin_critical_coupling_forward} shows that the onset of synchronization in a star-graph increases with the system size. This
dependence was also numerically observed in the original paper by G\'omez-Garde\~nez et al.~\cite{gomez2011explosive}. 

\begin{figure}[!t]
\begin{center}
\includegraphics[width=0.7\linewidth]{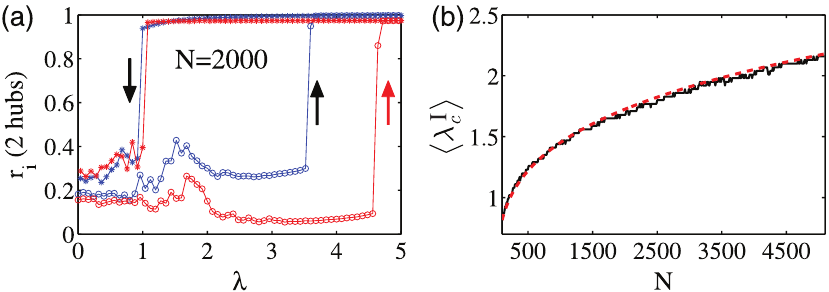}
\end{center}
\caption{(a) The local order parameter $r_i$ as a function of coupling $\lambda$ of two hubs selected from a BA network with $N = 2\times 10^3$ nodes and average degree $\left\langle k \right\rangle = 2$. Red curve: hub with degree $K_1 = 39$; 
blue curve: hub with degree $K_2 = 24$. (b) Critical coupling $\left\langle \lambda_c^{\rm{I}} \right\rangle$ of the forward propagation of $\lambda$ as a function of network size $N$. $\left\langle \cdot \right\rangle$ denotes an average over  50 different network realizations. Adapted from~\cite{zou2014basin}. Copyrighted by the American Physical Society.}
\label{fig:zou_basin_fig2}
\end{figure}

As previously mentioned, the analysis of the dynamics in star-graphs has the potential to qualitatively describe the emergence of explosive 
synchronization in SF networks, since the latter can be regarded
as a collection of interconnected hubs~\cite{coutinho2013kuramoto,zou2014basin}. Hubs and their low-degree neighbors form locally star-like structures creating, in this way, local locking manifolds as in Eq.~\ref{eq:zou_basin_locking_manifold}. In fact, the result in Eq.~\ref{eq:zou_basin_critical_coupling_forward} can be used 
to estimate the critical coupling of SF networks with degree 
distributions $P(k) \sim k^{-\gamma}$. The maximum degree for a SF network with the
critical exponent $\gamma$ scales as $k_{\max} \propto N^{\frac{1}{\gamma-1}}$~\cite{newman2010networks}. Therefore, using 
Eq.~\ref{eq:zou_basin_critical_coupling_forward} and considering an 
average of a network ensemble, one obtains the scaling relation
\begin{equation}
\left\langle \lambda_c^{\rm{I}} \right\rangle \propto  N^{\frac{1}{2(\gamma-1)}}.
\label{eq:zou_basin_critical_coupling_SF}
\end{equation} 
In order to sustain this hypothesis, Fig.~\ref{fig:zou_basin_fig2}(a) shows the local order parameter $r_i$ of two hubs in a SF network generated by the BA model with $N = 2000$ and $\left\langle k \right\rangle = 2$. As we can see, the parameters $r_i$ also undergo discontinuous phase transitions as a function of $\lambda$ and also agree with the limit $\lambda_c^{\rm{D}}\rightarrow \omega$, for $K\gg 1$ in Eq.~\ref{eq:zou_basin_critical_coupling_backward}. Furthermore, in Fig.~\ref{fig:zou_basin_fig2} (b) we see a good agreement between the scaling relation predicted in Eq.~\ref{eq:zou_basin_critical_coupling_SF} and the simulation results.
Noteworthy, alternative solutions for the synchronization of star graphs with frequency-degree correlations were recently 
proposed in~\cite{vlasov2015ExplosiveSyncIsDiscontinuous} using the Wattanabe-Strogatz approach~\cite{watanabe1993integrability,watanabe1994constants,pikovsky2008partially,pikovsky2011dynamics}, and in~\cite{jiang2015LowDimensionalES}, where the OA theory was employed. The corresponding stability analysis revealed that
only positive correlation between the frequencies and degrees yields discontinuous transitions in star-graphs. In particular, 
it was demonstrated that, for the forward propagation of coupling $\lambda$, after reaching the critical coupling $\lambda_c^{\rm{I}}$ the asynchronous state loses its stability, being replaced by the fixed point related 
to the synchronized state~\cite{vlasov2015ExplosiveSyncIsDiscontinuous}.  

Before reviewing other different types of correlations between topology and intrinsic dynamics introduced
in the literature, it is interesting to analyse situations in which the system is perturbed by properties
found in real applications. First, we consider when the condition $\omega_i = k_i$ is slightly disturbed 
by random fluctuations. Zou et al.~\cite{zou2014basin} also investigated the dynamics of star-graphs whose phases of the peripheral nodes
evolve according to 
\begin{equation}
\dot{\theta}_i = \omega + \zeta_i + \lambda\sin(\theta_H - \theta_i),\; (i=1,...,K), 
\label{eq:zou_basin_eq_of_motion_star_quenched}
\end{equation} 
where $\zeta_j$ is a variable randomly drawn from a uniform distribution in the range $\left[-\varepsilon,\varepsilon\right]$.
Note that only the hub is not subjected to a random frequency perturbation. Regarding the backward propagation 
of coupling $\lambda$, it was verified that for $\lambda > 
\lambda_c^{\rm{D}}$ the locking manifold $M_a$ is not 
destroyed by such random fluctuations, once the magnitude 
of fluctuation $\varepsilon$ is small enough (see region I 
in Fig.~\ref{fig:zou_basin_fig1}(b)). On the other hand, 
for values of coupling $\lambda < \lambda_c^{\rm{D}}$ the 
manifold $M_a$ is destroyed leading the peripheral nodes 
to oscillate incoherently at different frequencies, as 
seen in branch III in Fig.~\ref{fig:zou_basin_fig1}(b). 
The comparison with the dynamics in the absence 
of disorder is shown in branch II of the same figure. 
\begin{figure}[!t]
\begin{center}
\includegraphics[width=0.7\linewidth]{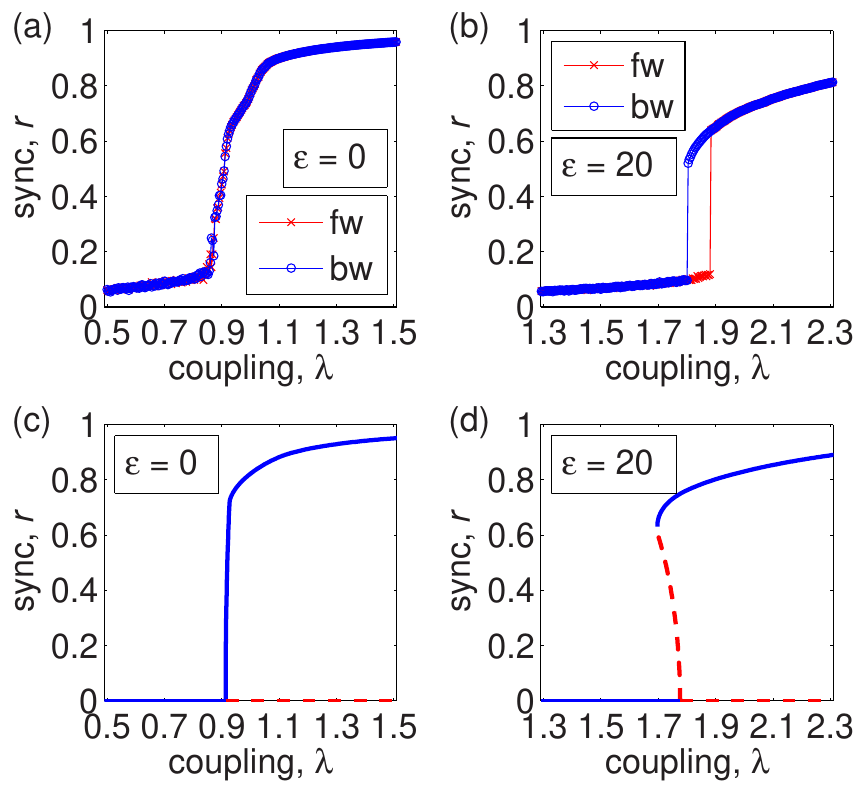}
\end{center}
\caption{ Synchronization diagram of the forward (fw) and backward (bw) propagation for SF networks with $N = 1000$ for (a) $\varepsilon = 0$ and (b) $\varepsilon = 20$. Panels (c) and (d) correspond to the theoretical curves of simulations in (a) and (b), respectively. All networks have $\gamma = 3.5$ and $k_{\min} = 10$. Reprinted with permission~\cite{skardal2014disorder}. Copyright 2014 by the American Physical Society.}
\label{fig:skarlda_disorder_fig1}
\end{figure}

The effect of addition of disorder to disturb the 
frequency assignment $\omega_i = k_i$ was further 
analyzed by Skardal and Arenas~
\cite{skardal2014disorder} for more heterogeneous 
structures, namely for SF  and stretched 
exponential networks with a degree distribution given 
by $P(k) \propto k^{b - 1}\exp[- \left(k/\mu 
\right)^b ]$. As shown in~
\cite{coutinho2013kuramoto}, SF networks with $
\gamma > 3$ no longer exhibit discontinuous phase 
transitions. However first-order transitions can be 
induced for cases in which the phase transitions are 
continuous under the condition $\omega_i = k_i$ if 
$\varepsilon$ is large 
enough~\cite{skardal2014disorder}. The effect of 
addition of disorder
on the order parameter $r$ can be seen in Fig.~\ref{fig:skarlda_disorder_fig1} for a network with $
\gamma = 3.5$. For $\varepsilon = 0$ (Fig.~\ref{fig:skarlda_disorder_fig1}(a)) the transition is 
continuous, as predicted in the theoretical curve in 
Fig.~\ref{fig:skarlda_disorder_fig1}(c). Yet, when $
\varepsilon$ is increased to $\varepsilon = 20$, a 
discontinuous phase transition with the presence of 
hysteresis emerges, as it is seen in the simulation in 
Fig.~\ref{fig:skarlda_disorder_fig1}(b) and the 
respective theoretical curve in Fig.~
\ref{fig:skarlda_disorder_fig1}(d). It is
interesting to note also that explosive 
synchronization as an effect
of inclusion of quenched disorder only arises for 
strong values of $\varepsilon$. More precisely, for 
less heterogeneous networks, even with perfect 
correlation between frequencies and degrees, explosive 
synchronization is not observed. The disorder 
introduced in the system
must overcome this homogeneity in the network structure so that the order parameter exhibits discontinuity. However, the larger the strength of disorder $\varepsilon$, the weaker the influence of degrees in the overall frequency $\omega_i = k_i + \zeta_i$. Therefore, if the magnitude of fluctuation $\varepsilon$ is sufficiently large, the frequency distribution is nearly uniform, leading to an analogous case as earlier considered by Paz\'o~\cite{pazo2005thermodynamic}. 

\begin{figure}[!t]
\begin{center}
\includegraphics[width=0.7\linewidth]{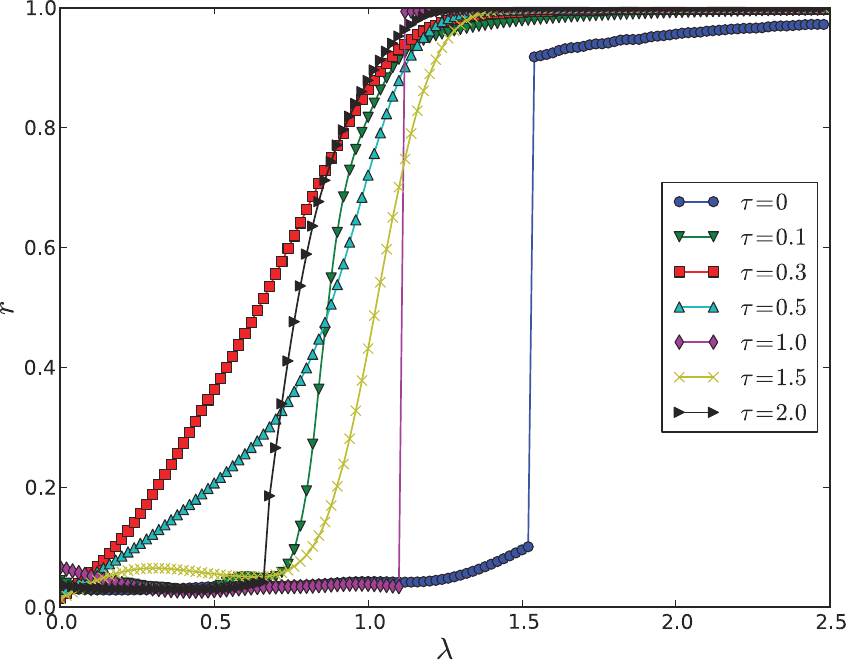}
\end{center}
\caption{ The order parameter $r$ vs. coupling strength $\lambda$ for SF networks with $\gamma = -3$ and $N = 10^3$ oscillators considering different time delays $\tau$. Reprinted with permission from~\cite{peron2012explosive}. Copyright
2012 by the American Physical Society.}
\label{fig:peron_delay_fig1}
\end{figure}

Another scenario found in many real-world systems is 
that
the interaction between the dynamical units may not 
occur instantaneously, but rather 
only after a time delay~\cite{lakshmanan2011dynamics} 
(see also Sec.~\ref{sec:time_delay}).  
Peron and Rodrigues~\cite{peron2012explosive} 
numerically investigated the Kuramoto model in time-delayed networks 
\begin{equation}
\dot{\theta}_i(t) = \omega_i + \lambda \sum_{j=1}^N A_{ij}\sin\left[\theta_j(t-\tau)  - \theta_i(t) \right], 
\label{eq:peron_delay_equation}
\end{equation}
where $\omega_i = k_i $ and $\tau$ is the time delay. 
The synchronization diagram considering BA networks is 
shown in Fig.~\ref{fig:peron_delay_fig1}. 
It is interesting to observe how the coherence of the 
populations of oscillators as a function of coupling 
dramatically changes when a time delay is included in 
the system. Even small values of $\tau$ are able to 
decrease the critical coupling, enhancing in this way 
the synchronization. Furthermore, besides enabling the 
network to reach the synchronous state for lower 
coupling strengths, the inclusion of a time delay also 
modifies the type of the phase transition. For 
instance, for $\tau = 1$, 
a discontinuous transition is obtained (similarly as in 
the original case with $\tau = 0$),  while for $
\tau=2$, the system
exhibits a continuous transition. It is also worth 
noting that some choices of $\tau$, namely $\tau = 0.1$ 
and $0.5$, yield 
non-usual scaling of the order parameter for lower 
values of the coupling strength. The authors in~
\cite{peron2012explosive} analytically studied the 
dynamics in star-graphs, which qualitatively explains 
the dependency of the order parameter observed
in Fig.~\ref{fig:peron_delay_fig1}. However, the mean-field solution of the model 
(\ref{eq:peron_delay_equation}) for degree-correlated 
frequencies in SF networks is still open. In 
fact, the detailed investigation of such a problem 
could yield insights into real applications in which 
the hysteric behavior can be avoided by the addition of 
time delay in the communication of the dynamical units. 

\subsection{Other kinds of correlations}

Naturally the results by G\'omez-Garde\~nes et al.~\cite{gomez2011explosive} raised many questions
regarding the dynamics of Kuramoto oscillators in networks. Of particular interest is the problem of whether other types of correlations between topology and intrinsic dynamics of the oscillators also lead to discontinuous synchronization transitions. 

\begin{figure}[!t]
\begin{center}
\includegraphics[width=0.7\linewidth]{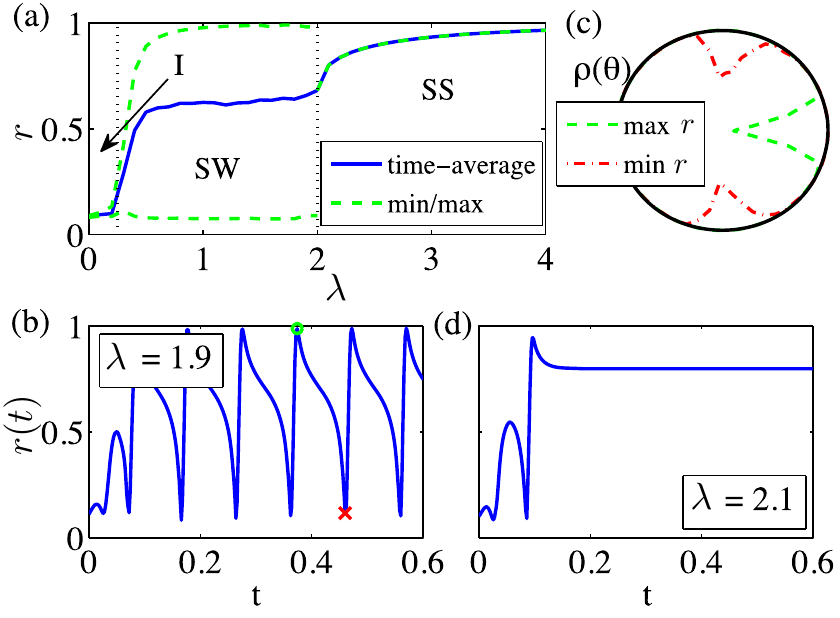}
\end{center}
\caption{ Numerical analysis of the Kuramoto model in an ER network with correlation between frequencies and degrees given by Eq.~\ref{eq:skardal_epl_correlation_function} with $N = 10^3$ and $\left\langle k \right\rangle = 10$. (a) Time-averaged order parameter $R$ as a function of coupling $\lambda$. (b) Order parameter $R$ as a function of time for $\lambda = 1.9$ and (c) the phase distribution $\rho(\theta)$ for points in red and green in panel (b). (d) Order parameter as a function of $t$ for $\lambda = 2.1$. Here, the acronyms I, SW and SS stand for incoherent, standing wave and stationary state, respectively. Adapted with permission from~\cite{skardal2013effects}.}
\label{fig:skarlda_epl_fig1}
\end{figure}

For instance, Skardal et al.~\cite{skardal2013effects} analyzed the Kuramoto model in uncorrelated networks where the frequencies and degrees are subject to the 
following joint probability distribution
\begin{equation}
P(k,\omega) = \frac{P(k)}{2} \left[ \delta(\omega - a k^b) + \delta(\omega + a k^b\right)],
\label{eq:skardal_epl_correlation_function}
\end{equation}
i.e., the frequency of oscillator $i$ is $\omega_i = \pm a k_i^{b}$. For the original case in which $a = b = 1$ and only positive frequencies 
are considered, the correlation between frequencies and degrees in~\cite{gomez2011explosive} is 
recovered. The model subjected to correlation between frequencies and 
degrees imposed by Eq.~\ref{eq:skardal_epl_correlation_function} exhibits 
significant different properties compared to the one in~\cite{gomez2011explosive}. Specifically, Fig.~\ref{fig:skarlda_epl_fig1}(a) shows the order parameter $r$ (Eq.~\ref{eq:restrepo_r_rn}) as a
function of coupling $\lambda$ for an ER network. One can identify three different regimes, namely (i) incoherent state, (ii) standing wave and (iii) stationary state~\cite{skardal2013effects}. State (i) is characterized by the 
absence of synchronization for $\lambda< \lambda_1$. In the (ii) regime, 
a spontaneous drift of the order parameter $r$ appears resulting 
in a time dependent behavior observed in Fig.~\ref{fig:skarlda_epl_fig1}(b) (the phase distribution $\rho(\theta)$ for two extreme values of $r$ can be seen in Fig.~\ref{fig:skarlda_epl_fig1}(c)). On the other hand, 
for $\lambda>\lambda_2$ the (iii) regime is reached for which the order parameter $r$ is time-independent, as shown in Fig.~\ref{fig:skarlda_epl_fig1}(d). 

Besides the standing wave state induced by the bimodal distribution in Eq.~\ref{eq:skardal_epl_correlation_function}, the connectivity pattern of the entrained oscillators strongly varies depending on the exponent $b$. More specifically, for $b < 1$ (sublinear) high degree nodes will form a
synchronized component in the network, while the drifting oscillators will be
mostly composed of low degree nodes. This scenario inverts for $b > 1$ (super-linear), i.e., the synchronized cluster is now dominated by low degree nodes. For the linear correlation regime, $b = 1$, the stationary state is characterized by a full locking, i.e., 
the absence of drifting oscillators. 

In order to get analytical insights into this behavior, it is convenient to define the order 
parameters for positive and negative frequencies as $r_i^{\pm}e^{i\psi^{\pm}} = \sum_{\omega_j\lessgtr 0}A_{ij}e^{i\theta_j}$
and $r^{\pm} = \frac{1}{N}\sum_{j=1}^N r_j^{\pm}/k_j^{\pm}$, where $k_j^{\pm}$ is the degree of node $j$ due to connections with nodes with
positive (resp. negative) frequencies. Using these definitions, the equations
of motion can be decoupled as 
\begin{equation}
\dot{\theta}_i = \omega_i + \lambda \left[r^{+}_i \sin(\psi_i^{+} - \theta_i) + r^{-}_i \sin(\psi_j^{-} - \theta_i) \right].
\label{eq:skardal_epl_decoupled_eq_of_motion}
\end{equation}
Setting the rotating frame $\phi_i = \theta_i - \Omega t$ and defining $\psi_i^{\pm} = \pm \Omega t$, the equations of motion can be rewritten as
\begin{equation}
\dot{\phi}_i = (\omega_i - \Omega) - \lambda r^{+}_i \sin\phi_i - \lambda r_i^{-}\sin(\phi_i + 2\Omega t).
\label{eq:skardal_epl_decoupled_eq_of_motion_rot_frame}
\end{equation}
Imposing the phase-locked solution $\dot\phi_i = 0$, we get the following relations~\cite{skardal2013effects}
\begin{equation}
r^{+}=\frac{1}{\left\langle k\right\rangle }\int_{2\left|a k^{b}-\Omega\right|\leq\lambda r^{+}k}P(k)k\sqrt{1-\frac{4\left(a k^{b}-\Omega\right)^{2}}{\left(\lambda r^{+}k\right)^{2}}},
\label{eq:skardal_epl_Rp}
\end{equation}
\begin{equation}
\Omega=a\frac{\int_{2\left|a k^{b}-\Omega\right|\leq\lambda r^{+}k}P(k)k^{b}dk}{\int_{2\left|a k^{b}-\Omega\right|\leq\lambda r^{+}k}P(k)dk},
\label{eq:skardal_epl_Omega}
\end{equation}
which are also satisfied by replacing $r^{+} \rightarrow r^{-}$~\cite{skardal2013effects}. Furthermore, for the SS regime, i.e. for $\Omega = 0$, Eq.~\ref{eq:skardal_epl_Rp} reduces to 
\begin{equation}
r=\frac{1}{\left\langle k\right\rangle }\int_{a k^{b}\leq\lambda r k}P(k)k\sqrt{1-\left(\frac{a k^{b}}{\lambda rk}\right)^{2}}dk.
\label{eq:skardal_effect_epl_R}
\end{equation} 
From Eq.~\ref{eq:skardal_effect_epl_R} it becomes clear that the three possible regimes are induced in dependence on the parameter $b$. Specifically, for $b > 1$ oscillators with degree $k\leq \left(\frac{\lambda r}{a}\right)^{\frac{1}{b - 1}}$ become entrained by the mean-field, whereas the oscillators with higher degrees become drifting. For $b < 1$ the range of degree of the synchronized oscillators is given by $k \geq \left( \frac{a}{\lambda r}\right)^{\frac{1}{1-b}}$. On the other hand, if $b = 1$, the dependence on $k$ in Eq.~\ref{eq:skardal_effect_epl_R} is lost and all oscillators become locked, once the critical coupling is reached and $\Omega=0$. 

\begin{figure}[!t]
\begin{center}
\includegraphics[width=1.0\linewidth]{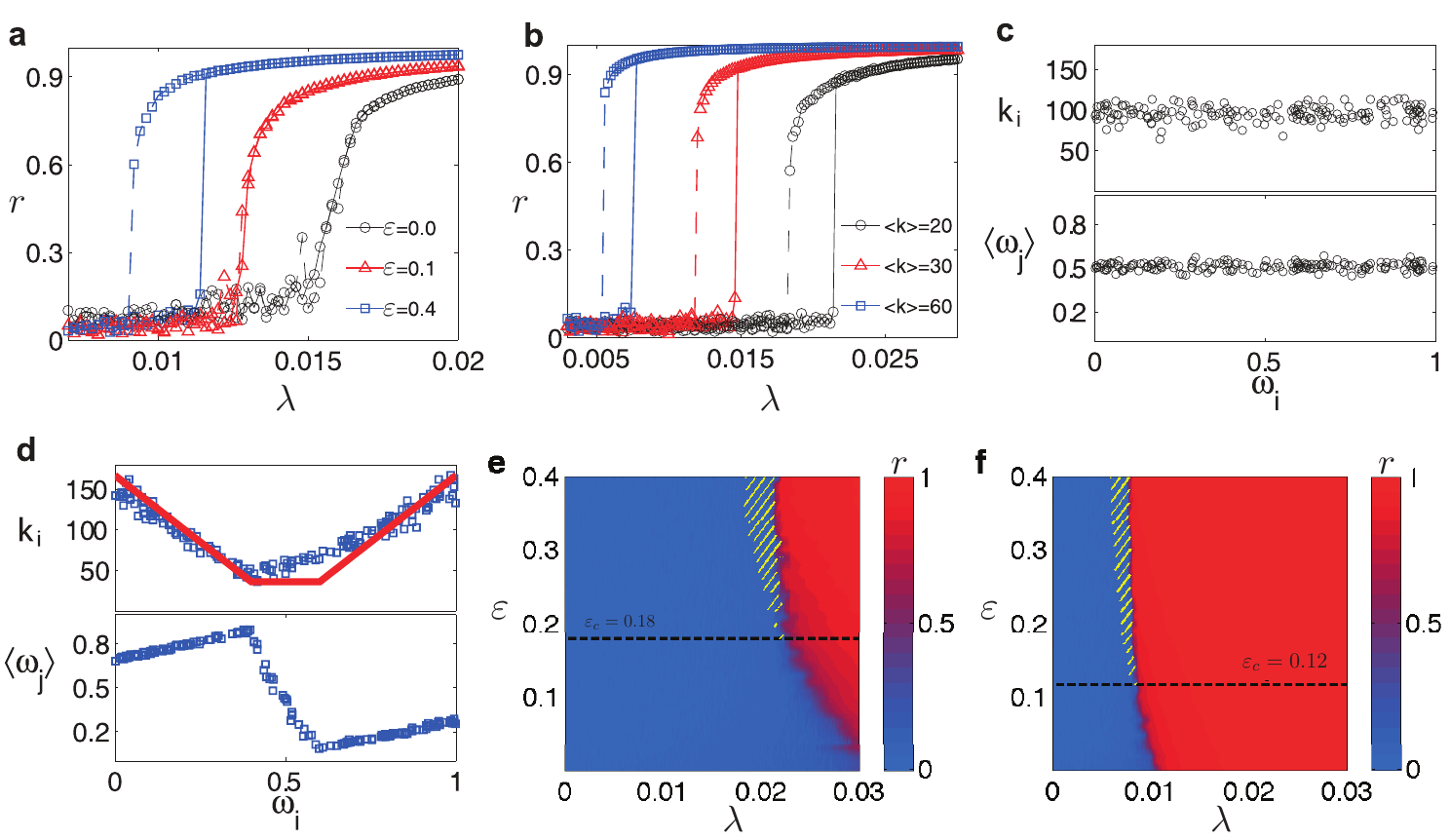}
\end{center}
\caption{ Synchronization diagrams for networks satisfying condition in 
Eq.~\ref{eq:leyva_scirep_condition_1} for (a) $\left\langle k 
\right\rangle = 40$ and different values of threshold $\gamma$, (b) 
$\varepsilon = 0.4$ and varying average degree $\left\langle k \right\rangle$. 
Solid (dashed) lines stand for the forward (backward) propagation of 
coupling $\lambda$. (c) Scatter plots of the degrees $k_i$ against 
frequencies $\omega_i$ in a network constructed under condition in 
Eq.~\ref{eq:leyva_scirep_condition_1} with $\left\langle k \right\rangle = 
100$ and frequency mismatch $\varepsilon = 0$; and (d) $\varepsilon = 0.4$.  
Parameter space $\lambda-\varepsilon$ coloured according the order parameter 
$r$ for (e) $\left\langle k \right\rangle = 20$ and (f) $\left\langle k 
\right\rangle = 20$. All networks have $N = 500$ oscillators. Reprinted by permission from Macmillan Publishers Ltd: Scientific Reports~\cite{leyva2013explosivescirep}, copyright 2013.}
\label{fig:leyva_scirep_fig1}
\end{figure}

The emergence of discontinuous synchronization transitions has been 
described as an effect exclusively due to the microscopic correlation
between dynamics and local topology. This now poses the question
of whether abrupt transitions can be observed in networks where 
the constraint $\omega_i \propto k_i $ does not hold and what would 
be the necessary conditions for the existence of such dynamical behavior in networks. 
One of the first steps towards the understanding of this problem was given by
Leyva et al.~\cite{leyva2013explosivescirep}. The authors remarked that, in order to obtain a
discontinuity in the order parameter as a function of coupling, one
should avoid a smooth emergence of a giant synchronous components. With this hypothesis Leyva et al. elaborated the condition 
for frequency assignment stating that every pair of nodes should satisfy~\cite{leyva2013explosivescirep}
\begin{equation}
|\omega_i - \omega_j|>\varepsilon_c.
\label{eq:leyva_scirep_condition_1}
\end{equation}
More specifically, the condition (\ref{eq:leyva_scirep_condition_1}) 
inhibits the formation of small synchronous clustering by imposing 
a gap in the frequency mismatch or, in other words, a frequency dissortativity. Figure~\ref{fig:leyva_scirep_fig1} shows simulations with ER networks
created following Eq.~\ref{eq:leyva_scirep_condition_1}. As we can see 
in Fig.~\ref{fig:leyva_scirep_fig1}(a), for sufficient large $\varepsilon$ the transition 
becomes discontinuous, a phenomenon not observed for the same 
network topology under the constraint $\omega_i = k_i$~\cite{gomez2011explosive,peron2012determination,coutinho2013kuramoto}.
Moreover, it is interesting to note in Figs.~\ref{fig:leyva_scirep_fig1}(c) and (d) that a correlation between frequencies and degrees naturally emerges by imposing the frequency mismatch beforehand. Figs.~\ref{fig:leyva_scirep_fig1} (e) and (f) further show the calculation of the 
critical mismatch $\varepsilon_c$ for $\left\langle k \right\rangle = 20$ and $60$, respectively.
Interestingly, it is possible to relax the condition in Eq.~\ref{eq:leyva_scirep_condition_1}, while preserving
the discontinuity of the order parameter. More specifically, it was shown that condition (\ref{eq:leyva_scirep_condition_1}) can be softened to
\begin{equation}
|\omega_i  - \left\langle  \omega_j \right\rangle |> \varepsilon_c, 
\label{eq:leyva_scirep_condition_2}
\end{equation}
where $\left\langle \cdots \right\rangle$ stands for the average over the neighbours of node $j$~\cite{leyva2013explosivescirep}. Thus, Eq.~\ref{eq:leyva_scirep_condition_2} imposes that two nodes $i$ and $j$ will be connected only if the difference between frequency $\omega_i$ overcomes the local mean-field of the neighborhood of the node $j$. As in the case of Eq.~\ref{eq:leyva_scirep_condition_1}, a spontaneous emergence of correlation between frequencies and degrees is observed~\cite{leyva2013explosivescirep}. Remarkably, the condition in Eq.~\ref{eq:leyva_scirep_condition_2} imposed in~\cite{leyva2013explosivescirep} is exactly the same condition $|\omega_i - \left\langle \omega_j \right\rangle|> \omega_c$ analytically found to obtain first-order transitions in star-graphs~\cite{coutinho2013kuramoto}.

\begin{figure}[!t]
\begin{center}
\includegraphics[width=0.6\linewidth]{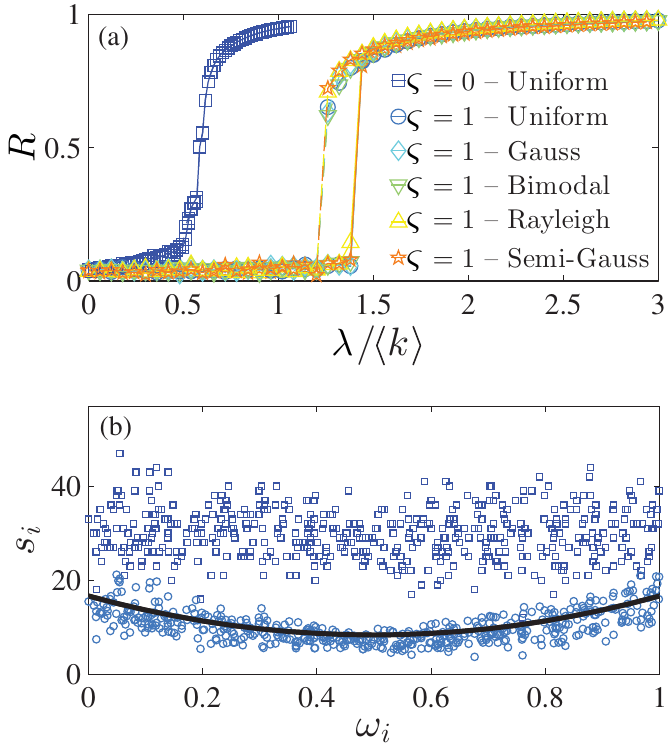}
\end{center}
\caption{(a) Order parameter $R$ (Eq.~\ref{eq:KM_ORDER_PARAMETER}) as a function of $\lambda$ for the model defined in Eq.~\ref{eq:leyva_weighted_eq_mov} considering $N = 500$, $\left\langle k \right\rangle = 30$ and different values of exponent $\varsigma$ and different choices for frequency distribution $g(\omega)$ in the range $[0,1]$. Solid (dashed) lines correspond to the forward (backward) propagation of coupling $\lambda$. (b) Scatter plots of the nodes' strength $s_i = \sum_j A_{ij} \left|\omega_i - \omega_j \right|^{\varsigma}$ vs. natural frequency $\omega_i$, where dark blue squares and circles correspond to the cases $\varsigma = 0$ and $\varsigma = 1$, respectively, for uniform frequency distribution in panel (a). Solid line represents the analytical prediction $s \sim \left( \omega - \frac{\varepsilon}{2}\right)^2 + \frac{1}{4\varepsilon}$ (Eq.~\ref{eq:leyva_weighted_strength}) for $\varepsilon=1$. Adapted with permission from~\cite{leyva2013explosive}. Copyrighted by the American Physical Society.}
\label{fig:leyva_weighted_fig1}
\end{figure}

Motivated by these findings that explosive synchronization is achievable for any frequency distributions, Leyva et al.~\cite{leyva2013explosive}  further analysed the following model 
\begin{equation}
\dot{\theta}_i = \omega_i + \frac{\lambda}{\left\langle k \right\rangle}\sum_{j=1}^N \Gamma_{ij}^{\varsigma} \sin(\theta_j - \theta_i).
\label{eq:leyva_weighted_eq_mov}
\end{equation} 
The frequencies $\omega_i$ are distributed according to a frequency distribution $g(\omega)$ and the link weights 
are given by 
\begin{equation}
\Gamma_{ij}^{\varsigma} = A_{ij}\left|\omega_i - \omega_j \right|^{\varsigma}, 
\label{eq:leyva_weighted_gamma}
\end{equation}
The interactions between the oscillators depend now on the frequency mismatch, 
however without imposing any frequency dissortativity as in Eqs.~\ref{eq:leyva_scirep_condition_1} and~\ref{eq:leyva_scirep_condition_2}. Figure~\ref{fig:leyva_weighted_fig1}(a) shows the synchronization diagram calculated
by numerically evolving the equations (\ref{eq:leyva_weighted_eq_mov}) for several choices of distribution
$g(\omega)$ and considering an ER network with $N=500$ nodes and $\left\langle k \right\rangle = 15$. It is interesting to note that the transition is second-order for a uniform distribution $g(\omega)$ and $\varsigma=0$, whereas for $\varsigma = 1$ with the same $g(\omega)$ the transition becomes discontinuous. Furthermore, once the range of frequencies is kept fixed ($\omega \in [0,1]$), the dependence of the order parameter $r$ on $\lambda$ remains
unchanged. 

Introducing the weights $\Gamma_{ij}^{\varsigma}$ also naturally induces a correlation between the nodes strengths
$s_i$ and the natural frequencies $\omega_i$, as seen in Fig.~\ref{fig:leyva_weighted_fig1}(b). The correlation $s_i \sim \omega_i$ spontaneously emerges as a consequence of 
the weighting procedure~\cite{leyva2013explosive}, in contrast with the correlation between frequencies and degrees considered in~\cite{gomez2011explosive}, where 
such constraint is imposed in order to yield discontinuous transitions. In fact, the dependence of $s$ on $\omega$
can be explicitly obtained for a fully connected graph. Thus, considering the model
\begin{equation}
\dot{\theta}_i = \omega_i + \frac{\lambda}{N}\sum_{j=1}^N \left|\omega_i - \omega_j \right| \sin(\theta_j - \theta_i), 
\label{eq:leyva_weighted_eqmotion_2}
\end{equation}    
one can express the equations as~\cite{leyva2013explosive}
\begin{equation}
\dot{\theta}_i =\omega_i + \lambda r_i \sin(\psi_i - \theta_i), 
\label{eq:leyva_weighted_eqmotion_2_decoupled}
\end{equation}
where $r_i$ and $\psi_i$ are the amplitude and the phase of the local mean-field, respectively; i.e., 
\begin{equation}
r_i e^{i\psi_i} = \frac{1}{N} \sum_{j=1}^N \left|\omega_i - \omega_j \right| e^{i\theta_j}. 
\label{eq:leyva_weighted_local_mean_field}
\end{equation}
The stationary solution in the continuum limit gives $\omega = \lambda r(\omega) \sin[\theta(\omega) - \psi(\omega)].$ 
By defining the functions
\begin{equation}
F(\omega)= r(\omega) \sin\psi(\omega) =  \int g(y)\left|\omega-y\right|\sin\theta(y)dy\nonumber,
\end{equation}
\begin{equation}
G(\omega)= r(\omega) \cos\psi(\omega) =  \int g(y)\left|\omega-y\right|\cos\theta(y)dy\nonumber,
\end{equation}
Eq.~\ref{eq:leyva_weighted_eqmotion_2_decoupled} in the stationary state is rewritten as
\begin{equation}
\frac{2}{\lambda}g(\omega)\omega=G(\omega)\frac{d^{2}F(\omega)}{d\omega^{2}}-F(\omega)\frac{d^{2}G(\omega)}{d\omega^{2}}.
\label{eq:leyva_weighted_FG_equation}
\end{equation}
Noting that $G(\omega) \approx r s(\omega)$ when the system is close to the synchronous state and considering a
uniform frequency distribution $g(\omega)$ in the range $\omega \in [-\varepsilon/2,\varepsilon/2]$, the obtained strength $s$ is
\begin{equation}
s(\omega) = \varepsilon \left[ \left(\frac{\omega}{\varepsilon} \right)^2 + \frac{1}{4}\right], 
\label{eq:leyva_weighted_strength}
\end{equation}
which is the black curve depicted in Fig.~\ref{fig:leyva_weighted_fig1}(b). Moreover, the order parameter $r$ can also be obtained analytically 
determined in a similar way as previously. More specifically, it can be shown that in the case
of the model in Eq.~\ref{eq:leyva_weighted_eqmotion_2} for uniform frequency distribution defined in the range $\omega \in [-\varepsilon/2,\varepsilon/2]$, the order parameter is given by
\begin{equation}
r=\int_{-\frac{\varepsilon}{2}H^{-1}(\lambda r)}^{\frac{\varepsilon}{2}H^{-1}(\lambda r)}g(\omega)\sqrt{1-\left[\frac{1}{\lambda r}H\left(\frac{2\omega}{\varepsilon}\right)^{2}\right]}d\omega, 
\label{eq:leyva_weighted_r}
\end{equation}
where $H(y)$ is the function
\begin{equation}
H(y)=\frac{4}{4+\pi}\left[\frac{y}{1+y^{2}}+\arctan(y)\right].
\label{eq:leyva_weighted_function_H}
\end{equation}
Solving Eq.~\ref{eq:leyva_weighted_r} one gets the full dependence of the parameter $r$ on the coupling $\lambda$. 


\begin{figure}[!t]
\begin{center}
\includegraphics[width=0.6\linewidth]{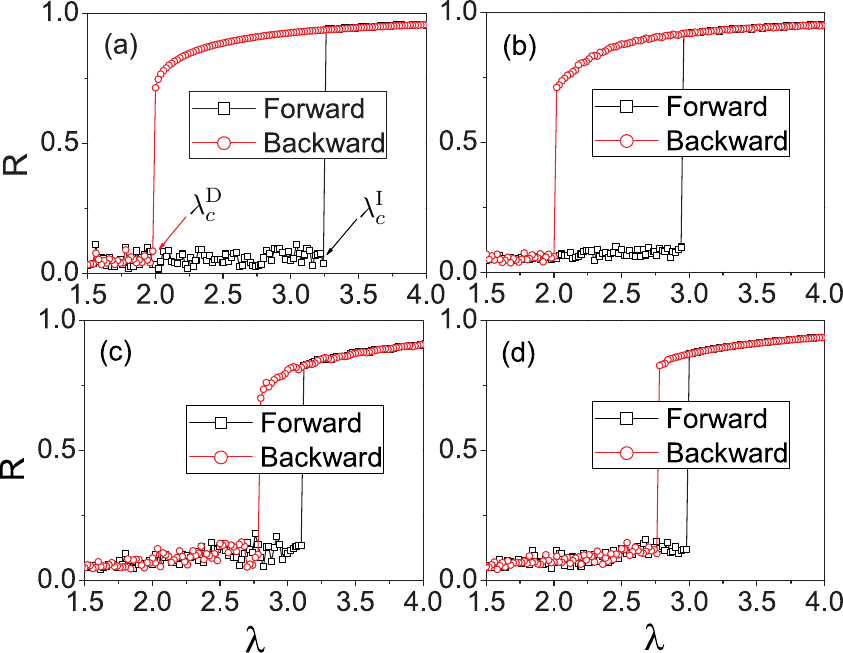}
\end{center}
\caption{Order parameter $R$ (Eq.~\ref{eq:KM_ORDER_PARAMETER}) as a function of coupling $\lambda$ of model defined in Eq.~\ref{eq:zhang_general_eq_of_motion} considering the fully connected graph with (a) Lorentzian and (b) Gaussian frequency distribution. Panels (c) and (d) correspond to the cases of an ER and SF network, respectively, with Lorentzian frequency distribution and average degree $\left\langle k \right\rangle = 6$. Adapted with permission from~\cite{zhang2013explosive}. Copyrighted by the American Physical Society.}
\label{fig:zhang_general_fig1}
\end{figure}
Analyzing the variations of the Kuramoto model discussed in this section we notice that having frequencies 
proportional to degree is just one of the possible mechanisms that can lead to abrupt 
transitions either in networks or in fully connected graphs. Another example is the model considered by 
Zhang et al.~\cite{zhang2013explosive}, which consists of a system governed by the following 
equations
\begin{equation}
\dot{\theta}_i = \omega_i + \frac{\lambda \left| \omega_i\right|}{k_i} \sum_{j=1}^N A_{ij} \sin(\theta_j - \theta_i), 
\label{eq:zhang_general_eq_of_motion}
\end{equation}
where $\omega_i$ are drawn from a distribution 
$g(\omega)$. Instead of directly correlating dynamics 
and topology, the 
model considered in~\cite{zhang2013explosive}
includes an asymmetric coupling between oscillators 
that is proportional to the individual natural 
frequencies. In Fig.~\ref{fig:zhang_general_fig1} the 
synchronization diagram of model by 
(\ref{eq:zhang_general_eq_of_motion}) is presented for 
different frequency distributions $g(\omega)$
and network topologies. Specifically, Fig.~\ref{fig:zhang_general_fig1}(a) and (b) consider 
oscillators in a fully connected graph with Lorentzian 
and 
Gaussian frequency distribution, respectively. As seen there, the nature of the transition is not affected by
the choice of $g(\omega)$ (see also~\cite{zhou2015ExplosiveSyncWithAsymmetricFreq} for investigations of model (\ref{eq:zhang_general_eq_of_motion}) considering asymmetric frequency distributions). The same effect is observed if the topology of the networks is varied, as depicted in the panels (c) 
and (d), where the authors considered ER and SF networks constructed through the CM.  
The only exception reported by the authors in~\cite{zhang2013explosive} in which the synchronization transition 
becomes continuous is obtained for $g(\omega) = \sqrt{\frac{2}{\pi}} \exp\left(-\frac{\omega^2}{2} \right)$ and $\omega>0$. 
Noteworthy, nonlinear frequency-weighted couplings have also 
been considered~\citep{PhysRevE.83.066214,doi:10.1142/S0218127412502306}, where, besides asynchronous and synchronous states, oscillatory and chimera states were reported to occur depending on the weighting exponent. 

\subsection{The role of degree-degree correlations}

We saw that one of the key mechanisms to induce a discontinuous
synchronization transition networks of Kuramoto oscillators is 
the existence of sufficient large frequency mismatches between 
nodes and their neighbors (see Eqs.~\ref{eq:leyva_scirep_condition_1} 
and~\ref{eq:leyva_scirep_condition_2})~
\cite{leyva2013explosivescirep,coutinho2013kuramoto}. In fact, the 
original correlation between frequencies and degrees considered by G\'omez-Garde\~nez et al.~\cite{gomez2011explosive} fits this frequency 
dissortativity condition for SF networks. Based on 
the analysis of star-graphs and on the assumption that they are 
building blocks for SF networks, the fulfillment of this condition qualitatively explains the 
routes to explosive synchronization in such structures, as analysed in Sec.~\ref{subsec:explosive_sync_analytical_approaches}. Thus, these 
results suggest that the type of the synchronization transition might 
strongly depend on the assortative properties of the underlying 
network.  

Analyzing the aforementioned conditions for the occurrence of 
discontinuous transitions, one expects that the lower the assortativity 
coefficient of the network, the larger the hysteretic behavior of the order parameter as a function of the coupling strength. By 
investigating the synchronization of SF networks with a tunable assortativity coefficient using the algorithm described in~\cite{xulvi2004reshuffling}, it was shown in~\cite{li2013reexamination} that the 
hysteresis area
decreases in strongly dissortative networks. Apparently this is in contrast with the assumption that larger frequency 
mismatches, and consequently negative degree-degree correlations, contribute to enhance the irreversibility of the phase 
transition. However, as remarked in~\cite{zou2014basin}, SF networks can be considered as a 
collection of star graphs only for a sufficient low average degrees or
when the network possess locally-tree like structure, i.e., without 
the presence of loops. Otherwise, the approximation of considering 
SF network as interconnected star-graphs does not hold, explaining the apparent contradictory result obtained 
in~\cite{li2013reexamination}. In fact, for assortativity values $\mathcal{A}< -0.35$ the synchronization 
transition is no longer discontinuous for the considered networks. Furthermore, as also 
discussed in Sec.~\ref{subsec:assortative}, the critical coupling $\lambda_c$ was 
observed to be positively correlated with the network assortativity~\cite{li2013reexamination}.

\begin{figure}[!t]
\begin{center}
\includegraphics[width=0.8\linewidth]{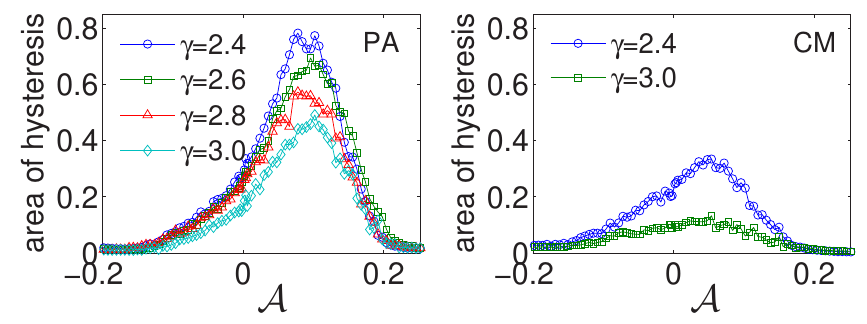}
\end{center}
\caption{Area of hysteresis defined in the synchronization diagram
by the forward and backward propagation of coupling strength as a function of the assortativity coefficient $\mathcal{A}$ for (left) growing SF networks constructed via the preferential attachment (PA) rule with initial attractiveness~\cite{dorogovtsev2000StructureOfGrowingNetworksPreferential} and (right) networks statically generated through the CM. Each point consists in an average over 10 different network realizations. Network parameters are $N = 10^3$ and $\left\langle k \right\rangle = 6$. Adapted with permission from~\cite{sendina2015effects}. Copyrighted by the American Physical Society.}
\label{fig:sendina_assortative_fig1}
\end{figure}

The irreversibility of 
the phase transition was further investigated in~\cite{sendina2015effects}. The hysteric behavior of the order parameter is maximized for positive 
values of the assortativity coefficient $\mathcal{A}$ in growing SF networks, as shown in Fig.~\ref{fig:sendina_assortative_fig1}(a) for different exponents $\gamma$. 
Surprisingly, networks 
statically constructed with the very same size $N$, average degree 
$\left\langle k \right\rangle$ and exponent $\gamma$  through the 
CM do not present the same dependency on 
$\mathcal{A}$ as
observed in Fig.~\ref{fig:sendina_assortative_fig1}(a)~\cite{sendina2015effects}. In this case, the hysteresis is maximized for 
slightly negative values of $\mathcal{A}$, besides 
having significant lower values (Fig.~\ref{fig:sendina_assortative_fig1}(b)). This interesting phenomenon was explained in~\cite{sendina2015effects} by studying the average betweenness centrality
among the highest connected hubs as a function of the degree-mixing. It 
was found that hubs belonging to networks constructed through growing 
processes exhibit significant higher betweenness centrality than those 
in networks generated by the CM. In other words, 
in growing SF networks, the hubs segregate more efficiently the low-degree 
nodes, whereas in the CM the highly connected nodes are 
more homogeneously distributed throughout the network. Thus, since 
explosive  synchronization is induced by the large frequency gaps between 
neighbors, frustrating in this way the paths to synchronization~\cite{sendina2015effects,leyva2013explosivescirep,gomez2007paths}, growing SF networks are more vulnerable to have paths frustrated by such 
gaps, explaining the higher irreversibility in the transition than in the 
CM. Furthermore, as the networks are rewired towards 
extreme values of positive or negative assortativity $\mathcal{A}$, other
topological properties, such as average shortest path length, clustering
coefficient and even modular structure~\cite{xulvi2004reshuffling}, are dramatically changed up to a point that the frequency gap is not sufficient to yield an abrupt formation of a synchronous giant component~\cite{sendina2015effects,leyva2013explosivescirep}.

Other works analyzed the effect of non-vanishing assortativity values on 
the synchronization transition under
frequency-degree correlations~\cite{liu2013effects,zhu2013criterion}. For instance, Liu et al.~\cite{liu2013effects} 
considered natural frequencies to be non-linearly correlated with degrees
in the form $\omega = k_i^{a}/b$, similarly as in Eq.~\ref{eq:skardal_epl_correlation_function}.
The authors found that in assortative networks the nodes no longer join abruptly 
a giant synchronized group, but instead a hierarchical synchronization transition settles in~\cite{liu2013effects}. 
This is in agreement with the properties described above. Noteworthy, similar dependencies of the nature of the
phase transition on degree-mixing properties were also observed in the synchronization of networks made 
up of FitzHugh-Nagumo oscillators with correlation between intrinsic dynamics and local topological properties~\cite{chen2013explosive}.

\subsection{Other works}

Further properties of synchronization in the presence 
of correlations between the intrinsic dynamics of the oscillators
and their local topology were investigated in~\cite{zhang2014explosive,hu2014exact,navas2015SynchronizationCentralityES} as well as the dynamics in other types of
networks, such as in modular~\cite{li2013effect} and in co-evolving~\cite{su2013explosive} ones. The effects of partially correlating frequencies and degrees were also investigated~\cite{pinto2015ExplosiveSyncPartialCorr}. Moreover, of particular
interest is the case analysed in~\cite{zhang2015explosive}. 
The authors posed the question whether explosive synchronization is achievable without the presence of any kind microscopic
correlation in the model. To investigate this issue,
networks with adaptive coupling following
\begin{equation}
\dot{\theta}_i = \omega_i + \lambda \kappa_i \sum_{j=1}^N A_{ij}\sin(\theta_j - \theta_i),
\label{eq:zhang_adaptive_eq_motion}
\end{equation}  
were considered, where $\kappa_i$ is the coupling parameter. A fraction $f$ of the oscillators is then chosen to have $\kappa_i = r_i/k_i$, where  $r_i$ is the local order parameter defined in Sec.~\ref{sec:traditional}, while for the other fraction $(1-f)$, $\kappa_i = 1$ is set. 
Investigating $r$ as a function of coupling $\lambda$, it was found that, after a critical fraction $f_c$ is crossed, the networks reach the synchronized state through a discontinuous transition, regardless the network topology or the frequency distribution $g(\omega)$. 
The same result is obtained when a single-layer network is replaced by
a two-layer network. Again, the parameter controlling the emergence of
discontinuity of $r$ is the fraction $f$ of oscillators with adaptive 
coupling $\kappa_i = r_i$. The findings in~\cite{zhang2015explosive} reinforced the idea that the necessary condition for the emergence of 
explosive synchronization is not the correlation between topology 
and intrinsic dynamics in networks nor a sufficient frequency 
mismatch among the oscillators, but rather the existence of any suppressive rule that impedes the formation of a giant synchronous component.  

Finally, it is also important to mention that the particular frequency assignment introduced by G\'omez-Garde\~nez et al.~\cite{gomez2011explosive} has been not only studied in networks of Kuramoto oscillators. In fact, the effects of degree-correlated frequencies were also investigated in networks made up of 
R\"ossler~\cite{leyva2012explosive_chaotic}, FitzHugh-Nagumo~\cite{chen2013explosive} and Stuart-Landau~\cite{bi2014explosive,chen2015self} oscillators, as well as in experiments
with chemo-mechanical oscillators~\cite{kumar2015ExperimentalEvidence}.


\section{Stochastic Kuramoto Model}
\label{sec:stochastic}

The traditional Kuramoto model under the influence of stochastic forces has 
been already extensively analyzed in many seminal papers, dating back the initial approach 
by Strogatz and Mirollo~\cite{strogatz1991stability} and others as~\cite{crawford1999synchronization} and~\cite{bonilla1993glassy}. The precise revision of the early works on 
the stochastic
Kuramoto model in the fully connected graph lies beyond the scope of this 
review. For such purpose we refer to~\cite{acebron2005kuramoto}. Here, nonetheless,  we focus on 
the discussion of the works which are particularly devoted to the analysis 
of the stochastic model in complex networks. Despite the vast literature 
on synchronization of stochastic Kuramoto oscillators~\cite{acebron2005kuramoto}, studies on complex topologies
have only been presented very recently~\cite{sonnenschein2012onset,sonnenschein2013networks,sonnenschein2013excitable,sonnenschein2013approximate,traxl2014general}. This fact comes as a surprise given the many possible scenarios and applications arising
from the interplay between network topology and stochastic 
dynamics. In this section we present the formulation of the stochastic Kuramoto model in networks and derive the corresponding
nonlinear Fokker-Planck equation (FPE) through which the critical coupling strength for the onset of synchronization is obtained. Furthermore, 
we analyse the low-dimensional behavior of the model using the
framework provided by the Gaussian Approximation (GA) technique as 
well as its generalization to excitable systems described by 
the Shinomoto-Kuramoto model. 

\subsection{Onset of synchronization}

Sonnenschein and Schimansky-Geier~\cite{sonnenschein2012onset} investigated the stochastic Kuramoto model by taking all-together the mean-field approaches
in uncorrelated networks~\cite{ichinomiya2004frequency} with the known 
results on the stability of the incoherent state in the fully connected graph~\cite{strogatz1991stability}. 
More specifically, they considered networks in the absence of degree 
correlations whose oscillators evolve according to the following
equations of motion
\begin{equation}
\dot{\theta}_i=\omega_{i}+\frac{\lambda}{N}\sum_{j=1}^{N}A_{ij}\sin\left(\theta_{j}-\theta_{i}\right)+\xi_{i}(t), 
\label{eq:sonne_onset_eq_mov}
\end{equation}
where the effect of noise is brought into the model by the terms $\xi_i(t)$, which account for the contribution of different
stochastic forces. Note the choice made by the authors to include the size of the network $N$ as a normalization term in order to obtain an intensive coupling. This approach differs from those ones adopted in most of the works discussed here, where the commonly adopted coupling is simply $\lambda_{ij} = \lambda$ $\forall i,j$. The terms $\xi_i(t)$ are assumed to be sources of Gaussian white noise satisfying
\begin{eqnarray}\nonumber
\left\langle \xi_{i}(t)\right\rangle  & = & 0,\\
\left\langle \xi_{i}(t)\xi_{j}(t)\right\rangle  & = & 2D\delta_{ij}\delta(t-t'), 
\label{eq:noise}
\end{eqnarray}
Using the annealed network approximation for the expected value of the adjacency matrix elements in uncorrelated networks, i.e., $\tilde{A}_{ij} = k_i k_j /\sum_m k_m $ (see the discussion in Sec.~\ref{subsec:early_works}) and using the order parameter defined in Eq.~\ref{eq:mean_field_annealed_order_parameter}, the equations (\ref{eq:sonne_onset_eq_mov}) can be decoupled as
\begin{equation}
\dot{\theta}_i=\omega_{i}+r\lambda\frac{k_{i}}{N}\sin(\psi-\theta_{i})+\xi_{i}(t), 
\label{eq:sonne_mean_field_1}
\end{equation}
which are precisely the same decoupled equations through mean-field as described in the
previous sections, with the exception of the noise terms $\xi_i(t)$ (Eq.~\ref{eq:noise}) and the choice for the normalization factor. The system (\ref{eq:sonne_mean_field_1}) governed by $N$ equations is equivalent to a Markov process characterized by the 
transition probability $\mathcal{P}(\boldsymbol{\theta},t| \boldsymbol{\theta}^0, t^0;\boldsymbol{\omega},\boldsymbol{k})$, where 
$\boldsymbol{\theta} = \left( \theta_1, ..., \theta_N\right)$ and $\boldsymbol{\theta}^0 = \left( \theta_1^0, ..., \theta_N^0\right)$  
are the vectors containing the phases at time $t$ and $t_0<t$, respectively. Likewise, $\boldsymbol{k} = \left(k_1, ..., k_N \right)$ and $\boldsymbol{\omega} = \left(\omega_1, ..., \omega_N \right)$ correspond to the vectors with degrees and frequencies of the nodes, respectively. Thus, 
the probability $\mathcal{P}$ satisfies the linear Fokker-Planck equation
\begin{equation}
\frac{\partial\mathcal{P}}{\partial t}=-\sum_{i=1}^{N}\frac{\partial}{\partial\theta_{i}}\left[\left(\omega_{i}\mathcal{P}+k_{i}\frac{\lambda}{N}\sum_{j=1}^{N}\frac{k_{j}}{\sum_{m=1}^{N}k_{m}}\sin(\theta_{j}-\theta_{i})\mathcal{P}\right)\right]+D\sum_{i=1}^{N}\frac{\partial^{2}\mathcal{P}}{\partial\theta_{i}^{2}},
\label{eq:sonne_fokker_planck_1}
\end{equation}
with the initial condition 
\begin{equation}
\mathcal{P}(\boldsymbol{\theta},t|\boldsymbol{\theta}^0,t^0;\boldsymbol{\omega},\boldsymbol{k}) = \delta^N(\boldsymbol{\theta} - \boldsymbol{\theta}^0).
\label{eq:sonne_initial_condition_P}
\end{equation}
The initial phase configuration $\boldsymbol\theta^0$ is assumed to be drawn 
independently from the frequencies $\boldsymbol\omega$ and degrees $\boldsymbol{k}$. Thus, 
the joint probability distribution of the initial phases, frequencies and degrees at time $t^0$
can be written as
\begin{equation}
\mathcal{P}(\boldsymbol\theta^0, t^0;\boldsymbol\omega, \boldsymbol{k}) = \prod_{i=1}^N P(\omega_i,k_i) \mathcal{P}(\theta_i^0,t^0).
\label{eq:sonne_joint_prob_initial_phase}
\end{equation}
The joint probability distribution $\mathcal{P}(\boldsymbol\theta,t;\boldsymbol\omega,\boldsymbol{k})$ describing the 
population for $t>t^0$ is obtained by integrating the joint probability $\mathcal{P}(\boldsymbol\theta,t|
\boldsymbol\theta^0,t^0;\boldsymbol\omega,\boldsymbol{k})\mathcal{P}(\boldsymbol\theta^0,t^0;\boldsymbol\omega,\boldsymbol{k})$ over the initial conditions, i.e., 
\begin{equation} {\mathcal
    P}\left(\boldsymbol{\theta},t;\boldsymbol{\omega},\boldsymbol{k}\right)=\int d{\theta_1^0}\ldots  d{\theta_N^0}\ \mathcal
    P\left(\boldsymbol{\theta},t|\boldsymbol{\theta}^0,t^0;\boldsymbol{\omega},\boldsymbol{k}\right)\prod_{i=1}^N
    P\left(\omega_i,k_i\right)\mathcal
    P\left({\theta_i^0},t^0\right)\,.
\label{eq:sonne_joint_prob_initial_phase_factorized}
\end{equation}
In order to further develop the description of the model, it is convenient to obtain 
the joint probability distribution $\mathcal{P}\left(\boldsymbol{\theta},t;\boldsymbol{\omega},\boldsymbol{k}\right)$
in a reduced formulation~\cite{sonnenschein2012onset}. In order to do so, we follow the approaches presented in~\cite{sonnenschein2012onset,crawford1999synchronization}. Let $\rho_n$ be the reduced joint probability
distribution  describing the phases $\theta_1, ..., \theta_n$, frequencies $\omega_1, ..., \omega_n$ and degrees $k_1, ..., k_n$, where $n<N$. Its relation with $\mathcal{P}\left(\boldsymbol{\theta},t;\boldsymbol{\omega},\boldsymbol{k}\right)$
is given by~\cite{crawford1999synchronization,sonnenschein2012onset}
\begin{eqnarray}\nonumber
\rho_{n}\left(\theta_{1},\ldots,\theta_{n},t;\omega_{1},\ldots,\omega_{n},k_{1},\ldots,k_{n}\right) & = & \int d \theta_{n+1}\ldots d \theta_{N}\int d \omega_{n+1}\ldots d \omega_{N}\\
 & \times & \int d k_{n+1}\ldots d k_{N}\ \mathcal{P}\left(\boldsymbol{\theta},t;\boldsymbol{\omega},\boldsymbol{k}\right).
\end{eqnarray}
For the case of $n=1$ oscillator, considering that all oscillators are statistically equivalent, the distribution $\rho_1$ is obtained by integrating Eq.~\ref{eq:sonne_fokker_planck_1}, which yields~\cite{crawford1999synchronization,sonnenschein2012onset}
\begin{eqnarray}\nonumber
\frac{\partial\rho_{1}}{\partial t} & = & -\frac{\partial}{\partial\theta_{1}}\omega_{1}\rho_{1}+D\frac{\partial^{2}\rho_{1}}{\partial\theta_{1}^{2}}\\
 &  & -\frac{\partial}{\partial\theta_{1}}\left[\frac{\lambda}{N}\frac{k_{1}\left(N-1\right)}{\sum_{m}k_{m}}\int d \theta_{2}\int d \omega_{2}\int d k_{2}\sin\left(\theta_{2}-\theta_{1}\right)\right.\\
 &  & \left.\times k_{2}\rho_{2}\left(\theta_{1},\theta_{2},t;\omega_{1},\omega_{2},k_{1},k_{2}\right)\right].
\end{eqnarray}
Note that the reduced probability distribution $\rho_1$ is related to the distribution $\rho_2$, which is hierarchically related
to the other distributions $\rho_n$ ($n>2$) by successively integrating Eq.~\ref{eq:sonne_fokker_planck_1}~\cite{crawford1999synchronization,sonnenschein2012onset}. For large populations $N \rightarrow \infty$, the correlation between two oscillators vanishes~\cite{crawford1999synchronization,sonnenschein2012onset}, i.e., 
\begin{equation}
  \rho_2\left(\theta_1,\theta_2,t;\omega_1,\omega_2,k_1,k_2\right)=\rho_1\left(\theta_1,t;\omega_1,k_1\right)\rho_1\left(\theta_2,t;\omega_2,k_2\right)\,.
\end{equation}
Hence, recalling that $\rho_1(\theta_1,t;\omega_1,k_1) = \rho_1(\theta_1,t|\omega_1,k_1)P(\omega_1,k_1)$, one 
gets a nolinear Fokker-Planck equation for the one-oscillator density $\rho_0$
\begin{equation}
\frac{\partial\rho_{1}(\left.\theta_{1},t\right|\omega_{1},k_{1})}{\partial t}=-\frac{\partial}{\partial\theta_{1}}\left[v(\theta_{1},t)\rho_{1}(\left.\theta_{1},t\right|\omega_{1},k_{1})\right]+D\frac{\partial^{2}\rho_{1}(\left.\theta_{1},t\right|\omega,k_{1})}{\partial\theta_{1}^{2}}
\label{eq:sonne_onset_fokker_planck_one}, 
\end{equation}
where the phase velocity is given by
\begin{equation}
v(\theta_1,t) = \omega_1 + r\tilde\lambda k_1\sin(\psi - \theta_1), 
\label{eq:sonne_onset_velocity}
\end{equation}
where $\tilde{\lambda} = \lambda/N$. The dependency on $\rho_1(\theta_1,t|\omega_1,k_1)$ is introduced by the order parameter
\begin{equation}
r e^{i\psi(t)} = \frac{1}{\left\langle k \right\rangle} \int_0^{2\pi}d\theta_2 \int_{-\infty}^{\infty}d\omega_2 \int_{k_{\min}}^{\infty} dk_2 e^{i\theta_2}\rho_1(\left.\theta_2, t \right| \omega_2, k_2)k_2P(\omega_2,k_2).
\label{eq:sonne_onset_order_parameter}
\end{equation}
From now on, the indices can be omitted so that the system 
is described by the variables $\theta$, $\omega$ and $k$ and the 
distribution $\rho$ satisfying $\int_0^{2\pi} \rho(\left.\theta,t \right|\omega,k)d\theta = 1$, $\forall$ $\omega,k$ and $t$. In order to derive
the critical coupling for the onset of synchronization, in~\cite{sonnenschein2012onset} the authors generalized the approach by Strogatz and Mirollo for the network case. The asynchronous state is defined by the incoherent solution 
\begin{equation}
\rho_0(\left. \theta \right| \omega, k) = \frac{1}{2\pi}, \; \forall \omega, k, t.
\label{eq:sonne_onset_incoherent_dist}
\end{equation}
Considering a small perturbation to the incoherent 
state
\begin{equation}
\rho(\left. \theta, t \right| \omega, k) = \frac{1}{2\pi} + \epsilon \delta \rho(\left. \theta, t \right| \omega,k),\; \epsilon \ll 1,
\label{eq:sonne_perturbation_incoherent_state}
\end{equation}
yields the following normalization condition
\begin{equation}
\epsilon \int_0^{2\pi} \delta \rho(\left. \theta, t \right|, \omega, k) d\theta =0.
\label{eq:sonne_normalization_cond_perturbation}
\end{equation}
Inserting Eq.~\ref{eq:sonne_perturbation_incoherent_state} into the nonlinear FPE~\ref{eq:sonne_onset_fokker_planck_one} and linearizing in the lowest order in $\epsilon$ gives
\begin{equation}
\frac{\partial\mathcal{\delta}\rho}{\partial t}=-\omega\frac{\partial\delta\rho}{\partial\theta}+\frac{\tilde{\lambda}k}{2\pi}\delta r\cos(\psi-\theta)+D\frac{\partial^{2}\mathcal{\partial\delta\rho}}{\partial\theta^{2}},
\label{eq:sonne_onset_FPE_delta_rho}
\end{equation}
where $\delta r$ is calculated by using $\delta \rho$ in Eq.~\ref{eq:sonne_onset_incoherent_dist}. Since $\delta \rho$ is a $2\pi$-periodic function, it is convenient 
to seek for solutions according to
\begin{equation}
\delta\rho(\left.\theta,t\right|\omega,k)=\frac{1}{2\pi}\sum_{m=1}^{\infty}\left[c_{m}(\left.t\right|\omega,k)e^{im\theta}+c_{m}^{*}(\left.t\right|\omega,k)e^{-im\theta}\right].
\label{eq:sonne_onset_fourier}
\end{equation}
Substituting Eq.~\ref{eq:sonne_onset_fourier} into~\ref{eq:sonne_onset_FPE_delta_rho} and noticing that
$c_0(\left. t \right| \omega,k)$ vanishes due to Eq.~\ref{eq:sonne_normalization_cond_perturbation} yields 
\begin{equation}
\delta re^{i\psi}=\frac{1}{\left\langle k\right\rangle }\int_{-\infty}^{\infty}d\omega'\int_{k_{\min}}^{\infty}dk'c_{1}^{*}(\left.t\right|\omega',k')k'P(\omega',k'),
\label{eq:sonne_delta_r}
\end{equation}
which multiplying by $e^{-i\theta}$ and taking the real part leads to
\begin{equation}
\delta r\cos(\psi-\theta)=\frac{1}{2\left\langle k\right\rangle }\left(\int_{-\infty}^{\infty}d\omega'\int_{k_{\min}}^{\infty} dk' c_{1}(\left.t\right|\omega',k')k'P(\omega',k')\right)e^{i\theta}+\textrm{c.c},
\label{eq:sonne_delta_r_cos}
\end{equation}
where c.c. denotes the complex conjugate. Finally, inserting Eq.~\ref{eq:sonne_onset_fourier}-\ref{eq:sonne_delta_r_cos} into Eq.~\ref{eq:sonne_onset_FPE_delta_rho} 
and grouping the coefficients proportional to $e^{i\theta}$, we obtain the amplitude equation for the fundamental model $c_1(t|\omega,k)$~\cite{sonnenschein2012onset}
\begin{equation}
\frac{\partial c_{1}}{\partial t}=-(D+i\omega)c_{1}+\frac{\tilde{\lambda}k}{2\left\langle k\right\rangle }\int_{-\infty}^{\infty}d\omega'\int_{k_{\min}}^{\infty}dk'c_{1}(\left.t\right|\omega',k')k'P(\omega',k').
\label{eq:sonne_eigenvalue_eq_c1}
\end{equation}
Seeking for solutions in the form~\cite{sonnenschein2012onset,strogatz1991stability} 
\begin{equation}
c_{1}(\left.t\right|\omega,k)\equiv b(\omega,k)e^{\sigma t},\;\; \sigma \in  \mathbb{C}
\end{equation}
one obtains the following eigenvalue equation
\begin{equation}
\sigma b=-(D+i\omega)b+\frac{\tilde{\lambda}k}{2\left\langle k\right\rangle }\int_{-\infty}^{\infty}d\omega'\int_{k_{\min}}^{\infty}dk'b(\omega',k')k'P(\omega',k')
\end{equation}
The term $b$ in the right-hand side can be written as
\begin{equation}
b(\omega,k)=\frac{Bk}{\sigma+D+i\omega},
\end{equation}
where $B$ is determined through the self-consistent equation
\begin{equation}
B=\frac{\tilde{\lambda}}{2\left\langle k\right\rangle }\int_{-\infty}^{\infty}d\omega'\int_{k_{\min}}^{\infty} dk' \frac{Bk'^{2}}{\sigma+D+i\omega}P(\omega',k').
\label{eq:sonne_onset_B_self_consistent}
\end{equation}
Supposing that the dependency on $\omega$ of the function $P(\omega,k)$ is unimodal and even, one can show that
the eigenvalue $\sigma$ is implicitly determined by~\cite{sonnenschein2012onset,strogatz1991stability} 
\begin{equation}
1=\frac{\tilde{\lambda}}{2\left\langle k\right\rangle }\int_{-\infty}^{+\infty}d\omega'\int_{k_{\min}}^{\infty} dk' \frac{(\sigma+D)k'^{2}}{(\sigma+D)^{2}+\omega'{}^{2}}P(\omega',k').
\label{eq:sonne_onset_eq_for_sigma}
\end{equation}
It is important to mention that the eigenvalue $\sigma$
must satisfy $\sigma > -D$ so that the right-hand side
of Eq.~\ref{eq:sonne_onset_eq_for_sigma} is always positive. Furthermore, if $\sigma > 0$ the fundamental
mode is linearly unstable, letting the order parameter
to increase exponentially as $r(t) \sim r_0 e^{\sigma t}$~\cite{sonnenschein2012onset,strogatz1991stability} . 
At $\sigma = \sigma_c = 0$ the stability of the incoherent is lost, leading to the following  
critical coupling for the onset of synchronization
\begin{equation}
\lambda_{c}=2N\left\langle k\right\rangle \left[\int_{-\infty}^{+\infty}d\omega'\int_{k_{\min}}^{+\infty}dk'\frac{Dk'^{2}}{D^{2}+\omega^{2}}P(\omega',k')\right]^{-1}.
\label{eq:sonne_critical_coupling}
\end{equation}

\begin{figure}[!t]
\begin{center}
\includegraphics[width=1.0\linewidth]{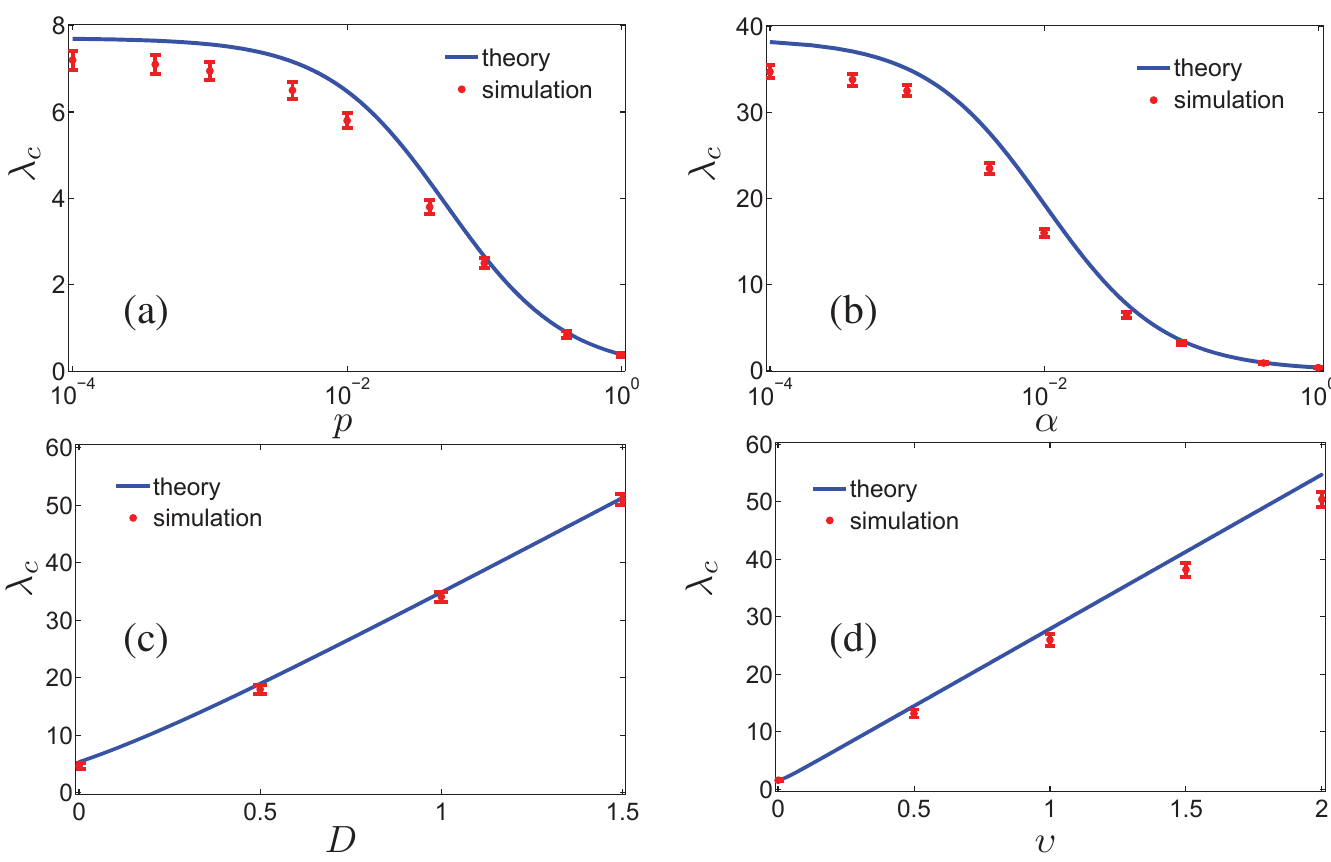}
\end{center}
\caption{ Critical coupling $\lambda_c$ as a function of (a) probability $p$ in the \emph{dense small-world model} considering noise strength $D = 0.05$, Gaussian frequency distribution with standard deviation $\upsilon = 0.2$ and parameter $\alpha = 0.05$; (b) density parameter $\alpha$ for $D = 0.05$, $\upsilon = 0.2$ and $p=0.01$; (c) noise strength $D$ considering $\alpha = 0.05$, $p=0.01$ and $\upsilon = 0.2$; and (d) standard deviation $\upsilon$ for $\alpha = 0.05$, $p=0.01$ and $D=0.05$. Adapted with permission from~\cite{sonnenschein2012onset}. Copyrighted by the American Physical Society.}
\label{fig:sonnenschein_onset_fig1}
\end{figure}

To test the validity of the above expression for the critical coupling, the authors in~\cite{sonnenschein2012onset} considered networks
following the so-called dense small-world model. In this topology, each node is connected to its
$K$ nearest neighbors in both directions in the ring, where $K$ is defined as 
\begin{equation}
K = \left\lfloor \frac{\alpha}{2}(N-1) \right\rfloor, \; 0\leq \alpha \leq 1. 
\label{eq:sonne_onset_K_definition}
\end{equation} 
The variable $\alpha$ tunes the fraction of the total number of nodes $(N-1)/2$ to which
each node is coupled. Starting with a ring network and employing this connection 
strategy leads to a regular network with the degree distribution $P_{\rm{local}}(k) = \delta_{k,2K+1}$. 
The small-world character of the model is included by the introduction of shortcuts. In contrast 
to the standard SW model, the shortcuts are not created through the rewiring of already
existing edges, but rather by the inclusion of new connections. More specifically, besides the 
$2K + 1$ nearest neighbors that a given node is connected to, a new connection is created
between the other $N - K - 1$ nodes with probability $p$. In the case $K=0$, the generated
structures are ER random networks for $0<p<1$, whereas $p=1$ yields the fully connected graph. The final 
network has thus the following degree distribution
\begin{equation}
P(k) = \binom{N - 2K - 1}{k - 2K -1} p^{k-2K-1}\left[1 - p \right]^{N-k}, \textrm{ for }k>2K.
\label{eq:sonne_onset_Pk_dense_SW}
\end{equation}
The merit of using the dense small-world world as a substrate to study  synchronization of Kuramoto oscillators is the fact that the number of edges linearly scales with the number
of nodes, justifying the normalization factor in Eq.~\ref{eq:sonne_onset_eq_mov}.  

The comparison between the result Eq.~\ref{eq:sonne_critical_coupling} and numerical simulations
is shown in Fig.~\ref{fig:sonnenschein_onset_fig1} as a function of several parameters of the model and considering Gaussian
frequency distribution. In particular, in Fig.~\ref{fig:sonnenschein_onset_fig1}(a) and (b) noise strength $D$ and variance of 
$g(\omega)$ are kept fixed as the network topology is varied. As it is seen, the denser the networks, the
better the agreement with the theoretical prediction, which is expected from the MFA. 
Noteworthy, the critical coupling $\lambda_c$ of the simulated networks is evaluated through FSS analysis, analogously as discussed in Sec.~\ref{sec:traditional}, where it is assumed
that the order parameter $r$ has the scaling form given by Eq.~\ref{eq:hong2002_scaling}.
Interestingly, the same scaling with system's size 
observed in the fully connected graph~\cite{son2010ThermalFluctuationEffectsOnFSS,choi2013ExtendedFSS} was found in the stochastic Kuramoto model in the dense SW network~\cite{sonnenschein2012onset}, namely $\beta \approx 1/2$
and $\bar{\nu} \approx 2$. Thus, by calculating the synchronization 
curves for different network sizes, one is then able to numerically 
evaluate the critical coupling strength $\lambda_c$ 
(Fig.~\ref{fig:sonnenschein_onset_fig1}). Moreover, a slight 
disagreement is also observed for strong standard deviation of the frequency 
distribution in Fig.~\ref{fig:sonnenschein_onset_fig1}(d), which is analogous to the discrepancy 
observed for strongly heterogeneous 
networks~\cite{sonnenschein2012onset}.



Sonnenschein et al.~\cite{sonnenschein2013networks} also extended the analysis developed in~\cite{sonnenschein2012onset} 
to the case in which the diversity of frequencies is related to the network topology through the joint distribution $P(\omega,k)$. The equations of motion are given by
\begin{equation}
\dot{\theta}_i = \omega_i + \frac{\lambda}{N^q}\sum_{j=1}^N A_{ij}\sin\left[\theta_j(t)  - \theta_i(t) \right] + \xi_i (t), 
\label{eq:sonnenschein_freqdispersion_eq_motion}
\end{equation}
where $q$ is a scaling parameter proportional to the 
number of links in the network. Again, the presence of stochastic forces is accounted for by the terms $\xi_i(t)$ which are considered to be sources of Gaussian white noise, satisfying Eq.~\ref{eq:noise}.
The natural frequency distribution is assumed to be unimodal and even
with standard deviation $\sqrt{\left\langle \omega^2_i \right\rangle - \left\langle \omega_i \right\rangle^2} = \upsilon_n(k_i) $, where
\begin{equation}
\upsilon_n (k_i) = \upsilon_0 \left(\frac{k_i}{\left\langle k \right\rangle}\right)^n, \; n \in \mathbb{R}, 
\label{eq:sonne_epjst_noisy_correlate_freq_deg}
\end{equation}
with $\upsilon_0$ being a constant parameter. The main motivation of the model in Eq.~\ref{eq:sonnenschein_freqdispersion_eq_motion} resides in the fact that a direct 
correlation between frequencies and local connectivity might be hard to be observable 
in real-world networks~\cite{sonnenschein2013networks}. Yet, the process of network formation in real-world 
systems can in fact yield nodes whose dynamical properties are intrinsically related
to the topological structure. This interplay between structure and dynamics
is observed, e.g. in neuronal networks, where the neurons connectivity
is related to the balance of inhibition and excitation~\cite{sonnenschein2013networks}.

Assuming the joint probability distribution $P(\omega,k) = g(\omega|k)P(k)$, where the conditional
distribution of frequencies $g(\omega|k)$ for a given degree $k$ is given by
\begin{equation}
g_{\rm{gauss}}(\omega | k) = \frac{1}{\sqrt{2\pi\upsilon_n(k)}}e^{-\frac{1}{2}\frac{\omega^2}{\upsilon_n^2(k)}},
\label{eq:sonnenschein_freqdispersion_ggauss}
\end{equation}
enables us to show using Eq.~\ref{eq:sonne_critical_coupling} that 
\begin{equation}
\lambda_{c,{\rm {gauss}}}=2\sqrt{\frac{2}{\pi}}\upsilon_{0}N^{q}\langle k\rangle^{1-n}\left\langle k^{2-n}\textrm{erfc}\left(\frac{D}{\sqrt{2}\upsilon_{n}(k)}\right)\exp\left(\frac{D^{2}}{2\upsilon_{n}(k)^{2}}\right)\right\rangle ^{-1}.
\label{eq:sonnenschein_freqdispersion_critical_coupling_gauss}
\end{equation}
In the absence of noise, i.e., $D = 0$, the previous equation is reduced
to
\begin{equation}
\lambda_{c,{\rm {gauss}}}(D=0)=2\sqrt{\frac{2}{\pi}}\upsilon_{0}N^{q}\frac{\langle k\rangle^{1-n}}{\langle k^{2-n}\rangle}.
\label{eq:sonnenschein_dispersion_critical_coupling_gauss_D_0}
\end{equation}
It is interesting to note that, if ER networks are considered, for $n=1$ and $2$ the critical coupling coupling $\lambda_{c,{\rm {gauss}}}(D=0)$ 
has an inverse dependence on the average degree $\left\langle k \right\rangle$. The same result was obtained for linear correlation between 
frequencies and degrees in~\cite{peron2012determination,coutinho2013kuramoto}. Although, in this
model, it is the frequency dispersion that is correlated with local topology, 
it also leads to similar effects as the ones observed when the frequencies
are directly proportional to degrees. For $n>0$, the hubs will potentially have higher frequencies due to the large dispersion, whereas
low degree nodes will have narrower distributions. This frequency assignment will contribute to the increase of the critical coupling for the onset of synchronization, since higher couplings are required to entrain the hubs in the mean-field. Finally, for $n<0$, hubs and low degree nodes switch their role, making the network to reach synchronization
for lower coupling strengths. Noteworthy, this interplay between the nodes degree and the emergence of a synchronous component was also observed in the model described in Sec.~\ref{sec:explosive_sync} with joint distribution given by Eq.~\ref{eq:skardal_epl_correlation_function}. 

Furthermore, a curious behavior  can be observed if one considers different system-size scaling in the equations of motion. More specifically, changing $N^q \rightarrow \mathcal{N}(k)$ in Eq.~\ref{eq:sonnenschein_freqdispersion_eq_motion} for two different cases, $\mathcal{N}(k) = \left\langle k \right\rangle$ and $\mathcal{N}(k) =  k $, one can show that the critical couplings in the absence of noise ($D=0$) obey
\begin{equation}
\lambda_{c,\left\langle k \right\rangle} = C \upsilon_0 \frac{\left\langle k \right\rangle^{2-n} }{\left\langle k^{2-n} \right\rangle},\; \lambda_{c,k} = C \upsilon_0 \frac{\left\langle k \right\rangle^{1-n} }{\left\langle k^{1-n} \right\rangle}, 
\label{eq:sonne_dispersion_critical_couplings_different_scalings}
\end{equation} 
regardless of the frequency distribution adopted, where $\lambda_{c,\left\langle k \right\rangle}$ and $\lambda_{c,k} $ correspond to $\mathcal{N}(k) = \left\langle k \right\rangle$ and $\mathcal{N}(k) =  k $, respectively; and $C$ is a proportionality constant, which is different for each frequency distribution. Interestingly, the dependency on the network topology vanishes for some values of $n$. In contrast with the critical coupling in Eq.~\ref{eq:sonnenschein_dispersion_critical_coupling_gauss_D_0} for scaling $N^q$. Note also that the critical couplings related to the scaling terms $\mathcal{N}(k) = \left\langle k \right\rangle$ and $\mathcal{N}(k) =  k $ correspond
to the shifted version of each other with respect to the correlation power $n$, i.e., $\lambda_{c,\left\langle k \right\rangle}(n) = \lambda_{c,k}(n+1)$.

\subsection{Gaussian approximation}
\label{subsec:gaussian_approximation}


The onset of synchronization of the stochastic Kuramoto model in
networks can also be derived through a different technique, namely the so-called
Gaussian-Approximation~\cite{zaks2003noise,sonnenschein2013approximate} (GA).  
The framework provided by the GA allows a suitable dimension reduction, though it is not
exact, in contrast to the OA theory~\cite{ott2008low}, which 
is, however, not applicable to stochastic systems~\cite{sonnenschein2013approximate}.  Before discussing the stochastic Kuramoto model on networks, let us first briefly introduce the results uncovered through the GA for the fully connected graph. 

The equations for the model with identical natural frequencies $\omega_0 = 0$ are 
\begin{equation}
\dot{\theta}_i = \xi_i(t) + \frac{\lambda}{N}\sum_{j=1}^N \sin(\theta_j - \theta_i).
\label{eq:sonen_approximate_eq_of_motion}
\end{equation}
As analysed in the previous section, in the thermodynamic limit $N \rightarrow \infty$, the above system 
can be described by the density $\rho(\theta,t)$, which satisfies the nonlinear Fokker-Planck
equation
\begin{equation}
\frac{\partial \rho}{\partial t} = D \frac{\partial^2 \rho}{\partial^2 \theta} - \frac{\partial}{ \partial \theta}\left[\lambda r \sin(\psi - \theta) \rho \right], 
\label{eq:sonne_approximate_nonlinear_fokker_planck}
\end{equation}
where $r$ and $\psi$ in the continuum limit are
\begin{equation}
r(t) e^{i\psi(t)} = \int_0^{2\pi} d\theta e^{i\theta} \rho(\theta,t).
\label{eq:sonne_approximate_r_continuum_limit}
\end{equation}
The Fourier series expansion of $\rho$ yields
\begin{equation}
\rho(\theta,t) = \frac{1}{2\pi} \sum_{n = -\infty}^{+\infty} \hat{\rho}_n(t) e^{-in\theta},
\label{eq:sonne_approximate_fourier_series}
\end{equation}
where $\rho_0 = 1$ and $\rho_{-n} = \rho_n^{*}$, since $\rho(\theta,t)$ is a real function. Using the
inverse Fourier transform one obtains
\begin{equation}
\hat{\rho}_n(t) = \int_0^{2\pi} d\theta \rho(\theta,t) e^{in\theta} \equiv c_n(t) + is_n(t), 
\label{eq:sonne_approximate_inverse_transform}
\end{equation}
which for $n=1$ yields the order parameter definition in Eq.~\ref{eq:sonne_approximate_r_continuum_limit}. Substituting
Eq.~\ref{eq:sonne_approximate_fourier_series} into the nonlinear Fokker-Planck equation (Eq.~\ref{eq:sonne_approximate_nonlinear_fokker_planck}) and grouping the exponential terms leads to the infinite 
chain of coupled complex-valued equations for the Fourier coefficients given by
\begin{equation}
\frac{\dot{\hat{\rho}}_n}{n} = \frac{\lambda}{2} \left(\hat{\rho}_{n-1}\hat{\rho}_1 - \hat{\rho}_{n+1}\hat{\rho}_{-1} \right) - D n \hat{\rho}_n, \; n=1,2,...,\infty.
\label{eq:sonne_approximate_chain}
\end{equation}
This equation provides a complete description and, due to the rapidly decay of the coefficients $\hat{\rho}_n$ as $n$ increases, it is possible to get relatively accurate results by truncating the system at sufficient large $n$~\cite{sonnenschein2013approximate}. However, a thorough bifurcation analysis of the system in Eq.~\ref{eq:sonne_approximate_chain}
would be difficult or even unfeasible depending on the value of $n$. Thus, in order to 
find a closure scheme for the infinite set of equations (\ref{eq:sonne_approximate_chain}), we assume that the phases 
of the oscillators are distributed according to a Gaussian distribution with mean $\mu(t)$ and variance $\upsilon^2(t)$. 
In this way the coefficients $c_n$ and $s_n$ in Eq.~\ref{eq:sonne_approximate_inverse_transform} become~\cite{sonnenschein2013approximate}
 \begin{equation}
 \begin{aligned}
 c_n(t)&=\mathrm e^{-n^2\upsilon^2(t)/2}\cos\left[n\mu(t)\right],\\
 s_n(t)&=\mathrm e^{-n^2\upsilon^2(t)/2}\sin\left[n\mu(t)\right], 
 \end{aligned}
\label{eq:sonne_approximate_cn_sn}
\end{equation}
leading to the following Fourier coefficients
\begin{equation}
\left| \hat{\rho}_n(t) \right| = e^{-n^2\upsilon^2(t) /2}. 
\label{eq:sonne_approximate_rho_n_gaussian}
\end{equation}
Changing the variables, one can further show that the mean $\mu$ and
the variance $\upsilon^2$ evolve according to~\cite{sonnenschein2013approximate}
\begin{equation}
\dot\mu = 0\mbox{ and }\dot{\upsilon^2}=2D+\lambda\left(\mathrm e^{-2\upsilon^2}-1\right),
\label{eq:sonne_approximate_dot_sigma}
\end{equation}
which can be explicitly solved: 
\begin{equation}
\begin{aligned}
\upsilon^2(t)=&\frac{1}{2}\ln\left[\frac{\lambda}{\lambda-2D}+\mathrm e^{-2t(\lambda-2D)}
\left(\mathrm e^{2\upsilon^2(0)}-\frac{\lambda}{\lambda-2D}\right)\right],
\label{eq:sonne_approximate_sigma2}
\end{aligned}
\end{equation}
for $\lambda \neq D $. Note that in the long-time limit $t \rightarrow \infty$ the variance
asymptotically converges to
\begin{equation}
\upsilon^2(t\rightarrow \infty) = \frac{1}{2} \ln \left(\frac{\lambda}{\lambda - 2D} \right).
\label{eq:sonne_approximate_var_asymp}
\end{equation}
According to the equation above, for strong couplings $\lambda \rightarrow 
\infty$, the variance tends to vanish $\upsilon^2 \rightarrow 0$, characterizing 
the complete synchronous state. Interestingly, the
critical coupling obtained through Eq.~\ref{eq:sonne_approximate_var_asymp}, 
$\lambda_c = 2D$, is the same
well-known valued exactly calculated in~\cite{strogatz1991stability}. The time evolution of the order parameter $r(t)$ can also be derived thanks to the closure scheme provided 
by the GA. More specifically, all coefficients $c_n$ and $s_n$ depend on 
$c_1$ and $s_1$. Therefore, using Eq.~\ref{eq:sonne_approximate_chain} 
the following system of equations is obtained~\cite{sonnenschein2013approximate,zaks2003noise}
 \begin{equation}
 \begin{aligned}
   \dot c_1 &=-Dc_1 +\frac{\lambda c_1 }{2}\left\{1-\left[\left(c_1 \right)^2+\left(s_1 \right)^2\right]^2\right\},\\
   \dot s_1 &=-Ds_1 +\frac{\lambda s_1 }{2}\left\{1-\left[\left(c_1 \right)^2+\left(s_1 \right)^2\right]^2\right\}.
 \end{aligned}
 \label{eq:sonne_approximate_dot_c1_dot_s1}
 \end{equation}
Noting that $c_1 = r$ and $s_1 = 0$, one derives 
\begin{equation}
\dot r = \frac{r}{2} \left[ \lambda\left(1 - r^4 \right) -2D\right], 
\label{eq:sonne_approximate_dot_r}
\end{equation}
for which in the stationary regime $\dot r = 0$ and neglecting the term $r^4$ near the onset of synchronization also leads to 
the critical coupling $\lambda_c = 2D$.

The generalization to the network case can be done straightforwardly, although the dimensionality reduction is not as drastic as in the fully
connected graph. The population of oscillators is described by the phase
distribution $\rho(\theta,t|k)$, i.e., the dependency on a given degree $k$
is introduced. With this definition, the coherence inside
each population of oscillators with the same degree $k$ is calculated by 
\begin{equation}
  r_k(t)\mathrm e^{i\psi_k(t)}=\int_{0}^{2\pi} d \theta'\ \mathrm
  e^{i\theta'}\ \mathcal \rho\left(\theta',t|k\right).
  \label{eq:sonne_approximate_orderp_k}
\end{equation}
The total order parameter is then given by
\begin{equation}
 r(t)\mathrm e^{i\psi(t)}= \frac{1}{\langle k\rangle} \int dk' P(k')k'r_{k'}(t)\mathrm e^{i\psi_{k'}(t)}, 
 \label{rkthetak}
\end{equation}
Using the GA one is able to show that 
the closure scheme yields
\begin{eqnarray*}
\dot{r}_{k} & = & \frac{1-\left(r_{k}\right)^{4}}{2N\left\langle k\right\rangle }\lambda k\left\langle k'r_{k'}\cos(\psi_{k'}-\psi_{k})\right\rangle_k -r_{k} D\\
\dot{\psi}{}_{k} & = & \frac{\left(r_{k}\right)^{-1}+\left(r_{k}\right)^{3}}{2N\left\langle k\right\rangle }\lambda k\left\langle k'r_{k'}\sin(\psi_{k'}-\psi_{k})\right\rangle_k,
\end{eqnarray*}
where $\langle ... \rangle_k = \int dk ... P(k)$ is the average over 
the degree distribution $P(k)$. Note that the populations of different degrees are coupled through the
averages $\langle ... \rangle_k$. Assuming $\psi_k = 0$ $\forall k$, the evolution of the total order parameter is given by
\begin{equation}
    \dot r^{g}=\frac{\lambda}{2N\langle k \rangle}\left\langle k'^2\left[1-\left(r_{k'}^g\right)^4\right]\right\rangle_k r^g-Dr^g.
 \label{r_networks}
\end{equation}
In the stationary state, neglecting the terms $r_{k'}^4$ leads to the critical coupling
\begin{equation}
\lambda_c = 2DN \frac{\left\langle k \right\rangle}{\left\langle k^2 \right\rangle}
\label{eq:sonne_approximate_critical_coupling_networks}
\end{equation}
which is the same result as obtained with Eq.~\ref{eq:sonne_critical_coupling}
for identical oscillators.

The GA technique was also employed in the analysis of heterogeneous networks made up of stochastic Shinomoto-Kuramoto active rotator oscillators~\cite{sonnenschein2013excitable}. The model proposed in~\cite{shinomoto1986phase} consists 
of a slight modification of the original Kuramoto model that encompasses 
the properties of excitable systems. Excitable behavior and collective oscillations are found in many different systems, including physical, chemical and biological ones, with a wide range of applications \cite{LiGarNeiSchi04}. Excitability in a given system is characterized by the presence of a stable equilibrium of its units, which can be replaced 
by a non-monotonic behavior when the system is subjected to sufficiently strong perturbations.  

The equations of the networks considered in~\cite{sonnenschein2013excitable} are given by
\begin{equation}
  \dot{\theta}_i(t)=1-a\sin\left(\theta_i(t)\right)+\frac{\lambda}{N}\sum_{j=1}^{N}A_{ij}\sin\left(\theta_j(t)-\theta_i(t)\right)+\xi_i(t),
\label{eq:sonne_excitable_model}
\end{equation}
where $a$ is the excitability threshold of the model. 
For a system with only one oscillator without the presence of noise, $
\dot\theta_i = 1 - a\sin\theta_i$, an excitable behavior is achieved 
for $|a|>1$ and the stationary solution is $\theta_i = \arcsin
\left(a^{-1}\right)$. On the other hand, if $|a|<1$, the system is in 
an oscillatory regime with frequency $\sqrt{1 - a^2}$. Noteworthy, 
contrary to the standard Kuramoto model, 
phase $\theta_i$ does not rotate uniformly, the slowest and fastest points are near $\phi = \pi/2$ and $3\pi/2$, respectively~\cite{shinomoto1986phase,kurrer1995noise,LiGarNeiSchi04,sonnenschein2013excitable}.
As before, the network is described by the one-oscillator probability density $\rho(\theta,t|k)$, where $\rho(\theta,t|k)d\theta$ gives the fraction of oscillators with 
phase between $\theta$ and $\theta + d\theta$ at time $t$ for a given 
degree $k$. The Fourier expansion of $\rho$ leads to
\begin{equation}
\rho(\theta,t|k) = \frac{1}{2\pi} \sum_{n=-\infty}^{\infty} \hat{\rho}_n (t|k) e^{-in\theta}, 
\label{eq:sonne_excitable_fourier}
\end{equation}
where the coefficients, which now are also defined for a given degree $k$, can be written as
\begin{equation}
\hat{\rho}_n(t|k) = c_n(t|k) + is_n(t|k),\; n=1,2,...,\infty.
\label{eq:sonne_excitable_cn_sn}
\end{equation}
Substituting Eq.~\ref{eq:sonne_excitable_fourier} in the corresponding 
nonlinear Fokker-Planck, one can show that for every $k$ the following 
infinite chain of coupled complex-valued equations determines the 
evolution of the coefficients $\rho_n(t|k)$: 
\begin{equation}
\begin{aligned}\frac{\dot{\hat{\rho}}_{n}(t|k)}{n}= & \frac{\tilde{\lambda}k}{2\langle k\rangle}\left(\hat{\rho}_{n-1}(t|k)\left\langle \hat{\rho}_{1}(t|k')k'\right\rangle _{k}-\hat{\rho}_{n+1}(t|k)\left\langle \hat{\rho}_{-1}(t|k')k'\right\rangle _{k}\right)\\
 & +\frac{a}{2}\left(\hat{\rho}_{n-1}(t|k)-\hat{\rho}_{n+1}(t|k)\right)-(Dn-i)\hat{\rho}_{n}(t|k).
\end{aligned}
\label{eq:sonne_excitable_chain}
\end{equation}
Similarly as before, it is assumed that the phases of 
each subpopulation formed by nodes with the same degree $k$ follow a Gaussian distribution with mean $\mu_k(t)$ and variance $\upsilon^2_k(t)$. Thus, in the thermodynamic limit $N\rightarrow\infty$ one can show
that in the GA the coefficients $c_n$ and $s_n$ in Eq.~\ref{eq:sonne_excitable_cn_sn} are given by
\begin{equation}
c_n(t|k)=\mathrm e^{-n^2\upsilon_k^2(t)/2}\cos\left(n\mu_k(t)\right),\ \ s_n(t|k)=\mathrm e^{-n^2\upsilon_k^2(t)/2}\sin\left(n\mu_k(t)\right).
\label{eq:sonne_excitable_cn_sn_GA}
\end{equation}
Applying the closure, scheme the reduced system of equations for the variables $\mu_k(t)$ and $\upsilon_k(t)$ is obtained~\cite{sonnenschein2013excitable}
\begin{equation}
\begin{aligned}
\dot \mu_k=&1-\mathrm e^{-\upsilon_k^2/2}\cosh\upsilon_k^2\\
&\times\left[a\sin \mu_k-\frac{\tilde{\lambda} k}{\langle k\rangle}\left(\langle k'\mathrm e^{-\upsilon_{k'}^2/2}\sin 
\mu_{k'}\rangle_k \cos \mu_k-\langle {k'}\mathrm e^{-\upsilon_{k'}^2/2}\cos \mu_{k'}\rangle_k\sin \mu_k\right)\right],\\
\dot \upsilon_k^2=&2D-2\mathrm e^{-\upsilon_k^2/2}\sinh\upsilon_k^2\\
&\times\left[a\cos \mu_k+\frac{\tilde{\lambda} k}{\langle k\rangle}\left(\langle k'\mathrm e^{-\upsilon_{k'}^2/2}\sin 
\mu_{k'}\rangle_k\sin \mu_k+\langle {k'}\mathrm e^{-\upsilon_{k'}^2/2}\cos \mu_{k'}\rangle_k \cos \mu_k\right)\right].
\end{aligned}
\label{eq:sonne_exictable_reduced_system_k}
\end{equation}
Again, the dimensionality reduction in the network case is not as severe as for the dynamics in the fully connected graph. The number of coupled equations one is left with is equal to twice the number of different degrees in the network, if the oscillators are identical. Generalizations for the Ott-Antonsen also share this property~\cite{barlev2011dynamics,ji2014low,restrepo2014mean}, in fact some approaches yield reduced systems with similar size as the original ones~\cite{barlev2011dynamics}. Nevertheless, the impact of network heterogeneity can be comprehensively analysed for some special topologies, which can then be used
as insights to the understanding of the dynamics of more sophisticated structures. In particular, for regular networks, Eqs.~\ref{eq:sonne_exictable_reduced_system_k} are reduced to
\begin{equation}
\dot \mu_k=1-a\mathrm e^{-\upsilon_k^2/2}\cosh\upsilon_k^2\sin \mu_k,\quad
\dot \upsilon_k^2=2D-2\mathrm e^{-\upsilon_k^2/2}\sinh\upsilon_k^2\left[a\cos \mu_k+\tilde{\lambda} k\mathrm e^{-\upsilon_k^2/2}\right].
\label{eq:sonne_excitable_reduced_system_regular_networks}
\end{equation}     
\begin{figure}[!t]
\begin{center}
\includegraphics[width=0.9\linewidth]{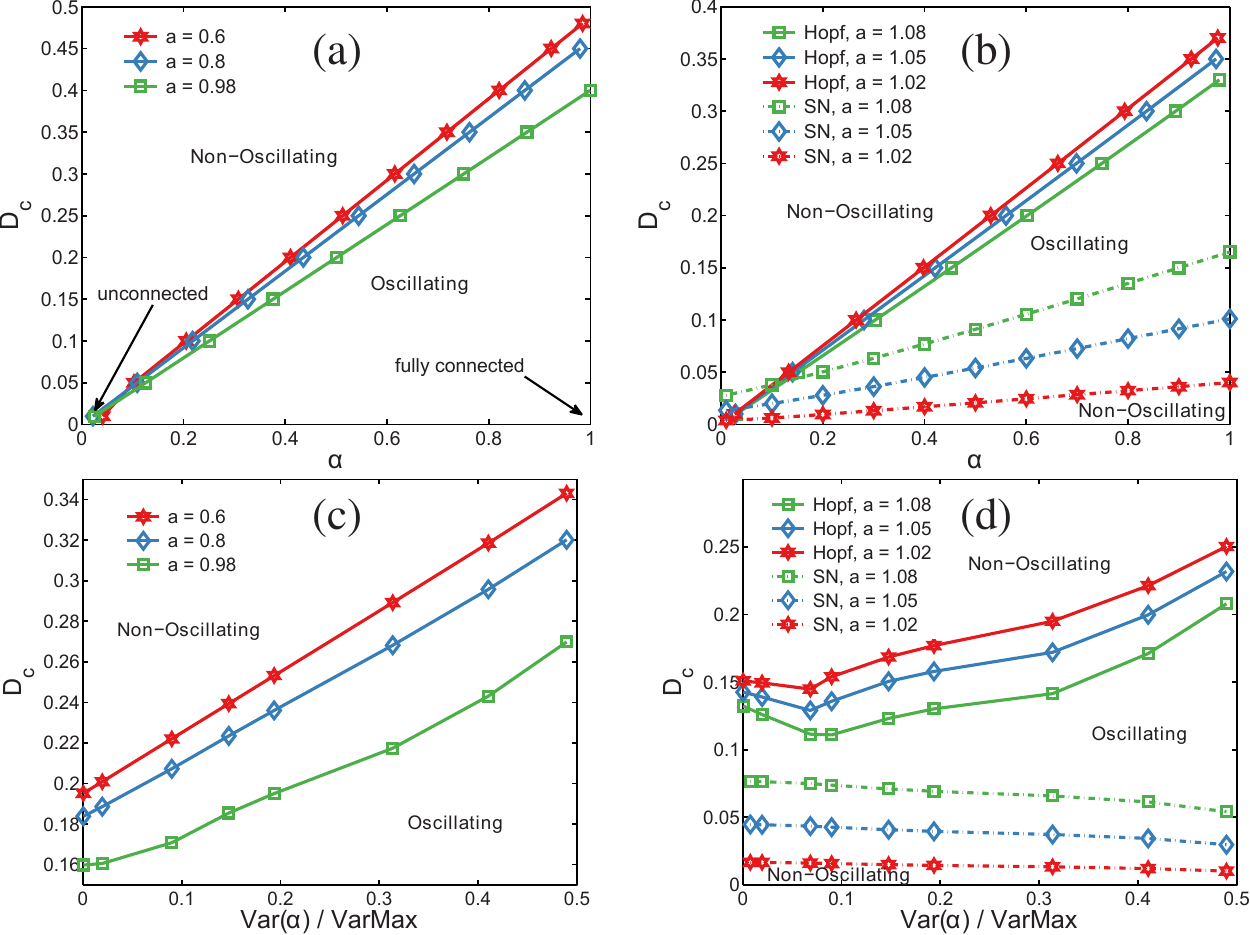}
\end{center}
\caption{ Critical noise strength $D_c$ as a function of connectivity fraction $\alpha$ of regular networks for excitation threshold (a) $a<1$ and (b) $a>1$ determined through numerical bifurcation analysis of the reduced system in Eq.~\ref{eq:sonne_excitable_binary_networks_Pk}. Panels (c) and (d) show $D_c$ as a function of $\textrm{Var}(\alpha)/\textrm{VarMax}$ of the reduced system in Eq.~\ref{eq:sonne_excitable_reduced_system_binary_networks}  for binary networks considering $a<1$ and $a>1$, respectively; where $\left\langle \alpha \right\rangle = 0.4$ and $\textrm{Var}(\alpha)/\textrm{VarMax} = 0.25$. In all panels coupling strength is set to $\lambda=1$ and SN indicate saddle-node bifuracations. Adapted with permission from
Springer Science+Business Media~\cite{sonnenschein2013excitable}.}
\label{fig:sonnenschein_excitable_fig1}
\end{figure} 
As shown in~\cite{sonnenschein2013excitable}, system (\ref{eq:sonne_excitable_reduced_system_regular_networks}) has 
exactly the same bifurcation diagram on the plane $a-D$ as the correspondent system for the fully connected graph, 
the equations only differ by a rescaling of the effective mean-field $\lambda \rightarrow k\lambda/N$. For sufficient 
large values of noise intensity $D$, changes in the excitability threshold $a$ have no effect on the system's stability. On 
the order hand, for moderate values of $D$, the oscillators can transit through Hopf and saddle-node bifurcation between 
regimes of synchronous and asynchronous firing and coherent oscillations with the absence of 
excitability~\cite{sonnenschein2013excitable,zaks2003noise}. Although the regular network exhibits a perfectly homogeneous 
topology, it still serves as a special model to analyze how the density of connections impacts on the such transitions. 
Let $D_c$ denote the bifurcation value of the noise strength - or also called as critical noise strength -, Fig.~\ref{fig:sonnenschein_excitable_fig1}(a) and (b) show 
its dependence on the variable $\alpha = k/N$, which quantifies  the density of the regular network, for different values of $a$. 
The regions ``Non-oscillating'' and ``Oscillating'' denote the asynchronous and synchronous regimes, respectively. Moreover, 
the state in which the active rotators fire coherently and the state in which they are synchronized without firing are treated 
indiscriminately as the ``Oscillating'' state in Fig.~\ref{fig:sonnenschein_excitable_fig1} in order to purely investigate the impact of the network topology on the
overall dynamics.  
As it is seen, the bifurcation point $D_c$  is almost linearly related with the parameter $\alpha$, in a way that 
the variation of network density does not lead to novel phenomena in the network dynamics; the effective
mean-field is only rescaled by such a modification in the topology~\cite{sonnenschein2013excitable}. Furthermore, 
for $a<1$ (Fig.~\ref{fig:sonnenschein_excitable_fig1}( (a)) only Hopf bifurcations are observed, whereas for $a>1$ (Fig.~\ref{fig:sonnenschein_excitable_fig1} (b)) saddle-node bifurcations also
emerge~\cite{sonnenschein2013excitable}. 

Topological heterogeneity can be included by considering a network
with only two different degrees.  The random binary network provides the simplest heterogeneous degree distribution, which is characterized by the degrees $k_1$ and $k_2$ (and the respective
densities $\alpha_1 = k_1/N$ and $\alpha_2 = k_2/N$) given by
\begin{equation}
P(k) = p \delta_{k,k_1} + (1-p)\delta_{k,k_2}
\label{eq:sonne_excitable_binary_networks_Pk}
\end{equation}
By fixing the average density $\left\langle \alpha \right\rangle$ (or equivalently the average degree $\left\langle k \right\rangle$) and varying the $\alpha_1$ and $\alpha_2$, it is possible to tune $\textrm{Var}(\alpha)$ up to the upper bound $\textrm{VarMax} = 0.25$ so that the bifurcation points can be analysed as a function of the disparity of connections between the subpopulations. Using the degree distribution in Eq.~\ref{eq:sonne_excitable_binary_networks_Pk}, the reduced system in Eq.~\ref{eq:sonne_exictable_reduced_system_k} can be rewritten for the binary network case as
\begin{equation}
 \begin{aligned}
\dot \mu_1=&1-\mathrm e^{-\upsilon_1^2/2}\cosh(\upsilon_1^2)\left[a\sin(\mu_1)+\frac{\lambda\alpha_1\alpha_2 (1-p)}{p\alpha_1+(1-p)\alpha_2}\sin(\mu_1-\mu_2)\mathrm e^{-\upsilon_2^2/2}\right] \\
\dot \upsilon_1^2=&2D-2\mathrm e^{-\upsilon_1^2/2}\sinh(\upsilon_1^2)\\
&\times\left[a\cos(\mu_1)+\frac{\lambda\alpha_1}{p\alpha_1+(1-p)\alpha_2}\left(\alpha_1 p\mathrm e^{-\upsilon_1^2/2}+\alpha_2 (1-p)\mathrm e^{-\upsilon_2^2/2}\cos(\mu_1-\mu_2)\right)\right] \\
\dot \mu_2=&1-\mathrm e^{-\upsilon_2^2/2}\cosh(\upsilon_2^2)\left[a\sin(\mu_2)+\frac{\lambda\alpha_1\alpha_2 p}{p\alpha_1+(1-p)\alpha_2}\sin(\mu_2-\mu_1)\mathrm e^{-\upsilon_1^2/2}\right] \\
\dot \upsilon_2^2=&2D-2\mathrm e^{-\upsilon_2^2/2}\sinh(\upsilon_2^2)\\
&\times\left[a\cos(\mu_2)+\frac{\lambda\alpha_2}{p\alpha_1+(1-p)\alpha_2}\left(\alpha_2 (1-p)\mathrm e^{-\upsilon_2^2/2}+\alpha_1 p\mathrm e^{-\upsilon_1^2/2}\cos(\mu_2-\mu_1)\right)\right], 
 \end{aligned}
\label{eq:sonne_excitable_reduced_system_binary_networks}
\end{equation}
where $\alpha_1 = k_1/N$ and $\alpha_2 = k_2/N$ are the connectivity fraction of each subpopulation. Fig.~\ref{fig:sonnenschein_excitable_fig1} (c) 
and (d) show the dependence of the critical noise strength $D_c$ on the variance $\textrm{Var}(\alpha)$. For $a<1$, 
$D_c$ exhibits similar dependence on $\textrm{Var}(\alpha)/\textrm{VarMax}$ as on $\alpha$ in regular networks, i.e., 
almost linear growing with Hopf bifurcations connecting the 
states. However, for $a>1$, the critical noise strength $D_c$ 
associated to the saddle-node bifurcation decreases as the degree 
distribution becomes more heterogeneous, enlarging the oscillating 
region. Another interesting
behavior is that the Hopf bifurcation lines are no longer monotonic 
as a function of $\textrm{Var}(\alpha)$. As seen in Fig.~
\ref{fig:sonnenschein_excitable_fig1}(d), 
$D_c$ reaches minimal values for slight heterogeneous networks, a 
phenomenon induced by the presence of a small fraction 
of highly connected nodes~\cite{sonnenschein2013excitable}.  

The active rotator model in the absence of noise in uncorrelated 
networks was studied in~\cite{tessone2008global} in the presence of 
attractive and repulsive couplings strengths. The 
authors verified that the network heterogeneity, along with the 
repulsive couplings, induces global firing in the ensemble, an 
effect that is akin to the collective firings 
promoted by noise and quenched disorder in globally coupled 
oscillators~\cite{tessone2007theory}. Strangely, despite its rich 
dynamics, the Shinomoto-Kuramoto model has been rarely tackled in 
heterogeneous networks in a way that many aspects of the influence 
of the connectivity pattern on the population coherence and excitability 
remain to be investigated. For instance, Sonnenschein et al.~
\cite{sonnenschein2014cooperative} 
recently investigated the Shinomoto-Kuramoto model in the fully 
connected graph where the natural frequencies and individual coupling 
strengths are correlated. More specifically, a system made up of two equally sized populations
of oscillators, each one characterized by the pairs $(\omega_1,
\lambda_1)$ and $(\omega_2,\lambda_2)$, was studied. Interestingly, it was found 
that global excitability is only reached when the excitable units 
have smaller coupling strengths than the self-oscillatory ones~
\cite{sonnenschein2014cooperative}.  This particular frequency-coupling correlation naturally motivates further generalizations of 
the model, such as in modular and mutilayer networks~
\cite{kivela2014multilayer,boccaletti2014structure}, in which other 
interesting dynamical patterns can possibly arise by the 
introduction of network connectivity. 

Besides the theoretical approaches described above, other aspects of the dynamics of the stochastic Kuramoto model were numerically addressed~\cite{khoshbakht2008phase,esfahani2012noise,traxl2014general,yanagita2012design,sheshbolouki2015FeedbackLoopsDestructive,tirabassi2015inferring}. In particular, Traxl et al.~\cite{traxl2014general} developed a numerical framework and systematically analyzed the influence
of noise and coupling strength on the maximum degree of synchronization of several networks; including fully connected, random, modular and real topologies. For all the cases considered, the authors reported that the maximum degree of synchronization 
is suitably fitted by a 2-dimensional sigmoidal function  linearly 
related with the noise and coupling strength~\cite{traxl2014general}.
  

\section{Second-order Kuramoto model}
\label{sec:second_KM}

The first-order Kuramoto model is analytically tractable and exhibits a large variety of synchronization phenomena in different contexts (see Sec.~\ref{sec:traditional}).  
However, it approaches too fast the partial synchronized state compared to  experimental observations and an infinite coupling strength is required to achieve perfect synchronization~\cite{acebron_adaptive_1998}.  
The  model with frequency adaptation, where both phase and frequency evolve in time and having synchronization slowed down by inertia, can solve such problems. 
Therefore, the second-order Kuramoto model (or the Kuramoto model with inertia) has been extensively investigated. 
The model with inertia was firstly explored in biological contexts. 
In the early 90's, motivated by the fact that 
certain species of fireflies, like Pteroptyx malaccae, 
are able to achieve perfect synchronization even for a 
stimulating frequency different from their intrinsic 
frequency, Ermentrout~\cite{ermentrout_adaptive_1991} 
proposed a model with frequency adaptation which has the 
ability to mimic such a perfect synchrony between coupled 
oscillators. 
Strogatz~\cite{strogatz_nonlinear_2001} and later Trees 
et al.~\cite{trees_synchronization_2005} showed that the same  model can be obtained from capacitively shunted junction 
equations to study synchronization in disordered arrays of 
Josephson junctions. Moreover Filatrella et al.~\cite{filatrella_analysis_2008} derived this model from 
the classical swing equation to study synchronization in power grids (see Sec.~\ref{sec:applications}). 

In this section, we will explain the second-order Kuramoto 
model from different points of view.  
From a methodological standpoint, as the focus of this 
section, we first explain the collective behavior of a set 
of coupled Kuramoto model with inertia using a mean-field 
analysis~\cite{tanaka_first_1997,tanaka_self-synchronization_1997} and  substantially extend the theory 
to the second-order Kuramoto model with frequency-degree 
correlation~\cite{ji2013cluster}. Subsequently, we 
quantify the stability of networks of coupled oscillators 
in terms of the basin of attraction of the synchronized 
state against large perturbations using the new concept of basin stability ($\mathcal{BS}$)~\cite{menck_how_2014}.
In particular, we focus on the interplay between $\mathcal{BS}$ and underlying structures. 

\subsection{Second-order Kuramoto model in complex networks}

The second-order Kuramoto model consists of an ensemble of coupled phase oscillators, $\theta_i$ for $i = 1, \dots, N$,  whose dynamics are governed by 
\begin{equation}
\ddot{\theta}_i = - \alpha\dot{\theta}_i + \omega_i + \lambda\sum\limits^{N}_{j=1} A_{ij} \sin{(\theta_j-\theta_i)},
\label{Eq:SKM_complex_networks}
\end{equation}
where $\alpha$ is the dissipation parameter. If  the left part of this equation is set to zero, we recover the first-order Kuramoto model. In comparison to the first-order Kuramoto model, the second-order mode exhibits a region of bistability in the bifurcation diagram, which will be shown below. 

\subsubsection{Illustration}
\label{SKM_illustration_one_node}

In order to get insights into the fundamental dynamics of Eq.~\ref{Eq:SKM_complex_networks}, let us consider the one-node model, where one node is connected to a grid and whose dynamics follows 
\begin{equation}
\ddot{\theta} = - \alpha\dot{\theta} + \omega + \lambda \sin{(\theta_L - \theta)},
\label{Eq:SKM_infinite_bus_bar}
\end{equation}
where $\theta_L$ is the phase of the large system. 
The grid is regarded to be infinite in the sense that its state can not be affected by the node's dynamics. Hence, here we set $\theta_L \equiv 0$, without loss of generality. Such a model also depicts the governing dynamics of the driven pendulum~\cite{strogatz_nonlinear_2001}, Josephson junctions~\cite{strogatz_nonlinear_2001} and the one-machine infinite bus system of a generator in a power-grid~\cite{chiang_direct_2011}. The corresponding governing dynamics can also depict a two-node model with the same bifurcation diagram~\cite{rohden_self-organized_2012}. 

We can define the energy function $E(\nu,\omega)$ for model (\ref{Eq:SKM_infinite_bus_bar})  as~\cite{chiang_direct_2011}
\begin{equation}
E(\theta,\nu) = E_k(\nu) + E_p(\theta),
\label{Eq:SKM_infinite_bus_bar_energy_function}
\end{equation}
where $\nu \equiv \dot{\theta}$ and the kinetic and potential energy are given by$E_k(\nu) = \frac{\nu^2}{2}$ and $E_p(\theta) = - \omega \theta - \lambda \cos{(\theta)}$, respectively. 
In the absence of damping and external driving, i.e. $\alpha=0$ and $\omega=0$, by introducing a dimensionless time $\tau = \sqrt{\lambda} t$, Eq.~(\ref{Eq:SKM_infinite_bus_bar}) becomes
\begin{equation}
\frac{d^2 \theta}{d \tau^2} = - \sin{\theta}. 
\end{equation}
Such system has two fixed points within the range $[0,2\pi)$. 
One fixed point $(\theta^*,\nu^*) = (\pi,0)$ is a saddle. 
The other one is located at $(\theta^*,\nu^*) = (0,0)$. The origin is a nonlinear center, as the system is reversible and conservative with the energy function $E(\theta, \nu) = \nu^2/2 - \cos(\theta) = \mathrm{constant}$. A local minimal energy is located at the fixed point with $E(0,0) = -1$~\cite{strogatz_nonlinear_2001}. Small orbits around the center are small oscillations, called librations.  The obits grow with an increase in $E$ until $E=1$ along with the heteroclinic trajectories linking the saddles. With further increases in $E$, i.e. $E>1$, the system starts oscillating periodically over or below the heteroclinic trajectories~\cite{strogatz_nonlinear_2001}. 

Furthermore, via including a linear damping to this system, i.e. $\alpha>0$ but with $\omega=0$, 
the system therein has one stable fixed point at $(\theta^*,\nu^*) = (0,0)$ and one saddle $(\theta^*,\nu^*) = (\pi,0)$~\cite{strogatz_nonlinear_2001}. 
With small damping, vibrations start converging to the stable fixed point.  The energy decreases monotonically along the trajectories with the rate $\frac{d E(\theta, \nu)}{d t} = \frac{d ( \nu^2/2 - \lambda\cos(\theta))}{d t} = -\alpha \nu^2$, except for the fixed points with $\nu^*=0$. 

\begin{figure}
\begin{center}
\includegraphics[width=0.7\linewidth]{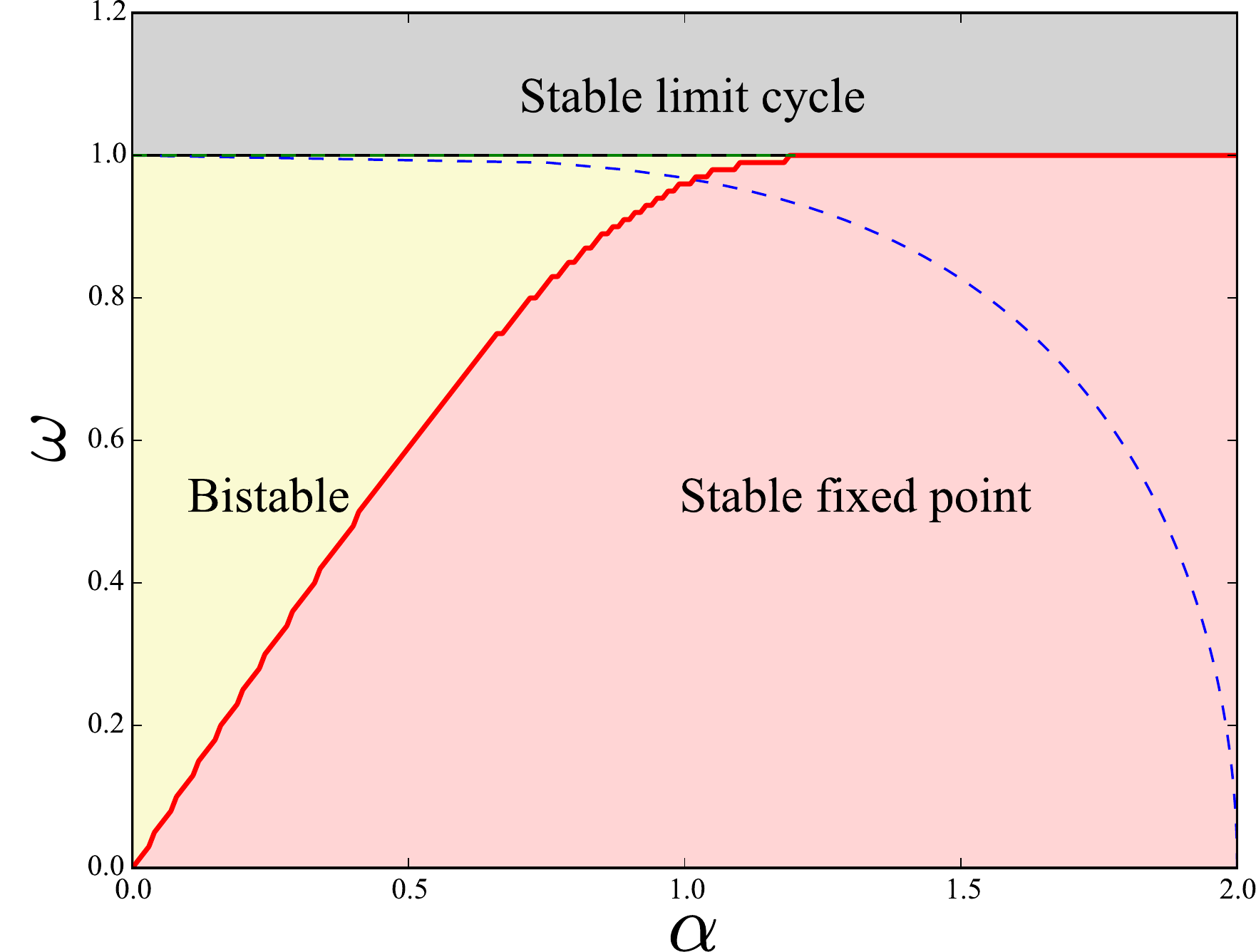}
\end{center}
\caption{Bifurcation diagram of the one-node model~(\ref{Eq:SKM_infinite_bus_bar}). In the grey region only the stable limit cycle exists. The red region indicates the globally stable fixed point. In the region of bistability, stable fixed point and stable limit cycle coexist. The blue dashed line is plotted at $\alpha^2 = 4 \sqrt{1-\omega^2}$. Here we set $\lambda =1$.}
\label{Fig:SKM_infinite_bus_bar_parameter_space}
\end{figure}

Finally, the third case is the complete original system~(\ref{Eq:SKM_infinite_bus_bar}) with damping as well as external driving, i.e. $\alpha>0$ and $\omega>0$. For convenience, one could either set $\lambda =1$ or introduce a dimensionless time $\tau = \sqrt{\lambda}t$. The bifurcation diagram is shown in Fig.~\ref{Fig:SKM_infinite_bus_bar_parameter_space}. 
For $\omega > 1$, all rotations converge to a unique and stable limit cycle and no fixed points are available in the region of the stable limit cycle. 
For $\omega<1$, two fixed points comprise a saddle and a sink, satisfying $\nu^*=0$ and $\sin(\theta^*)=\omega$. The linear stability of the fixed points is determined by the Jacobian matrix
\[ J = \left( \begin{array}{cc}
0 & 1  \\
-\cos(\theta^*) & -\alpha   \end{array} \right)\] 
with two eigenvalues 
\begin{equation}
\sigma_{1,2} =  \frac{-\alpha \pm \sqrt{\alpha^2 - 4 \cos(\theta^*)}}{2}.
\end{equation}

The fixed point with $\cos(\theta^*) = - \sqrt{1-\omega^2}$ with $\sigma_1>0$ and $\sigma_2<0$ is a saddle. The other fixed point with $\cos(\theta^*) =  \sqrt{1-\omega^2}$ is stable due to its real part of both eigenvalues $\mathrm{Re}(\sigma_{1,2})<0$. Moreover, it is a stable node for $\alpha^2 - 4 \sqrt{1-\omega^2}>0$ and a stable spiral, otherwise (separated by the blue dashed line in Fig.~\ref{Fig:SKM_infinite_bus_bar_parameter_space}). 

There are three types of bifurcations in the bifurcation diagram in Fig.~\ref{Fig:SKM_infinite_bus_bar_parameter_space}. 
For small $\alpha$ and suppose that we start from $\omega>1$, the system rotates periodically. We slowly decrease $\omega$ and, at some critical value $\omega_c<1$, rotations merge with the saddle and are destroyed in a homoclinic bifurcation. Its critical value is determined by 
\begin{equation}
\omega_c = \frac{4 \alpha}{\pi},
\label{Eq:SKM_homoclinic_bifurcation_line}
\end{equation}
 as $\alpha \rightarrow 0$ based on the Melnikov's analysis~\cite{strogatz_nonlinear_2001}. 
 This bifurcation line separates the region of bistability 
 and the region of globally stable fixed point. 
Noteworthy, this result can also be obtained by employing Lyapunov's second method~\cite{manik_supply_2014}.
For large $\alpha$, we slowly decrease $\omega$ from $\omega>1$. The rotations are destroyed by an infinite-period bifurcation and fixed points appear. 
Suppose that we start from a stable fixed point for small $\alpha$ and slowly 
increase $\omega$, two fixed points collide and annihilate 
each other in a saddle-node bifurcation with $\omega_c=1$. 
  
The node can either converge to a fixed point or oscillate periodically in the region of bistability. 
Menck et al.~\cite{menck_how_2014} proposed a way to approximate the curve of the stable cycle. For $\lambda=0$, the model~(\ref{Eq:SKM_infinite_bus_bar}) has one stable limit cycle with frequency $\nu(t) = \omega / \alpha$ and phase $\theta(t) = \nu(t) t + \theta(0)$. For large $\lambda$, a similar solution is still fulfilled with the average frequency $\left \langle \nu \right\rangle = \int\limits^{T}_{t=0} \nu(t) dt/T \approx \omega / \alpha$ and $\theta(t) \approx \left\langle \nu \right\rangle t + \theta(0)$, where $T$ is a long integrating period. To derive an expression of the instant frequency, the frequency $\nu(t)$ is assumed to fluctuate around the average frequency $\left\langle \nu \right\rangle$ as  $\nu(t) = \left\langle \nu \right\rangle + f(t)$, where $f(t)$ remains to be solved. Inserting this into Eq.~(\ref{Eq:SKM_infinite_bus_bar}) yields 
\begin{equation}
\dot{f} = -\alpha f - \lambda \sin(\left \langle \nu \right\rangle t + \theta(0)),
\end{equation}
which has one special solution as follows
\begin{equation}
f(t)  = - \frac{\alpha \lambda}{\left\langle \nu \right \rangle^2 + \alpha^2} \left(\sin\left(\left\langle \nu \right \rangle t +\theta(0)\right) - \frac{\left\langle \nu \right \rangle}{\alpha} \cos\left(\left\langle \nu \right \rangle t + \theta(0)\right)\right).
\label{Eq:One_node_version_f}
\end{equation}
For $\left\langle \nu \right \rangle \gg \alpha$ and via inserting $\left\langle \nu \right \rangle \approx \omega/\alpha$ into Eq. (\ref{Eq:One_node_version_f}), the instant frequency therein is approximated by~\cite{menck_how_2014} 
\begin{equation}
\nu(t) \approx \omega/\alpha + \frac{\alpha\lambda}{\omega} \cos(\omega t/\alpha  + \theta(0)). 
\label{Eq:SKM_instant_frequency_approximation}
\end{equation}
Equation~(\ref{Eq:SKM_instant_frequency_approximation}) is in good agreement with numerical results~\cite{menck_how_2014,ji_basin_2014-1}. 
Integrating Eq.~(\ref{Eq:SKM_instant_frequency_approximation}) yields~\cite{menck_how_2014}
\begin{equation}
\theta(t) \approx \omega t/\alpha  + \frac{\alpha^2\lambda}{\omega^2} \sin{(\omega t/\alpha  + \theta(0)) + \theta(0)}. 
\end{equation} 
If $\omega^2 \gg \alpha^2\lambda $, $\theta(t) \approx \omega t/\alpha + \theta(0)$ which is consistent with the previous assumption.

\subsubsection{Mean-field theory without noise}
\label{SKM_Mean_field_theory_without_noise}

In the previous subsection, we explain the basic dynamics of the one-node model, in this and next subsections, we will explain  networked oscillators without and with noise, respectively. 

Tanaka et al.~\cite{tanaka_self-synchronization_1997,tanaka_first_1997} were the first to analyze 
the collective behavior of a set of coupled Kuramoto model 
with finite (large) inertia. Due to the finite inertia, 
the system exhibits a first-order phase transition, with 
discontinuous jumps and hysteresis, different from the 
second-order transition obtained from  the Kuramoto model 
without inertia. 
In particular, the mean-field framework is provided 
for investigating the hysteretic behavior in the large 
network size and validated theoretical results with a 
uniform, bounded intrinsic frequency distribution. 
To compare with the case of without inertia, the following general form of the second-order Kuramoto model was considered~\cite{tanaka_self-synchronization_1997,tanaka_first_1997}
\begin{equation}
m\ddot{\theta}_i = - \alpha\dot{\theta}_i + \omega_i + \frac{\lambda}{N} \sum\limits^{N}_{j=1}  \sin{(\theta_j-\theta_i)},
\label{Eq:SKM_TL_1997}
\end{equation}
where $m\ddot{\theta}_i$ denotes the inertia of the $i$-th oscillator. After a suitable coordinate transformation via replacing the coupling term by the mean-field quantities $(R,\psi)$ and assuming $\alpha=1$,  Eq.~(\ref{Eq:SKM_TL_1997}) is rewritten as
\begin{equation}
m\ddot{\theta}_i = - \dot{\theta}_i + \omega_i + \lambda R \sin{(\psi-\theta_i)}. 
\label{Eq:SKM_TL_1997_mean_field}
\end{equation}
In the stationary state, the set of oscillators is split into one subgroup of oscillators locked to the mean phase and the other subgroup of drifting oscillators whirling over (or below depending on the sign of $\omega$) the locked subgroup. Therefore, the overall phase coherence $R$ sums two certain coherence $R_\mathrm{lock}$ and $ R_\mathrm{drift}$, contributed by these two subgroups, respectively, i.e., 
\begin{equation}
R = R_\mathrm{lock} + R_\mathrm{drift}. 
\label{Eq:r_lock_plus_drift}
\end{equation}
To detect  multistability of the system, two kinds of simulations are considered: increasing (I) and decreasing (D) adiabatically the coupling strength $\lambda$~\cite{tanaka_self-synchronization_1997,tanaka_first_1997}, respectively. Note that the analytical process is similar to that of the first-order Kuramoto model (see Sec.~\ref{sec:traditional}) but  with different synchronization conditions.  
(I) When  $\lambda$ increases from a small value, the phase coherence $R^\mathrm{I}$ persists around a small fluctuation due to effects of network sizes until a critical coupling strength, denoted by $\lambda^\mathrm{I}_c$. Above $\lambda^\mathrm{I}_c$, the system jumps to a weakly synchronized state. $R^{\rm{I}}$ increases with further increases in $\lambda$ and then saturates to a constant value for sufficiently large coupling strengths. 
(D)  When $\lambda$ decreases from a sufficiently large value, the system is initially in the strongly synchronized state and $R^\mathrm{D}$ remains nearly constant until a critical coupling strength, denoted by $\lambda^\mathrm{D}_c$. Beyond this threshold, the system jumps back to the  incoherent state. Hysteretic behaviors therein are observed. Two critical coupling strengths $\lambda^\mathrm{I}_c$ and $\lambda^\mathrm{D}_c$ are almost the same for small $m$  (e.g. $m=0.95$) but its difference enlarges for large $m$ (e.g. $m=2.0$ or $6.0$)~\cite{tanaka_first_1997}. Noteworthy, 
 $\lambda^\mathrm{D}_c$ is the same as that of the first-order Kuramoto model~\cite{tanaka_self-synchronization_1997}.  
Different dynamical regimes are observed~\cite{tanaka_first_1997,tanaka_self-synchronization_1997}, including the incoherent state (IS), the weakly synchronized state (WSS), the strongly synchronized state (SSS), a transition state from WSS to SSS and vice-versa.

In case ($\mathrm{I}$), all oscillators initially drift 
around its own natural frequency $\omega_i$. With 
increasing $\lambda$, oscillators with a small natural 
frequency below the threshold $\omega_{\mathrm{I}}$, i.e. 
$|\omega_i| < \omega_{\mathrm{I}}$, start being attracted 
to the locked group. With further increases in 
$\lambda$, $\omega_{\mathrm{I}}$ enlarges, oscillators 
with a large natural frequency become synchronized and the 
phase coherence $R^\mathrm{I}$ increases. 
For sufficiently large coupling, $\omega_{\mathrm{I}}$ 
exhibits plateaus and $R^\mathrm{I}\approx 1$. 
If the inertia is rather small, i.e. $\frac{1}{\sqrt{m
\lambda R^\mathrm{I}}} \ll 1$, the homoclinic bifurcation 
is tangent to the line 
(\ref{Eq:SKM_homoclinic_bifurcation_line}). Using   
Melnikov's method one is able to obtain $
\omega_{\mathrm{I}} = \frac{4}{\pi}\sqrt{\frac{\lambda 
R^{\mathrm{I}}}{m}}$~\cite{tanaka_self-synchronization_1997}. During this process, a secondary 
synchronization of drifting oscillators is observed for 
larger $m$. This phenomenon was confirmed in~\cite{olmi_hysteretic_2014}, where  
the synchronized motions were  validated by comparing the evolution of the 
instantaneous frequency $\nu_i(t) = \dot{\theta}_i$ of the 
secondary synchronized oscillators in random networks and 
also in the Italian high-voltage power grid.

In case ($\mathrm{D}$), initially almost all oscillators are locked to the mean phase $\psi$ if the initial coupling strength $\lambda$ is large enough, and $R^\mathrm{D} \approx 1$. With decreasing $\lambda$ further, locked oscillators are desynchronized and start whirling when their natural frequency exceeds the threshold $\omega_{\rm{D}}$, i.e. $|\omega_i| > \omega_{\rm{D}}=\lambda R^\mathrm{D}$, where a  saddle node bifurcation occurs. Therefore, given the synchronization boundary $\omega_{\mathrm{I}}$ and $\omega_D$, the contribution to the locked coherence follows~\cite{tanaka_first_1997,tanaka_self-synchronization_1997,olmi_hysteretic_2014}
\begin{equation}
R^\mathrm{I, D}_\mathrm{lock}= \lambda R^\mathrm{I, D} \int^{\theta_{{\mathrm{I}},{\mathrm{D}}}}_{-\theta_{\mathrm{I,D}}} \cos^2{\theta} g(\lambda R^\mathrm{I, D} \sin\theta) d \theta,
\label{Eq:SKM_Tanaka_r_lock}
\end{equation}
where $\theta_{\mathrm{I}} = \sin^{-1}(\omega_{\mathrm{I}}/(\lambda R^\mathrm{I}))$ and $\theta_{\mathrm{D}} = \sin^{-1}(\omega_{\mathrm{D}}/(\lambda R^\mathrm{D}))$.

The phase coherence from drifting oscillators takes the same form as in the first-order Kuramoto model~\cite{acebron2005kuramoto} and is given by~\cite{tanaka_first_1997,tanaka_self-synchronization_1997,olmi_hysteretic_2014}
\begin{equation}
R^\mathrm{I, D}_\mathrm{drift}= \int_{|\omega|>\omega_{\mathrm{I,D}}} \int e^{i\theta} \rho_{\mathrm{drift}}(\theta,\omega) g(\omega) d \theta d \omega,
\label{Eq:SKM_Tanaka_r_drift}
\end{equation}
where $\rho_{\mathrm{drift}}$ is the density of the drifting oscillators with the phase $\theta$ and frequency $\omega$. 
 $\rho_{\mathrm{drift}}(\theta,\omega)$ is proportional to $|\nu|^{-1}$, i.e. $\rho_{\mathrm{drift}}(\theta,\omega) = \frac{\hat{\omega}}{2 \pi}|\nu|^{-1}$, where $\hat{\omega}$ is the frequency of the periodic solution of $\theta$~\cite{tanaka_self-synchronization_1997}. Expanding the cosine function in terms of the Bessel functions and applying the Poicare-Lindstead method~\cite{tanaka_self-synchronization_1997}, Eq.~(\ref{Eq:SKM_Tanaka_r_drift}) can be simplified as 
 \begin{equation}
 R^\mathrm{I, D}_\mathrm{drift}= \int_{|\omega|>\omega_{\mathrm{I,D}}} \int^{\hat{T}}_0 \cos(\theta(t,\omega)) g(\omega) d t d \omega,
 \end{equation}
where $\hat{T} = \frac{2\pi}{\hat{\omega}}$ is the period of the running periodic solution of $\theta$. This theoretical framework can be extended to a general distribution of $g(\omega)$ with extended tails to solve the self-consistent equations of $R$ analytically~\cite{tanaka_first_1997}.

\begin{figure}
\begin{center}
\includegraphics[width=0.7\linewidth]{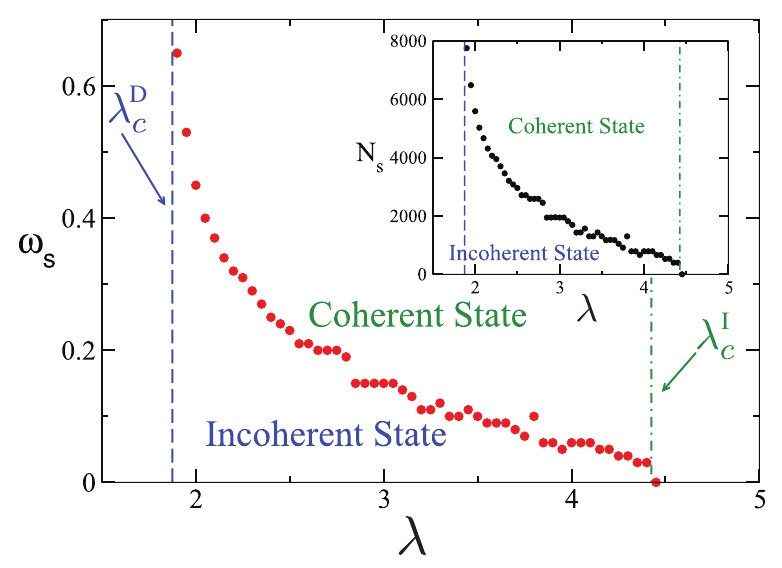}
\end{center}
\caption{Minimal $\omega_S$ inducing the system to the coherent state with respective to the coupling strength $\lambda$.  The inset plots the minimal number of nodes with $|\omega_i|<\omega_S$ as a function of $\lambda$. The green dot-dashed and blue dashed lines indicate the increasing $\lambda^\mathrm{I}_c$ and decreasing  $\lambda^\mathrm{D}_c$ coupling thresholds, respectively. Adapted with permission from~\cite{olmi_hysteretic_2014}. Copyrighted by the American Physical Society.}
\label{Fig:Olmi_hysteretic_2}
\end{figure}

Now, consider the same dynamics~(\ref{Eq:SKM_TL_1997}) with 
natural frequencies   distributed according to the Gaussian 
distribution $g(\omega) = \frac{1}{\sqrt{2 \pi}} e^{-\omega^2/2}$  
\cite{olmi_hysteretic_2014}. To numerically investigate the 
hysteretic loop of the locked natural frequency threshold as a 
function of $\lambda$ within the range $ 
[\lambda^\mathrm{D}_c, \lambda^\mathrm{I}_c]$, one can record the 
maximal locked natural frequency $\omega_{\rm{I}}$ ($\omega_{\rm{D}}$) with 
respect to the  increasing (decreasing) coupling strength 
$\lambda$  in steps $\delta \lambda$ in order to uncover the hysteresis. Note that the 
results are independent of  the step 
size~\cite{olmi_hysteretic_2014}. 
Recalling that within the interval $[\lambda^\mathrm{D}_c, \lambda^\mathrm{I}_c]$, incoherent and strongly synchronized  states coexist~\cite{tanaka_first_1997,tanaka_self-synchronization_1997}. Such  results can also be validated via perturbing incoherent states~\cite{olmi_hysteretic_2014}. In particular, initially the system is in the asynchronous state.  At each $\lambda$,  all the oscillators with $|\omega_i| < \omega_S$  are locked to the mean phase, and then the mean phase coherence $r$ is recorded after a long transient time, where $\omega_S$ is an artificial threshold. To identify the interval $[\lambda^\mathrm{D}_c, \lambda^\mathrm{I}_c]$, at each  $\lambda$, via increasing $\omega_S$ from small to high values, the lowest value of $\omega_S$ is taken as minimal $\omega_S$ when a coherent state is observable, otherwise $\omega_S=0$ it is set~\cite{olmi_hysteretic_2014}. Using this process, the critical coupling interval can also be reproduced (Fig.~\ref{Fig:Olmi_hysteretic_2}).   These results  give insights on the probability of the system retaining to the coherent or incoherent states~\cite{menck_how_2013,menck_how_2014}.

\begin{figure}
\includegraphics[width=\linewidth]{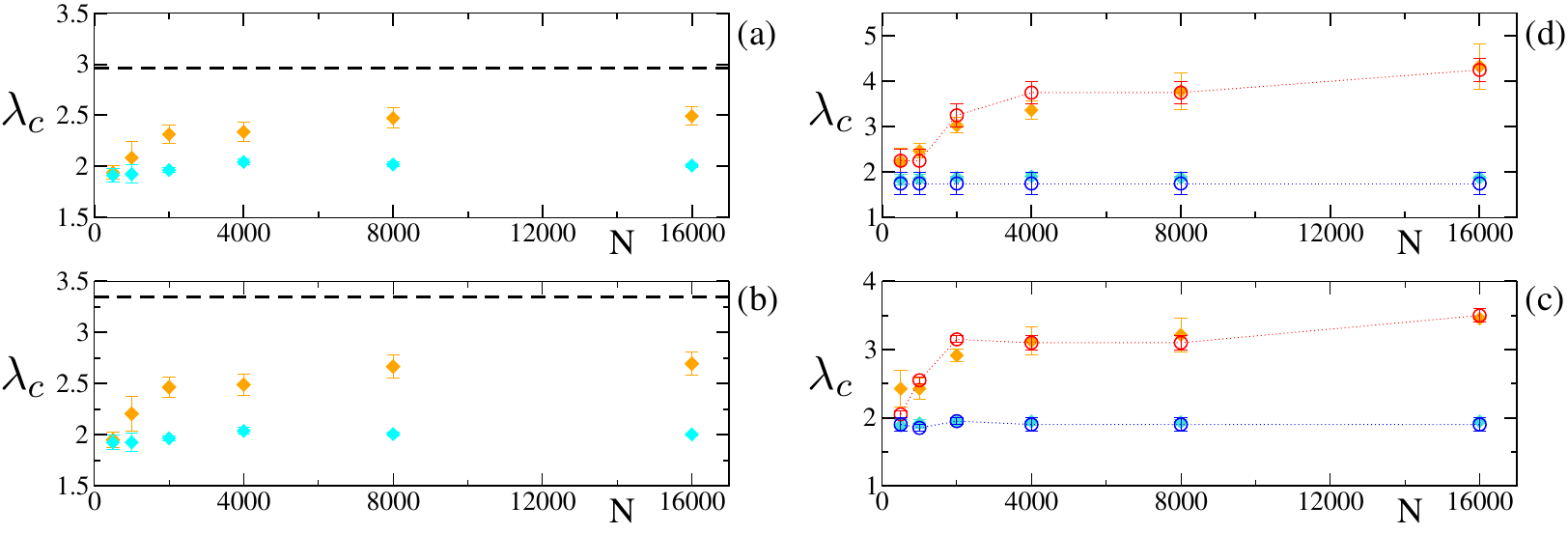}
\caption{Critical couplings $\lambda^\mathrm{I}_c$ and $\lambda^\mathrm{D}_c$ versus the network size $N$ with different values of $m$ in Eq.~(\ref{Eq:SKM_TL_1997}): (a) $m=0.8$, (b) $m=1$, (c) $m=2$ and $m=6$. In each panel, upper points indicate the value of $\lambda^\mathrm{I}_c$ and lower points indicate the value of $\lambda^\mathrm{D}_c$. In panels (a) and (b) with small $m$, the dashed black lines indicate the onset of collective synchronization obtained from Eq.~\ref{Eq:SKM_onset_synchronization_gaussian}. Adapted with permission from~\cite{olmi_hysteretic_2014}. Copyrighted by the American Physical Society.}
\label{Fig:Olmi_hysteretic_6}
\end{figure}

Additionally,  the finite size of the system can also affect the thresholds of the hysteretic transition given  the value of the inertia (Fig. \ref{Fig:Olmi_hysteretic_6}).  
In comparison to the results obtained from the MFA in the limit of large network sizes $N$, $\lambda^{{\mathrm{I}}}_c$ 
increases steadily with  $N$ and keeps constant 
for large inertia values. 
Interestingly, in the case of decreases in $\lambda$,  $\lambda^
\mathrm{D}_c$ remains nearly constant. 
For increasing $\lambda$, a good agreement of the 
critical coupling between  simulations of large networks and the 
MFA is achieved at small $m$ but not 
at large $m$. The underlying mechanisms can be understood by the emergence of the secondary synchronization of drifting 
oscillators, which was firstly observed  in~\cite{tanaka_first_1997}. 
Clusters of whirling oscillators (the secondary synchronization transition) are formed  with an increase in the inertia were also observed in \cite{olmi_hysteretic_2014}, and the emergence of the clusters can also be observed in  random regular networks and in the high-voltage power grid in Italy.    
Note that in regular networks, where each degree is constant $k_i = k_c$ $\forall i$, the hysteretic 
loop enlarges with respect to  increases in the 
fraction of connected links $k_c / N$.

Additionally to the research on hysteresis, the linear stability of the incoherent solution of the Kuramoto model with inertia with different natural frequency distributions was rigorously analyzed~\cite{acebron_synchronization_2000}. 
 The critical coupling $\lambda_c$, where the system exhibits a first-order transition from non-synchronized to synchronized states, follows~\cite{gupta_nonequilibrium_2014}
\begin{equation}
\frac{1}{\lambda_c} = \frac{\pi g(0)}{2} - \frac{m}{2} \int^{\infty}_{-\infty} \frac{g(\omega)}{1+m^2 \omega^2} d\omega,
\label{Eq:SKM_onset_synchronization}
\end{equation}
where $g(\omega)$ is unimodal with width $\Delta$. Without inertia, i.e. $m \rightarrow 0$, Eq.~(\ref{Eq:SKM_onset_synchronization}) reduces to the exact formula of the onset of collective synchronization of the first-order Kuramoto model (Eq.~\ref{eq:KM_CRITICAL_COUPLING}). For a Lorentzian $g(\omega)$, $\lambda_c$ is given by the following relation~\cite{olmi_hysteretic_2014} 
\begin{equation}
\lambda_c = 2 \Delta (1+ m \Delta),
\end{equation}
which is consistent with the results obtained in~\cite{acebron_synchronization_2000}. 
For a Gaussian distribution and with a rather small $m$, the first corrective terms are determined by~\cite{olmi_hysteretic_2014}
\begin{equation}
\lambda_c = 2 \Delta \sqrt{\frac{2}{\pi}} \left \lbrace 1 + \sqrt{\frac{2}{\pi}} m \Delta + \frac{2}{\pi} m^2 \Delta^2 + \sqrt{\left(\frac{2}{\pi}\right)^3 - \frac{2}{\pi}} m^3\Delta^3 \right\rbrace + O(m^4\Delta^4).
\label{Eq:SKM_onset_synchronization_gaussian}
\end{equation}
With the increases in $m$ and $\Delta$ for the Lorentzian as well as the Gaussian distribution, it becomes harder and harder for the system to achieve complete synchronization~\cite{olmi_hysteretic_2014}.

When the natural frequency distribution $g(\omega)$ is unimodal, symmetric, and has zero mean, the mean phase could be taken as a constant, e.g. $\psi(t) \equiv 0$. In the continuum limit, where $N \rightarrow \infty$, the fluctuations of $r(t)$ vanish  and $r(t)$ therein is assumed to be constant, i.e. $r(t) \equiv r$. In this case, the system could be considered as a set of an one-node model. With respect to parameter values in the bifurcation diagram in Fig.~\ref{Fig:SKM_infinite_bus_bar_parameter_space}, one oscillator could be located in the region of the stable limit cycle, of the stable fixed point or of bistability with the coexistence of two  such stable solutions.

\subsubsection{Mean-field theory with noise}
Next, we briefly review the second-order Kuramoto model that includes noise:
\begin{equation}
\begin{split}
\dot{\theta}_i &= \nu_i\\
m\dot{\nu}_i &= - \nu_i + \omega_i + \lambda r   \sin{(\psi-\theta_i)} + \xi_i(t),
\end{split}
\label{Eq:SKM_noise}
\end{equation}
where $\xi_i(t)$'s are independent sources of Gaussian white noises, with $\left\langle \xi_i \right\rangle =0$ and correlation $\left\langle \xi_i(t) \xi_j(s) \right\rangle = 2 D \delta_{ij} \delta(t-s)$. 

When $N \rightarrow \infty$, for the one-oscillator probability density, Acebr\'on and Spigler~\cite{acebron_adaptive_1998} considered the evolution equation of $\rho(\theta, \nu, \omega, t)$ as
\begin{equation}
\frac{\partial \rho }{\partial t} = \frac{D}{m^2 } \frac{\partial^2 \rho}{\partial \nu^2} - \frac{1}{m } \frac{\partial}{\partial \nu} \left[(-\nu + \omega + \lambda r \sin(\psi - \theta))\rho \right] - \nu \frac{\partial \rho}{\partial \theta},
\label{Eq:SKM_noise_continuum_limit_rho}
\end{equation}
where $\rho$ is normalized with $\int^{\infty}_{-\infty} \int^{\pi}_{-\pi} \rho(\theta, \nu,\omega,0) d \theta d\nu =1$. 
For identical oscillators, $g(\omega) = \delta(\omega)$, via setting  the stationary solution $\rho(\theta, \nu) = \chi(\theta) \eta (\nu)$ and assuming that $\eta(\nu)$ is independent of $\lambda$ (motivated by numerical simulations~\cite{acebron_adaptive_1998}), the phase and frequency distributions are obtained explicitly from Eq. (\ref{Eq:SKM_noise_continuum_limit_rho}) and such results are validated by simulations~\cite{acebron_adaptive_1998}.   
In this case, the critical coupling from incoherence to coherence is independent of the inertia.  For bimodal distributions  of natural frequency, the inertia tends to destabilize incoherence and to harden the bifurcation from incoherence to synchronized states \cite{acebron_synchronization_2000}. In particular, in the absence of inertia, the bifurcation is supercritical, but it becomes subcritical with  increasing inertia. 

Instead of Eq.~(\ref{Eq:SKM_noise_continuum_limit_rho}), via averaging over the velocity $\nu(t)$ in the long-time limit, the Fokker-Planck equation for the probability distribution $\rho(\theta,\nu,\omega,t)$ can be reduced  into the Smoluchowski equation, which yields~\cite{hong1999NoiseEffectsSync,bonilla_chapman-enskog_2000} 
\begin{equation}
\frac{\partial \rho(\theta, t)}{\partial t} = \frac{\partial}{\partial \theta} \left[ \left( \frac{\partial V(\theta)}{\partial \theta} \rho(\theta) + D \frac{\partial \rho(\theta)}{\partial \theta} \right) \left( 1+ m \frac{\partial^2{ V(\theta)} }{\partial \theta^2 }\right) \right],
\label{Eq:SKM_noise_smoluchowski}
\end{equation}
with the washboard potential $V(\theta) \equiv - \lambda r \cos(\theta) - \omega \theta$. For $D=0$, by analyzing the stationary state of Eq.~\ref{Eq:SKM_noise_smoluchowski}, the self-consistent equation is obtained~\cite{hong_phase_2000} 
\begin{equation}
r = \left( \frac{\pi}{2} - \frac{m}{2}\right) g(0 ) \lambda r  + \frac{4}{3} m g(0) (\lambda r)^2 + \frac{\pi}{16}  g^{''}(0) (\lambda r)^3 + O(\lambda r)^4. 
\end{equation}
In the presence of the inertia ($m  \neq 0$), drifting oscillators as well as locked oscillators contribute to the phase coherence and the resulting quadratic term of the order $(\lambda r)^2$ induces hysteresis in the bifurcation diagram as observed before~\cite{tanaka_first_1997,tanaka_self-synchronization_1997}.  The hysteresis is reduced with the presence of  noise. The critical coupling strength increases monotonically with the increase of $D$~\cite{hong_phase_2000}. Moreover, via analyzing the  power spectrum of the phase velocity,  in contrast with synchronization suppressed by noise, the response of the phase velocity to the external driving is enhanced by a certain amount of noise~\cite{hong_phase_2000}. Meanwhile, two related results were obtained: a consistent two-term Smoluchowski approximate equation in the limit of small inertia and  the amplitude equation for an O(2)-symmetric Takens-Bogdanov bifurcation at the tricritical point of a standard Kuramoto model using  the Chapman-Enskog method~\cite{bonilla_chapman-enskog_2000}.  

Recently, Gupta et al.~\cite{gupta_nonequilibrium_2014} investigated coupled oscillator systems with inertia and noise in different situations.  The dynamics studied in~\cite{gupta_nonequilibrium_2014} written in the dimensionless form follows
\begin{equation}
\begin{split}
\dot{\theta}_i &= \nu_i \\
\dot{\nu}_i &= - \frac{1}{\sqrt{m}} \nu_i + \delta \omega_i + r   \sin{(\psi-\theta_i(t))} + \xi_i(t), 
\end{split}
\label{Eq:SKM_Gupta_dynamics}
\end{equation}
where $\delta$ is the width of the frequency distribution $g(\omega)$. Given a realization of $g(\omega)$, where the set of oscillators consists of $N_1$ oscillators with frequencies $\omega_1$ and $N_2$ oscillators with frequencies $\omega_2$. 
When $\delta=0$, the stationary solution of $\rho(\theta,\nu)$ is
 \begin{equation}
 \rho_\mathrm{st}(\theta,\nu) \propto \mathrm{exp} \left [ -\left(\nu^2/2 - r \cos(\theta) \right) /D\right],
 \end{equation}
which corresponds to the canonical equilibrium and the stationary solution of the order parameter  $r$ is determined by a self-consistent equation. 
For $\delta \neq 0$, the incoherent stationary state of $\rho_\mathrm{inc}(\theta,\nu,\omega)$  is 
\begin{equation}
\rho_\mathrm{inc}(\theta,\nu,\omega) = \frac{1}{(2\pi)^{3/2} \sqrt{D}} \mathrm{exp}\left[(-\left( \nu - \delta \omega \sqrt{m}\right)^2)/(2D) \right]. 
\end{equation}
Interestingly, in terms of a linear stability analysis of the incoherent state,   the stability threshold $\delta^\mathrm{inc}$ for the incoherent state in different situations can be obtained~\cite{gupta_nonequilibrium_2014}. With $m \approx 0$ at fixed $D$,  $\delta^\mathrm{inc}(m,D)$ satisfies 
\begin{equation}
 2 = \int^{\infty}_{-\infty} \frac{D g(\omega) d \omega}{ D^2 + \omega^2 (\delta^\mathrm{inc}(0,D))^2  }. 
\end{equation}
If $D \approx D_c = 1/2 $ then $\delta^{\rm{inc}}(m,D) \approx 0$. 
 When $D \approx 0$ at fixed $m$, $\delta_{\mathrm{noiseless}}^\mathrm{inc}(m,D)$ satisfies 
 \begin{equation}
 1 = \frac{\pi g(0)}{ 2 \delta^\mathrm{inc}_\mathrm{noiseless}} - \frac{m}{2} \int^{\infty}_{-\infty} \frac{g(\omega) d \omega}{ 1 + m^2 (\delta^\mathrm{inc}_\mathrm{noiseless})^2 \omega^2}. 
 \end{equation}
Komarov et al.~\cite{komarov_synchronization_2014} studied a generic model in the presence of inertia, noise, and phase shift, i.e.
\begin{equation}
m_i\ddot{\theta}_i = - \dot{\theta}_i + \omega_i + \lambda r   \sin{(\psi-\theta_i - \varphi)} + \xi_i(t),
\label{Eq:SKM_noise_Komarov}
\end{equation}
 where the inertia $m_i$ is distributed according to the density function $f(m)$, and $\varphi$  is the phase shift. Rich phenomena emerge via considering various symmetry and asymmetry distributions of $f(m)$ and $g(\omega)$, e.g., the derivation of an exact solution of the self-consistent solution of the order parameter shows nontrivial phase transitions to synchrony due to correlations between natural frequencies and the moments of inertia~\cite{komarov_synchronization_2014}. 

Note that the recent review by Gupta et al.~\citep{gupta_kuramoto_2014} provided a general mean-field analysis framework of the second-order Kuramoto model with noise, focusing on the equilibrium and out-of-equilibrium aspects of its dynamics from a statistical physics point of view.

\subsubsection{Frequency-degree correlation}
Let us now turn to effects of network topologies on dynamics.  
As illustrated in Sec.~\ref{sec:explosive_sync}, the correlation between the dynamics and the structure can induce the emergence of dynamical abrupt phase transitions.  
In this case, the natural frequency distribution  becomes asymmetric.  
Basnarkov and Urumov~\cite{basnarkov2008kuramoto} investigated the first-order Kuramoto model with natural frequencies distributed according to a unimodal asymmetric function and showed that a first-order phase transition occurs if the distribution has a sufficiently large flat section.  
The MFA method~\cite{tanaka_self-synchronization_1997,tanaka_first_1997} (shown in the section~\ref{SKM_Mean_field_theory_without_noise})  is provided for the second-order Kuramoto  model~(\ref{Eq:SKM_complex_networks}) with symmetric frequency distribution, but the method for the model with asymmetric distribution is still open.

Ji et al.~\cite{ji2013cluster, ji_analysis_2014} substantially extended the first-order Kuramoto model with frequency-degree correlation as discussed in Sec.~\ref{sec:explosive_sync} to the Kuramoto model with inertia.  By considering  $\omega_i$ of each oscillator $i$ proportional to its degree with zero mean, i.e. $\omega_i = B(k_i - \left\langle k \right\rangle)$ so that $\sum_i \omega_i=0$, the original dynamics becomes 
\begin{equation}
\ddot{\theta}_i = - \alpha\dot{\theta}_i + B(k_i - \left\langle k \right\rangle) + \lambda\sum\limits^{N}_{j=1} A_{ij} \sin{(\theta_j-\theta_i)},
\label{Eq:CES_orignal_dynamics}
\end{equation}
where $B$ is a proportionality constant that weights the influence of the local structure on the natural frequencies.
%
When $N \rightarrow \infty$, and in uncorrelated networks, after the transformation via replacing the coupling term by the imaginary term of the continuum-limit version of $r$ (Eq.~\ref{eq:order_parameter_peron_rodrigues1}), Eq.~(\ref{Eq:CES_orignal_dynamics}) becomes  
\begin{equation}
\ddot{\theta} = -\alpha \dot{\theta} + B(k-\left\langle k \right\rangle) + k\lambda r \sin(\psi - \theta),
\label{Eq:CES_mean_field}
\end{equation}
where the subscript $_i$ is dropped in the continuum limit.

In the mean-field version (\ref{Eq:CES_mean_field}), each oscillator appears to be uncoupled from the others but interacts through the mean-field properties $(r, \psi)$ and the phase $\theta$ is pulled toward  $\psi$ by the coupling strength $k\lambda r$. 
Natural frequencies are proportional to degrees, and since the degree distribution is not necessary symmetric, $\psi$ cannot be set as a constant, but rather oscillates periodically. 
 Here, we assume that $r$ is at the steady state, otherwise, complex phenomena could occur, e.g. secondary synchronization~\cite{tanaka_first_1997,olmi_hysteretic_2014}. To derive sufficient conditions for synchronization, for convenience, a new rotating reference is defined as $\phi = \theta - \psi$. Substituting this into Eq.~(\ref{Eq:CES_mean_field}) yields
\begin{equation}
\ddot{\phi}=-\alpha\dot{\phi}+B[k-\left\langle k \right\rangle-C(\lambda r)]- k\lambda r\sin \phi,
\label{Eq:CES_phi_new_reference}
\end{equation}
where $C(\lambda r) \equiv (\ddot{\psi} + \alpha\dot{\psi})/B$.   
In this case, each oscillator can be treated separately and behaves independently, i.e. either synchronizes to the mean-field or runs periodically with frequency given by Eq.~(\ref{Eq:SKM_instant_frequency_approximation}), which further depends on the parameter combination that includes the dissipation coefficient $\alpha$, the new natural frequency $B[k-\left\langle k \right\rangle-C(\lambda r)]$ and the new coupling strength $k\lambda r$.

Provided that nodes with degree   within the range $[k_1, k_2]$ are synchronized, i.e. $\dot{\phi}=0$ and $\ddot{\phi}=0$, their phases are  $k$-dependent with $\phi= \arcsin\left(\frac{B\left(k-\left\langle k\right\rangle -C(\lambda r)\right)}{k\lambda r}\right) $ and the density function $\rho(\phi|k)$ can be rewritten as $\rho(\phi|k)=\delta\left[\phi-\arcsin\left(\frac{B\left(k-\left\langle k\right\rangle -C(\lambda r)\right)}{k\lambda r}\right)\right] 
\mbox{ for }k\in\left[k_{1},k_{2}\right]$.
After substituting the density function into the definition of the order parameter (see  Sec. 2), the locked order parameter $r_\mathrm{lock}$ follows
\begin{eqnarray}
r_{\mathrm{lock}} = \frac{1}{\left\langle k\right\rangle }\int_{k_{1}}^{k_{2}}\int_0^{2\pi} P(k)ke^{i\phi(t)}
  \delta{\left[\phi-\arcsin{\left(\frac{B(k-\langle k\rangle-C(\lambda r))}{k\lambda r}\right)}\right]}d\phi dk\nonumber,\\
\label{Eq:CES_r_lock_complex}
\end{eqnarray}
and its real term becomes 
\begin{equation}
r_{\mathrm{lock}} =\frac{1}{\left\langle k \right\rangle}\int_{k_1}^{k_2}\left. kP(k)\sqrt{1-\left(\frac{B\left(k-\left\langle k\right\rangle -C(\lambda r)\right)}{k\lambda r}\right)^2} \right. dk . 
\label{Eq:CES_r_lock}
\end{equation} 

\begin{figure}
\begin{center}
\includegraphics[width=1.0\linewidth]{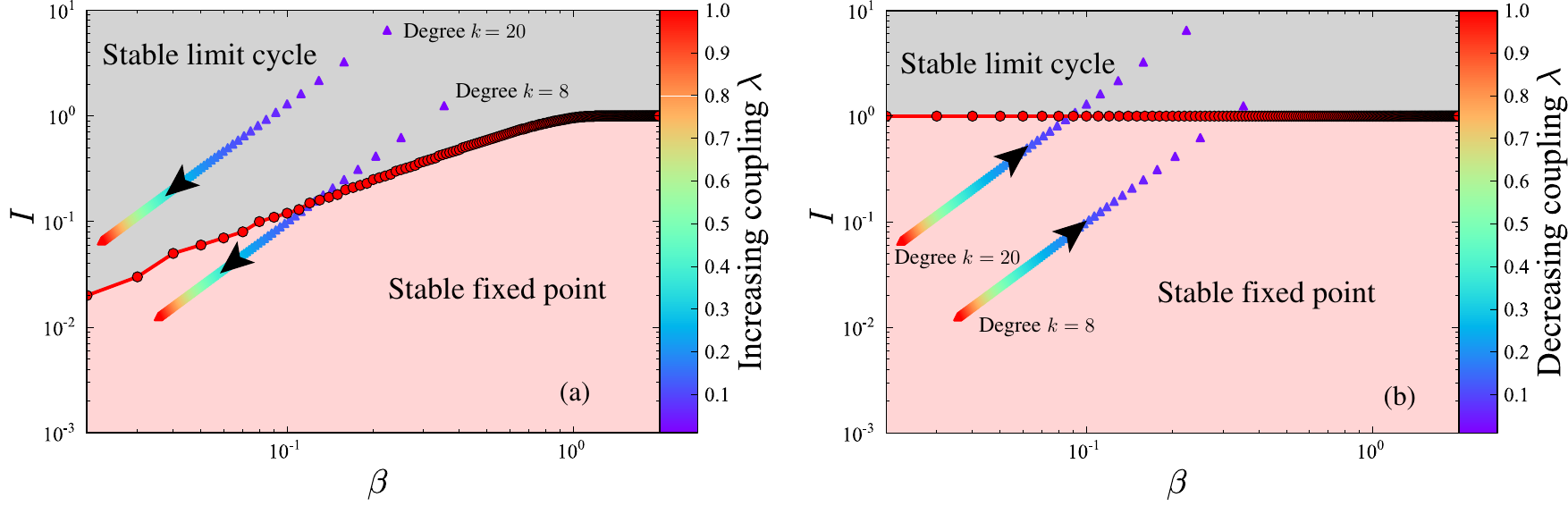}
\end{center}
\caption{Parameter space of one-node model for the increasing coupling strength (a) and for the decreasing coupling strength (b). The red (dark) region indicates the existence of the stable fixed point (the stable limit cycle). 
Adapted with permission from~\cite{ji_analysis_2014}. Copyrighted by the American Physical Society.}
\label{Fig:ji_analysis_2014_1}
\end{figure}

On the other hand, nodes with $k\in k_{\mathrm{drift}} \equiv \left[k_{\min},k_1\right]\cup \left[k_2,k_{\max}\right]$ are drifting, where $k_{\min}$ denotes the minimal degree and $k_{\max}$  the maximal degree.  
These drifting nodes rotate with the period $\hat{T}$ and the frequency $\hat{\omega} = \frac{2\pi}{\hat{T}}$ in the stationary state. 
As the density
$\rho_{\rm{drift}}( \phi,t|k)$ is proportional to $|\dot{\phi}|^{-1}$~\cite{tanaka_first_1997,tanaka_self-synchronization_1997} and
$\oint \rho_{\rm{drift}}(\phi|k) d\phi =\int^{\hat{T}}_0 \rho_{\rm{drift}}(\phi|k) \dot\phi dt =1 $, we get 
$\rho_{\rm{drift}}(\phi|k)=\hat{T}^{-1}|\dot\phi|^{-1}=\frac{\hat{\omega}}{2\pi}|\dot\phi|^{-1}$.
Therefore, the drifting order parameter $r_\mathrm{drift}$ becomes
\begin{equation}
r_{\mathrm{drift}}=\frac{1}{2\pi \left\langle k \right\rangle}\int_{k \in k_{\mathrm{drift}}}\int^{\hat{T}}_0 kP(k) \hat{\omega} |\dot\phi|^{-1} e^{i\phi(t)} \dot{\phi}dt dk.
\label{Eq:CES_r_drift_1}
\end{equation}
As nodes with negative (positive) natural frequency oscillate over (under) the locked group, one can assume that $\dot{\phi}<0$ for $k \in \left[ k_{\min},k_1\right]$ and $\dot{\phi}>0$ for $k \in \left[k_2,k_{\max}\right]$ without loss of generality.
A perturbation approximation of the self-consistent equations enables us to get a series expression of the periodic solution $\phi (t)$ using the Poincare-Lindstead method and approximate $\cos(\phi(t))$ using Bessel functions~\cite{tanaka_self-synchronization_1997}. After performing some manipulations on Eq.~(\ref{Eq:CES_r_drift_1}) motivated by~\cite{tanaka_first_1997}, one gets the final solution of the real part of $r_{\text{drift}}$ as follows
\begin{equation}
r_{\mathrm{drift}} = \left(-\int_{k_{\min}}^{k_1} + \int_{k_2}^{k_{\max}} \right)\frac{-rk^2\lambda \alpha^4 P(k)}{B^3\left[k - \left\langle k \right\rangle -C(\lambda r)\right]^3\left\langle k\right\rangle} dk. 
\label{Eq:CES_r_drift_3}
\end{equation}
The self-consistent equation of $r$ sums the contribution $r_\mathrm{lock}$ (\ref{Eq:CES_r_lock}) from oscillators locked to the mean-field and the contribution $r_\mathrm{drift}$ (\ref{Eq:CES_r_drift_3}) from the rest, i.e. $r= r_\mathrm{lock} + r_\mathrm{drift}$.  
To solve this self-consistent equation, three parameters remain to be solved:  constant $C$ and the range of the degree of synchronized nodes  $[k_1,k_2]$. 
Considering the complex order parameter summing Eqs.~(\ref{Eq:CES_r_lock_complex}) and (\ref{Eq:CES_r_drift_1}) and
following a similar procedure to express $\int_0^{\hat{T}} \cos\phi(t) dt$~\cite{tanaka_self-synchronization_1997} for the integral $\int_0^{\hat{T}} \sin\phi(t) dt$ in its imaginary term, we yield the self-consistent equation 
\begin{eqnarray}\nonumber
0 & = & \frac{1}{\left\langle k\right\rangle }\int_{k_{1}}^{k_{2}}kP(k)\frac{B(k-\langle k\rangle-C(\lambda r))}{k\lambda r}dk\\
 &  & +\frac{1}{2\left\langle k\right\rangle }\left(\int_{k_{\min}}^{k_{1}}+\int_{k_{2}}^{k_{\max}}\right)\frac{rk^{2}\lambda\alpha^{2}P(k)}{B^{2}\left[k-\left\langle k\right\rangle -C(\lambda r)\right]^{2}}dk
  \label{Eq:CES_C}
\end{eqnarray}
Given the variable $\lambda r$, $C(\lambda r)$ can be taken as a function of $k_1$ and $k_2$.  

In order to determine these quantities, for notational simplicity, we set $\beta \equiv \alpha / \sqrt{k\lambda r}$ and 
$I \equiv B(k - \left\langle k \right\rangle - C(\lambda r))/(k\lambda r)$. 
As depicted in panel (a) of Fig.~\ref{Fig:ji_analysis_2014_1}, initially all nodes are in the region of the stable limit cycle with increasing $\lambda$ until the onset of synchronization $\lambda^\mathrm{I}_c$. At $\lambda^\mathrm{I}_c$, the homoclinic bifurcation occurs and nodes within the synchronization boundary start synchronizing to the mean-field. For small value of $\beta$, $[k^\mathrm{I}_1, k^\mathrm{I}_2]$ must fulfil two conditions: $\frac{\left|B\left(k - \left\langle k \right\rangle - C(\lambda r)\right)\right|}{k\lambda r}\leq 1$ and $ \frac{\left|B\left(k - \left\langle k \right\rangle - C(\lambda r)\right)\right|}{k\lambda r}\leq \frac{4\alpha}{\pi \sqrt{k\lambda r}}.$

\begin{figure}
\begin{center}
\includegraphics[width=1.0\linewidth]{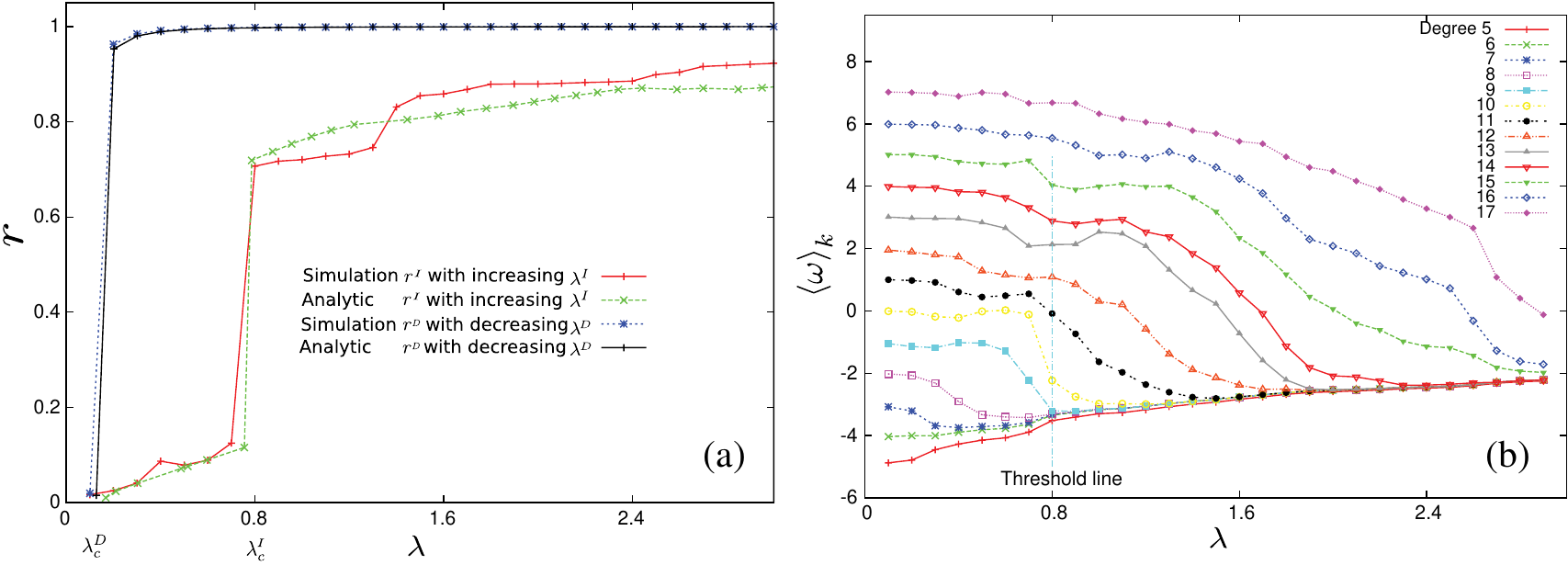}
\end{center}
\caption{Analytical (in blue) and numerical (in red) results of the order parameter $r$ with increasing and decreasing strength (a) and $C(\lambda r)$ with increasing coupling strength (b) for synchronization diagrams.   Adapted with permission from~\cite{ji2013cluster}. Copyrighted by the American Physical Society.}
\label{Fig:ji_analysis_2}
\end{figure}


For  decreasing $\lambda$, nodes start from the phase-locked synchronous state (panel (b) in Fig.~\ref{Fig:ji_analysis_2014_1}). With  decreasing in $\lambda$, nodes reach the asychronous state at  $\lambda^\mathrm{D}_c$, at which a saddle-node bifurcation occurs.  Therefore, the phase-locked oscillators satisfy  $\frac{\left|B\left(k - \left\langle k \right\rangle - C(\lambda r)\right)\right|}{k\lambda r}\leq 1$ and the synchronization boundary follows $[k_1^\mathrm{D}, k_2^\mathrm{D}] \equiv \left[\frac{\left\langle k \right\rangle + C(\lambda r)}{1 + \frac{\lambda r}{B}} , \frac{\left\langle k \right\rangle + C(\lambda r)}{1 - \frac{\lambda r}{B}} \right]. $
With the above synchronization boundary $[k_1,k_2]$, the self-consistent equation of the order parameter can be obtained as a sum of Eqs.~\ref{Eq:CES_r_lock} and \ref{Eq:CES_r_drift_3} as a function of $C$, $k_1$ and $k_2$. 
Additionally, the imaginary term of the complex order parameter (\ref{Eq:CES_C}) should then be used to obtain the dependence $C=C(\lambda r)$. The comparison between analytical results of the order parameter and simulations are shown in Fig.~\ref{Fig:ji_analysis_2} (a).
 
To further uncover the first-order phase transition [Fig.~\ref{Fig:ji_analysis_2} (a)], the average frequency $\left\langle \omega \right\rangle_k$ of nodes with the same degree $k$ (here called cluster) is visualized as a function of the coupling strength $\lambda$, and its calculation follows  
 \begin{equation}
\left\langle \omega\right\rangle _{k}=\sum_{[i|k_{i}=k]}\langle \dot{\theta}_{i}\rangle_{t}/(NP(k)), 
 \label{Eq:CES_average_frequency_k} 
 \end{equation}
  where 
$\langle \dot{\theta}_i \rangle_t=\int^{t+T}_t\dot\theta_i(\tau)dt / T$. 
Unlike explosive synchronization in~\cite{gomez2011explosive} and as discussed in Sec.~\ref{sec:explosive_sync} (see Fig.~\ref{fig:gomezgardenez_explosive_fig2}), where all nodes synchronize abruptly at the same coupling, a new phenomenon was found. Oscillators join the synchronous component grouped into clusters of nodes with the same degree, where small degree nodes form the synchronous component simultaneously, whereas high degree nodes synchronize successively (see Fig.~\ref{Fig:ji_analysis_2}). This phenomenon was termed \textit{cluster explosive synchronization}~\cite{ji2013cluster, ji_analysis_2014}.

\subsubsection{Further works}

Additionally, research has been done related to   models with time delay~\cite{kachhvah_time_2014}, with periodic driving~\cite{hong_inertia_1999} and chimera states~\cite{jaros_chimera_2015,olmi_intermittent_2015}. 
 
(i) Time delay: Finite inertia and time delay are related in some systems, e.g., finite-time interval  required for transforming information in a superconducting junction network~\cite{park_synchronization_1997} and in coupled nonlinear electronic
circuits~\cite{ramana_reddy_experimental_2000}. The interplay between inertia and time delay was analytically and numerically investigate~\cite{hong_spontaneous_2002}, where the emergence of spontaneous phase oscillation without external driving was reported. Such spontaneous oscillation was found to suppress synchronization and its frequency was observed to decrease when inertia and time delay decreased. Moreover, the phase diagram  can be analytically obtained (in the three-dimensional space of inertia, time delay and coupling strength), where oscillatory and  stationary states appear.

Another recent contribution studied the onset of synchronization in a star network of the second-order Kuramoto model with frequency-degree correlation in the presence of a time delay. The model   is governed by~\cite{kachhvah_time_2014}
\begin{equation}
m\ddot{\theta}_i = - \dot{\theta}_i + \omega_i + \frac{\lambda}{N} \sum\limits^{N}_{j=1}  \sin{(\theta_j(t-\tau)-\theta_i(t))},
\label{Eq:SKM_time_delay}
\end{equation}
where $\omega_i = k_i$ and $\tau$ denotes the time delay. 
To address the effects of $\tau$ on the collective synchronization, a mean-field analysis was conducted following the same analytical process of~\cite{peron2012explosive}, yielding a good approximation between numerical results and the MFA. However, the results need to be further investigated to include the influences of the inertia $m$ on the onset of synchronization. 


(ii) External periodic forcing: To understand  effects of inertia on the  collective synchronization in a set of coupled oscillators with external driving, consider the following dynamics~\cite{hong_inertia_1999}
\begin{equation}
m\ddot{\theta}_i = - \dot{\theta}_i + \omega_i + \lambda r  \sin{(\psi-\theta_i(t))} + I_i \cos{(\Omega t)},
\label{Eq:SKM_external_driving}
\end{equation}
where $I_i \cos{(\Omega t)}$ describes the periodic driving on the $i$-th oscillator. In the absence of inertia ($m =0$), only oscillators locked to the external driving contribute to the collective synchronization. Otherwise, with inertia ($m \neq 0$), drifting oscillators as well as oscillators locked to the external driving contribute to the collective synchronization. It was shown analytically that  the inertia tends to suppress synchronization~\cite{hong_inertia_1999}. 

(iii) Chimera states: Jaros et al.~\cite{jaros_chimera_2015} illustrated different types of chimera states in the Kuramoto model with inertia and its dynamics follows
\begin{equation}
m\ddot{\theta}_i = - \dot{\theta}_i + \frac{\lambda}{2 L +1} \sum^{i+L}_{j = i-L} \sin{(\theta_j(t)-\theta_i(t) - \varphi)}, 
\end{equation}
where $\varphi$ indicates the phase lag and each node is linked to its $L$ nearest neighbors to the left and to the right.  A different type of spatiotemporal pattern was defined, termed \textit{imperfect chimera states}, where a small number of oscillators escapes from the synchronized cluster. Moreover, by varying  $\varphi$,  the transition of chimera states from coherence to incoherence was observed. 

Additionally, note that effects of assortative mixing on synchronization are worth to being investigated. 
We saw in Sec.~\ref{subsec:assortative} that the combination of mean-field theory and the OA ansatz on the first-order Kuramoto model with frequency-degree correlation can yield a good approximation to discover the influence of topology on collective synchronization~\cite{restrepo2014mean}. It would be interesting to combine recent developments on the low-dimensional
behavior of the second-order Kuramoto model~\cite{ji2014low} with the approach in~\cite{restrepo2014mean} in order to further analyse the recent results
obtained on assortative networks of second-order Kuramoto oscillators~\cite{peron2015EffectsOfAssortativeMixing}. 
Moreover, effects of shortcuts in SW networks on synchronization remain to be further explored~\cite{sasaki_hysteretic_2015}.

\subsection{Basin stability}

In the last decades, much research effort has been devoted to explore how the synchronizability of  a network of coupled oscillators depends on network topology~\cite{pecora_master_1998,arenas2008synchronization}, but from a local perspective, related to  spectral properties of the underlying structure. 
The seminal work~\cite{wiley_size_2006} initiated a new line of research by proposing a new stability approach that is related to the size of the basin of attraction for a synchronous state. Two additional questions were posed: How likely will a network fall into sync, starting from random initial conditions? And how does the likelihood of synchronization depend on the network topology? Alternatively,  the first question can be addressed differently \cite{menck_topological_2012, menck_how_2013}: How likely will a network return to the synchronous state after even large random perturbations? Substantially,  the correlation was investigated between basin stability and the network architecture in the second-order Kuramoto model~\cite{menck_how_2014}. 
In this subsection, we review the basin stability formalism and the main results obtained so far. 
\subsubsection{Basin stability formalism}

Traditional linear stability is too local to adequately quantify in many applications how stable a state is~\cite{menck_how_2013}. A new concept was defined, termed $\textit{basin stability}$ ($\mathcal{BS}$), which quantifies the likelihood that a system will retain a desirable state after even large perturbations \cite{menck_topological_2012,menck_how_2013}. Basin stability is non-local, nonlinear and easily applicable to high-dimensional systems, even with fractal basin boundaries. It is related to the volume of the basin of attraction. 
The concept of $\mathcal{BS}$ can be easily applied to high-dimensional systems as complex networks~\cite{menck_topological_2012,menck_how_2013, menck_how_2014}.

To quantify how stable a synchronous state of the networked Kuramoto model with inertia (\ref{Eq:SKM_complex_networks}) is against large perturbations depending on network topologies,  the basin stability $\mathcal{BS}_i$ at each node $i$ is defined as~\cite{menck_how_2013,menck_how_2014}
\begin{equation}
\mathcal{BS}_i = \int \chi(\theta_i,\nu_i) \rho_{\rm{p}}(\theta_i,\nu_i) d \theta_i d\nu_i,
\label{Eq:Basin_stability_definition}
\end{equation}
 with $\theta_j(0) = \theta^*_j   $ and $     \nu_j(0) =0 $ for all $  j \neq i $
where $\chi(\theta_i,\nu_i)$ is an indicator function with $\chi(\theta_i,\nu_i) = 1$ if $(\theta_i, \nu_i)$ belongs to its basin of attraction of the synchronous state, and $\chi(\theta_i,\nu_i) = 0$, otherwise. $\rho_{\rm{p}}$  is a perturbation density function with the normalization condition $\int \rho_{\rm{p}}(\theta_i,\nu_i) d\theta_i d\nu_i =1$. 
$\theta_j(0) = \theta^*_j$ and $ \nu_j(0) =0 $ for all $  j \neq i$ indicate that initially all nodes are  in the synchronous state except $i$. The value of $\mathcal{BS}_i$ at node $i$ expresses the likelihood that the system returns to the synchronous state after $i$ having been subjected to (large) perturbations. Specifically, $\mathcal{BS}_i=0$ when the node $i$ is unstable, and $\mathcal{BS}_i =1$ when $i$ is globally stable. 

Numerically, $\mathcal{BS}_i$ is estimated by means of a Monte-Carlo method. More specifically, the dynamics is integrated independently for $M_i$ different initial conditions drawn according to $\rho_{\rm{p}}$, one can count the number $S_i$ of initial conditions at which the system converges to the synchronous state and calculate $\mathcal{BS}$  as~\cite{menck_how_2014}
\begin{equation}
\mathcal{BS}_i = \frac{S_i}{M_i}. 
\label{Eq:Basin_stability_numerically}
\end{equation}
This is a repeated Bernoulli experiment, and thus the standard error $e$ of $\mathcal{BS}$ follows
\begin{equation}
e = \frac{\sqrt{\mathcal{BS}_i (1-\mathcal{BS}_i)}}{M_i}, 
\end{equation}
which turns out to be independent of the system's dimension, making $\mathcal
BS$ easily applicable to high-dimensional systems~\cite{menck_how_2014}.  

\subsubsection{Basin stability approximation}

\begin{figure}
\begin{center}
\includegraphics[width=\linewidth]{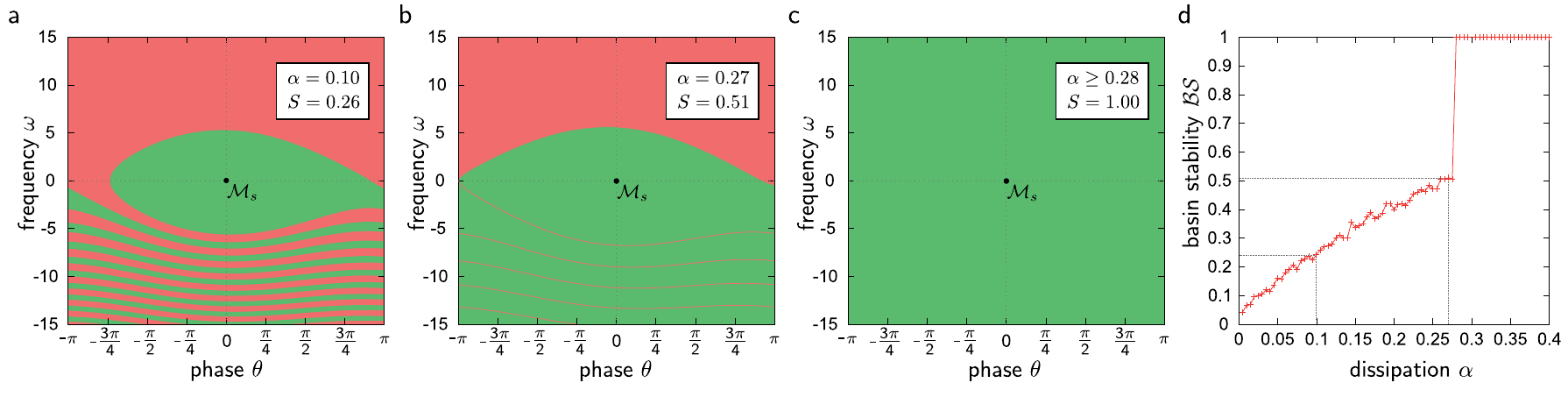}
\end{center}
\caption{State space of the model~(\ref{Eq:SKM_infinite_bus_bar}) with $\alpha=0.1$ (a), $\alpha=0.27$ (b) and $\alpha=0.3$ (c) at $\omega =1$ and $\lambda=8$. The basin of attraction of the stable fixed point $\theta^*$ is colored green, and that of the stable limit cycle is colored red. (d) Basin stability $\mathcal{BS}$ obtained from Eq.~(\ref{Eq:Basin_stability_numerically}) of the synchronous state versus the the damping parameter $\alpha$. Here $\rho(\theta,\nu)$ is a uniform distribution restricted to a lox $(\theta,\nu) \in [-\pi,\pi] \times [-100,100]$. 
Adapted from~\cite{menck_topological_2012}. }
\label{Fig:menck_topological_2012_1}
\end{figure}

As an illustration, we consider the one-node model (\ref{Eq:SKM_infinite_bus_bar})~\cite{menck_topological_2012,menck_how_2014},
The basic dynamics of the one-node model is shown in Sec.~\ref{SKM_illustration_one_node}. Here we focus on evaluating the volume of the attracting basin of the synchronous state. 
In Fig.~\ref{Fig:menck_topological_2012_1}, the basin of attraction of the synchronous state with respect to $\alpha$ is plotted in panels (a-c)  and its volume increases with $\alpha$ until the critical value $\alpha_c = \frac{\pi \omega}{4 \sqrt{\lambda}}$ obtained from~(\ref{Eq:SKM_homoclinic_bifurcation_line}), at which the homoclinic bifurcation occurs (shown in the bifurcation diagram Fig.~\ref{Fig:SKM_infinite_bus_bar_parameter_space}). Correspondingly, basin stability $\mathcal{BS}$ increases with $\alpha \in (0,\alpha_c)$ from Fig.~\ref{Fig:menck_topological_2012_1} (a) to (b) and then persists at $\mathcal{BS}=1$ with $\alpha>\alpha_c$ as shown in Fig.~\ref{Fig:menck_topological_2012_1}(d). Considering a simple case where $\rho_{\rm{p}}(\theta,\nu)$  is uniformly distributed within the region of $(\theta,\nu) \in [-\pi,\pi] \times [-\epsilon,\epsilon]$ with $\epsilon=100$ and $\mathcal{BS}$ therein is proportional to the volume of the basin of attraction of the fixed point. 

Note that in Fig.~\ref{Fig:menck_topological_2012_1}(d), basin stability $\mathcal{BS}$ increases almost linearly with $\alpha$ in the interval $(0,\alpha_c]$. Provided that the linear correlation between $\mathcal{BS}$ and $\alpha$ validates within the region of bistability given other values  $\omega$ and $\lambda$, Ji et al. \citep{ji_onset_2015} estimated $\mathcal{BS} (\alpha,\omega,\lambda)$ with respect to $\alpha$, $\omega$ and $\lambda$. 

Firstly, when the system is weakly dissipated with $\alpha \approx 0$ denoted by $\alpha_0$, the volume of the fixed point's basin ensembles all the librations and the basins' boundary is tangent to the stable manifold of the saddle $(\theta^*_2,\nu^*_2)$. Observed numerically that the basin is symmetrical. 
According to the energy function~(\ref{Eq:SKM_infinite_bus_bar_energy_function}), the upper $\nu_+$ and the lower $\nu_-$ boundaries are approximated by $\nu_\pm \approx \pm \sqrt{2 (\omega \theta + \lambda \cos(\theta) + E(\theta^*_2,\nu^*_2))}$. Therefore, the approximation of  $\mathcal{BS} (\alpha_0, \omega, \lambda)$  at weakly dissipated system follows \citep{ji_onset_2015}
\begin{equation}
\mathcal{BS} (\alpha_0, \omega, \lambda) \approx \int^{\theta^*_2}_{\theta_0} (\nu_+ - \nu_-)/(4\epsilon\pi) d \theta,
\label{Eq:Basin_stability_approximation_alpha_0}
\end{equation}
where $4\epsilon\pi$ is taken for normalization and $\theta_0$ is the left joint point between the lower and upper basin boundary, satisfying $0= \sqrt{2 (\omega \theta_0 + \lambda \cos(\theta_0) + E(\theta^*_2,\nu^*_2))}$ with $\theta_0 \neq \theta^*_2$. 

At  $\alpha_c$, the upper basin boundary is tangent to the curve of the limit cycle. Thanks to the approximation of the limit cycle curve~(\ref{Eq:SKM_instant_frequency_approximation}), the upper basin boundary follows $\omega/\alpha_c + f(\theta,\alpha_c, \omega, \lambda) $, where $f(\theta,\alpha_c, \omega, \lambda) = \frac{\lambda \alpha^2 }{\alpha^4 + \omega^2} (\frac{\omega}{\alpha} \cos(\theta) - \alpha \sin(\theta))$. As numerically observed that the state space under the upper line belongs to the attracting basin of the stable fixed point. Therefore, at the onset  $\alpha_c$, $\mathcal{BS} (\alpha_c, \omega, \lambda)$ is approximated as  \citep{ji_onset_2015} 
\begin{equation}
\mathcal{BS} (\alpha_c, \omega, \lambda) \approx \int^{\pi}_{-\pi} \frac{\omega/\alpha_c + f(\theta,\alpha_c, \omega, \lambda) + \epsilon}{4 e\epsilon\pi}. 
\end{equation}

Therefore, given the linear correlation between $\mathcal{BS}$ and $\alpha \in [\alpha_0,\alpha_c]$, $\mathcal{BS}$ is defined as a function of  $\alpha$, $\omega$ and  $\lambda$   \citep{ji_onset_2015}
\begin{equation}
\mathcal{BS}(\alpha, \omega, \lambda) \approx \frac{\mathcal{BS}(\alpha_0,\omega, \lambda) - \mathcal{BS}(\alpha_c,\omega, \lambda)}{\alpha_0-\alpha_c} (\alpha - \alpha_0)
+ \mathcal{BS}(\alpha_0,\omega, \lambda), 
\label{Eq:basin_stability_omega_alpha_lambda}
\end{equation}
for $\alpha \in [\alpha_0,\alpha_c]$
and  $\mathcal{BS}(\alpha, \omega, \lambda) =1$ for $\alpha > \alpha_c$. 

By definition Eq. (\ref{Eq:Basin_stability_definition}), $\mathcal{BS}$ depends on the choice of the probability density of random perturbations $\rho_{\rm{p}} (\theta,\nu)$.  
The formula of $\mathcal{BS}$ Eq. (\ref{Eq:basin_stability_omega_alpha_lambda}) is derived given the linear correlation between $\mathcal{BS}$ and the dissipation parameter $\alpha$ when the perturbations are uniformly distributed within a region,  which was  firstly considered for convenience.   The Gaussian distribution $\rho_{\rm{p}}(\theta,\nu) = \frac{1}{(2 \pi)^{3/2}b} \mathrm{exp} \left( -\frac{\nu^2}{2b^2}\right)$ was also considered in~\cite{menck2014wires}. Given parameter values in the region of bistability, the dependence of $\mathcal{BS}$ on the width $b$ of $\rho$ was investigated. At small $b$, $\mathcal{BS}$ persists at a high value. With the increases in $b$ beyond a critical value, $\mathcal{BS}$ decreases sharply and then remains at almost small constant values  with high values of $b$.

Basin boundaries can be either smooth, e.g. intricately intertwined shapes in panel (a) and (b) of Fig.~\ref{Fig:menck_topological_2012_1} or fractal~\cite{menck2014wires}. Fractal basin boundaries can strongly influence the predictability of which attractor a system eventually converge. Suppose that the measurement of initial conditions has an uncertainty $\varepsilon$, if the initial conditions close to fractal basin boundaries, it is uncertain to which attractor the system will converge. This causes some trouble when estimating $\mathcal{BS}$ in the long-term behaviour. The simple damped pendulum with periodic driving was considered and it was shown that  $\mathcal{BS}$ estimation is robust, i.e. $\mathcal{BS}$ can be applicable to nonlinear dynamical systems even with fractal basin boundaries~\cite{menck2014wires}.

\subsubsection{Correlation with network topology}

\begin{figure}
\begin{center}
\includegraphics[width=0.8\linewidth]{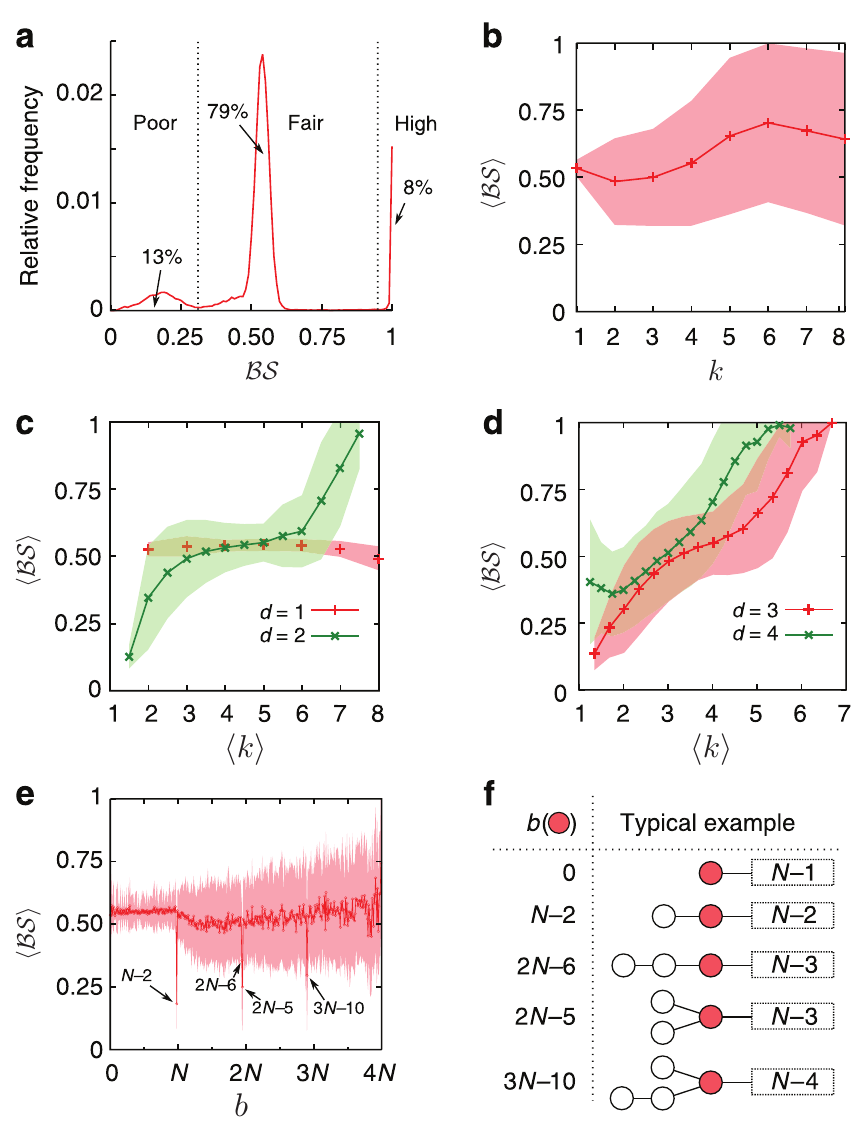}
\end{center}
\caption{Relation between single-node basin stability $\mathcal{BS}$ and network topologies. (a) The histogram of $\mathcal{BS}_i$ for all nodes in the assemble. Most nodes have fair basin stability delimited by two dotted lines. Average basin stability $ \left \langle \mathcal{BS} \right\rangle$ with respect to degree (b), average degree of neighbors of nodes with degree $k=1$ and $k=2$ (c) and that of nodes with degree $k=3$ and $k=4$ (d). $ \left \langle \mathcal{BS} \right\rangle$ with respect to shortest-path betweenness $b$ (d). Some illustrations of nodes with certain distinct values of $b$. Adapted from~\cite{menck_how_2014}.}
\label{Fig:menck_how_2014_2}
\end{figure}

To reveal the relation between stability and network structure, in \cite{menck_how_2014,menck_topological_2012} an ensemble of $1,000$ randomly generated networks with average degree $\left\langle k \right\rangle =2.7$ was statistically studied. Furthermore, two types of natural frequency are selected with $N/2$ nodes with positive natural frequency $\omega_i = +1$ and $N/2$ nodes with negative natural frequency $\omega_i = -1$. The histogram of the single-node basin stability $\mathcal{BS}$ is given in Fig.~\ref{Fig:menck_how_2014_2} showing that most nodes have a fair value of $\mathcal{BS}$. More interesting and special are nodes with poor basin stability ($\mathcal{BS} < 0.3$) resp. high basin stability ($\mathcal{BS} > 0.95$).  To uncover these special nodes, the authors  averaged basin stability $ \left \langle \mathcal{BS} \right\rangle$ of all nodes having the same degree $k$ in Fig~\ref{Fig:menck_how_2014_2}(b). Unlike initial expectations, nodes with higher $k$ do not have larger  $ \left \langle \mathcal{BS} \right\rangle$. A more insightful characteristic is shown in Fig.~\ref{Fig:menck_how_2014_2}(c-d). Average  $\mathcal{BS}$ increases with the average degree $\left\langle k \right\rangle_k$ of neighbors of nodes with degree $k \geq 2$ except nodes with $k =1$. The main conclusions are  obtained from the characteristic shown in  Fig.~\ref{Fig:menck_how_2014_2}(e): $ \left \langle \mathcal{BS} \right\rangle$ as a function of the shortest path and betweenness $b$. Most $b$'s have no certain effects on $ \left \langle \mathcal{BS} \right\rangle$. Interestingly, certain $b$'s, indicated by arrows, have extremely low values of $\mathcal{BS}$. Fig.~\ref{Fig:menck_how_2014_2}(f) sketches some examples of typical nodes in red with such betweenness $b$. These nodes are the so-called \textit{dead ends} or \textit{dead trees} and strongly diminish the stability of the network~\cite{menck_how_2014}.  

The stability of a system with varying network topologies was also investigated given a bipolar natural distribution \cite{menck_how_2014}. 
Actually, for fixed networks, the stability also varies with the topological localization of $\omega_i$. For example, both the values of $\omega_i$ of a given node $i$ and its topological localization affects phase coherence in the networked Kuramoto model with a bipolar distribution of $\omega_i$. Furthermore, synchronization can be enhanced when nodes are surrounded by nodes with opposite $\omega_i$~\cite{buzna2009synchronization}. Therefore, it is interesting to investigate how the localization of generators and consumers influences the stability of power grids. 

\begin{figure}
\begin{center}
\includegraphics[width=0.7\linewidth]{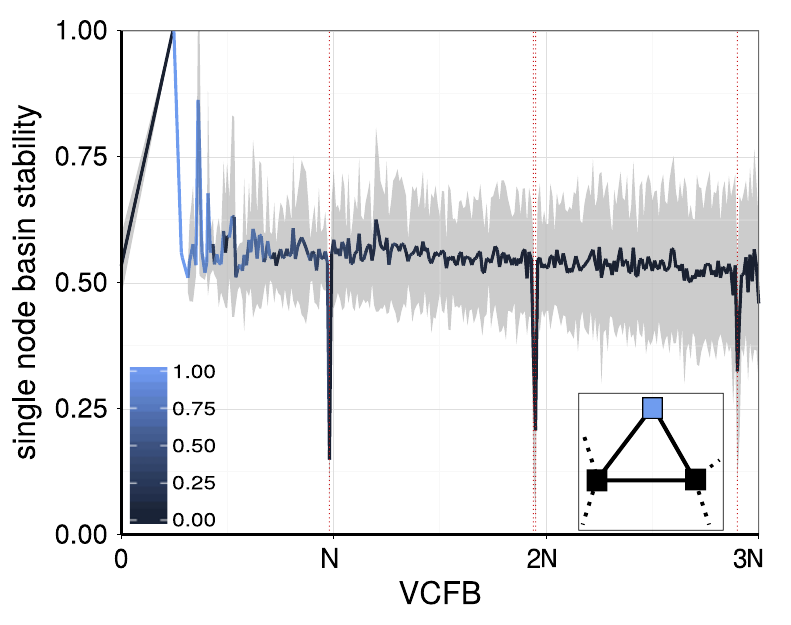}
\end{center}
\caption{Single-node basin stability versus the vertex current flow betweenness (VCFB). The four red dotted lines indicates poor basin stability of dead tree gateways. The line's color indicates the share of nodes in detours with the respective betweenness value. The shade indicates the standard derivation. The inset sketches a detour motifs with the detour node colored in blue. From~\cite{schultz_detours_2014}}
\label{Fig:schultz_detours_2014_3} 
\end{figure}

Schultz et al.~\cite{schultz_detours_2014} further investigated the relationship between stability and topological properties using a random growth model and other spatially embedded infrastructure
networks~\cite{schultz_random_2014}. Such a model reproduces various network characteristics and provides a wide range of network topologies. The dynamics  is based on the classical Kuramoto model with inertia but with weighted coupling strength as follows~\cite{schultz_detours_2014}
\begin{equation}
\ddot{\theta}_i = - \alpha\dot{\theta}_i + \omega_i + \sum\limits^{N}_{j=1} \lambda_{ij} \sin{(\theta_j-\theta_i)},
\label{Eq:Detours_SKM_k_ij}
\end{equation}
where $\lambda_{ij}$ is determined by the voltage amplitudes and generator constants, and it is also related to the link weights, precisely $\lambda_{ij} \propto \frac{1}{X_{ij}}$, where $X_{ij}$ are determined by the entries of the reactance matrix. A statistical analysis using a Monte-Carlo rejection method was used to 
investigate the influence of network motifs, including four-size motifs, dead tree gateways and \textit{detours}, on the stability of networks.
Dead tree gateways are termed to refer an ensemble of nodes within dead trees, e.g. red nodes in Fig.~\ref{Fig:menck_how_2014_2}(f). 
 Detour nodes are nodes in triangles with very low shortest-path betweenness~\cite{schultz_detours_2014}. Characteristics of nodes on detours are with degree two and with unit value of the clustering coefficient. If networks are resistance, these nodes take a significant amount of transformations. 
 In comparison to~\cite{menck_topological_2012,menck_how_2014}, where only poor $\mathcal{BS}$'s are detected in terms of shortest-path betweenness,  downward and upward peaks of the curve of single-node basin stability were observed with respect to the vertex current flow betweenness (VCFB) as shown in Fig.~\ref{Fig:schultz_detours_2014_3}. The appearance of detour nodes seems to prevent poor local basin stability and the identification of them is very important for enhancing the whole network stability. 
Moreover, due to costly simulations, e.g. $100,000 \times 500$ times simulations used in~\cite{menck_how_2014}, the strategy to estimate that $80\%$ of nodes are neither dead tree gateways nor detour nodes shows a good agreement with numerical results~\cite{schultz_detours_2014}. 

From the above various analysis, we can learn that it could be possible to uncover specific correlation between stability and local structures given a specific system with  fixed configurations, but the results can be varied depending on the stability methods. It should be noted, however, that
$\mathcal{BS}$ is a first-order method. In order to deepen the understanding of stability in different contexts, e.g., in partial synchronization, more subtle definitions should be considered. Progress in this direction has been made recently~\cite{hellmann2015survivability,mitra2015IntegrativeQuantifierMultiStability}.

\section{Optimization of synchronization}
\label{sec:optimization}

The observation that the level of synchronization depends on networks structure has motivated questions about the optimal topology to reach the highest synchronization level. Gleiser and Zanette~\cite{Gleiser06} considered the Kuramoto model, in which the dynamics of each oscillator is described by the equations (\ref{eq:KMNETWORKS}) with $\lambda_{ij} = \lambda/k$ (see Sec.~\ref{sec:traditional}).
On a time interval of length $T$, the average oscillation frequency is defined by $\omega_i^{\rm{eff}} = \frac{1}{T} \int_t^{t+T} \dot{\theta_i}(t') dt'$.
If the coupling between two oscillators is sufficiently strong, the synchronous state $\omega_i^{\rm{eff}} = \omega_j^{\rm{eff}}$ is reached for $T\to \infty$.

The network adaptation is performed by the following procedure~\cite{Gleiser06}. After the time interval $T$ is past, a node $i$ is selected at random. The difference  $\delta_{ij} = |\omega_i^{\rm{eff}} - \omega_j^{\rm{eff}}|$ is calculated for all $j \neq i$. The oscillator $k$ for which $\delta_{ik}$ is minimum is selected. If $k$ is a neighbour of $i$, the network is not changed. Otherwise, it is selected, amongst the neighbors of $i$, the node $l$ for which $\delta_{il}$ is maximum. Then, the connection between $i$ and $l$ is replaced by a link between $i$ and $k$. After this update, the process is repeated until a stationary state is attained. For practical purposes, the authors assumed that the procedure is executed only if the maximal difference between the average frequencies of oscillator $i$ and all its neighbours $m$ is higher than a defined threshold.

By considering a random homogeneous network, the authors verified that the connections are changed in order to create and keep connections between oscillators that are more likely to become synchronized~\cite{Gleiser06}. In addition, they verified that the clustering coefficient is changed during the adaptive process. For small values of order parameter, the clustering coefficient is almost unchanged. The same behavior is verified for strong coupling strengths. However, for intermediate values, the clustering presents a maximum, which indicates that the resulting network acquires a more structured (less random) organization. The analysis of the mean distance, reinforce that the network structure moves from a random topology to a SW network~\cite{Gleiser06}. In fact, the authors verified that the network evolution creates cluster of densely connected nodes, which are formed by mutually synchronized oscillators. The occurrence of these clusters decreases the mean distance of the network. 

To complement the analysis by Gleiser and Zanette~\cite{Gleiser06}, 
Brede~\cite{Brede08} studied the relation between the native frequency of 
the oscillators and the organization of optimized networks of Kuramoto 
oscillators. He defined that a network $A_1$ is more synchronizable than 
a network $A_2$ if the order parameter $r_{A_1}(\lambda) > r_{A_2}
(\lambda)$ for all $\lambda$. To optimize the network structure, he 
proposed the following procedure. For a fixed value of the coupling 
strength $\lambda^*$, a network is adapted by a rewiring in the 
coupling's network configuration. More specifically, a random chosen link 
is rewired and this rewiring is accepted if the average fitness 
$\overline{r}(t,\lambda^*) = (1/\Delta t)\sum_{t = 
T_{\rm{rel}}}^{T_{\rm{rel}} + \Delta t} r(t,\lambda^*)$ is increased. 
$T_{\rm{rel}}$ is the number of time steps for relaxation. This process 
is repeated until no move was accepted during $2\sum_{ij}A_{ij}$ 
interactions. It was verified that to obtain optimal networks, native 
frequencies of adjacent oscillators must be anti-correlated~\cite{Brede08}. \textcolor{black}{Furthermore, if the network structure remains fixed and optimization is performed by swapping the natural frequencies of adjacent oscillators instead of by link rewiring, a positive correlation between frequencies and degrees emerges, yielding even higher levels of synchronization (see Sec.~\ref{sec:explosive_sync}).} Therefore, the level of synchronization 
depends not only on the network structure, but also on the assignment of 
frequencies to nodes~\cite{Brede08} \textcolor{black}{(a similar study as in~\cite{Brede08} has been recently carried out by considering a simulated annealing method~\cite{freitas2015synchronization})}. Moreover, he verified that the optimized networks 
are very small, homogeneous and no cliquish. This result is in contrast 
with those in~\cite{Gleiser06}, in which the authors verified that 
networks are characterized by a high level of cliquishness and large 
average distances. This analysis was further extended 
in~\cite{brede2010SynchronizationTransitionsCorrelatedOscillators}, where it was demonstrated that the 
previous observations are also valid in the thermodynamic
limit. He confirmed that the synchronization properties depend on both 
the coupling topology and network structure. 

In a subsequent paper, Brede~\cite{Brede082} considered the same method as in~\cite{Brede08} to analyze local synchronization and the onset of synchronization by adapting the network structure. However, a measure of synchronization given by a combination of the global order parameter and the measure of pairwise coherence of links as a local measure of synchronization was assumed~\cite{gomez2007paths},
\begin{equation}
r_{\mathrm{link}} = \frac{1}{E}\sum_{k,l} \Big | \lim_{\Delta T \to \infty} \frac{1}{\Delta T} \int_{\tau_r}^{\tau_r + \Delta T} e^{i(\phi_l(t)-\phi_k(t))dt}  \Big |,
\end{equation}
where $E$ is the total number of links, $\tau_r$ represents the relaxation time (see Sec.~\ref{subsec:relaxation}) and $\Delta T$ is the time over which the coherence between adjacent oscillators is measured. This measure was used to detect community structures in networks~\cite{arenas2008synchronization,arenas2006SynchronizationRevealsTopological,arenas2006synchronizationProcessesInComplexNetworks,arenas2007synchronizationAndModularityInComplexNetworks}. Notice that a network can present $r_{\mathrm{link}}  \simeq 1$ even when $r \ll 1$. The synchronization of the network is measured by
\begin{equation}
\mathcal{F}(\lambda^*) = b r(\lambda^*) + (1-b) r_{\mathrm{link}} (\lambda^*),
\end{equation} 
where $b$ is a constant that balances the contribution of the global and local synchronization. The author considered $b = 1/2$ in most of the analysis presented in~\cite{Brede082}. 
 
\begin{figure}[!t]
\begin{center}
\includegraphics[width=0.9\linewidth]{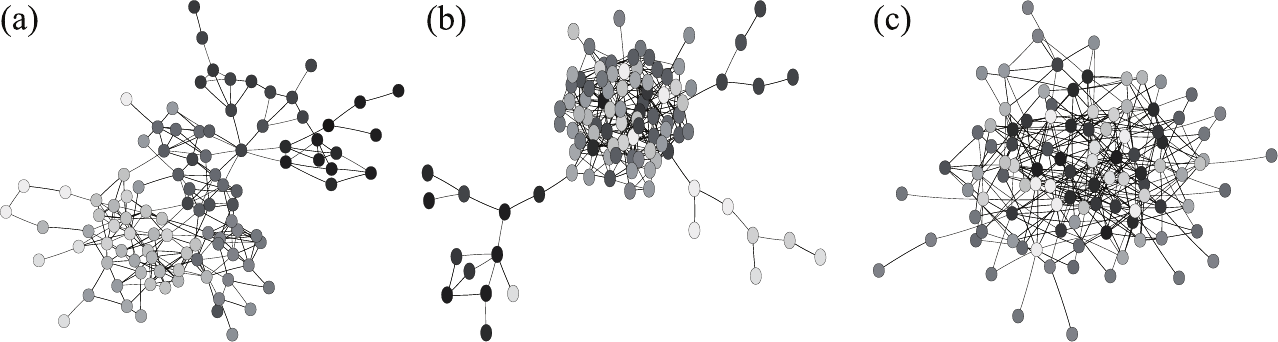}
\end{center}
\caption{Example of optimal networks constructed for (a) $\lambda^* = 0.05$ 
(low coupling strength), (b) $\lambda^* = 0.15$ (intermediate coupling 
strength) and (c) $\lambda^* = 0.40$ (large coupling strength). The color 
of the nodes denotes the natural frequency, where white and black 
correspond to $\omega_i = - 1$ and $\omega_i = 1$, respectively. Adapted 
with permission from Springer Science+Business Media~\cite{Brede082}.}
\label{fig:opt1}
\end{figure}

By considering ER networks as the starting topology for the 
adaptation process, the connections were rewired in order to 
increase $\mathcal{F}$. Brede \cite{Brede082} verified that 
depending on the coupling $\lambda^*$, considered as a fixed 
parameter during the network adaptation, the optimized network 
presents different properties. As we can see in Fig. 
\ref{fig:opt1}, for small coupling, the optimized network 
presents modular organization. These modules are formed by nodes 
presenting similar natural frequencies. As the coupling strength 
is increased, the community organization disappears 
\textcolor{black}{increasing also the anti-correlation between 
adjacent natural frequencies.} In addition, Brede verified that 
networks adapted from large coupling have a later onset of 
synchronization, but reach the fully synchronized state rapidly.
\textcolor{black}{Noteworthy, optimization 
of both local and global optimization also yields correlation 
between frequencies and degrees. Interestingly, in some sense anticipating the phenomenon of explosive synchronization~\cite{gomez2011explosive} discussed in Sec.~
\ref{sec:explosive_sync}, abrupt transitions 
to the synchronized state were reported in \cite{Brede082} for 
networks optimized for large $\lambda^*$, and in~\cite{fan2009EnhancementSynchronizability}, where networks with higher correlations between frequencies and degrees exhibited
lower critical couplings for the onset of synchronization.}

In another paper, Brede investigated optimization in directed networks~\cite{brede083}. In this case, he verified that optimized networks present homogeneous in-degree and skewed out-degree distributions. Directed networks were also analyzed by Zeng et al.~\cite{zeng2012manipulating}, where the authors changed only a fixed number of links in order to shorten the convergence time to synchronization on directed networks.

The topology optimization of a network of non-identical oscillators was also studied by Carareto et al.~\cite{carareto2009optimized}. Their main goal was to obtain a complete synchronization with the smallest possible coupling strength $\lambda$. The authors verified that optimized networks exhibit strong anti-correlation between natural frequencies of adjacent oscillators, as in~\cite{Brede08}. The verification that synchronization is enhanced when nodes are surrounded by neighbors of the opposite frequency was also observed for networks presenting symmetric bipolar distribution of natural frequencies~\cite{buzna2009synchronization}. Carareto et al.~\cite{carareto2009optimized} also verified that this anti-correlation and degree homogeneity are not conflicting, proposing a solution for the heterogeneity paradox~
\cite{motter2005NetworkSyncDiffParadoxHeterogeneity}.

Weighted networks were also studied in terms of optimization. Tanaka and Aoyagi \cite{tanaka2008optimal} considered the Kuramoto model in weighted networks, i.e.,
\begin{equation}
\dot{\theta}_i = \omega_i + \sum_i A_{ij} w_{ij} \sin(\theta_j -\theta_i),
\end{equation}
where $w_{ij}$ is the coupling strength between oscillators $i$ and $j$. The optimization was performed with respect to the coupling strength and natural frequency via assuming the constraint $\sum_{ij} \lambda_{ij}^\alpha = \lambda_{\mathrm{total}}^\alpha$, where $\alpha$ and $\lambda_{\mathrm{total}}$ are constants, and symmetric connections. 
By considering the steepest gradient of $r^2$, where $r$ is the order parameter, it was found theoretically update rules for the coupling strength and the natural frequency. Through the optimization procedure, they found that stronger weights are assigned to connections between pairs of oscillators with very different natural frequencies. The authors also verified that for a large $\lambda_{\mathrm{total}}$, the system of oscillators presents two natural frequencies, whereas for small $\lambda_{\mathrm{total}}$, there is a convergence of frequencies to one single value.

Several other papers further analyzed how the network organization can be 
adjusted to reach the optimal synchronization. The Kuramoto model with noise 
was addressed by Yanagita et al.~\cite{yanagita2012design}. Kelly and 
Gottwald~\cite{kelly2011topology} introduced an energy-like measure to quantify 
the correlation of frequencies with the same magnitude but opposite signs. They 
proposed an algorithm for minimization of this energy and, therefore, obtained 
optimized networks. The proposed method is computationally fast to generate 
optimized networks. Skardal et al.~\cite{skardal2014optimal} derived  a 
synchrony alignment function to measure synchronization, which can be used to 
optimize networks. They verified that synchronization is advanced by an 
alignment of the frequencies with the most dominant Laplacian eigenvectors. A 
matching between the heterogeneity of frequencies and network structure also 
improves the synchronization. Finally, we point out that optimization strategies
considering external forcing were investigated in~\cite{sendina2009entraining,li2015synchronizing,skardal2015ControlCoupledOscillator}.


\section{Applications}
\label{sec:applications}

The Kuramoto model was firstly conceived with the aim 
to qualitatively explain how populations of oscillators
fall into synchronization. However, since its formulation 
40 years ago many real-world applications were reported, a fact 
that not even Kuramoto himself could 
foresee~\cite{kuramotovideo}. Now, his
model has become relevant for the understanding of the dynamics
of real complex systems which are internally organized into 
topologies way different than the regular ones long studied 
in theoretical physics.   
In this section we review some applications of the model in 
different fields, such as engineering, neuroscience, physics and even seismology. We should remark 
though that this is not about a comprehensive and complete review
of real examples of synchronization, but rather about the 
description of recent applications of the Kuramoto model to real-world phenomena putting emphasis on the interplay between 
structure and dynamics. Regarding the former purpose we refer the 
reader to the classical texts in the field~\cite{pikovsky2003synchronization,acebron2005kuramoto,arenas2008synchronization}.      

\subsection{Power-grids}

Power-grids, as one of the largest man-made networks, are a typical example to study collective behavior of networked elements, and its research has become  increasingly important to its stability and network  design  for physicists and engineers~\cite{mallada2011improving,rohden_self-organized_2012,lozano2012RoleOfNetworkTopology,dorfler_synchronization_2013,dorfler2014synchronization,nardelli2014ModelsForTheModernPG,menck_how_2013,motter2013SpontaneousSynchronyInPowerGridNetworks,dewenter2015LargeDeviationProperties,kim2015CommunityConsistencyBasinStability}.

\begin{figure}[!t]
\begin{center}
\includegraphics[width=0.85\linewidth]{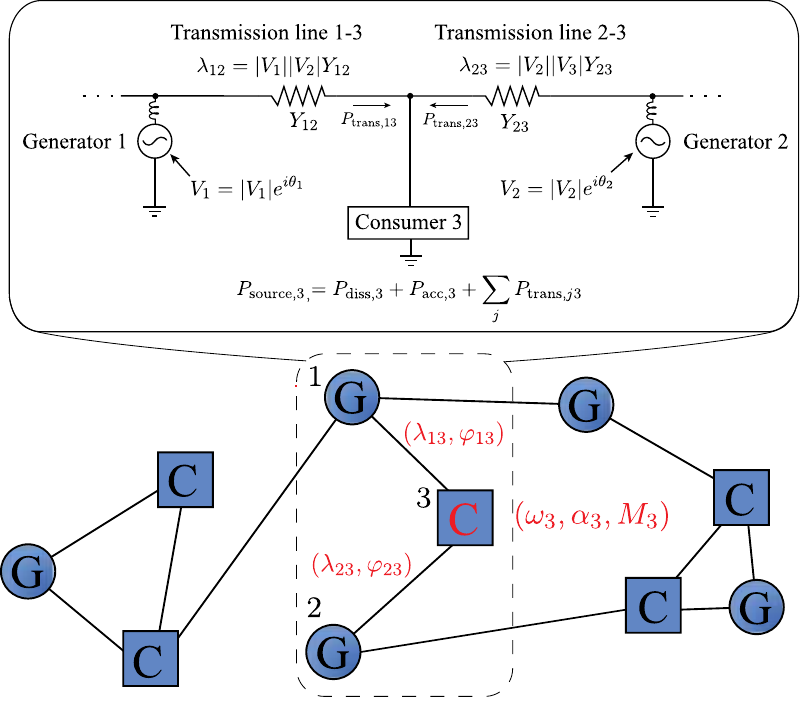}
\end{center}
\caption{Illustration of the reduced model for power grid dynamics (Eq.~\ref{Eq:power_grid_orignal_model}) highlighting the two possible roles played by the nodes, namely as generators ($G$) or consumers ($C$).}
\label{fig:illustration_powergrid}
\end{figure}

Consider an undirected and weighted AC power network, with $N$ generators and reduced admittance matrix $\mathbf{Y}$, 
at each generator $i$, the voltage phasor is accounted for by $V_i = |V_i| e^{i\theta_i}$  with the phase $\theta_i$ and magnitude $|V_i|$. 
Between connected generators $i$ and $j$, the allowed maximum power transferred is denoted by  $\lambda_{ij} = |V_i| |V_j| Y_{ij}$ as the coupling weight  and  the energy loss along the transmission is accounted for by  $\varphi_{ij}$ with $\varphi_{ii}=0$. 
The governing dynamics as \textit{the network-reduced power system model}  at generator $i$ is given by \citep{machowski1997power,dorfler_synchronization_2013}  
\begin{equation}
M_i \ddot{\theta}_i = - \alpha_i \dot{\theta}_i  + \omega_i +  \sum\limits^{N}_{j=1}\lambda_{ij}A_{ij}\sin(\theta_j-\theta_i + \varphi_{ij} ), 
\label{Eq:power_grid_orignal_model}
\end{equation}
where $M_i$ is the inertia coefficient, $\alpha_i$ is the damping coefficient, $\omega_i$ is the effective power input to $i$, and $\varphi_{ij}$ plays the role of phase shift. Figure~\ref{fig:illustration_powergrid} illustrates the network representation of a power-grid modelled by (\ref{Eq:power_grid_orignal_model}).

If the generator properties, voltage magnitudes and line reactances are assumed to be the same, and 
all lines are lossless, i.e. $\varphi_{ij}=0$,  one can reproduce the original second-order dynamics (\ref{Eq:SKM_complex_networks}). 
 Let us review first the derivation of the model via simplifying the swing equation.  
To mimic essential properties of the nonlinear dynamics of a population of $N$ interconnected   dynamical units in power grids, we consider a power grid model on coarse scales~\cite{filatrella_analysis_2008}. 
The state of each unit (machine) $i = 1,\dots N $ is determined by its phase angle $\phi_i(t)$ and its velocity $d{\phi}_i(t)/dt$. Each unit rotates with the same frequency $\Omega = 2 \pi \times 50$ Hz or $\Omega = 2 \pi \times 60 $ Hz, thus
\begin{equation}
\phi_i(t) = \Omega t + \theta_i(t),
\label{Eq:SKM_Derivation_phi}
\end{equation}
where $\theta_i$ indicates the phase difference to the reference phase $\Omega t$. 

During the rotation, the dissipated power is given by $P_{\mathrm{diss},i}=K_D(\dot{\phi}_i)^2,$
where $K_D$ is a friction coefficient.  The kinetic energy follows $E_{\mathrm{kinetic},i} = I_i\dot{\phi}_i^2/2$ and the accumulated kinetic power is given by $P_{\mathrm{acc},i}=\frac{d E_{\mathrm{kinetic},i}}{dt},
$
where $I_i$ is the moment of inertia. If a power flow between machines $i$ and $j$ exists, the power transmission is proportional to the sine of the phase difference, i.e.  $\sin{(\phi_j-\phi_i)}$ and the transmitted power follows $P_{\mathrm{trans}, ij}=-P_{\mathrm{max},ij} \sin{(\phi_j-\phi_i)},
$
where $P_{\mathrm{max},ij}$ is the maximal capacity of the transmission line. If there is no power flow, $P_{\mathrm{trans}, ij}=0$.

The power source $P_{\mathrm{source},i}$, which is fed into each machine $i$, has to be met by the sum of the power transmitted within the grid plus the accumulated and dissipated power, i.e., $P_{\mathrm{source},j}=P_{\mathrm{diss},j}+P_{\mathrm{acc},j}+ \sum\limits^{N}_{i=1} P_{\mathrm{trans},ij}.$
Inserting Eq.~\ref{Eq:SKM_Derivation_phi} and the expressions of powers into the condition for energy conservation 
one gets~\cite{filatrella_analysis_2008,witthaut_braesss_2012,rohden_self-organized_2012}
\begin{equation}
I_j \Omega \ddot{\theta}_j = P_{\rm{source},j} - K_D \Omega^2 - 2K_D \Omega \dot{\theta}_j + \sum_{i=1}^N P_{\max,ij}\sin(\theta_i - \theta_j),
\label{Eq:SKM_Derivation_balance_2}
\end{equation} 
where it was assumed $|\dot{\phi}_j| \ll \Omega$ so that the terms $\dot{\theta}_i^2$ and $\dot{\theta}_i\ddot{\theta}_j$ can be neglected.
For the sake of simplicity, considering that the moment of inertia $I_i$ and the line maximal capacity $P_{\mathrm{max},ij}$ are the same for all elements of the grid; defining $\omega_j \equiv (P_{\mathrm{source},j}-K_D\Omega^2)/(I\Omega)$, $\alpha \equiv 2K_D/I$ and $\lambda_{ij} \equiv P_{\mathrm{max},ij}/(I\Omega)$, finally yields the original dynamics in Eq.~(\ref{Eq:SKM_complex_networks})~\cite{filatrella_analysis_2008,witthaut_braesss_2012,rohden_self-organized_2012,ji2013cluster} (Fig.~\ref{fig:illustration_powergrid}). 
The power-grid model (\ref{Eq:power_grid_orignal_model}) with
multiple time constants, nonhomogeneous coupling and nonuniform phase 
shifts was explored in~\cite{dorfler_synchronization_2012}.


The dynamics is simple enough to be analytically determined, yet sufficiently complex to exhibit various phenomena with insightful and comprehensive explanations.  
The bifurcation diagram of the one-node model is shown in Fig. \ref{Fig:SKM_infinite_bus_bar_parameter_space} and that of a two-node subsystem consisting of one generator and one consumer shows quantitatively the same behavior.  They all capture the essential feature of a real power grid, i.e., the coexistence of self-organized operation and power outage (see Sec. \ref{sec:second_KM} of the second-order Kuramoto model).   
In this case, networks exhibit instability even without overloads \citep{manik_supply_2014}.  

In~\citep{dorfler_synchronization_2013} a condition for synchronization on any arbitrary network was derived rigorously in terms of network topology and grid parameters . 
However, the model is based on simplifying (and unrealistic) assumptions, which makes its applications difficult. 
In order to fill this gap and further analyze more realistic networks, high-order models have been recently considered~\citep{weckesser2013impact,schmietendorf_self-organized_2014,auer2015ImpactOfModelDetail}. One of the issues pursued by these approaches is the precise understanding of effects of network structures on the collective behavior 
of power-grids. For instance, the British power-grid exhibits a bistable regime where normal operation and power outage coexist, as long as it contains a two-node topology \citep{rohden_self-organized_2012} (see Sec. \ref{sec:second_KM}).   
The system could jump from self-organized synchronization to power outage under large perturbations. Further evidences were provided in various topologies, e.g., regular, random, and SW networks \citep{rohden_impact_2014}.   

A possible source of perturbation that could cause power outrages is the temporary resetting of the values of power
generation or consumption $\omega_i$, breaking the conservation
condition and thus leading the system out of the synchronized state~\cite{mallada2011improving,rohden_self-organized_2012,lozano2012RoleOfNetworkTopology,dorfler_synchronization_2013,dorfler2014synchronization,menck_how_2013,motter2013SpontaneousSynchronyInPowerGridNetworks}.   
Futhermore, analyzing the single-node basin stability (see Sec. \ref{sec:second_KM}), it was found that, under the influence of even large perturbations, dead ends, dead trees and detour motifs play an important role in the grid stability \citep{menck_how_2014, schultz_detours_2014}. With respect to different nodes, basin stability as a function of the coupling strength was investigated, and it was found that the corresponding transition varies accordingly. In terms of the transition over an even large time window, nodes have a low community consistency, which yields that the basin stability transition  is not sufficient to detect communities  but is able to capture  typical characteristics  of individual  nodes \citep{kim2015CommunityConsistencyBasinStability}.  
 
Cascading failures in the power supply are not only directly caused by large perturbations on single elements, but can also be induced by adding or removing links, e.g. the 2006 European blackout~\cite{netz2006ReportStatus,bundes2007ReportFederal,utce2006b}.  
Interestingly, including new lines to a grid under normal operation does not always promote synchrony. In fact, it can potentially lead the network to operate incoherently and possibly induce power outages, a process that is similar to the Braess's paradox in traffic networks   \citep{witthaut_braesss_2012}. Moreover, local overloads could be resulted by adding a remote link, which induces nonlocal failures \citep{witthaut_nonlocal_2013}.

Instead of adding links, a drastic change of electric power supply, the number of wind parks and other renewable energy sources could be increased, decentralizing generators and thus posing the question of how the self-organized synchronization and its stability would change correspondingly~\citep{rohden_self-organized_2012}.  Combing renewable energy sources in the energy transition raises novel issues on the stability and design of power-grid networks. For instance, by replacing large centralized generators by small distributed ones, the number of crucial links, which if damaged desynchronizes the network, decreases, and therein more decentralized grids become more robust to topological failures \citep{rohden_self-organized_2012}. 
  In terms of dynamical perturbations, more decentralized grids become less robust \citep{rohden_self-organized_2012}. 
Numerical simulations on various network topologies support that  in decentralized grids, the system reaches its stable state for lower transmission line capacities, and thus decentralization favors power grids \citep{rohden_impact_2014}. 
Effects of temporal energy feed-in fluctuations induced by renewable sources, e.g. solar plants~\cite{schmietendorf_self-organized_2014}, on synchronization and stability remain open.   
A rewiring algorithm by implementing a simple hill-climb method on switching edges was proposed to enhance synchronization \citep{pinto_synchrony-optimized_2014}. Actually, given even that the network topology is fixed, localization  of natural frequencies could affect synchronization. For example, in symmetric bipolar population networks, it was shown that  synchronization  is  promoted 
when nodes are surrounded by nodes of different natural frequency  \citep{buzna2009synchronization} (see also Sec.~\ref{sec:optimization}).

Due to the increasing fraction of renewable energy sources, stable operation suffers from fluctuations of temporally evolving generations. To stabilize the normal operation, it was proposed to adapt generation via feedback control, termed as decentral smart grid control, where a local price is firstly determined by the local frequency as a function of time intervals and time delays, and then the frequency is varied according to this price  \citep{schafer_decentral_2015}.  The suggested control method promotes grid stability for, e.g., a sufficiently large averaging interval \citep{schafer_decentral_2015}.  
Additional to renewable sources, plug-in electric vehicles serve as distributed energy and increase power grid transient stability \citep{gajduk_improving_2014}. 

Various power-grid models could be derived depending on assumptions adopted for the generators equations (\ref{Eq:power_grid_orignal_model}),  and the validity and appropriateness of them depend on specific purposes \citep{nishikawa_comparative_2015}.  
Taking into account that the electrodynamical behavior yields a third-order model with voltage dynamics (extended from the classical second-order Kuramoto model (\ref{Eq:SKM_complex_networks})), where the third  characteristic, voltage, is included besides phase and frequency~\citep{schmietendorf_self-organized_2014}. In comparison to the second-order Kuramoto model (\ref{Eq:SKM_complex_networks}), this shows different stability behavior due to the evolution of voltage dynamics and should provide more realistic features of real power grids. 
Furthermore, considering voltage dynamics could yield even higher dimensional model  \citep{weckesser2013impact}. For an overview of modern achievements
and current open problems on the interdisciplinary 
research on power-grid dynamics we refer the reader
to two recently published open issues~\cite{heitzig2014InterdisciplinaryChallenges,FocusisueNJP} and 
to the survey by D\"orfler and Bullo~\cite{dorfler2014synchronization}, which offers 
an comprehensive review on applications of networks
of phase oscillators to technological systems.


\subsection{Neuronal networks}
\label{subsec:neuroscience}

Synchronization phenomena play a prominent role in neuroscience. In fact, a plenty of evidences shows that synchronization is the key process through which information is processed in cortical areas of the brain~\cite{pikovsky2003synchronization,gray1994synchronous}. In particular, experiments in mammalian  brains point out that different spatial patterns of synchronous firing are observed in the visual cortex when different visual stimuli are presented~\cite{gray1994synchronous}. Further support 
for the hypothesis that collective behavior of groups of neurons is a general property 
of neuronal systems is found in the sensoriomotor cortex~\cite{mackay1997synchronized}. In this case, experiments
show that synchronous oscillations emerge in field potential recordings with well defined frequencies whose amplitude and spatial coverage crucially depend on the motor task being executed~\cite{gray1994synchronous,mackay1997synchronized}. 

In order to better understand the underlying mechanisms that lead
to neural synchronization, the neuroscience community has been invested 
great effort on the computational and theoretical study of neural-
mass models coupled through structural substrates provided by 
neuroanatomical networks~\cite{bullmore2009ComplexBrainNetworks,breakspear2010GenerativeModelsCorticalOscillations,cabral2014ExploringTheNetworkDynamics}. However, the analysis of highly 
accurate models in the physiologically point of a view can be rather 
challenging due to their complexity. On the other hand, in some 
situations, the models can be simplified to coupled phase 
oscillators akin to the Kuramoto model, which still yields a rich 
and non-trivial dynamics while being neurobiologically plausible~\cite{breakspear2010GenerativeModelsCorticalOscillations}. For instance, Wilson-Cowan oscillators are generally used
in the modeling neuronal dynamics of cortical regions~
\cite{wilson1973mathematical}. Using a phase reduction technique (see~\cite{nakao2015PhaseReductionApproach} for a recent review on phase reduction theory)
one can show that the 
Wilson-Cowan model leads to a Kuramoto-like interaction. 
Furthermore, another feature 
that must be taken into account is that the oscillators are
spatially embedded, a condition that inherently induces 
distance dependent coupling between the oscillators.
These concepts were recently illustrated in~\cite{sadilek2015physiologically}, where
the Wilson-Cowan model was shown, under the conditions 
that make the phase-reduction valid, to be equivalent to
a Kuramoto model in a two-layered network subjected to time-delayed coupling. In this model, three macroscopic cortical
dynamical states were found, namely incoherence (related to background activity), high synchronization (associated to epileptic behaviors) and also chaotic states, which 
turn out to be related to resting-state activities~\cite{sadilek2015physiologically}. 
 
Neurons in the visual cortex respond accordingly to visual stimuli with different orientations. Thus, when a specific stimulus with a particular orientation is given, neurons that respond to the specific stimuli 
synchronize. Furthermore, evidences show that neurons in the fourth layer of the cortex have distance dependent connections with other neurons in a heterogeneous fashion~\cite{bullmore2009ComplexBrainNetworks}. These aspects of the neuron dynamics and the connectivity pattern of neurons in the visual cortex were taken altogether in~\cite{tauro2014ModellingSpatialPatternsVisualCortex}, where the angle of synchronization of neurons was investigated by using Kuramoto oscillators 
coupled in SF networks embedded in 2D Euclidean spaces~\cite{rozenfeld2002ScaleFreeNetworksOnLattices,ben2003GeographicalEmbeddingSFNetworks}.  
Interestingly, despite its simplicity, the model was able to reproduce the emergence of clustered and stripped patterns akin to those ones observed in real experiments in infant macaques~\cite{obermayer1994development} and monkeys~\cite{obermayer1993geometry}. Moreover, by tuning the balance between inhibitory and excitatory couplings the model in~\cite{tauro2014ModellingSpatialPatternsVisualCortex} was even able to obtain patterns that qualitatively resemble experiments with ferrets~\cite{coppola1998unequal}. 

%

Another utility of the Kuramoto model is to probe 
the topology of networks derived from neuronal data. As 
seen in Sec.~\ref{sec:different_topologies}, transient dynamics can reveal the 
underlying hierarchical organization of the network   
through the routes to synchronization~\cite{arenas2006SynchronizationRevealsTopological,arenas2007synchronizationAndModularityInComplexNetworks,gomez2007paths,kim2010StructuralPropertiesSynCluster,gomez2011EvolutionMicrodMeso,stout2011LocalSyncComplexNetCoupledOsc}. Such an approach was 
employed in the analysis of the cortical network of 
cats in~\cite{gomez2010FromModularToCentralized}, where it was verified that a 
synchronization transition between local and global states 
controlled by highly connected regions, such as the visual 
and auditory cortex, emerges. Interestingly, similar 
results as those ones got with the Kuramoto model in~\cite{gomez2010FromModularToCentralized} can be obtained with models that encompass more 
sophisticated properties of neuronal dynamics. A similar 
strategy was carried out in~\cite{honey2008dynamical} with the particular 
objective to characterize how lesions could affect the
overall dynamics of the macaque cortical network. Lesions 
were modeled as perturbations on the synchronized and 
the impact of such disturbances was assessed by the time
required for the nodes to return to the locked state. As 
shown in~\cite{honey2008dynamical}, the most robust nodes, i.e., the ones that 
presented shorter relaxation times, were reported to 
correspond to highly connected cortex areas, with the 
exception of the region V4, which turns out to be an intra-cluster hub. These results are in agreement with those 
presented  in~\cite{moreno2004synchronization} regarding the relaxation time of nodes 
in SF networks. The effects of lesions in the local and global dynamics of cortical networks were further explored in~\cite{vavsa2015EffectsOfLesions}, however with a different methodology. Specifically, instead
of considering lesions in the cortex as random perturbations in the phases, the removal of nodes in the cortical network was adopted as such. In this case, the strongest impact in the networks'
global dynamics is achieved when nodes with higher eigenvector centrality are removed, inducing also metastability in the system~\cite{vavsa2015EffectsOfLesions}. 

Throughout this review, we saw the impact of different topological 
characteristics on synchronization mainly considering 
networks generated through random models (see especially Sec.~\ref{sec:traditional} 
and~\ref{sec:different_topologies}). It is interesting to note that most of these findings 
were investigated in parallel in the neuroscience community by 
considering instead networks whose structure are derived 
from real data sets~\cite{schmidt2015KuramotoModelSimulationOfNeuralHubsAndDynamic,honey2008dynamical,gomez2010FromModularToCentralized}. The questions posed in these studies 
are similar to those tackled among the physicists, e.g. the role 
played by hubs, communities, degree-degree correlations, etc; 
but with the aim of identifying the neurobiological importance of such properties. For instance, Schmidt et al.~\cite{schmidt2015KuramotoModelSimulationOfNeuralHubsAndDynamic} investigated the 
potential role played by hubs in the brain dynamics by analysing the 
Kuramoto model in anatomical networks obtained through functional 
magnetic resonance (fMRI). One of the reported findings is that 
hubs have higher synchronization between themselves, a fact that 
is a consequence of high assortativity and modularity exhibited by 
these networks. Similar results were also observed for the Kuramoto model simulated on the macaque~\cite{honey2008dynamical} and cat~\cite{gomez2010FromModularToCentralized} cortical networks. 

It is also important to mention the variation of the Kuramoto 
model analyzed by Cabral et al.~\cite{cabral2011RoleLocalNetworkOscillationsResting} that 
incorporates heterogeneous time delay and stochastic fluctuations, i.e.: 
\begin{equation}
\dot{\theta}_i = \omega_i + \lambda \sum_{j=1}^N A_{ij}\sin\left[ \theta_j(t - \tau_{ij}) - \theta_i(t)\right] + \xi_i(t), 
\label{eq:cabralmodel}
\end{equation}
where $\tau_{ij}$ is set to be proportional to the physical distance 
between nodes in the functional brain network. In~
\cite{cabral2011RoleLocalNetworkOscillationsResting}, the authors 
used the model in the gamma frequency range ($>30\rm{Hz}$) to 
generate time series to serve as input in the Balloon-Windkessel 
hemodynamic model~\cite{friston2003dynamic}. Remarkably, the time 
series obtained through (\ref{eq:cabralmodel}) were able to 
accurately describe the emergence of slow neural activity 
fluctuations empirically measured in real networks~
\cite{cabral2011RoleLocalNetworkOscillationsResting}. Temporal and 
spatial synchronization patterns from fMRI were also reproduced 
for low frequencies between 0.01-0.13$\rm{Hz}$~
\cite{ponce2015resting}, a range  for which the time delays 
between the oscillators can be neglected, since the delay time 
scale in the cortex area is much faster than the periods of the 
oscillators~\cite{ringo1994TimeIsOfTheEssence}.
Furthermore, the variant of the Kuramoto model (\ref{eq:cabralmodel}) 
proved itself to be relevant in the study of criticality 
in brain dynamics~
\cite{botcharova2014MarkersOfCriticalityPhaseSync}, besides also 
reproducing moment-to-moment fluctuations of phase differences of 
resting states in magnetoencephalography (MEG) oscillations 
recorded during finger movement experiments~
\cite{botcharova2015RestingStateMEGOscillations}.
Other aspects of dynamical criticality in human brain functional 
networks were studied in~\cite{kitzbichler2009broadband} by 
comparing times series generated with the Kuramoto model and 
fMRI data recorded from normal subjects under resting conditions.    


Other recent works have investigated further aspects of the relation between structure and dynamics of phase oscillators in anatomical and functional cortical networks~\cite{dumas2012anatomical,yan2013emergence,watanabe2013rich,cabral2014ExploringMechanimsSpontaneous,cabral2014ExploringTheNetworkDynamics,schmidt2014DynamicsOnNetworks,tassef2014MouseHairExpression,maistrenko2007MultistabilityKMSynapticPlasticity}, which includes studies on the effects of remote synchronization~\cite{vuksanovic2014FunctionalConnectivity,nicosia2013RemoteSynchronizationReveals,gambuzza2013AnalysisRemoteSynchronization},
a toolbox for large-scale simulations of brain dynamics using different models~\cite{sanz2015MathematicalFrameworkLargeScale}
and discussions about the limitations of the above mentioned approaches~\cite{daffertshofer2011InfluenceOfAmplitudeConnectivity}.


\subsection{Networks of disordered Josephson junctions}

A locally coupled Kuramoto model with a second-order time derivative (LKM2) can be derived from a coupled resistively and capacitively shunted junction equations (RCSJ) for an underdamped ladder with periodic boundary conditions \cite{trees_synchronization_2005}. Phase synchronizations on these two models are investigated in terms of the Kuramoto order parameter and the degree of frequency synchronization defined as 
\begin{equation}
f  = 1- \frac{s_V(\lambda)}{s_V(0)},
\end{equation}
where $s_\nu(\lambda)$ denotes the standard derivation of the $N$ time-average voltage $\left\langle V_i\right\rangle$ as a function of the coupling strength $\lambda$ as follows
\begin{equation}
s_V(\lambda) = \sqrt{\frac{\sum^{N}_{i=1}\left[ \left\langle V_i\right\rangle - (1/N) \sum^N_{j=1} \left\langle V_j\right\rangle \right]^2}{N-1}}.
\end{equation}
When $f=1$ all $N$ junctions' frequencies are signaled and $f=0$ when neighboring rung junctions are uncoupled. 
In the absence of any coupling i.e. $\lambda=0$, phase value of each junction follows 
\begin{equation}
\phi_i (t) = A + B e^{-\tau/\beta_c}  + \omega_i \tau,
\end{equation}
where $\tau$ is dimensionless time unit given by $\tau = 2 e\left\langle I_c \right\rangle/ \hbar \left\langle R^{-1}\right\rangle$, with $\left\langle I_c \right\rangle$ the average critical current and $\left\langle R \right\rangle$ the average resistance. Moreover, $\beta_c$ is the McCumber parameter given by $\beta_c = 2e\left\langle I_c \right\rangle\left\langle C \right\rangle/ \hbar \left\langle C\right\rangle^2$, where $\left\langle C \right\rangle$ is the average capacitance of the system. $A$ and $B$ are constants that depend on the initial conditions. 
In the uncoupled limit we have $d\phi_i/d\tau = \omega_i$, thus $\omega_i$ can be regarded as the voltage across junction $i$ in the uncoupled regime.  

Good agreements between LKM2 and RCSJ are achieved for phase synchronization as well as frequency synchronization, but the LKM2 is easier to be solved and understood compared to the RCSJ \cite{trees_synchronization_2005}. On SW networks, shortcut links enhance synchronization on ladder arrays, but in two-dimensional arrays, the effects on synchronization are marginally restricted~\cite{trees_synchronization_2005}.

\subsection{Seismology}

An interesting and promising recent application of the 
Kuramoto model comes from the modeling of earthquakes. 
Scholz~\cite{scholz2010LargeEarthquakeTriggering} was the 
first to postulate that the Kuramoto model can be 
applicable to explain the sequencing of nearby faults after 
a large earthquake. This is supported by the hypothesis that 
a seismically active fault can be regarded as a limit cycle 
relaxation oscillator. Specifically, when a given fault on 
the earth's crust accumulates stress until a certain 
threshold an earthquake occurs, which reduces the stress in 
this particular fault and, at the same time, distributes the 
energy released contributing to increase the stress in the 
nearby faults and then triggering new events~\cite{scholz2010LargeEarthquakeTriggering}. Although 
this process resembles a pulse-coupled dynamics, many 
insights can be gained from the study of phase oscillators. 
However, considering synchronization of faults in a globally 
coupled topology seems quite unrealistic, since the spread 
of energy is made through local fault-fault interactions, making 
the network approach to the problem much more suitable. In order to model 
this process, Vasudevan et al.~\cite{vasudevan2015earthquake} considered 
a directed network constructed from the sequencing of earthquake events 
provided by the Incorporated Institutions for Seismology (IRIS) for the 
time span between 1970 and 2014. In this network, each node corresponds 
to an earthquake event and a directed edge departing from node $i$ and 
reaching node $j$ exists if earthquake $i$ triggered another one in node 
$j$. Furthermore, the phases of the model were considered to evolve   
according to~\cite{vasudevan2015earthquake}
\begin{equation}
\dot{\theta}_i = \omega_0 + \lambda \sum_{j=1}^N e^{-\kappa d_{ij}} \sin(\theta_j - \theta_i + \varphi),
\label{eq:vasudevan_model_earthquake}
\end{equation}
where $\varphi$ accounts for the non-instantaneous energy transmission, 
$d_{ij}$ is the shortest path length between 
two fault points $i$ and $j$ in the earthquake events grid and $\kappa$ is the strength of the non-local coupling. Interestingly, it was found that the dynamics in the earthquake network supports the occurrence of chimera states, which might be a result from the interplay between 
phase-lag interaction and the geodetic constraints in the earthquake zones~\cite{vasudevan2015earthquake}. The study of sequencing of earthquakes using phase models is still in its infancy and much remains to be explored, specially concerning the investigation of  effects of other parameters in the model, such as heterogeneity in the natural frequencies and the inclusion of heterogeneous time-delays between the oscillators. Another promising direction is the consideration of other models that incorporate excitable behavior, e.g. the Shinomoto-Kuramoto model (see Sec.~\ref{sec:stochastic}).  

Finally, we would like to mention that the Kuramoto
model in complex networks has been considered in many other applications, such as machine learning~\cite{miyano2007data,novikov2014oscillatory,miyano2008collective,pluchino2008CommunitiesRecognition,hong2012SynchronizationBasedApproach,miyano2012DeterminingAnomalous}, characterization of financial market networks~\cite{peron2011CollectiveBehaviorFinancialMarkets}, modelling of groups of animals in motion~\cite{leonard2012DecisionVersysCompromise}, oscillatory dynamics of cell networks~\cite{kori2012StructureCellNetworks}, logistics~\cite{lammer2006DecentralisedControl,fujiwara2011synchronization,donner2008multivariate} and opinion dynamics~\cite{pluchino2005changing,pluchino2006OpinionDynamicsAndSyncCollab,bruggeman2013solidarity}, besides being employed in the methods
for community detection in networks based on phase oscillators discussed in Sec.~\ref{subsec:community}. 

\section{Conclusions and perspectives}
\label{sec:conclusions}

In this article we have reviewed recent advances in the study 
of Kuramoto oscillators in complex networks. We have focused our 
analysis on how the local and global dynamics are influenced by the network 
connectivity pattern. As we saw, in the last years, the Kuramoto model has 
been scrutinized and many different scenarios were considered
since the first works addressing synchronization in networks were
published. Here we sum up these contributions and commenting what are most interesting and promising open problems for future research.

We started by analysing the dynamics in standard network models (Sec.~\ref{sec:traditional}), i.e. SW, random and 
SF topologies; which were the subject 
of the early works that established the 
foundations for many further studies. In particular, we presented 
necessary approximations to treat the problem, including the most employed 
MFA. Interestingly, although being unquestionably seminal contributions, it was 
evident that the first works 
investigating the Kuramoto model in complex networks were surrounded 
by questions that the inclusion of the heterogeneous connections had raised. 
Some of these questions, unfortunately, still remain without an answer. 
In particular, the disagreement between the finite value observed in 
simulations and the prediction of a vanishing critical coupling for the onset of synchronization by the MFA in SF networks 
is one of the most challenging puzzles that persisted from the early 
investigations. However, despite the fact that Kuramoto 
oscillators have long been studied in these traditional topologies, there are also recent fundamental new results on the dependence on the system 
size and on the relaxation dynamics of the model. In the former topic, we 
showed the importance of considering sample-sample fluctuations for the 
correct estimation of the scaling exponents. Regarding the relaxation dynamics, we saw the non-intuitive phenomenon of 
requiring larger
relaxation times to reach the stationary state in the SW regime.
This is a surprising
effect giving the many results by previous studies showing that 
synchronization is attainable for weaker couplings in comparison 
to regular 
topologies. 
It is worth mentioning that the study of the time-dependent 
dynamics was also greatly benefited thanks to the Ott-Antonsen theory, 
whose application was exemplified in the study of the relaxation 
dynamics of SF networks. 

\textcolor{black}{In Sec.~\ref{sec:different_topologies} we moved one step further in the level 
of topological description of networks  
by introducing a new set of structural features that are absent 
in uncorrelated ones.} More specifically, we analyzed synchronization of 
Kuramoto oscillators in network models that mimic, to some extent, 
connectivity patterns which are often encountered in real-world networks, 
such as clustering, modular organization and degree-degree 
correlations. Initial numerical investigations exhibited that the increasing
of clustering has a negative effect on the network synchronization. In other 
words, networks with a high occurrence of triangles reach lower levels 
of synchronization in comparison with networks with the same degree 
distribution at the some coupling strength. However, as we have discussed, 
as the clustering is increased when employing stochastic rewiring algorithms, 
several other properties in the network are changed, even though the degree
distribution remains fixed. The alternative approach is then to consider
the aforementioned models, which, besides being able to generate particular
topological properties of interest, further allow analytical tractability. 
Possibly here lies one of the most fruitful directions to be explored 
in future works, i.e. the development of analytical techniques that go beyond 
the MFAs for locally tree-like networks. 

\textcolor{black}{After investigating the effects of structural properties of real systems on network dynamics, in Sec.~\ref{sec:general_coupling} we reviewed effects of dynamical features that provide a more realistic modelling of special classes of systems. Specifically, we studied effects of time-delayed couplings, which are present in the synchronization of systems in which the speed of signal transmission is comparable to 
oscillators' periods, and scenarios where the couplings (and the network structure) is time-dependent. Regarding the former, we observed 
that the dynamics is significantly changed compared to the case where the interaction is instantaneous. In particular, new dynamical 
states were observed to emerge. However, while the impact of delay on synchronization is well understood in small populations, there is still a lack of results on the time-delayed dynamics of large heterogeneous networks. There are interesting topics for new research in this direction, such as progress can be achieved by applying the OA theory in networks in order to thoroughly evaluate the influence of delay on the dynamics as well as the relation with the network structure. The same is valid for research on adaptive topologies. This is one of the branches in the research of the Kuramoto model in networks that is in its infancy and much remains to be done.}

Section~\ref{sec:explosive_sync} reviewed one of the most active topics in the 
last years within the research of phase oscillators in networks: the 
correlation between natural frequencies and degrees. This strategy
to assign frequencies to the network yielded the first report of 
discontinuous synchronization phase transitions in SF networks and, consequently, attracted the attention of many researchers. However, 
as later shown, 
this particular mechanism turned out to be not the only one to 
lead to such an abrupt emergence of collective behavior in networks. The key point is the existence of a 
sufficiently
large gap between the natural frequencies of connected nodes, 
a condition
that is automatically satisfied in SF networks if $\omega_i 
\sim k_i$. The 
effect of other correlations between intrinsic dynamical 
characteristics and local topological properties could be an 
interesting topic for future research.

Section~\ref{sec:stochastic} addressed the Kuramoto
model in networks in the presence of noise. Intriguingly, despite 
the popularity of the model in the last decade, its stochastic
version still has many scenarios to be explored. For instance, 
we presented the derivation of the nonlinear 
Fokker-Planck equation and thereby obtained the onset of 
synchronization. 
However, a quantitative analysis of the interplay 
between the stochastic dynamics and the structure of SF was 
still not tackled. In particular, it would be interesting to 
combine the approaches
developed in Sec.~\ref{sec:stochastic} and those in Sec.~\ref{subsec:finite_size_effects} in order to 
explicitly determine the dependence with the system size. Another  interesting direction is to analyze how the relaxation time in 
networks is affected by stochastic forces. For this task, the 
OA ansatz used to uncover the temporal dynamics (see Sec.~\ref{subsec:relaxation}) 
is no longer applicable. Conversely, as shown in Sec.~\ref{subsec:gaussian_approximation}, a 
dimension reduction is still possible through a Gaussian 
approximation, which has been recently 
providing insightful results into the time dependent 
behavior of the stochastic Kuramoto and Shinomoto-Kuramoto 
model~\cite{sonnenschein2013approximate,sonnenschein2013excitable,sonnenschein2014cooperative,sonnenschein2015collective}. Regarding the latter, we saw that interesting 
phenomena can arise in the excitable dynamics as a consequence
of the inclusion of heterogeneous connectivity patterns. Other
scenarios are naturally appealing to consider in future 
studies, such as the inclusion of delay and degree-degree 
correlations. In particular, it would be interesting, for 
instance, to verify whether the transitions between 
periodic and quasiperiodic states induced by 
degree correlations~\cite{restrepo2014mean} and chaos generated by frequency
correlations~\cite{skardal2015frequency} still persist in the presence of noise. 
Moreover, the stochastic Kuramoto model subjected to 
attractive and repulsive couplings~\cite{sonnenschein2015collective,hong2011KuramotoModelPositiveNegative,hong2011ConformistsContrarians,hong2012MeanFieldBehavior,louzada2012HowToSuppressSync,zanette2005synchronization} could possibly 
reveal further peculiarities in the dynamics in networks.  

In Sec.~\ref{sec:second_KM} we have reviewed studies on the second-order 
Kuramoto model. Such a variation of the original model was
firstly conceived in order to encompass frequency adaptation, 
a property observed in synchronization of biological 
rhythms, such as exhibited by fireflies and circadian rhythms.
We started by first providing a brief overview of the recent
developments on the model in the fully connected graph 
with and without the influence of stochastic fluctuations. We followed up by 
considering the model in uncorrelated networks, where the
mean-field treatment presented in Sec.~\ref{subsec:early_works} was applied 
to the case with inertia. Despite the progress achieved in the last years in understanding the effects of inertia~\cite{acebron2005kuramoto}, the theory discussed still has some fundamental limitations. For instance, in order to 
obtain the synchronization conditions for locked oscillators we assumed sufficiently
small values for the damping coefficient so that Melnikov's 
method can be applied. For large values of mass and damping 
the approximations are no longer valid and 
results to fill this gap are still missing. Furthermore, the 
stochastic Kuramoto model with inertia in networks also remains 
largely unexplored. An interesting approach to tackle this problem would be to 
consider dimension reduction techniques such as the Gaussian 
approximation in Sec.~\ref{subsec:gaussian_approximation} or even the one recently introduced in~\cite{gottwald2015ModelReduction}. It is also
worth mentioning that the second-order Kuramoto model was brought
to higher attention lately partially due to recent 
interest in the synchronization of power grid systems. We 
strongly believe that this branch in the field of network 
synchronization will develop even further. In particular, much 
effort has been put into the characterization of the stability
of such systems using new set of complexity measures. For 
instance, the aforementioned basin stability has been 
successfully applied to quantify the robustness of power-grid 
networks against large perturbations to the synchronized state.  

\textcolor{black}{Methods to optimize synchronization were examined in Sec.~\ref{sec:optimization}. Here the main question addressed is which type of topology or frequency assignment maximizes the collective behavior among the oscillators
given certain constraints. This topic is particular appealing for various
applications, since typically the design of real systems
is surrounded by constrains that arise due to a limited amount of resources. In this regard, much can be learned from investigations 
presented in Sec.~\ref{sec:optimization}. Moreover, studies on 
optimization of any kind of dynamics in networks heavily rely on expensive
computational tasks. However, interesting analytical approaches have
been taken concerning optimization of networks consisting of Kuramoto oscillators~\cite{skardal2014optimal,gottwald2015ModelReduction,pinto2015optimal}. Interestingly, the study of optimal frequency assignments can be benefited from the results discussed in Sec.~\ref{sec:explosive_sync}, since
the correlation between frequencies and degrees has been reported to provide an optimal scenario for the emergence of synchronization. In the light of these results, it is therefore very likely that more work will attempt to better understand conditions which significantly improve network synchronization. }

\textcolor{black}{
Section~\ref{sec:applications} is devoted to review some 
applications of the Kuramoto model in real systems. There we discussed how insights gained in the long studied question of 
how topology shapes dynamics (and vice-versa)
have been put into practice. For instance, thanks to the joint
approach of network characterization and application of non-linear
stability measurements, important findings were obtained regarding 
the dynamics of real power-grid networks. In particular, the contribution 
of particular types of subgraphs to the overall grid's stability was 
assessed through the new concept of basin stability, whereby the non-intuitive phenomenon 
that stability is threatened by dead-ends and dead-trees in the topology 
was uncovered. These findings allied with extensions to higher-order power-grid models have great potential to become guidelines in the development 
of new policies in power-grid design and management. Furthermore, it is becoming 
clear that, despite its simplicity, the Kuramoto model plays an important
role in the characterization and modelling of the dynamics in cortical 
networks. Understanding the neuronal dynamics in networks 
obtained from real data using the model variations discussed in 
Sec.~\ref{sec:applications} can potentially pave the way for studies
with more sophisticated models.  
These and other examples showed in Sec.~\ref{sec:applications} along all 
the results reported in this paper might also motivate applications of the Kuramoto model in other fields. }


The relation between dynamics and structure of the Kuramoto model 
in complex networks was scrutinized over the last years. Much 
has been done since the first works addressing synchronization in 
networks came out. However, there are still many gaps to be 
filled and other scenarios to be explored. The combinations of 
different topologies and variations of the model dynamics are 
countless. Furthermore, the myriad of works 
published in the last decade on synchronization of 
phase oscillators in networks, as well as other dynamical 
processes, was a natural consequence of the growing interest in 
trying to understand the topological organization of real complex 
systems. Nowadays, we observe an analogous movement that  
aims at characterizing the structure of 
multilayer networks, a fact that makes us to expect a huge number of works devoted to the analysis of dynamical processes 
in these structures.


Finally, we believe that the present review complements previous 
works offering a guide on structural 
aspects of the dynamics of Kuramoto oscillators for new 
researchers in the field.


\section{Acknowledgements}
\label{sec:acknowledgements}

We are indebted with B. Sonnenschein, E. R. dos Santos, P. Schultz, C. Grabow, M. Ha and C. Choi for insightful and helpful discussions.
T.P. acknowledges FAPESP (No. 2012/22160-7 and No. 2015/02486-3) and IRTG 1740.
P.J. thanks founding from the China
Scholarship Council (CSC). 
F.A.R. acknowledges CNPq (Grant No. 305940/2010-4) and FAPESP (Grant No. 2013/26416-9). J.K. would like to acknowledge IRTG
1740 (DFG and FAPESP).

\appendix
\renewcommand*{\thesection}{\Alph{section}}

\section{Complex Networks}
\label{sec:appendix}
\begin{figure*}[!t]
\begin{center}
\subfigure[]{\includegraphics[width=0.4\linewidth]{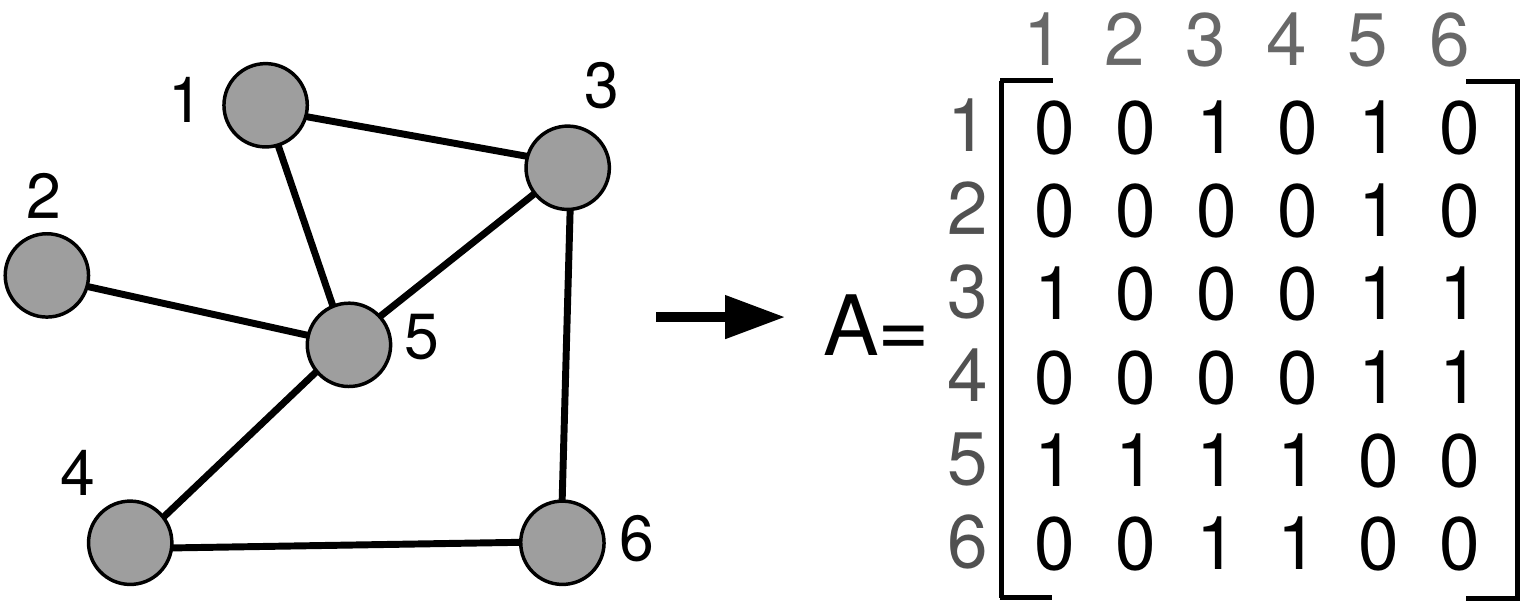}} \hspace{0.5cm}
\subfigure[]{\includegraphics[width=0.4\linewidth]{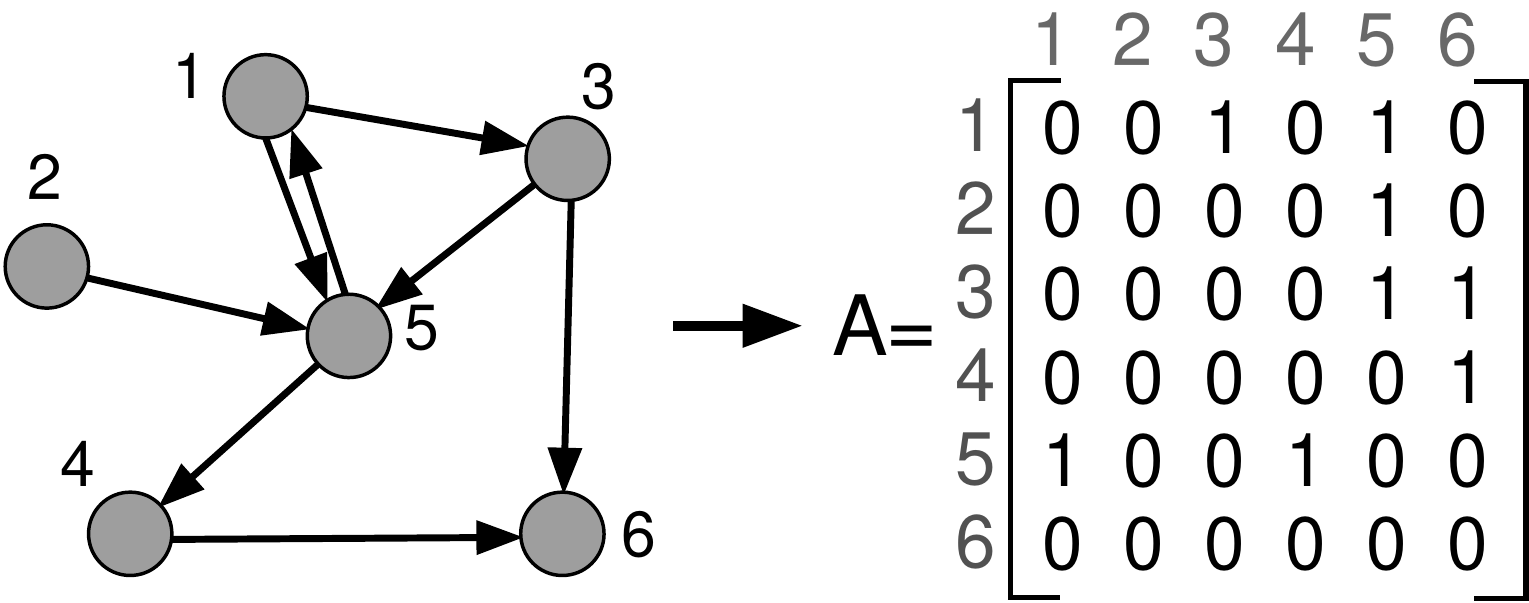}}
\end{center}
\caption{Graphs can be represented by adjacency matrices. In (a) we have a example of a undirected graph and (b) a directed graph. In (a), the matrix elements $A_{ij} = 1$ if there is a connection between the nodes $i$ and $j$ or $A_{ij} = 0$ otherwise. In (b), the matrix elements $A_{ij} =1 $ if there is a directed connection (an arc) from node $i$ to node $j$. } \label{Fig:mapping}
\end{figure*}

The structure of complex network can be represented by a graph $G$ composed of an ordered pair of disjoint sets $G=(V,E)$, where $V$ is a set of elements called vertices (nodes) and $E$ is a subset of ordered pairs of distinct elements of $V$, called edges~\cite{newman2003StructureFunctionFunctionComplexNetworks, albert2002statistical}. If the network is directed, then the connections are called arcs (or directed links). While a protein-protein interaction network is undirected, a network of neuronal connections is typically directed~\cite{costa2011analyzing}. Mathematically, the organization of networks can be represented by the adjacency matrix $\mathbf{A}$, whose elements $A_{ij} = 1$ if there is a connection between nodes $i$ and $j$ or $A_{ij} = 0$ otherwise. In undirected networks, $\mathbf{A}$ is symmetric, whereas for directed networks, $\mathbf{A}$ is generally asymmetric. Figure~\ref{Fig:mapping} illustrates the mapping of a network in the respective adjacency matrix for the directed and undirected cases.

The organization of networks can be characterized by network measures~\cite{costa2007characterization}. The node degree $k_i$, which represents the number of edges connected to node $i$, is one of the most simple measures for local network characterization. For undirected networks it can be calculated as
\begin{equation}
  k_i = \sum_j A_{ij} = \sum_j A_{ji}.
\end{equation}
For instance, in Figure~\ref{Fig:mapping}(a), we have $k_2 = 1$,
$k_1=k_4=k_6= 2$, $k_3=3$ and $k_5 = 4$. 

The density of connections in a network can be quantified by its average degree, which is calculated by 
\begin{equation}
  \langle k \rangle = \frac{1}{N} \sum_i k_i = \frac{1}{N} \sum_{ij}
  A_{ij}.
\end{equation}

In the case of directed networks, in which the connections have directions, a node can be characterized by the out-degree, $k_i^\mathrm{out}$, which yields the number of outgoing edges, and the in-degree, $k_i^\mathrm{in}$, corresponding to the number of incoming edges~\cite{newman2010networks}. In terms of the adjacency matrix, we have
\begin{eqnarray}
  k_i^\mathrm{out} & = & \sum_j A_{ij} , \\
  k_i^\mathrm{in}  & = & \sum_j A_{ji} .
\end{eqnarray}
For instance, in Figure~\ref{Fig:mapping}(b), $k_1^\mathrm{in} = 1$ and $k_1^\mathrm{out}=2$.  The total degree of a node $i$ is defined as $k_i = k_i^\mathrm{in} + k_i^\mathrm{out}$.

In unweighted networks, the connections are binary, i.e., $A_{ij} = 1$ or $0$, indicating the presence or absence of an edge. However, some networks have weighted connections, which can indicates the strength of the interactions between pairs of nodes~\cite{Barrat04}. For a road network, for instance, the weight can specify the inverse of the distance between pairs of cities. Therefore, the closer two cities, the higher the strength of their interconnection. A weighted network can be completely
represented in terms of its weight matrix $\mathbf{W}$, so that each element $W_{ij}$ expresses the weight of the connection from vertex $i$ to vertex $j$. In addition, instead of the node degree, we use the node strength of $i$, $s_i$, which is defined as the sum of the weights of the corresponding edges~\cite{Barrat04}. In terms of the weight matrix, for an directed and weighted network, the in and out-strength is defined as~\cite{Barrat04} 
\begin{eqnarray}\label{Eq:strength}
  s^\mathrm{out}_i & = & \sum_{j} W_{ij}\label{strength_out}, \\
  s^\mathrm{in}_i  & = & \sum_{j} W_{ji}\label{strength_in}.
\end{eqnarray}

By using the concept of degree, it is possible to characterize the large-scale network organization by considering the degree distribution $P(k)$, which yields the probability that a node selected randomly presents the degree $k$. The second statistical moment of $P(k)$ quantifies the network level of heterogeneity and is related to several dynamical processed running on the top of complex networks~\cite{boccaletti2006complex}. Particularly, the critical coupling of the Kuramoto model depends inversely on the second moment of $P(k)$~\cite{ichinomiya2004frequency}. 

Networks whose degree distribution follows a power-law (or Pareto distribution~\cite{Newman05:CP}) are called scale-free networks~\cite{albert2002statistical}, i.e.,
\begin{equation} \label{Eq:power_law}
P(k) \sim k^{-\gamma},
\end{equation}
implying that the degree distribution appears as a straight line when plotted on a loglog graph. Such networks are characterized by an inhomogeneous topological organization, where most of the nodes present a small number of connections, whereas a small fraction of nodes is densely connected. These densely connected nodes, called hubs, play a fundamental role in the topology and dynamics of complex networks. 

Another fundamental concept in networks is related to distances, such as the number of edges along the path connecting two vertices~\cite{costa2007characterization}.  There are many important topological and dynamical properties related to distances, such as random walks, which are defined by any sequence of adjacent nodes. A path is a special type of walk, which does not allow repetition of nodes or edges. The paths with the minimum length between two nodes $i$ and $j$ are called the shortest paths~\cite{costa2007characterization}. Observe that it is possible to have more than one path with the shortest length connecting two nodes. This length, which is called geodesic distance, is henceforth represented as $d_{ij}$. Thus, it is possible to characterize the large scale organization of a network by averaging the geodesic distances between every pair of nodes. The average shortest path length is defined by
\begin{equation}
 \ell = \frac{1}{N(N-1)}\sum_{i \ne j} d_{ij}.
 \label{meandist}
\end{equation}
Observe that this definition has a limitation, since the value of $\ell$ diverges if two nodes do not belong to the same component, which yields $d_{ij}=\infty$. A possible approach to avoid this divergence is to consider only the nodes in the largest component while calculating the value of $\ell$. 

\begin{figure}[!t]
\begin{center}
\includegraphics[width=0.2\linewidth]{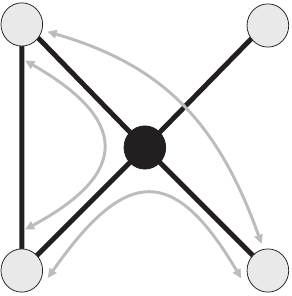}
\end{center}
 \caption{Illustration of the definition of transitivity. This network presents one triangle and eight connected triples (three of these triples are indicated by gray arrows). Therefore, in this network, we have $\mathcal{T} = 3/8.$}
\label{Fig:trans}
\end{figure}

Networks can also be characterized in terms of the presence of triangles (subgraphs composed by three fully connnected nodes)~\cite{newman2003StructureFunctionFunctionComplexNetworks}. The fraction of triangles can be determined by a measurement called transitivity (or global clustering coefficient). Transitivity measures the probability that the adjacent neighbors of a vertex are connected. For undirected unweighted networks, the transitivity is calculated as~\cite{newman2003StructureFunctionFunctionComplexNetworks,newman2010networks,costa2007characterization}
\begin{equation}
\mathcal{T}= \frac{3 N_{\triangle}}{N_3},
  \label{clustcoeff}
\end{equation}
where $N_{\triangle}$ corresponds to the number of triangles observed in a network and $N_3$ is the number of connected triples. Fig.~\ref{Fig:trans} illustrates the concepts of triangles and connected triples. The factor three in this equation is due to the fact that each triangle consists of three different connected triples, one with each of the vertices as central vertex, which assures that $0 \leq \mathcal{T} \leq 1$. By using the adjacency matrix, these properties can be calculated as
\begin{equation}
  N_\triangle  =  \sum_{k>j}^N \sum_{j>i}^N\sum_{i=1}^N A_{ij}A_{ik}A_{jk},
\end{equation}
\begin{equation}
    N_3  = \sum_{k>j}^N \sum_{j>i}^N\sum_{i=1}^N (A_{ij}A_{ik}+A_{ji}A_{jk}+A_{ki}A_{kj}).
\end{equation}

\begin{figure}[!t]
\begin{center}
\includegraphics[width=0.95\linewidth]{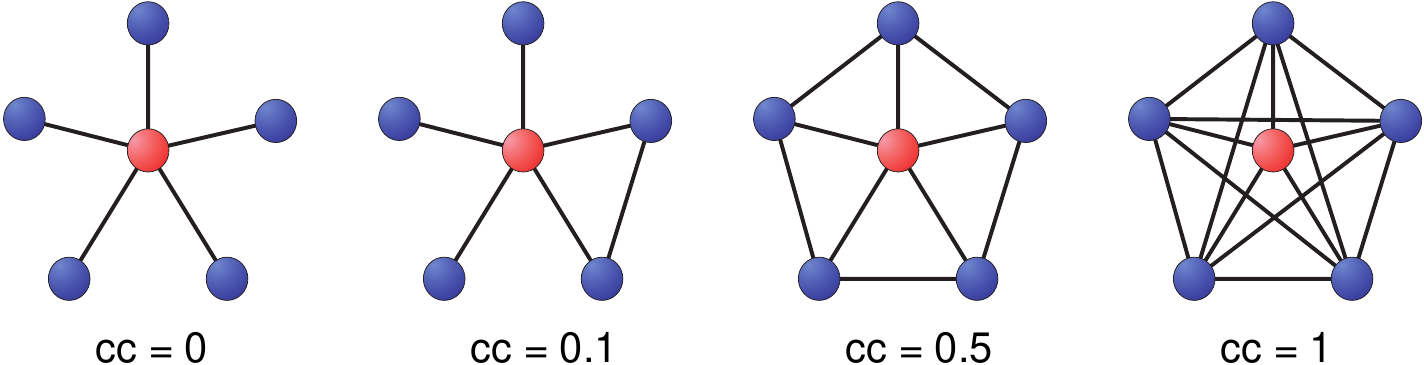}
\end{center}
 \caption{Examples of clustering coefficient calculated for the (central) red node.}
\label{Fig:cc}
\end{figure}

$\mathcal{T}$ is a measure for large scale network characterization. However, it is possible to define the local clustering coefficient of a node $i$, which quantifies how well the immediate neighbors of that node are interrelated~\cite{watts1998collective}. More specifically, the clustering coefficient is defined as
\begin{equation} \label{Eq:cc}
  cc_i = \frac{2e_i}{k_i(k_i-1)},
\end{equation}
where $e(i)$ is the total number of undirected edges interconnecting the immediate neighbors of $i$. Notice that $0 \leq cc(i) \leq 1$. The number of edges among the neighbors of a node $i$ can be calculated by
\begin{equation}
e_i = \frac{1}{2} \sum_{j=1}^N \sum_{k=1}^N A_{ij}A_{jk}A_{ki}.
\end{equation}
Figure~\ref{Fig:cc} illustrates the calculation of the clustering coefficient.

Networks can also be characterized in term of degree-degree correlations~\cite{boguna2003epidemic}, which can be analyzed in terms of the conditional probability $P(K'=k'|K=k) = P(k'|k)$, i.e., the probability that a vertex of degree $k$ is connected with a vertex of degree $k'$. In uncorrelated networks, $P(k'|k)$ does not depends on $k$ and the only relevant function for networks characterization is $P(k)$~\cite{boguna2003epidemic}. However, due to the difficulties in estimating the conditional probability of degree-degree connectivity in empirical data~\cite{newman2010networks}, researchers often use the assortativity coefficient $\mathcal{A}$ as a measure of degree-degree correlation. The correlation between a pair of random variables is calculated by taking into account the Pearson correlation coefficient. Thus, it is natural to recur to this approach to determine the correlations in networks~\cite{newman2003StructureFunctionFunctionComplexNetworks}, \emph{i.e.},
\begin{equation}
\mathcal{A} = \frac{
            \frac{1}{|E|} \sum_{j>i}k_i k_j A_{ij} -
            \left[ \frac{1}{|E|} \sum_{j>i}
                   \frac{1}{2} (k_i+k_j) A_{ij} \right]^2
           }
           {
            \frac{1}{|E|}\sum_{j>i}\frac{1}{2}(k_i^2+k_j^2) A_{ij} -
            \left[ \frac{1}{|E|}\sum_{j>i}
                   \frac{1}{2}(k_i+k_j) A_{ij}  \right]^2
           },
  \label{Eq:correlation}
\end{equation}
where $|E|$ is the number of edges. Note that $-1 \leq \mathcal{A} \leq 1$. In the case of $\mathcal{A} > 0$, vertices with similar degrees tend to connect with each other and the network is classified as assortative. On the other hand, if $\mathcal{A} < 0$, vertices with high degree tend to connect with low degree nodes, or vice versa, and the network is called disassortative. If $\mathcal{A}=0$ there is no correlation between the degrees. 

Centrality is a node property that can influence the network synchronization (e.g.~\cite{Hong04}). Although centrality is important for node characterization, there is no single definition of centrality in a complex network~\cite{Arruda014}. For instance, the concept of ``centrality'' can be related to the average shortest distances~\cite{girvan2002community}, network spectra~\cite{perra2008spectral} and random walks~\cite{estrada2008communicability}. A possible measure is given by the betweenness centrality~\cite{girvan2002community}, \emph{i.e.},
\begin{equation}
  b_u = \sum_{ij} \frac{\sigma(i,u,j)}{\sigma(i,j)},
  \label{betweenness}
\end{equation}
where $\sigma(i,u,j)$ is the number of shortest paths between vertices $i$ and $j$ that pass through vertex (or edge) $u$, $\sigma(i,j)$ is the total number of shortest paths between $i$ and $j$, and the sum is over all pairs $i,j$ of distinct vertices. Thus, a central node in a network should be crossed by many paths and therefore present a high value of betweenness centrality. This property can be verified in a city, as its nodes at center tend to present the highest traffic, i.e.\ to go from a region to another, it is necessary to pass through the center of the city. $b_u$ plays a fundamental role on network synchronization~\cite{Hong04}.

In addition to the degree-degree correlation, presence of triangles and heterogeneous structure, complex networks may display modular organization~\cite{girvan2002community, fortunato2010community}. The modules are groups whose vertices are more densely interconnected one another than with the rest of the network. These modules are called communities in network theory. Many networks present community organization~\cite{fortunato2010community}, e.g. social sciences, families, friendship circles, co-workers, scientific collaborations and villages are examples of communities. In the World Wide Web, pages related to the same or related topics tend to be connected, forming communities.

There are several methods to find communities in networks~\cite{fortunato2010community}. To quantify the quality of a particular division, most methods have adopted the modularity measure~\cite{newman2004findingAndEvaluatingCommunity}. This measurement is based on the comparison between the fraction of the edges that fall within the communities and the expected fraction whether edges were distributed at random. Mathematically,
\begin{equation}
Q = \frac{1}{2M} \sum_{i=1}^N \sum_{j=1}^N \left(A_{ij} -\frac{k_ik_j}{2M} \right)\delta(C_i,C_j),
\end{equation}
where $\delta(C_i,C_j)$ is the Kronecker delta, which is equal to one if $C_i=C_j$ and zero, otherwise. The term $(k_ik_j)/2E$ represents the expected number of edges between the vertices $i$ and $j$ in the configuration model. The best network division in communities yields the largest value of modularity. 

Several other measures can be considered for networks characterization~\cite{costa2007characterization}. They are used to understand the relation between networks structure and synchronization.

\subsection{Complex networks models}

Complex systems exhibit specific types of organization resulting from intrinsic rules and constraints that guide the system's evolution, e.g. the interconnection between components in power grids (whose nodes are power stations, transmission circuits, and substations) is to a great extent a consequence of the geographical distance between nodes~\cite{albert2004structural}. In other words, two geographically close nodes tend to be connected, while distant nodes are not likely to connect.  Therefore, the degree distribution of this type of network depends on the geographical node distribution~\cite{albert2004structural}.  For instance, if the distribution of nodes is uniform in the two-dimensional space, it is expected that the degree distribution
will be close to a Poisson distribution. On the other hand, in collaboration networks, composed by scientists connected according to coauthorships of scientific works, the most connected researchers tend to attract more scientists for collaborations (knwon as the rich get richer paradigm). In this case, the degree distribution follows a power law.

Since different rules result in different topologies, some basic models have been proposed to generate networks with a given degree distribution and other topological properties. These models are fundamental to generate networks in which it is possible to control some properties, such as $\left\langle k \right\rangle$ and $\mathcal{T}$. As following, we present the most famous models of complex networks. 

\subsubsection{Random Graphs}

A simple stochastic model for the generation of random graphs was proposed
by the mathematicians Paul Erd\H{o}s and Alfred R\'{e}nyi in
1959~\cite{renyi1959random}. A random graph (ER) is generated by starting with $N$ disconnected nodes and then connecting each pair of nodes according to a fixed probability $p$. Therefore, this process is a type of $N^2$ realization of a Bernoulli process with probability of success $p$. Therefore, the number of connections follows a binomial distribution. Nevertheless, as most realizations of this model take into account large values of $N$ and small values of $p$, the degree distribution tends to a Poisson distribution (known as the law of rare events). For small $p$ and large $N$, we have
\begin{equation}
P(k) = \frac{N!}{(N-k)!k!} p^k (1-p)^k \simeq \frac{e^{-\langle k \rangle} \langle k
\rangle^k}{k!},
\end{equation}
where $\langle k \rangle = p(N-1)$ is the average degree of the network. Random graphs exhibit a second order percolation phase transition according $\left\langle k \right\rangle$. If $\langle k \rangle > 1$, there is an emergence of a giant component~\cite{bollobas1998random}. 

\subsubsection{Small-world networks}

The ER network is not suitable to model the vast majority of real-world systems. Indeed, it is difficult to find a system whose structure is similar to ER. In this way, in order to generate networks with a more structured topology, Watts and Strogatz suggested a model whose structure lies between a complete regular graph and a random network~\cite{watts1998collective}. The Watts-Strogatz small-world model (SW) generates networks with a large number of loops of size three, i.e.\ three nodes connected to each other (a loop).  This property is observed in cohesive networks, such as in society, where
two friends $A$ and $B$ present a high probability to share a common
friend $C$, as quantified by the clustering coefficient (Eq.~\ref{Eq:cc}).

\begin{figure}[!t]
\begin{center}
\includegraphics[width=0.9\linewidth]{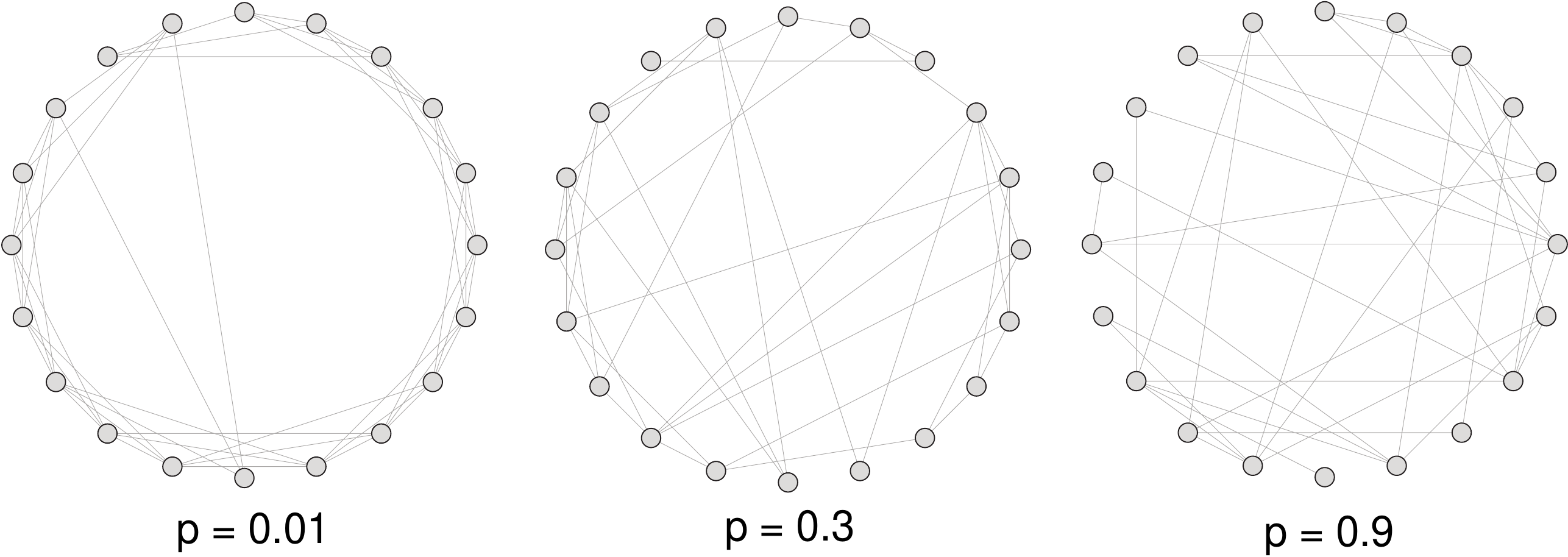}
\end{center}
 \caption{Small-world networks composed by $N = 20$ vertices and $\left\langle k \right\rangle = 3$. Different rewiring probabilities $p$ are
  taken into account.}
\label{Fig:ws}
\end{figure}

In order to construct a SW network, one starts with a ring of $N$ vertices, in wihch each vertex is connected to $\kappa$ nearest neighbors in each direction, totalizing $2\kappa$ connections. Next, we choose a vertex and the edge to its nearest
clockwise neighbour. With probability $p$, this edge is reconnected to a vertex chosen
uniformly at random over the entire ring. This process is repeated by moving clockwise around the ring, considering each vertex at time until one lap is completed. When $p = 0$ we have an ordered lattice with clustering coefficient, but with long average shortest paths. When $p\rightarrow 1$, the network becomes a random graph with small shortest distances between nodes but few loops. Notice that for $p=1$ a network does not become an
ER network. Figure~\ref{Fig:ws} presents SW networks generated with different values of the probability $p$. 

\subsubsection{Scale-free networks}

The model of Watts and Strogatz overcomes the lack of structure present in random graphs while providing an elegant theory to explain the presence of cycles of order three, as well as the dependence between the average shortest path and the network size. However, since this model does not generate networks with power-law degree distribution, it is not suitable to represent many real-world networks, such as the Internet, biological networks and World Wide Web~\cite{costa2011analyzing}. In 1999, Barab\'{a}si and Albert~\cite{barabasi1999emergence} proposed a simple algorithm to construct networks, whose degree distribution follows a power-law. This model, known as SF Barab\'{a}si-Albert model (BA), is based on two principles: (i)
growth and (ii) preferential attachment. The network is generated starting with a set of $m_0$ connected vertices; afterwards, at each step of the construction the network grows with the addition of a new vertex. For each new vertex, $m$ new edges are
inserted between the new vertex and some previous vertex. The vertices which receive the new edges are chosen following a linear preferential attachment rule, i.e.\ the probability of the new vertex $i$ to connect with an existing vertex $j$ is proportional to the degree of $j$,
\begin{equation}
\mathcal{P}(i\rightarrow j) = \frac{k_j}{\sum_n k_n}.
\end{equation}
This rule indicates that the most connected nodes present a high probability to receive new connections. 

\subsubsection{Configuration model}

\begin{figure}[!t]
\begin{center}
\includegraphics[width=0.3\linewidth]{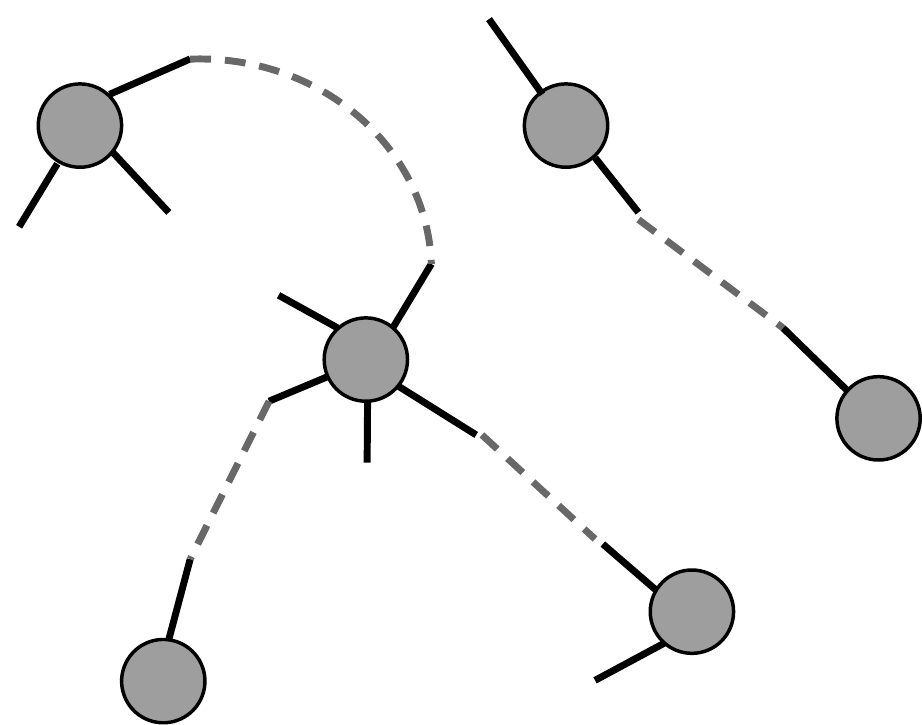}
\end{center}
 \caption{Illustration of the configuration model. Stubs are selected uniformly and connected, forming a network with a defined degree sequence. Dashed lines indicate the connections between pairs of stubs.}
\label{Fig:configuration}
\end{figure}

ER, SW and BA models generate networks with a well defined degree distribution. However, in some cases, the degree distribution of real-world networks is not homogeneous and the coefficient of the power law is not the same as that observed in the BA model. Therefore, it is important to have a model that enables the construction of a network with an arbitrary degree distribution, while preserving the other network properties as random.

The configuration model is a model of random graph with a defined degree sequence $\boldsymbol{k} = \{k_1,k_2,\ldots, k_n\}$~\cite{newman2001random,molloy1995critical,bender1978asymptotic}, where $k_i$ represents the number of connections of node $i$. To construct the network, initially, each node $i$ receives a total of $k_i$ stubs (half edges). Then, at each time step, two stubs (half-edges) selected uniformly are connected. This process is repeated until all stubs are connected. In Figure~\ref{Fig:configuration} we can see an illustration of the configuration model algorithm. Notice that self-loops and multi-edges are allowed in this model. At the end of the process, we have a network with a defined degree distribution.

\bibliographystyle{elsarticle-num}
\bibliography{references2}

\end{document}